\documentclass [12pt]{mn_thesis}

\usepackage[dvips]{pstcol}
\usepackage{epsfig}
\usepackage{subfigure}
\usepackage{amsmath}
\usepackage{rotating}
\definecolor{green}{rgb}{0.1,0.7,0.1}

\newcommand{\beq}{\begin{equation}}
\newcommand{\eeq}{\end{equation}}

\newcommand{\dslash}{\not{\hbox{\kern-2pt $\partial$}}}
\newcommand{\pslash}{\not{\hbox{\kern-2.3pt $p$}}}
 \newtoks\nslashfraction
 \nslashfraction={.13}
 \newcommand{\nslash}[1]{\setbox0\hbox{$ #1 $}
   \setbox0\hbox to \the\nslashfraction\wd0{\hss \box0}/\box0 }

\begin{document}
\normalspacing

\title{Study of Radiative Decays of Psi(2S) Mesons}
\author{Kaiyan Gao}
\adviser{Ronald Poling}
\graduationdate{December}{2008}

\begin{titlepage} \maketitle \end{titlepage}

\begin{copyrightpage} \makecopyright \end{copyrightpage}

\begin{acknowledgements} There are many people to whom I should show my appreciation and gratitude. 
\ This dissertation is completed in the CLEO collaboration, where 
I learned and experienced a lot in experimental high energy physics.
I would like to express my appreciations and gratitude to the people 
who helped me in various ways.

First I would like to thank my thesis adviser, Ron Poling for his 
support over the years. \
His patience, insight and advice have provided me guidance through 
the process of selecting, designing and performing the analysis. \ 
His enthusiasm and passion for physics and way of cooperating with
others certainly provide me a good model of a successful physicist. \ 
It was a great pleasure to work with him and have him as my adviser.

Pete Zweber cooperate with me on the project that involved in this 
analysis. \ From him I have learned a lot of background knowledge 
about this analysis and general analysis strategies. \ 
Datao Gong provided a lot of guidance and advice at the beginning of 
this project. \ He shared with me his experience of the previous analysis 
that invoked the motivations of this project. \ He also provided 
great technical helps during the early period of this analysis. \ 
During the first year I worked at Wilson Lab of Cornell University, my 
friend and colleague Selina Li helped me to know the hardware system, 
to get familiar with the collaboration, and to discuss analyses. \ 
I would also like to thank all other current and past CLEO colleagues 
who work or previously worked at the University of Minnesota, 
Yuichi Kubota, Dan Cronin-Hennessy, Jon Urheim, Alex Smith, 
Chris Stepaniak, Brian Lang, Alexander Scott, Tim Klein and Justin Hietala 
for their advices and helps.

I would like to thank all other people in the CLEO collaboration for their 
helps and discussions in many aspects, such as hardware, software, and 
analysis.

I would like to thank Larry McLerran, who was my first year graduate 
adviser, for his support and guidance during the earlier years.

I would like to thank my family members for their love and support through 
the years. \ My father, Chongshou Gao, a theoretical particle physicist, 
was the first person who told me the importance of experiments in physics. \ 
His enthusiasm for physics inspired my interests for physics. \ My mother, 
Baiqing Xie, a professor in computer science, showed me the 
importance of hard work and scientific methods. \ 
My brother, Kaizhong Gao, a theoretical condense matter physicist, helped 
me with his experience
doing research both during and outside of his graduate 
studies. \ During my difficult time, they all show their endless support 
to me. \ I would also like to thank my unborn son, who has not been given
his name yet, for the pleasures he has brought to me during the last few 
months of this dissertation.

Last but not least,
I would like to specially thank my husband, You Lin, for his love, 
friendship, patience, and encouragement through my career. \ I always
have wonderful time with him for discussing about all kinds of topics 
we can think of. \ 
As a theoretical condense matter physicist, he shares with me his 
thoughts about physics, phylosophy and other areas. \ His love makes
everyday of my life bright and brings smile on my face.
 \end{acknowledgements}

\begin{dedication} To my family.
 \end{dedication}

\begin{abstract} We studied the decay $\psi(2S) \to \gamma~\eta_c(2S)$ with
$25.9 \times 10^{6}$ $\psi(2S)$ events collected with the CLEO-c detector.
\ No $\psi(2S) \to \gamma~\eta_c(2S)$ decays were observed in any of the
eleven exclusive $\eta_c(2S)$ decay modes studied. \
The product branching fraction upper limits were determined for all
modes. \ The 90\% confidence level upper limit of
${\cal B}(\psi(2S) \to \gamma~\eta_c(2S)) < 7.4 \times 10^{-4}$
was obtained.

 \end{abstract}

\contentspage

\tablelistpage

\figurelistpage

\mainbody
\doublespacing

\chapter{Introduction}
\label{chap:introduction}

This dissertation is devoted to the study of the radiative decay 
of charmonium meson $\psi(2S)$ created in $e^{+}e^{-}$ annihilations 
at the center of mass energy of $3.7~{\rm GeV}$. \ Specifically, we 
have studied the radiative decay $\psi(2S) \to \gamma \eta_{c}(2S)$
using CLEO-c detector by reconstructing the final states of 
$\eta_{c}(2S)$ decays exclusively. \ To provide the background for 
the research I start with the fundamental knowledge of particle 
physics.

Since the start of human history, the composition of the world has always 
been a subject of intense interest. \ From thousands of years ago, 
when the name atom was given, to recent decades, experimental observations 
and theoretical studies have allowed us to understand the fundamental nature 
of materials more and more deeply. \ The latest and the most successful 
model of the fundamental particles that compose the universe is the Standard 
Model, which began with the introduction of quarks by Murray Gell-Mann and 
George Zweig in 1964. \ The observation of the top quark completed the quark 
sector of the Standard Model 
\cite{PhysRevLett.74.2422,PhysRevLett.74.2626} and provided strong 
support for the model. \ It is not a complete description, however, as 
demonstrated by recent observations of new physics beyond the Standard 
Model, which may greatly expand our knowledge about the universe. 

\section{The Standard Model}
\label{sec:standardmodel}

Nowadays it is common knowledge that all materials are composed of
atoms and molecules. \ Atoms are formed from elementary particles:  
electrons, protons and neutrons. \ Since the discovery of the first
elementary particle, the electron, in 1897 by J. J. Thomson at the Cavendish 
Laboratory of Cambridge University, hundreds of elementary particles have
been found and various theoretical models have been built to explain the
properties and interactions of these particles. \ 
For over forty years, the Standard Model has been developed as a framework 
for understanding and studying these elementary particles. \ Experimental 
measurements have demonstrated that the Standard Model provides an 
excellent description and have provided increasingly precise 
determinations of many unpredicted parameters.

\subsection{Quarks and Leptons}
Within the Standard Model, the particles with no known substructure are 
categorized in three types: quarks, leptons and gauge bosons. \ The quarks 
and leptons are the building blocks of composite particles, and are 
fermions with spin $s=\frac{1}{2}$. \ 
The gauge bosons are the mediators of the fundamental interactions and 
have whole-integer spin.

Hadrons are composite particles made of quarks. \ There are six kinds 
of quarks and they are characterized according to their ``flavor'': $u$, 
$d$, $c$, $s$, $t$, and $b$ (up, down, charm, strange, top, and 
bottom quarks) with charges (in units of the electron charge) of 
$\frac{2}{3}$, $-\frac{1}{3}$, $\frac{2}{3}$, $-\frac{1}{3}$, 
$\frac{2}{3}$, and $-\frac{1}{3}$, respectively. \ The flavor quantum 
numbers are the third component of isospin $I_{3}$ of $u$ and $d$ quarks, 
the charm of $c$ quarks, 
the strangeness of $s$ quarks, 
the topness of $t$ quarks, 
and the bottomness of $b$ quarks. 

\newcommand{\thickhline}{\noalign{\hrule height 1.2pt}}
The properties of the quarks are summarized in Table~\ref{table:quarks}. \
Quarks have not been found to 
exist independently and are only found as constituents of composite 
particles. \ The explanation of this is that quarks have another quantum 
number, color, so named because of similarity with color theory of 
visible light. \ Because net color 
is not observed in nature, the quarks are ``confined'' into the composite 
particles called hadrons, which are colorless. 
\begin{table}[htbp]
\caption[Properties and additive quantum numbers of the quarks]
{\label{table:quarks}Properties and additive quantum numbers of the quarks: 
mass ($M$), electric charge ($Q$), isospin 3-component ($I_{3}$), 
Charm ($C$), strangeness ($S$), topness ($T$) and bottomness($B$). \ Three 
generations of quarks are separated by thick lines.}
\begin{center}
\begin{tabular}{|l|c|c|c|c|c|c|c|}
\thickhline
Flavor & $M ({\rm GeV}/c^2)$ & $Q$ & $I_{3}$ & $C$ & $S$ & $T$ & $B$ \\ \thickhline
$u$ - up   & $3~{\rm MeV}/c^2$ & $+\frac{2}{3}$ & $+\frac{1}{2}$ & 0 & 0 & 0 & 0 \\ \hline
$d$ - down & $6~{\rm MeV}/c^2$ & $-\frac{1}{3}$ & $-\frac{1}{2}$ & 0 & 0 & 0 & 0 \\ \thickhline
$c$ - charm  & $1.24~{\rm GeV}/c^2$ & $+\frac{2}{3}$ & 0 & $+1$ & 0 & 0 & 0 \\ \hline
$s$ - strange & $95~{\rm MeV}/c^2$ & $-\frac{1}{3}$ & 0 & 0 & $-1$ & 0 & 0 \\ \thickhline
$t$ - top    & $172~{\rm GeV}/c^2$ & $+\frac{2}{3}$ & 0 & 0 & 0 & $+1$ & 0 \\ \hline
$b$ - bottom & $4.2~{\rm GeV}/c^2$ & $-\frac{1}{3}$ & 0 & 0 & 0 & 0 & $-1$ \\ \thickhline
\end{tabular}
\end{center}
\end{table}

There are two kinds of hadrons. \ Baryons consist of three quarks and are 
therefore fermions, e.g. a proton $p \sim uud$ and a neutron $p \sim ddu$. \ 
Mesons consist of one quark and one antiquark and are therefore bosons, e.g.
$\pi^{+} \sim u{\bar d}$ and $K^{+} \sim u{\bar s}$. \ Hadrons are particles 
that participate in strong interactions directly.

The Standard Model leptons are the electron ($e^-$), muon ($\mu^-$) and 
tau ($\tau^-$), each with charge of $-1$, 
and their corresponding chargeless neutrinos ($\nu _e$, 
$\nu _{\mu}$ and $\nu _ {\tau}$). \ Some of the properties of the leptons 
are given in Table~\ref{table:leptons} \ Each lepton type ($e^-$, $\mu^-$, 
$\tau^-$) is associated with a conserved lepton number. \ Recent evidence 
of oscillations among neutrino flavors and nonzero neutrino masses 
provides the first evidence for physics beyond the Standard Model 
\cite{PhysRevLett.81.1562,araki:081801,michael:191801,ashie:101801}.
\begin{table}[htbp]
\caption[Properties of the leptonss]
{\label{table:leptons}Properties of the leptons. \ Three generations of  
leptons are separated by thick lines.}
\begin{center}
\begin{tabular}{|l|c|c|}
\thickhline
Flavor                        & Mass & Electric Charge       \\ \thickhline
$\nu_{e}$ - electron neutrino & $<2~{\rm eV}/c^2$     & 0    \\ \hline
$e$ - electron                & $0.511~{\rm MeV}/c^2$ & $-1$ \\ \thickhline
$\nu_{\mu}$ - muon neutrino   & $<0.19~{\rm MeV}/c^2$ & 0    \\ \hline
$\mu$ - muon                  & $106~{\rm MeV}/c^2$   & $-1$ \\ \thickhline
$\nu_{\tau}$ - tau neutrino   & $<18.2~{\rm MeV}/c^2$ & 0    \\ \hline
$\tau$ - tau                  & $1.78~{\rm GeV}/c^2$  & $-1$ \\ \thickhline
\end{tabular}
\end{center}
\end{table}

Tables~\ref{table:quarks} and \ref{table:leptons} show the natural grouping 
of the quarks and leptons in three generations. \ Why there are three 
generations is one of the deepest mysteries of particle physics.

For each quark and lepton there is a corresponding antiquark and antilepton, 
and as described above the quarks come in three colors. \ The physical 
properties of the quarks and leptons from different generations are identical 
except for the masses of the fermions.

\subsection{Intermediate Particles and Fundamental Interactions}

The Standard Model describes the electromagnetic, weak and strong 
interactions. \ It does not incorporate the fourth of the fundamental 
interactions, gravitation. \ Each interaction has its corresponding 
``force carrier'', particles that are called gauge bosons. \ 
These mediators are all bosons with spin $s=1$. \ 
Massless photons ($\gamma$) are chargeless
and associated with the electromagnetic interaction, which has infinite 
interaction range. \ Massive $W^{\pm}$'s and $Z^0$'s are associated 
with the weak interaction, which has a very short range. \ They are 
self-interacting and the $W^{\pm}$'s have electric charge of $\pm 1$, 
while $Z^0$'s are neutral. \ Eight massless and chargeless gluons ($g$) 
are associated with the strong interaction, which is also short-ranged 
because the gluons carry color. \ As gluons have eight different colors, 
they not only interact with quarks but also interact with themselves.

\subsection{Parameters of the Standard Model}
A fundamental feature of the weak interaction in the Standard Model is 
quark mixing. \ The weak eigenstates of quarks are not the same as their 
mass eigenstates. \ The mixing of flavors and generations induced by weak 
interactions is parameterized by the Cabibbo-Kobayashi-Maskawa (CKM) 
matrix, $\mathbf{V}$, a $3 \times 3$ unitary matrix, operating on quark 
mass eigenstates, as shown in Equation~\ref{ckmmatrix}.
\begin{equation}
\label{ckmmatrix}
\left(\begin{array}{ccc}
        d'\\
        s'\\
        b'
        \end{array}\right) = \left( \begin{array}{ccc}
                                     V_{ud}&V_{us}&V_{ub}\\
                                     V_{cd}&V_{cs}&V_{cb}\\
                                     V_{td}&V_{ts}&V_{tb}
                                     \end{array} \right)
\left(\begin{array}{ccc}
        d\\
        s\\
        b
        \end{array}\right).
\end{equation}
The 2008 Nobel Prize in physics was awarded to M.~Kobayashi and T.~Maskawa 
for this formulation, ``the discovery of the origin of the broken symmetry 
which predicts the existence of at least three families of quarks in nature''.
Without flavor mixing, quarks could only decay within their generation. \ 
We would expect the heavier member to decay 100\% of the time to its 
lighter partner ($c \to s$, $t \to b$) and the lighter one to be stable. 
\ Since the CKM matrix is unitary, the nine elements 
are reducible to four independent parameters. \ In the standard 
parameterization of the matrix, three angles ($\theta_{12}$,
$\theta_{23}$, $\theta_{13}$) and one phase ($\delta_{13}$) represent the 
four independent parameters.

In experimental high energy physics, in addition to the properties of 
particles such as masses, charges and spins, the fundamental parameters 
of the CKM matrix must also be measured. \ These parameters are reflected 
in the decay widths (lifetimes), branching fractions, and other detailed 
properties of the particles and their decays. \ Redundant measurements of 
the CKM parameters are important and necessary as they are a powerful test 
on the viability of the Standard Model.

Based on various measurements, the current ranges for the magnitude of 
the CKM matrix elements have been compiled by the Particle Data Group 
\cite{PDBook2006} and are given in Equation~\ref{ckmmatrixnum}:
\begin{equation}
\label{ckmmatrixnum}
V = 
\left( \begin{array}{ccc}
        0.9741 \to 0.9756&0.219 \to 0.226&0.0025 \to 0.0048\\
        0.219 \to 0.226&0.9732 \to 0.9748&0.038 \to 0.044\\
        0.004 \to 0.014&0.037 \to 0.044&0.9990 \to 0.9993
        \end{array} \right) .
\end{equation}
The fact that the coupling between different generations of the quarks 
is small can be seen from the small values of the non-diagonal elements.

\subsection{Symmetries and Conservation Laws} 

If a system or phenomenon remains unchanged under a transformation of 
one or more physics variables, then the system or the phenomenon is 
regarded to have symmetry with respect to this transformation. \ 
According to Noether's Theorem \cite{Noether}, any symmetry of a 
physical system under a transformation that does not explicitly 
depend on time has a corresponding conservation law. \ The interactions 
and decays of particles are governed by conservation laws.

One type of conservation law is called an exact conservation law. \ The 
most commonly used conservation laws, conservation of energy, 
conservation  of linear momentum, conservation of angular momentum, and 
conservation of electric charge are all exact conservation laws. \ 
Conservation of energy is associated with symmetry under time 
translation. \ Conservation of linear momentum is a mathematical 
consequence of symmetry under continuous translation in space. \ 
Conservation of angular momentum corresponds to the continuous 
rotational symmetry of space. 

Another type of conservation law is an approximate conservation law. \ 
Such a law holds only under some particular situations.

\subsubsection{Isospin}

Isospin symmetry is an important internal symmetry that was introduced 
early in the development of nuclear and particle 
physics. \ The concept of isospin came from the similarity 
between neutrons and protons. \ Neutrons and protons have the same spin 
of $\frac{1}{2}$ and nearly the same mass, but neutrons are chargeless and 
protons have $+1$ unit of electric charge. \ They behave nearly identically 
under the 
strong interactions but differently in weak and electromagnetic 
interactions. \ By introducing isospin space, the abstract internal space 
associated with isospin quantum number $I$, neutrons and protons can be 
regarded as a doublet of the quantum state with $I = \frac{1}{2}$. \ 
The algebra of isospin is the same as that of angular momentum and isospin 
is also an additive quantum number. \ Therefore 
protons and neutrons have different projections on the third dimension in 
isospin space, i.e. $I_{3} = \frac{1}{2}, -\frac{1}{2}$ for the proton and 
neutron, respectively. \ Similarly, pions have isospin of $1$, so by 
following the algebra of angular momenta, they have three possible
third components of isospin, $I_{3} = -1, 0, 1$ and the corresponding pion  
states are denoted as $\pi^{-}$, $\pi^{0}$, $\pi^{+}$. \ The third component 
of isospin is chosen by convention based on the charge of the particle, e.g. 
the proton has charge $+1$ and is assigned the positive value of $I_{3}$.

Isospin symmetry is an approximate symmetry describing the strong 
interaction. \ The conservation of isospin requires that a system 
remains unchanged in strong interactions, i.e. the isospin $I$ and its 
third component $I_{3}$ are conserved. \ Similar to the relationship 
between the conservation of angular momenta and the rotational 
symmetry of space, the conservation of isospin comes from the 
rotational invariance of strong interactions in isospin space. \
Conservation laws of isospin does not hold in weak interactions.

\subsubsection{Parity}

Parity ($\mathcal{P}$) is a discrete (non-continuous) operation and is 
defined as the inversion of space coordinates:
\begin{equation}
\mathbf{x} \to - \mathbf{x}.
\end{equation}
According to its definition, parity satisfies $\mathcal{P}^{2}[\psi]=\psi$. \ 
Therefore, it has two eigenvalues, $P = \pm 1$. \ 
Under $\mathcal{P}$, the signs of both position $\mathbf{x}$ and 
linear momentum $\mathbf{p}$ are reversed, but angular momentum, defined as
$\mathbf{L} = \mathbf{x} \times \mathbf{p} = \mathbf{L}$ remains 
unchanged. \ Thus $\mathbf{x}$ and $\mathbf{p}$ have 
parity of $-1$ and $\mathbf{L}$ has parity of $+1$. \ 
The classical variables, the time, the energy, 
the angular momentum of a particle, and masses, charges, coupling 
constants, etc. satisfy $P = +1$. \ The variables with $P = -1$ 
include the position, the velocity, the linear momentum of a particle, 
and electric field. 

The invariance under the $\mathcal{P}$ transformation is associated with  
conservation of the parity quantum number. \ Parity is conserved in strong 
and electromagnetic interactions.

Parity is a multiplicative quantum number, i.e. the parity of a 
multiparticle state is the product of the parities of all component 
particles. \ The wave function of a particle is characterized by its
intrinsic parity, which is the parity quantum number $P$ of the particle. \ 
Basic nonrelativistic quantum mechanics demonstrates that 
if the orbital angular momentum of a system is $L$, then the parity of 
the system is the product of $(-1)^{L}$ and the intrinsic parity. \ 
The intrinsic parities of particles are defined in accordance with the 
conservation law, and are assigned based on experiments and by convention. \ 
By convention, quarks have positive parity and antiquarks have negative 
parity. \ The intrinsic parity of a hadron can be determined by the 
strong interactions that produces it, or through decays not involving 
the weak interaction. \ The intrinsic parity of a neutral particle 
can be determined from experiments, e.g. $P(\pi^{0}) = -1$. \ 
Photons and gluons have $P(\gamma) = -1$ and $P(g) = -1$,
respectively. \ The intrinsic parity of a particle composed of a 
fermion and antifermion pair is $-1$ and the intrinsic parity of a 
particle composed of a boson and antiboson pair is $+1$.

\subsubsection{Charge Conjugation}

Another discrete transformation is the charge conjugation (${\mathcal C}$)
transformation, which turns a particle into its 
antiparticle. \ The charge conjugation operator ${\mathcal C}$ 
reverses the electric charge and all the internal quantum numbers, 
including lepton number, baryon number, $I_{3}$, strangeness, and the other
flavors.
The eigenvalue of ${\mathcal C}$, $C = \pm 1$, is called charge conjugation 
parity ($C$), similar to $P$ parity. \ Only truly neutral systems with all 
quantum charges and magnetic moment zero are eigenstates of charge parity. \ 
Photons and gluons both have $C$ parity $-1$. \ For a system that is 
composed of a particle and antiparticle pair, if the orbital angular
momentum is $L$ and the sum of the spin angular momenta of the two 
particles is $S$, the $C$ parity 
is $C = (-1)^{L+S}$. \ $C$ parity is also a multiplicative quantum 
number.

Invariance under the charge conjugation transformation leads 
to the conservation of charge parity and is called 
$C$-symmetry. \ $C$-symmetry holds in strong and electromagnetic 
interactions, and even gravitational interactions. \ However, weak 
interaction does not obey charge conjugation symmetry.

\subsubsection{Other Conservation Laws}

There are other conservation laws, e.g. conservation of strangeness and 
baryon number, that also govern elementary particle interactions. \ 
The strangeness $S$, defined as the number of strange antiquarks ${\bar s}$ 
minus the number of strange quarks $s$, is a quantum number 
introduced to describe decays of particles in strong, electromagnetic 
and weak interactions. \ It has been observed experimentally that the 
change of the strangeness quantum number satisfies $\Delta S = 0$ for 
strong and electromagnetic 
interactions and $|\Delta S| = 0, 1$ for weak interactions.

Baryon number is defined as one third of the difference between the 
number of quarks 
$q$ and the number of antiquarks ${\bar q}$, where the fraction one third 
comes from the fact that a baryon consists of three quarks or 
antiquarks. \ The conservation of baryon number holds in all interactions 
of the Standard Model. \ Lepton number is similarly conserved in all 
interactions.

More detailed descriptions and discussions about conservation laws can 
be found in any standard textbook, such as Ref.~\cite{Griffiths}.

\section{Charmonium States}

Quarkonium is a flavorless bound state which consists of a quark and
its own antiquark bound by strong interactions. \ Quarkonia include
charmonia and bottomonia. \ A charmonium meson($c\bar{c}$ 
meson) is the bound state of a charm quark and a charm antiquark, 
and a bottomonium meson ($b\bar{b}$ meson) is the bound
state of a bottom quark and a bottom antiquark. \ No toponium exists
because of the large mass of top quark and antiquark. \ The quark
would decay through electroweak interactions before a bound state
can form. \ Bound states of light quark and antiquark pairs are
normally not called quarkonium, partly because in the experiments
the states observed are usually the mixture of the light quark states. \ 
The large mass difference between the charm or bottom quark and the lighter
quarks prevents the charmonium and bottomonium states from mixing with each
other or with the lighter flavorless mesons. \
Charmonium and bottomonium are very important for studying the
strong interaction. \ The two charmonia involved in this dissertation 
are the $\psi(2S)$ (often written as $\psi^{\prime}$ or $\psi(3686)$) and 
$\eta_{c}(2S)$ (often written as $\eta_{c}^{\prime}$).

The charmonium resonances were first discovered in experiments in 1974 
\cite{PhysRevLett.33.1404, PhysRevLett.33.1406, PhysRevLett.33.1453},
known as the ``November Revolution,'' when the $J/\psi$ meson was 
independently discovered by a group at the Stanford Linear Accelerator 
Center and by a group at Brookhaven National Laboratory. \ The 
charmonium resonances were quickly interpreted as $c\bar{c}$ bound states 
and were very powerful in establishing the reality of quarks and the 
Standard Model.

Non-perturbative Quantum chromodynamics (QCD) is used to compute 
the properties of mesons. \ 
Similar to nonrelativistic models of the hydrogen atom, the 
motion of the quarks in a quarkonium state is nonrelativistic so 
they can be assumed to move in a static potential. \ 
One technique that can be effectively applied to quarkonium is 
application of quantum mechanics, with specific choices of the 
quark-antiquark potential. \ Such calculations predict the masses 
of quarkonia states, and can be directly tested with data. \ 
One of the candidate potentials is 
\cite{Eichten80}
\begin{equation}
V(r) = -\frac{\kappa}{r} + \frac{r}{a^{2}} ,
\end{equation}
where $r$ is the effective radius of the quarkonium state and $\kappa$ 
and $a$ are parameters. \ The first term is the Coulomb-type force that 
dominates at the short distances (asymptotic freedom). \ It corresponds 
to the potential induced by one-gluon exchange between the quark and its 
anti-quark, analogous to the Coulomb electromagnetic potential. \ The 
second term is linear in $r$ and produces quark confinement.

The charmonium states, as well as other elementary particles, are 
denoted with the spectroscopic notation $n^{2S+1}L_{J}$. \ The principal 
quantum number $n = 1, 2, 3 ...$  represents the ground state, and 
successive excited states of the particle. \ $S$ represents the spin of 
the particle, which for charmonium takes values $S = 0, 1$, since its 
quark and antiquark constituent are fermions and have spin projections 
of $\frac{1}{2}$ or $-\frac{1}{2}$. \ $L$ 
denotes the orbital angular momentum and the letters $S$, $P$, $D$ ...
stand for $L = 0, 1, 2,$ ... . \ $J$ is the total angular momentum with 
$J = |L + S|, |L + S| -1, ..., |L - S|$. \ Specific states are 
frequently denoted by their $J^{PC}$, where $P$ and $C$ are the parity 
and charge conjugation quantum numbers, respectively. \ For example, the 
two charmonia involved in this dissertation are $\eta_{c}(2S)$ and 
$\psi(2S)$. \ The $2S$ means that for these states, $n = 2$ and $L = 0$. \ 
In fact, $\eta_{c}(2S)$ denotes $1^{1}S_{0}$ and $\psi(2S)$ is 
$1^{3}S_{0}$, so $\eta_{c}(2S)$ has $S = 0$ and $\psi(2S)$ has $S = 1$. \ 
The total angular momenta are $J = 0$ for $\eta_{c}(2S)$ and 
$J = 1$ for $\psi(2S)$. \ According to the parity convention for 
quarks and antiquarks, both $\eta_{c}(2S)$ and $\psi(2S)$ have parity 
$-1$. \ However, the $C$ parity of $\eta_{c}(2S)$ is $+1$ and that of 
$\psi(2S)$ is $-1$ because $C = (-1)^{L+S}$. \ Thus the 
characterization of $\eta_{c}(2S)$ is $J^{PC} = 0^{-+}$ and that of 
$\psi(2S)$ is $J^{PC} = 1^{--}$.

\begin{figure}[htbp]
  \centering
  \includegraphics[width=.95\textwidth]{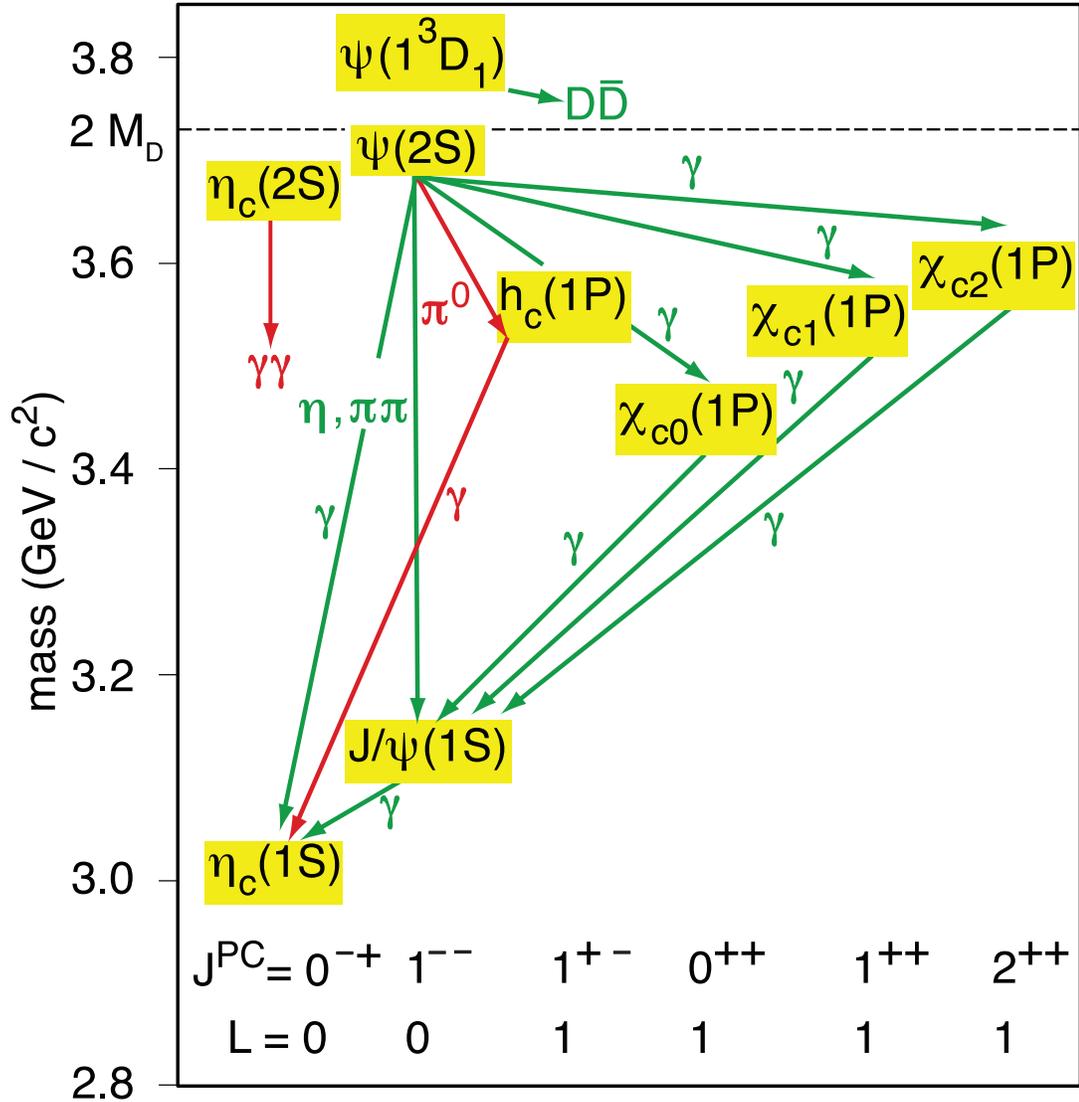}
  \caption[Charmonium resonances]{The spectrum of charmonium resonances. \ 
  The states are labeled with spectroscopic notation $n^{2S+1}L_{J}$, 
  where $n$ is the principal quantum number, $S$ represents the spin of 
  the particle and $S = 0, 1$, $L = S, P, D, ...$ denotes the orbital 
  angular momentum of $L = 0, 1, 2,$ ..., and $J$ is the total angular 
  momentum. \ In addition to this notation, parity ($P$) and charge 
  conjugation ($C$) are used in the notation $J^{PC}$.}
  \label{fig:charmstates}
\end{figure}

The charmonium state $\eta_{c}(2S)$ and $\psi(2S)$ are the first 
excited states of $\eta_{c}(1S)$ and $J/\psi(1S)$ respectively. \ 
Therefore the transition between $\eta_{c}(2S)$ and $\psi(2S)$ is 
similar to the transition between $\eta_{c}(1S)$ and $J/\psi(1S)$. \
The lowest energy states for mesons with quark-antiquark pairs have 
$S = 0$ and negative parity and are called pseudoscalar mesons, 
e.g. $\eta_{c}(1S)$ and $\eta_{c}(2S)$. \ For excited states of 
mesons the quark spins are parallel, so if a state has zero orbital 
angular momentum, then the total angular momentum of the particle is 
$1$. \ Such states are called vector mesons, e.g. $J/\psi$ 
and $\psi(2S)$. \ As the vector mesons have the same quantum numbers 
as photons (virtual photons), i.e. $J^{PC} = 1^{--}$, they can be 
produced directly in $e^{+}e^{-}$ collisions, in which virtual 
photons are formed. \ Other charmonium states can be observed in the 
transitions of $J^{PC} = 1^{--}$ states or in $p\bar{p}$ collisions. \ 
Figure~\ref{fig:charmstates} shows the spectrum of the charmonium 
resonances and the observed transitions among them. 

\section{$\eta_{c}(2S)$ Production at $\psi(2S)$}

\subsection{$\psi(2S)$ Decays}

\begin{figure}[htbp]
  \centering
  \subfigure[$\psi(2S) \to \gamma \to l^{+}l^{-}$]
    {\label{fig:feynmandiagram_a}
     \includegraphics[width=.48\textwidth]{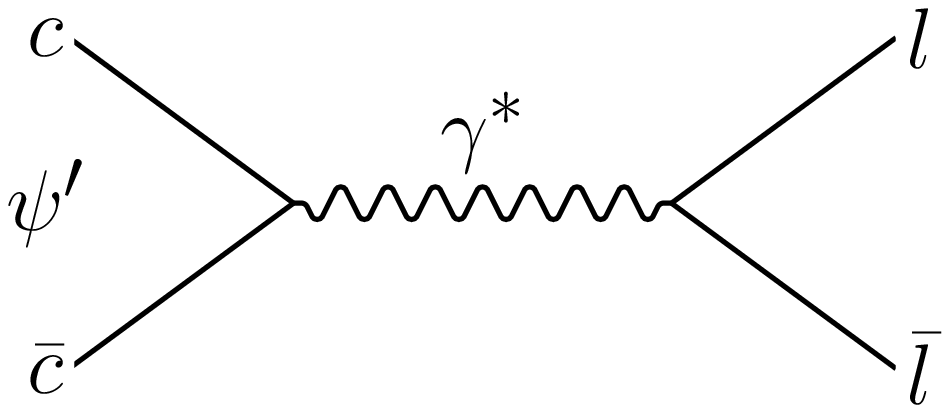}} 
  \subfigure[$\psi(2S) \to \gamma \to hadrons$]
    {\label{fig:feynmandiagram_b}
     \includegraphics[width=.48\textwidth]{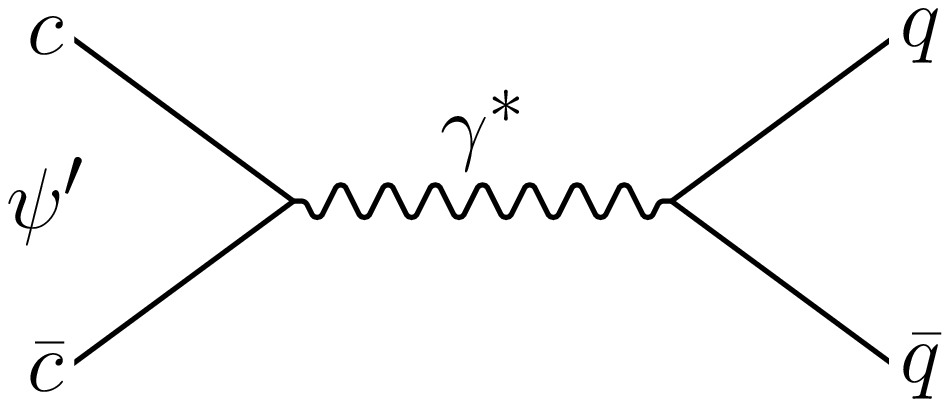}} 
  \subfigure[$\psi(2S) \to g + g + g \to hadrons$]
    {\label{fig:feynmandiagram_c}
     \includegraphics[width=.48\textwidth]{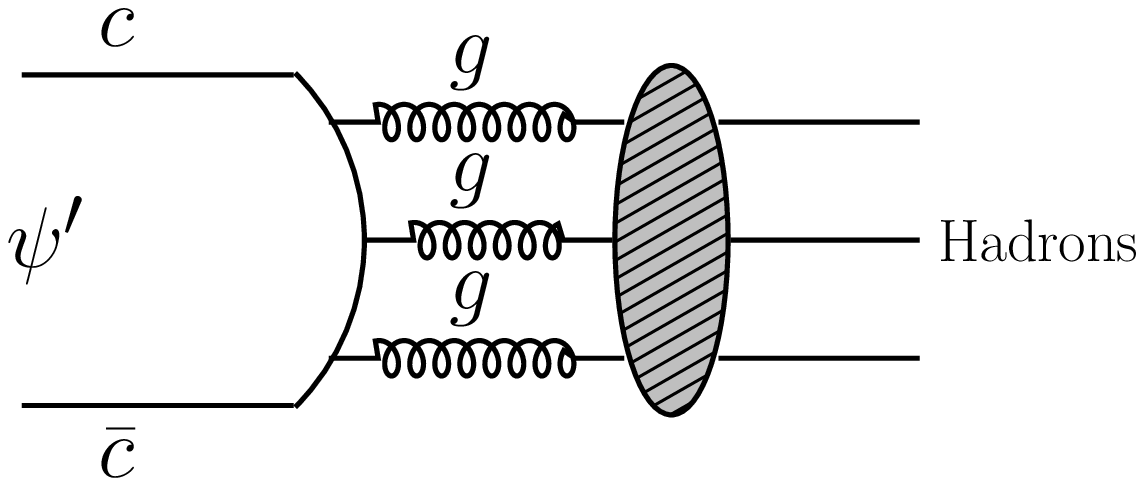}} 
  \subfigure[$\psi(2S) \to \gamma + g + g \to \gamma + hadrons$]
    {\label{fig:feynmandiagram_d}
     \includegraphics[width=.48\textwidth]{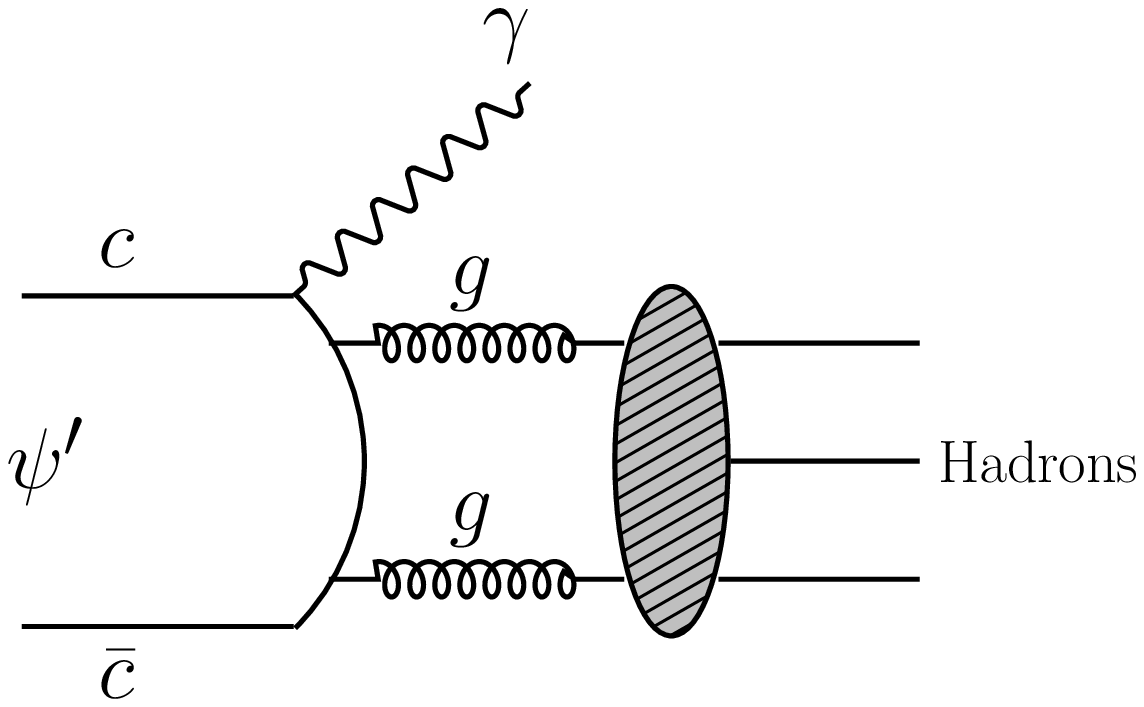}} 
  \caption[$\psi(2S)$ decays]{$c\bar{c}$ annihilation and decay 
  mechanisms for $\psi(2S)$ ($\psi^{\prime}$):  
  (a) leptonic decay through virtual photon; (2) hadronic decay through 
  virtual photon; (c) hadronic decay through three gluons; 
  (d) radiative decay through two gluons and a photon.}
  \label{fig:feynmandiagram}
\end{figure}

The charmonium states with $J^{PC} = 1^{--}$ have four possible ways 
to decay, as shown in Figure~\ref{fig:feynmandiagram}. \ In 
experiments, they are observed to be through leptonic decays, 
hadronic decays, and radiative decays. 
\begin{itemize}
\item[(a)] Leptonic decays (Figure~\ref{fig:feynmandiagram_a}): \ 
The quark and antiquark pair annihilate to create a virtual
photon and then the virtual photon produces a lepton and antilepton 
pair, i.e. $J/\psi,\psi(2S) \to \gamma^{*} \to l^{+}l^{-}$.
\item[(b)] Hadronic decays by the electromagnetic interaction 
(Figure~\ref{fig:feynmandiagram_b}): \ 
The quark and antiquark pair annihilate to create a virtual
photon and then the virtual photon produces a quark and antiquark  
pair. \ The quark-antiquark pair fragments into a final state of 
hadrons, i.e. $J/\psi,\psi(2S) \to \gamma^{*} \to {\rm hadrons}$. \ 
\item[(c)] Hadronic decays by the strong interaction 
(Figure~\ref{fig:feynmandiagram_d}): \ 
The quark and antiquark pair annihilate to three gluons, then 
the three gluons fragment into a final state of hadrons,
i.e. $J/\psi,\psi(2S) \to g + g + g \to {\rm hadrons}$.
\item[(d)] Radiative decays (Figure~\ref{fig:feynmandiagram_c}): \ 
The quark and antiquark pair annihilate to a photon and two 
virtual gluons, then the two gluons produce the final state of hadrons, 
i.e. $J/\psi,\psi(2S) \to \gamma + g + g \to \gamma + {\rm hadrons}$. \ 
\end{itemize}

The strong interaction dominates over the electromagnetic interaction, 
in spite of the suppression (OZI) of the three-gluon process. \
It is impossible for a charmonium state with $J^{PC} = 1^{--}$ to decay
to a single gluon as the particle is colorless but a gluon carries color
charge. \ Decay through two gluons is forbidden by $C$ conservation. \
For example, $97.85\%$ of $\psi(2S)$ mesons decay to hadrons through 
three gluons and only $1.73\%$ of $\psi(2S)$ mesons decays to hadrons 
through virtual photons.

\subsection{Radiative Decay of $\psi(2S) \to \gamma \eta_{c}(2S)$}

The meson $\psi(2S)$ has a mass of ($3684\pm0.034$)~MeV and can decay
radiatively to another $c\bar{c}$ meson, such as 
$\psi(2S)\to\gamma\eta_{c}(1S)$ and $\psi(2S)\to\gamma\chi_{c2}$. \ 
The radiative decay $\psi(2S)\to\gamma\eta_{c}(2S)$, that is the focus 
of this dissertation, has not been observed.

In our investigations, $\psi(2S)$ mesons are produced by the 
annihilation of electron-positron pairs at a center-of-mass energy 
of $3686~{\rm MeV}$. \ An unknown portion of the $\psi(2S)$ mesons 
would decay into a transition photon and a pseudoscalar $c\bar{c}$ 
meson $\eta_{c}(2S)$ through electromagnetic interaction. \ 
Some of the generated $\eta_{c}(2S)$ mesons would decay into 
hadrons, ultimately into detectable particles, like pions, kaons, and 
photons. \ The first step of the 
radiative decay is a two body decay, which means the particle 
$\eta_{c}(2S)$ could be observed and measured through the energy 
distribution of the transition photon.

The experimental challenge of the measurement for this decay channel is 
to detect the $\sim 48$~MeV radiative photons in an experimental 
environment with considerable background. \ Inclusive study is 
impractical for these low photon energies. \ 
Our alternative approach is to study the decay exclusively by 
reconstructing $\eta_{c}(2S)$ from its hadronic decay products and 
looking at the $\eta_{c}(2S)$ candidate mass for evidence of 
$\psi(2S) \to \gamma~\eta_{c}(2S)$. 

\subsection{Theoretical Background of $\psi(2S) \to \gamma~\eta_{c}(2S)$}
\label{sec:theory}

The purpose of this analysis is to study the direct magnetic dipole (M1) 
radiative transition $\psi(2S) \to \gamma~\eta_c(2S)$ by reconstructing
exclusive decays of the $\eta_c(2S)$. \ The partial width for a direct
M1 radiative transition between S-wave charmonium states is given by
\begin{equation}
\label{transitionrate}
\Gamma(^3S_1 \rightarrow~^1 S_0~\gamma) =
\frac{4\alpha}{3}\frac{e^2_c}{m^2_c}k^3I^2,
\end{equation}
where $\alpha$ is the fine structure constant, $e_c$ ($m_c$) is the
charge (mass) of the charm quark, and $k$ is the energy of the
transition photon. \ The factor $I$ incorporates the matrix element
for the spin-flip transition of the $c\overline{c}$ pair. \
The energy of the transition photon is determined by the mass difference
between the S-wave spin-triplet and spin-singlet charmonium states and
therefore is related to the hyperfine mass splitting between them.

The hyperfine mass splitting is defined as
\begin{equation}
\Delta M_{\rm{HF}}(nS) = \frac{32\pi\alpha_s}{9m^2_c}|\psi(0)|^2,
\end{equation}
where $n$ denotes the principal quantum number, $\alpha_s$ is the strong
coupling constant, and $\psi(0)$ is the wave function of the S-wave
spin-triplet state at the origin. \ Table~\ref{tab:theoryhf}
lists the theoretical predictions for the hyperfine splitting. \
The splitting of the first radially excited state is consistently less
than that for the ground state, reflecting coupled-channel effects
between the $\psi(2S)$ and the nearby open-charm threshold
\cite{Eichten04,Martin82,Eichten80}.

Table~\ref{tab:theorybf} lists the theoretical predictions for the
$\psi(2S) \rightarrow \gamma~\eta_c(2S)$ branching fraction. \ 
These include phenomenological models, including nonrelativistic 
potential models, effective field theory calculations, and QCD 
calculations implemented via lattice gauge theory (LQCD). \
The results fall in the range $(0.6-36.0)\times10^{-4}$.
\singlespacing
\begin{table}[htbp]
\caption[Predictions for the hyperfine mass splitting]
{Predictions for the hyperfine mass splitting between the first
two S-wave spin-triplet and spin-singlet charmonium states. \ The labels
$\Delta M_{\rm{HF}}(2S)$ and $\Delta M_{\rm{HF}}(1S)$ are defined as
$M(\psi(2S))-M(\eta_c(2S))$ and $M(J/\psi)-M(\eta_c(1S))$, respectively.}
\label{tab:theoryhf}
\medskip
\begin{center}
\begin{tabular}{ccc}
\hline
\hline
Ref. & $\Delta M_{\rm{HF}}(2S)$ [MeV] & $\Delta M_{\rm{HF}}(1S)$ [MeV] \\
\hline
Experiment\cite{PDBook2006}        & 48$\pm$4  & 116.5$\pm$1.2  \\
\hline
Potential Model\cite{Pumplin75}    & 92        & 119            \\
Potential Model\cite{Eichten81}    & 83        & 130            \\
Potential Model\cite{Buchmuller81} & 42-65     & 99            \\
Potential Model\cite{Moxhay83}     & 49        & 78             \\
Potential Model\cite{McClary83}    & 70        & 113            \\
Potential Model\cite{Grotch84}     & 72-109    & 121-196        \\
Potential Model\cite{Godfrey85}    & 60        & 130            \\
Perturbative QCD\cite{Pantaleone86} & 69.0      & 101.0          \\
Potential Model\cite{Gupta89}      & 71        & 116            \\
Potential Model\cite{Fulcher90}    & 204       & 139            \\
Potential Model\cite{Fulcher91}    & 86        & 117            \\
Potential Model\cite{Eichten94}    & 78        & 117            \\
HQET\cite{Zeng95}                  & 60        & 100            \\
Potential Model\cite{Gupta96}      & 68.0      & 117.8          \\
Potential Model\cite{Chen96}       & 54.2-82.5 & 106.1-128.3    \\
Potential Model\cite{Motyka98}     & 98        & 117            \\
BSLT\cite{Lahde02}                 & 38        & 102            \\
Quark Model\cite{Ebert03}          & 98        & 117            \\
Phenomenology\cite{Badalian03}       & 57$\pm$8  & ---          \\
Perturbative QCD\cite{Recksiegel04} & 37$\pm$35 & 88$\pm$26    \\
Potential Model\cite{Eichten04}      & 46.1      & 117*         \\
Quark Model\cite{Ebert05}            & 51        & 119          \\
Potential Model\cite{Barnes05}       & 42-53     & 108-123        \\
Potential Model\cite{Lakhina06}      & 38-41     & 80-108         \\
Potential Model\cite{Radford07}      & 66.9-88.4 & 115.22-117.06  \\
Quenched LQCD\cite{LQCDOkamoto02}    & 34$\pm$25              & 72.6$\pm$0.9   \\
Unquenched LQCD\cite{LQCDMcNeile04}  & -18$\pm$189,124$\pm$24 & 80$\pm$1,105$\pm$19   \\
\hline
\hline
\end{tabular}
\end{center}
*Input to theory, not a prediction.
\end{table}
\normalspacing
\begin{table}[htbp]
\caption[Predictions for $\psi(2S) \rightarrow \gamma~\eta_c(2S)$ 
branching fraction]
{Predictions for $\psi(2S) \rightarrow \gamma~\eta_c(2S)$ branching 
fraction. \ 
The energy of the transition photon is denoted by $k$.
The $\psi(2S)$ full width ($\Gamma(\psi(2S)) = 337~{\rm keV}$ from 
Ref.~\cite{PDBook2006}
is used to determine  ${\cal B}(\psi(2S) \rightarrow \gamma~\eta_c(2S))$, 
as described in the text.}
\label{tab:theorybf}
\medskip
\begin{center}
\begin{tabular}{ccc}
\hline
\hline
Ref. & $k$ [MeV] & ${\cal B}(\psi(2S) \rightarrow \gamma~\eta_c(2S))$ [$10^{-4}$] \\
\hline
Experiment\cite{PDBook2006}             & 48$\pm$4   & ---  \\
\hline
Potential Model\cite{Eichten80}    & 24.9(49.6) & 0.585(4.45) \\
Potential Model\cite{Zambetakis83} & 48*        & 3.1         \\
Potential Model\cite{Grotch84}     & 92         & 0.6-27.0    \\
Potential Model\cite{Godfrey85}    & 60         & 8.0         \\
Potential Model\cite{Zhang91}      & 92         & 13.4-20.2   \\
Potential Model\cite{Eichten02}    & 32         & 1.2         \\
Quark Model\cite{Ebert03}  & 32(91)     & 1.3(29)     \\
BSLT\cite{Lahde03}                 & 46         & 0.89        \\
Potential Model\cite{Barnes05}     & 48         & 5.0-6.2     \\
Potential Model\cite{Radford07}    & 66.9(88.4) & 12(36)       \\
Phenomenology\cite{Eichten07}      & 47.8       & 2.6$\pm$0.7 \\
\hline
\hline
\end{tabular}
\end{center}
*Calculated using mass values from 
Ref.~\cite{PDBook2006}.
\end{table}

\subsection{Estimates of Production and Decay Rates}
\label{ExpNumProdEvt}

The PDG does not provide enough information for the branching fractions 
of $\psi(2S) \to \gamma \eta_{c}(2S)$ and $\eta_{c}(2S) \to X$ for 
definite predictions of event rates in any experiment. \ We have made 
certain assumptions to estimate these values for our analysis.

An estimate of expected $\psi(2S) \to \gamma \eta_{c}(2S)$ rate can be 
based on the similarity between $\psi(2S) \to \gamma \eta_{c} (2S)$ 
and $J/\psi(1S) \to \gamma \eta_{c} (1S)$. \ We start from the 
transition rate between spin-0 and spin-1 S-wave states, which is given 
in Equation~\ref{transitionrate}. \ 
For estimate, we assume the 
 matrix element is 
the same for both decays and use the mass differences to calculate the 
energies of transition photons. \ For the $2S$ decay, we have 
\begin{equation}
\label{2sphotonenergy}
E_{\gamma}(\psi(2S) \to \gamma \eta_{c}(2S)) 
= \frac{M^2_{\psi(2S)} - M^2_{\eta_c(2S)}}{2M_{\psi(2S)}}
= (47.8 \pm 3.9)~{\rm MeV},
\end{equation}
where $M_{\psi(2S)}$ ($M_{\eta_c(2S)}$) is the mass of the 
$\psi(2S) = 3686.093 \pm 0.034$ MeV ($\eta_c(2S) = 3638 \pm 4$ MeV). \
Similarly, for the $1S$ decay, we have
\begin{equation}
\label{1sphotonenergy}
E_{\gamma}(J/\psi \to \gamma \eta_{c}(1S)) = (114.3 \pm 1.1)~{\rm MeV}.
\end{equation} 
From the branching fraction for $J/\psi \to \gamma \eta_{c}$, 
\begin{equation}
\label{bfjpsitogetac}
{\mathcal B}(J/\psi \to \gamma \eta_{c}) = ( 1.3 \pm 0.4 )\%
\end{equation}
and the full width of $J/\psi$, 
\begin{equation}
\label{widthjpsi}
\Gamma_{J/\psi} = 93.4 \pm 2.1~{\rm keV},
\end{equation}
we can obtain the partial width of $J/\psi \to \gamma \eta_{c}$: 
\begin{equation}
\label{widthjpsitogetac}
\Gamma(J/\psi \to \gamma \eta_{c}) = {\cal B}(J/\psi \to \gamma \eta_{c})
\times\Gamma(J/\psi)
= (1.21 \pm 0.37)~{\rm keV}.
\end{equation}
With the above information the partial width of 
$\psi (2S) \to \gamma \eta_{c} (2S)$ can be calculated as
\begin{eqnarray}
\Gamma(\psi(2S) \to \gamma \eta_{c}(2S))
& = & \Gamma(J/\psi \to \gamma \eta_{c}(1S))~
      \left(\frac{E_{\gamma}(\psi(2S) \to \gamma 
      \eta_{c}(2S))}{E_{\gamma}(J/\psi \to \gamma \eta_{c}(1S))}\right)^3
\nonumber \\
& = & (0.086 \pm 0.035)~{\rm keV}
\label{widthpsi2stogetac2s}
\end{eqnarray}
The full width of $\psi (2S)$ is
\begin{equation}
\Gamma_{\psi(2S)} = 283 \ {\rm keV}.
\end{equation}
Therefore, the branching fraction of $\psi(2S) \to \gamma \eta_{c}(2S)$
is estimated to be
\begin{equation}
\label{widthpsi2s}
{\cal B}(\psi(2S) \to \gamma \eta_{c}(2S)) = (2.6 \pm 1.0) \times 10^{-4} .
\end{equation}

To estimate the expected number of produced events for each decay mode
considered, we need estimates of the branching fractions of
$\eta_{c}(2S) \to X$, where $X$ is a particular final state. \ 
These are estimated by scaling from the
corresponding $\eta_{c}(1S)$ branching fractions:
\begin{eqnarray}
{\cal B}(\eta_{c}(2S) \to X)
& = & \left(\frac{\Gamma(\eta_c(1S))}{\Gamma(\eta_c(2S))} \right){\cal B}(\eta_{c}(1S) \to X) \nonumber \\
& = & \left(\frac{25.5 \pm 3.4~{\rm MeV}}{14 \pm 7~{\rm MeV}}\right){\cal B}(\eta_{c}(1S) \to X) \nonumber \\
& = & (1.8 \pm 0.9)\times{\cal B}(\eta_{c}(1S) \to X).
\label{modebfconvertion1sto2s}
\end{eqnarray}

\subsection{Previous Measurements of $\eta_{c}(2S)$ Productions and Decays}

In this analysis we are motivated to measure the mass of the $\eta_{c}(2S)$ 
from the decay channel $\psi(2S) \to \gamma \eta_{c}(2S)$, and improve 
the measurement of full width of the $\eta_{c} (2S)$. \ Previous 
measurements compiled by the PDG give an average of $3638 \pm 5~{\rm MeV}$ 
on the $\eta_{c}(2S)$ mass 
and $14 \pm 7~{\rm MeV}$ on the full width of $\eta_{c}(2S)$ in 2005 
\cite{PDBook2006}. \ The measurements that contributes to the average 
of PDG 2006 are from CLEO, 
$M_{\eta_{c}(2S)} = 3642.9 \pm 3.1 \pm 1.5~{\rm MeV}$ 
  and $\Gamma_{\eta_{c}(2S)}=6.3\pm12.4\pm4.0~{\rm MeV}$ 
  with 61 events in 2004 
\cite{cleo2004fusion, cbx03-20, cbx03-42}, 
from BaBar, $M_{\eta_{c}(2S)} = 3630.8 \pm 3.4 \pm 1.0~{\rm MeV}$ 
  and $\Gamma_{\eta_{c}(2S)} = 17.0 \pm 8.3 \pm 2.5~{\rm MeV}$ 
  with $112\pm24$ events in 2004 
  \cite{babar2004fusion},
and from Belle, $M_{\eta_{c}(2S)} = 3654 \pm 6 \pm 8~{\rm MeV}$ 
  with $39 \pm 11$ events in 2002 
\cite{PhysRevLett.89.102001}.
In the CLEO measurement, two-photon production of $\eta_{c}(2S)$, 
followed by $\eta_{c}(2S) \to K_{S}K\pi$ was studied with samples 
$13.6~{\rm fb}^{-1}$ of CLEO II/II.V data and $13.1~{\rm fb}^{-1}$ 
of CLEO III data. \ The result was a measurement of the ratio 
$R = [ \Gamma _{\gamma \gamma}(2S) \times {\mathcal B}(\eta_{c}(2S) 
\to K_{S}K\pi)] / [\Gamma _{\gamma \gamma}(1S)\times {\mathcal B}(\eta_{c}(1S)
\to K_{S}K\pi)] = 0.18 \pm 0.05({\rm stat}) \pm 0.02({\rm syst})$. \ 
The most recent average of measurements gives a small correction of 
the mass of $\eta_{c}(2S)$ from $3638 \pm 5~{\rm MeV}$ to 
$3637 \pm 4~{\rm MeV}$, but the width has remained unchanged since 2006 
\cite{PDBook2008}. \ 
In the BaBar measurement, the $\eta_{c}(2S)$ mass and width were measured 
with a sample of $88~{\rm fb}^{-1}$ collected at the $\Upsilon (4S)$. \ 
In 2006, BaBar inclusively studied the decay 
$B^{\pm} \to K^{\pm} \eta_{c}(2S)$ with a sample of
$210.5~{\rm fb}^{-1}$ collected at the $\Upsilon (4S)$, finding 
${\mathcal B}(B^{\pm} \to K^{\pm} \eta_{c}(2S)) 
= (3.4 \pm 1.8({\rm stat}) \pm 0.3({\rm syst})) \times 10^{-4}$, with a 
significance of $1.8~\sigma$ \cite{aubert:052002}. \ This result is 
listed in PDG 2008 \cite{PDBook2008}. \ In a recent BaBar measurement 
$B^{\pm} \to K^{\pm} \eta_{c}(2S)$ and $\eta_{c}(2S) \to K \bar{K}\pi$, 
where $K \bar{K}\pi = K_{S}K\pi + KK\pi^{0}$ was studied with a sample 
of $349~{\rm fb}^{-1}$ collected at the $\Upsilon (4S)$. \ They measured 
the ratio of 
$B^{\pm} \to K^{\pm} \eta_{c}(2S)$ and $\eta_{c}(2S) \to K \bar{K}\pi$
to the corresponding $\eta_{c}(1S)$ process. \ Using the result from the 
PDG 2008, this leads to 
${\mathcal B}(\eta_{c}(2S) \to K \bar{K}\pi) = (1.9 \pm 0.4({\rm stat})
\pm 0.5({\rm syst}) \pm 1.0({\rm br}))$. \ However, the mass 
resolutions of the two separate modes 
$\sigma(K_{S}K\pi) = 9 \pm 1~{\rm MeV}$ and 
$\sigma(KK\pi^{0}) = 20 \pm 9~{\rm MeV}$ are very different and the 
peaks of $\psi(2S)$ and $\eta_{c}(2S)$ overlap each other 
\cite{Aubert:2008kp}. \ 
In a recent Belle measurement, 
two-photon production of $\eta_{c}(2S)$, followed by 
$\eta_{c}(2S) \to 4 \pi$ or $\eta_{c}(2S) \to KK \pi \pi$, was studied 
with a sample of $395~{\rm fb}^{-1}$ collected at the $\Upsilon (4S)$, 
leading to results of 
$\Gamma _{\gamma \gamma}(2S) \times {\mathcal B}(\eta_{c}(2S) \to 4 \pi)
< 6.5~{\rm eV}$ and 
$\Gamma _{\gamma \gamma}(2S) \times {\mathcal B}(\eta_{c}(2S) \to KK\pi\pi)
< 5.0~{\rm eV}$ at 90\% confidence level \cite{Uehara2008}.

\section{Objectives and Organization of this Dissertation}

Information about $\eta_{c}(2S)$ production and decay is currently very 
limited. \ Measurements at the $\Upsilon (4S)$ energy are hampered by 
serious backgrounds and limited to only a few modes. \ Measurements in 
the cleaner environment of $e^{+}e^{-}$ annihilations near charm 
threshold have significant advantages. \ This is the principal 
motivation for the work described in this dissertation.

Starting in fall 2003, the CESR-c/CLEO-c program began an anticipated 
three year run though the actually run period was 4.5 years until the 
shut down on March 2, 2008. \ During this period, the CLEO-c detector 
was used to collect the following data:
\begin{itemize}
\item $572~{\rm pb}^{-1}$ of integrated luminosity on the $\psi(3770)$
(see Section~\ref{sec:luminosity})
\item about $27$ million $\psi(2S)$ decays
\item $21~{\rm pb}^{-1}$ of continuum below the $\psi(2S)$
\item $47~{\rm pb}^{-1}$ of scan data near $E_{CM} = 4170~{\rm MeV}$
\item $13~{\rm pb}^{-1}$ of data at $E_{CM} = 4260~{\rm MeV}$
\item about $600~{\rm pb}^{-1}$ of data at $E_{CM} = 4170~{\rm MeV}$ for 
$D_{s}$ physics
\end{itemize}
While the data samples were smaller than the original project goals, 
the CLEO-c project is regarded as a great success.

In this project, the decay channel $\psi(2S) \to \gamma \eta_{c}(2S)$ 
is investigated through the exclusive reconstruction of candidate events
in the CLEO-c $\psi(2S)$ data sample. \ A similar nearby decay channel,
$\psi(2S) \to \gamma \chi_{c2}$, has been studied for validation and 
comparison purpose, using
CLEO's previous measurement of this channel.

Based on the current information about the decay 
$\psi(2S) \to \gamma \eta_{c}(2S)$, we evaluated the project by 
making an estimate on the number of evens that can possibly produced. \ 

For the CLEO-c $\psi(2S)$ sample of 25.9 million events, the estimated
number of $\psi(2S) \to \gamma \eta_{c}(2S)$ events produced is
$N_{\rm prod} = 6700 \pm 2600$, where we have
used a branching fraction of $(2.6 \pm 1.0) \times 10^{-4}$. \ 
With this information, the expected numbers of produced events of
all considered modes are listed in Table~\ref{table:numofproduced}. \
The error in $B(\eta_{c}(2S)\to X)$
is dominated by $\Gamma(\eta_{c}(2S)) [50\%]$, and the error in
$B(\psi(2S)\to\gamma\eta_{c}(2S))$ is dominated by
$B(J/\psi\to \gamma\eta_{c}(2S)) [31\%]$.
\begin{table}[htbp]
\caption[Expected number of produced events in each mode]
{\label{table:numofproduced}Expected number of produced events
in each mode. \ As a reference the 1\% assumption of branching fraction
of a mode would provide a number of produced event of $(67 \pm 18)$.}
\begin{center}
\begin{tabular}{|l|c|c|c|c|}
  \hline
  Mode & $B(\eta_{c}(2S)\to X)$ &
         $B(\psi(2S)\to\gamma\eta_{c}(2S)) \times $ &
         $N$ produced \\
       & (\%) & $B(\eta_{c}(2S)\to X) (10^{-6})$ & \\ \hline
  $4\pi$
    & $2.2\pm1.3$ & $5.7\pm3.3\pm2.2$ & $147\pm84\pm57$ \\ \hline
  $6\pi$
    & $3.6\pm2.3$ & $9.5\pm5.9\pm3.7$ & $245\pm153\pm94$ \\ \hline
  $KK\pi\pi$
    & $2.7\pm1.8$ & $7.1\pm4.6\pm2.7$ & $184\pm121\pm71$ \\ \hline
  $KK\pi^{0}$
    & $2.1\pm1.1$ & $5.5\pm2.9\pm2.1$ & $142\pm77\pm55$ \\ \hline
  $K_{S}K\pi$
    & $4.2\pm2.2$ & $10.9\pm5.9\pm4.2$ & $283\pm152\pm109$ \\ \hline
  $\pi\pi\eta(\gamma\gamma)$
    & $2.3\pm1.5$ & $6.1\pm3.9\pm2.3$ & $157\pm100\pm60$ \\ \hline
  $\pi\pi\eta(\pi\pi\pi^{0})$
    & $1.4\pm0.9$ & $3.5\pm2.2\pm1.3$ & $91\pm58\pm35$ \\ \hline
  $\pi\pi\eta^{\prime}, \eta^{\prime}\to\pi\pi\eta(\gamma\gamma)$
    & $0.9\pm0.6$ & $2.3\pm1.5\pm0.9$ & $59\pm39\pm23$ \\ \hline
  $KK\eta(\gamma\gamma)$
    & $< 1.1$ & $< 2.9$ & $< 75$ \\ \hline
  $KK\eta(\pi\pi\pi^{0})$
    & $< 0.64$ & $< 1.7$ & $< 43$ \\ \hline
  $KK\pi\pi\pi^{0}$
    & -- & -- & -- \\ \hline
  $KK4\pi$
    & $1.8\pm1.2$ & $4.7\pm3.0\pm1.9$ & $121\pm78\pm47$ \\ \hline
  $K_{S}K3\pi$
    & -- & -- & -- \\ \hline
\end{tabular}
\end{center}
\end{table}

The expected number of produced $\eta_{c}(2S)$ was estimated assuming
a full width of $14\pm7$~MeV for $\eta_{c}(2S)$ according to PDG 
average \cite{PDBook2006}. \ We expected that the measurement can be 
improved with a larger CLEO-c sample based on these PDG values as 
discussed in Chapter~\ref{chap:analysis_etac2s}. 

From the above estimate we find that the measurement of the decay 
$\psi(2S) \to \gamma \eta_{c}(2S)$ is possible in spite of technical 
challenges and the lack of previous knowledge on the physical 
properties of $\eta_{c}(2S)$. \ 
If enough events can be observed we also plan to measure the partial 
width of $\psi(2S) \to \gamma \eta_{c}(2S)$. \ The only previous 
contribution to the measurement was given by CLEO, in 2004, which is 
$<0.2 \times 10^{-2}$ 
\cite{cleo2004fusion, cbx03-20, cbx03-42}.

Following this introduction, the remainder of this dissertation is 
organized as follows. \ 
The CESR-c accelerator and CLEO-c detector are described in detail in 
Chapter~\ref{chap:apparatus}. \ The analysis 
set up and the event-selection criteria, as well as the study of the 
related decay channel $\psi(2S) \to \gamma \chi_{c2}$ are included in 
Chapter~\ref{chap:analysis_etac2s}. \ 
In Chapter~\ref{chap:conclusions} the summary, discussions and 
conclusions of the investigation of the radiative decay 
$\psi(2S) \to \gamma~\eta_{c}(2S)$ are presented.

\chapter{Experimental Apparatus}
\label{chap:apparatus}

The analysis presented in this dissertation was based on data collected 
with the accelerator and detector at the Cornell University Laboratory 
of Elementary Particle Physics. \ The design, construction and operation 
of the apparatus has been carried out by the collaboration of hundreds 
of people. \ In this chapter the Cornell Electron Storage Ring (CESR-c) 
is described in Section \ref{sec:cesr} and the CLEO-c detector in Section 
\ref{sec:detector}. \ The trigger and data acquisition systems are 
described in Section \ref{sec:daq} and event reconstruction in Section 
\ref{sec:reconstruction}. \ Finally, Section \ref{sec:cleog} explains the 
simulation of the detector response for determination of the detection 
efficiencies and backgrounds for measurements of various physical 
processes. 

\section{Cornell Electron Storage Ring - CESR-c}
\label{sec:cesr}

The Cornell Electron Storage Ring running at charm threshold (CESR-c) is a 
symmetric electron-positron collider with a circumference of 768 meters. \ 
It is located at Wilson Synchrotron Laboratory, about forty feet under 
the track-and-field facility of Cornell University. \ Accelerators like 
CESR-c produce short-lived particles under controlled conditions, allowing 
more detailed studies than with particles from natural sources like cosmic 
rays. \ In CESR-c, electrons and 
positrons annihilate via the electroweak interaction to produce final 
states consisting of hadrons and leptons. \ These annihilations occur 
inside the CLEO-c detector, which measures the particles' momenta, 
energies and other properties.

\begin{figure}[htbp]
\begin{center}
\includegraphics[width=.9\textwidth]{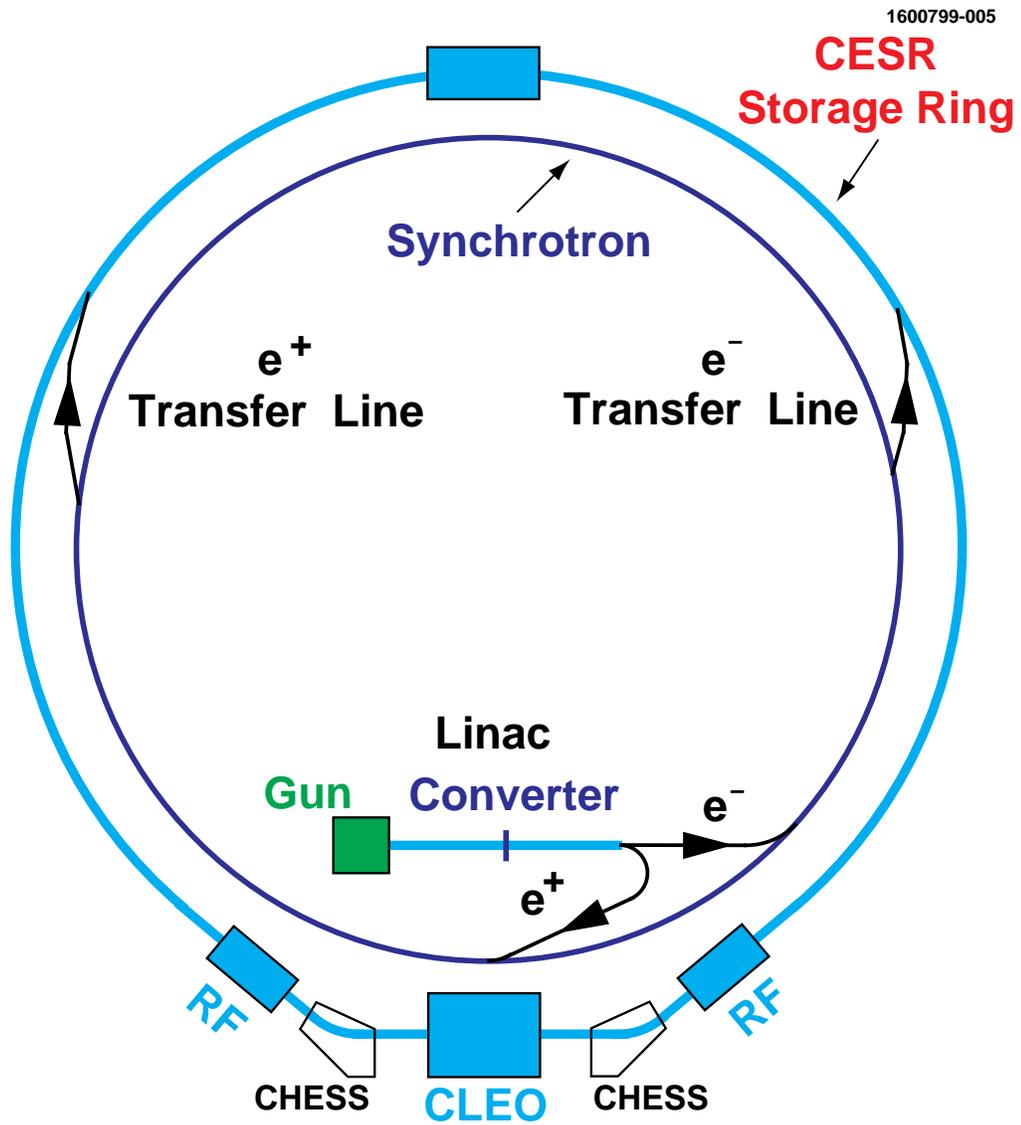}
\end{center}
\caption{\label{fig:cesr}{The CESR $e^+e^-$ collider.}}
\end{figure}

CESR-c is designed to produce collisions between electrons and positrons 
with center-of-mass energies between 3 and 5~GeV. \ 
For the data sample used for this dissertation, the beam energy was 
set at around 3.7~GeV, 
to produce $\psi(2S)$ charmonium mesons nearly at rest. \ CESR-c consists 
of three parts: the linear accelerator (Linac), the CESR synchrotron, and the 
storage ring, as shown in Figure~\ref{fig:cesr}. 

\subsection{Linac}

The Linac is the first part of the CESR-c, in which 
electrons and positrons are generated, collected, bunched and accelerated 
before they enter the synchrotron for further acceleration.

The electrons are boiled off of a hot filament by thermionic emission. \ 
They are accelerated to 150~keV by static electric fields and injected into
the 30-meter-long vacuum pipe which is the main Linac structure. \ The 
Linac has eight radio-frequency (RF) cavities, which generate oscillating 
electric fields that accelerate the electrons to an 
energy of 300~MeV before they are injected into the synchrotron.

The positrons are created by intercepting the 150~MeV accelerated electrons 
halfway down the Linac and directing them into a tungsten target. \ 
Showers of electrons, positrons and 
photons are generated from the target. \ Positrons are selected by a 
magnetic field and then accelerated along the rest of the Linac
to a final energy of 200~MeV for injection into the synchrotron.

\subsection{Synchrotron}

The synchrotron is a circular vacuum pipe that fills the CESR-c tunnel 
and consists of four RF accelerating cavities of 3-meter length and 
192 bending and focusing magnets. \ The RF cavities add energy to 
the electrons and positrons each time they pass 
through. \ Dipole magnets 
bend the trajectories of the electrons and positrons so that they can 
move along the circular path within the synchrotron. \ Quadrupole and 
other focusing magnets keep the electrons and positrons confined to 
trajectories near the axis of the beam pipe. \ As the radius of the 
circular motion of electrons and positrons is given by $R=p/qB$, in order 
to keep the particle beam bunches moving in synchrotron, the dipole magnetic 
field $B$ needs to be synchronized with the momentum increase of the 
particles. \ It takes about 1500 revolutions or 1/100 second to 
accelerate the particles to the desired beam energy, after which they 
are injected into the CESR-c storage ring. \ At the typical beam energy of 
2~GeV , the beam particles travel at 99.999997\% of the speed of light.

\subsection{CESR-c Storage Ring}

Once the electron or positron beam has been accelerated to the desired 
energy, it is injected into the storage ring which operates by the same 
principle as the synchrotron. \ The electrons and positrons travel in a 
closed orbit in a much better vacuum than the synchrotron. \ 
The storage ring contains 106 quadrupole focusing magnets and 86 dipole 
bending magnets, and in addition it also has sextupole and octupole 
magnets for very precise focusing of the beams. \ When the electrons 
and positrons circulate in the ring, they emit electromagnetic radiation 
that causes energy loss, which is called synchrotron radiation. \ 
The energy is restored by superconducting radio-frequency (RF) cavities 
that operate at a frequency of 500~MHz. \ These RF cavities are similar 
to those used in the Linac and synchrotron, except that they are used 
primarily to restore the energy of the particles, while those in the 
Linac and synchrotron are used to accelerate the particles.

The synchrotron radiation deposits energy in the vacuum chamber wall and 
the generated heat due to the radiation is carried away by circulating 
water. \ Some of the synchrotron radiation is used by the Cornell High 
Energy Synchrotron Source (CHESS) facility for X-ray research in the 
areas of physics, chemistry, biology, environmental science, and 
materials science.

\begin{figure}[htbp]
\begin{center}
\includegraphics[width=.9\textwidth]{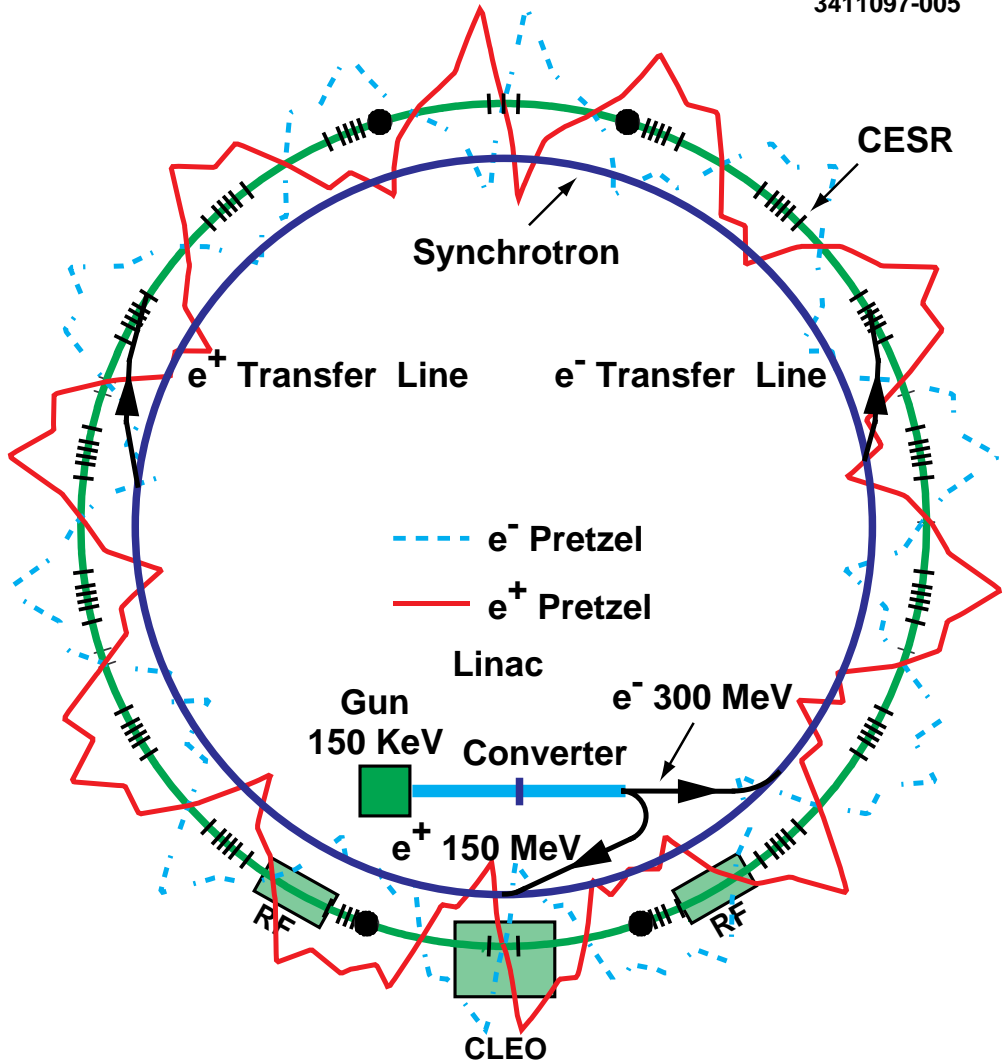}
\end{center}
\caption[Schematic of CESR showing the ``pretzel'' orbits.]
{\label{fig:pretzel}{Schematic of CESR showing the ``pretzel'' orbits which 
are used to separate the electron and positron beams at parasitic 
crossing locations.}}
\end{figure}

In standard operation CESR-c stores the particles in 45 bunches of 
electrons and 
45 bunches of positrons (configured as nine ``trains'' of five bunches), 
circulating in opposite directions. \ They are allowed to cross at one 
interaction region, where the CLEO-c detector is located. \ One ``fill'' 
of electrons and positrons lasts about one hour. \ Since both the electron 
and positron bunches are stored in the same ring, four electrostatic 
separators are used to set up ``pretzel'' orbits and ensure that the 
electrons and positrons miss each other when they pass through the unwanted 
intersecting locations, sometimes referred to parasitic crossings. \ Thus
the electrons and positrons collide only at the interaction point, which
is at the center of CLEO-c detector. \ Figure~\ref{fig:pretzel} shows a 
greatly exaggerated \cite{yellowbook}
schematic of the pretzel orbits. \ At the interaction point, the two beams 
are steered into each other and collide at a small crossing angle of 
approximately $\pm 2.6$~mrad ($\sim 0.15^{\circ}$).

\subsection{Luminosity}
\label{sec:luminosity}

The annihilation rate depends on the ``luminosity'' of the storage ring, 
which in turn depends 
primarily on the stored current and beam size. \ Instantaneous luminosity 
$\mathcal L$ is defined as the number of collisions for each cm$^2$ of 
cross section per second by the approximate equation
\begin{equation}
{\mathcal L} \equiv nf\frac{N_{e^+}N_{e^-}}{A},
\end{equation}
where $f$ is the frequency of revolution of the particles, $n$ is the
number of electron or positron bunches in the beam, $N_{e^-}$ and 
$N_{e^+}$ are the number of electrons and positrons in each bunch
respectively, and $A$ is the cross-sectional area of the beams. \ The 
total number of events for a particular process, $N$, is given by
\begin{equation}
N = \sigma_{i} \int{\mathcal L}dt,
\end{equation}
where $\sigma_{i}$ is the cross section for the process. \ The integral 
of the instantaneous luminosity, $\int{\mathcal L}dt$, is generally 
referred to as the integrated luminosity or luminosity. \ In CLEO-c,
the integrated luminosity for a given data sample is determined by a 
weighted average of the results from
three processes $e^{+}e^{-} \to e^{+}e^{-}$, $\mu^{+}\mu^{-}$, and 
$\gamma \gamma$, for which the cross sections are precisely determined by 
QED. \ Each of the three final states relies on different components
of the detector with different systematic effects \cite{cbx05-10}.  

\subsection{CESR-c Upgrades}

CLEO-c and CESR-c have been developed from the previous CLEO III detector
and CESR accelerator. \ For over 20 years beginning in 1979,
CESR collected electron-positron annihilation data at a center-of-mass 
energy of $\sim 10.5$~GeV for the study of $B$ mesons near threshold. \ 
For the CLEO-c and CESR-c project, CESR was proposed to run around charm 
threshold, $\sim$~3-5~GeV. \ Modifications to the accelerator and 
detector were necessary to accommodate the lower center-of-mass energy.

As the rate of the synchrotron radiation is proportional to $E^{4}$, the 
reduction of the beam energy greatly altered the beam dynamics. \ 
The decrease of the synchrotron radiation affects CESR performance through 
two important beam parameters: damping time, with which perturbations in 
beam orbits caused by injection and other transitions decay away, and 
horizontal beam size, or horizontal emittance, which measures the spread 
of particles in the bend plane. \ At decreased center-of-mass energy, 
these effects can limit the luminosity to unacceptable levels. \ In CESR-c, 
superconducting wiggler magnets were used to increase the synchrotron 
radiation \cite{yellowbook}, reduce the damping time, and increase the 
luminosity. \ A wiggler magnet 
is a series of dipole magnets with high magnetic fields. \ Each successive 
dipole has its direction of magnetic field flipped. \ When a particle 
passes the wiggler, it will oscillate and therefore emit additional 
synchrotron radiation while keeping the overall path of the particle in 
the ring the same. \ The damping time is decreased and the horizontal 
beam size is increased because of increase of the synchrotron radiation. \
CESR-c included twelve superconducting wiggler magnets for low-energy
running.

\section{CLEO-c Detector}
\label{sec:detector}

When an electron and positron collide and annihilate, the flash of 
energy results in the creation of new matter. \ The CLEO detector,
named for the one historically coupled to Caesar, was built to 
study these collisions, with details described in 
References~\cite{Kubota:1992ww,Hill:1998ea}.

The CLEO-c detector was the final stage of a series of upgrades of the CLEO 
detector since its commissioning in 1979. \ Beginning in 1989, the CLEO II 
detector was a productive source of physics research, due to its very good 
drift chamber based tracking system and outstanding electromagnetic 
calorimeter. \ In 1995, after a three-layer 
silicon strip vertex detector and a new beam-pipe were installed, and an 
argon-ethane gas mixture in the drift chambers was switched to 
helium-propane, CLEO II was upgraded to CLEO II.V. \ A few years later 
CLEO II.V was upgraded to CLEO III by adding a Ring Image Cherenkov (RICH) 
detector for particle identification and 
a new tracking system, which consisted of a new central drift chamber
surrounding a four-layer silicon strip vertex detector. \ The conversion
from CLEO III to CLEO-c in 2003 consisted of the replacement of the 
silicon vertex detector with a low-material gaseous tracking detector (ZD).

\subsection{CLEO-c Detector}

\begin{figure}[htbp]
\begin{center}
\rotatebox{270}{\includegraphics[width=.7\textwidth]{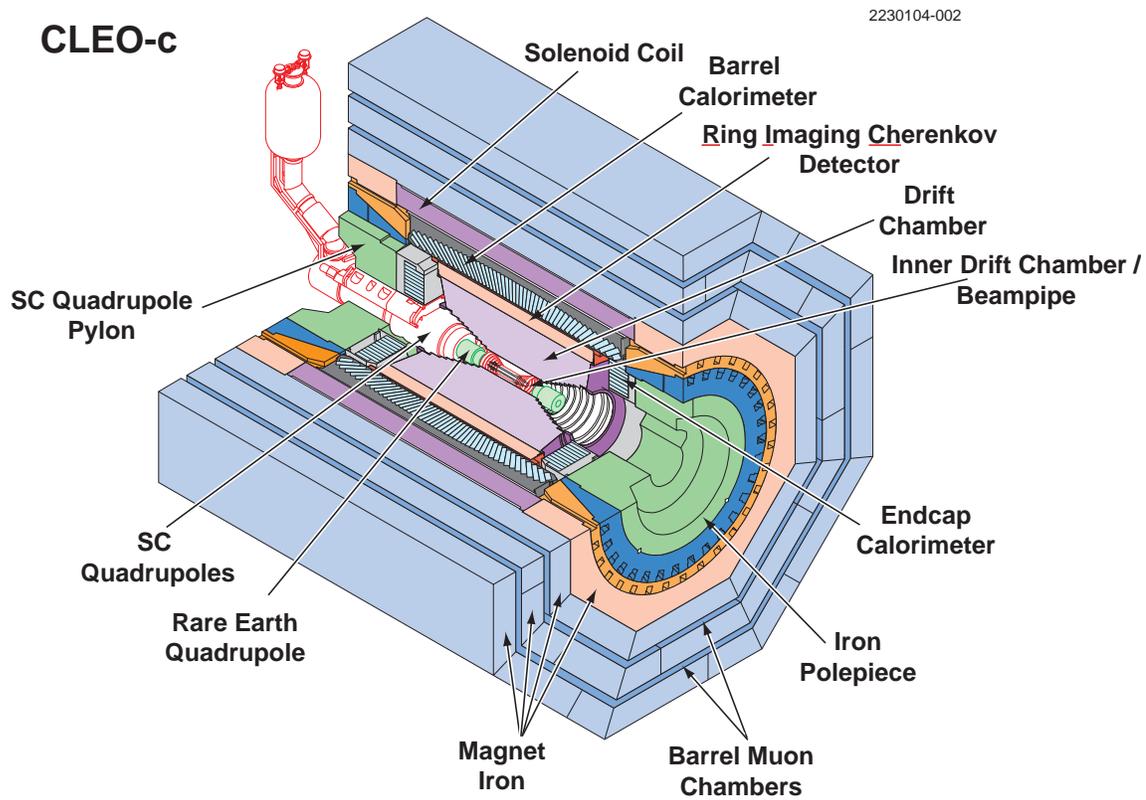}}
\end{center}
  \caption{\label{fig:cleoc} The CLEO-c detector.}
\end{figure}

The CLEO-c detector (Figure~\ref{fig:cleoc}) measures about 6 meters on 
a side, and is constructed of about 900,000 kilograms of iron and over 
25,000 individual detection elements. \ It is an approximately 
cylindrically symmetric detector aligned along the beam line, and 
covering about 93\% of the $4\pi$ solid angle. \ Electrons and positrons 
collide at the interaction point (IP) in the beam 
pipe, which is located at the center of the detector. \ The annihilation 
produces new particles, which decay quickly into long-lived or stable 
charged particles, electrons, muons, protons, pions, and kaons, and some 
neutral particles including photons. \ Particles pass through different 
sub-detectors of the CLEO-c detector, leaving measurable signals. \ The 
tracking system provides the precise measurement of the trajectories of 
charged particles and information on the rate of energy deposit in the 
material ($dE/dx$). \ The Ring Imaging Cherenkov (RICH) detector provides 
a velocity measurement for charged particle identification. \ The 
electromagnetic crystal calorimeter (CC) is used to measure the energies 
of electrons and photons. \ A superconducting solenoid provides a uniform 
magnetic field over the whole tracking system, RICH and CC. \ The muon 
detector is used to detect muons, which can penetrate all inner 
sub-detectors if they have sufficient momentum.

The whole operation of the experiment includes the constant calibrations, 
event data taking process and offline data analysis, which includes the
Monte Carlo studies and physics analysis.

\subsection{Tracking System}

The tracking system of CLEO-c consists of two drift chambers, which 
provide precisely measured space points for charged particles emerging 
from the interaction. \ These devices reside inside a superconducting 
solenoidal magnet that provides a uniform field of 1~T along the beam 
direction. \ The tracking system measures the helical trajectories of 
charged particles, and online and offline software reconstruct these 
trajectories to obtain precise measurements of particle momenta.

\subsubsection{Inner Drift Chamber - ZD}

The innermost sub-detector of the CLEO-c detector is the inner drift 
chamber (ZD), which is located right outside of the beam pipe. \ The 
major hardware modification from CLEO III to CLEO-c is the replacement of 
the Silicon Vertex Detector with the wire vertex drift chamber ZD. \ The 
silicon vertex chamber, as the tracking drift chamber for CLEO III, was used 
to provide extremely accurate track position measurements in $r - \phi$ 
and $z$ as close as possible to the interaction point for studies at 
the $\Upsilon (4S)$. \ However, when the center of mass 
energy was reduced from 10~GeV down to 3-5~GeV, 
the momentum distribution of the charged tracks is shifted down to lower 
values. \ Minimizing material is crucial at lower center-of-mass energies 
since multiple scattering dominates the momentum measurement error for 
low momentum tracks. \ As the tracks have lower velocities, the importance 
of vertexing is reduced because vertex separation for decays of particles 
like $D^{\pm}$ and $D^{0}$ are too small to be resolved with the silicon 
vertex detector. \ Since the extreme precision of the previous silicon 
vertex detector was not needed and the material was a disadvantage, it was 
replaced with the ZD chamber \cite{cbx01-20}.
\begin{figure}[htbp]
\begin{center}
\rotatebox{270}{\includegraphics[width=.7\textwidth]{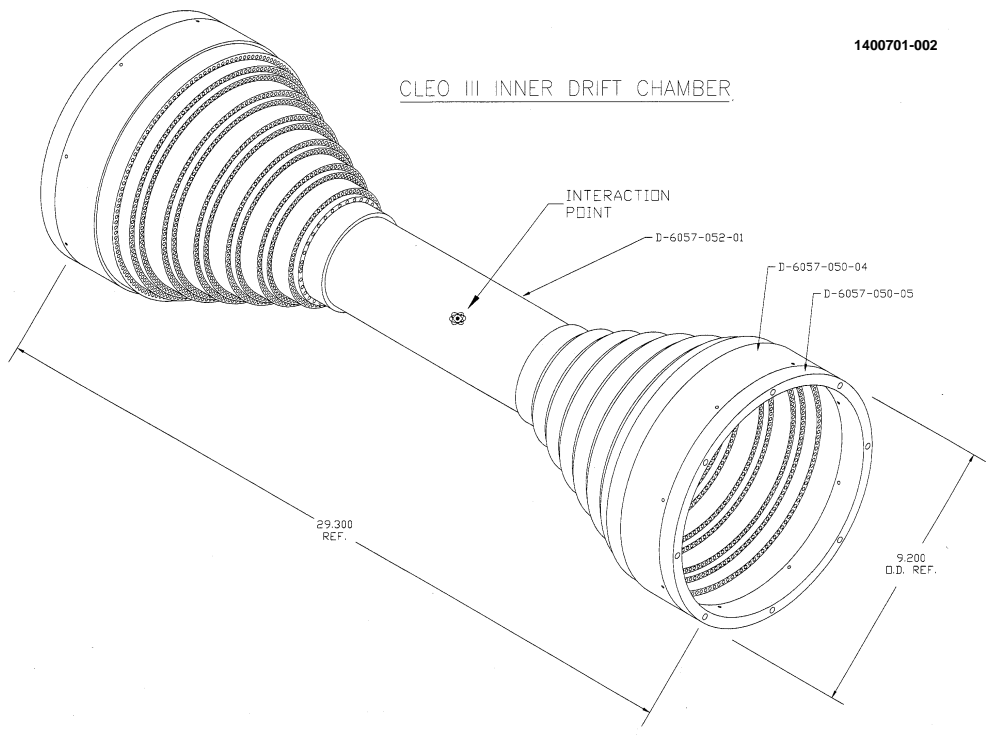}}
\end{center}
  \caption{\label{fig:zd} An schematic of ZD drift chamber.}
\end{figure}

The ZD is located just outside the beam pipe and covers radii from 
4.1~cm to 11.8~cm from the interaction point. \ An isometric view of 
the structure of the ZD is shown in Figure~\ref{fig:zd}. \ It supports 
gold-plated tungsten sense wires and gold-plated aluminum field wires. \ 
The sense wires are in six layers with the number of cells per layer 
ranging from 34 to 66 for a total of 300. \ Each cell measures $10$~mm 
and consists of a sense wire held at $+ 1900$~V, surrounded by field wires 
at ground. \ Cells share bordering field wires. \ The wires are oriented 
at a small (stereo) angle with respect to the beam axis, allowing 
measurement of $r$, $\phi$ and $z$. \ The stereo angle of the ZD ranges 
from $10.3^{\circ}$ in the innermost field wire layer to $15.4^{\circ}$ in 
the outermost field wire layer, with the stereo hyperbolic sag of 4~mm. \ 
The ``all stereo'' design optimizes the measurement of $z$ (position in the
beam direction), from which the inner drift chamber was given the name 
``Z Detector.'' \ The ZD's stepped structure provides spatial measurement 
of charged particles within $|\cos \theta| < 0.93$, where $\theta$ is the 
angle of the the particle with respect to the beam.

When a charged particle passes through a cell, it ionizes the atoms of 
gas filling. \ The ZD is filled with a very pure mixture of 60\% 
helium and 40\% propane (C$_{3}$H$_{8}$) which provides a very long 
radiation length ($\sim 330$~m) 
\cite{yellowbook}. \ The electrons 
produced from the ionized atoms are accelerated toward the sense wire 
and away from the field wires. \  As the electric field increases toward 
the sense wire, these primary electrons ionize other gas atoms and create 
an ``avalanche'' at the sense wire. \ The time of the electric pulse 
observed on the sense wire and the charge, which is related to the 
deposited energy, are recorded. \ This information is used to map out the 
trajectories of the charged particles and to fit tracks.

\subsubsection{Main Drift Chamber - DR}

Outside of the ZD detector is the main drift chamber (DR), a much larger 
device that measures particles out to a distance of 0.8~m from the 
interaction point. \ The main drift chamber has design similar to that 
of the ZD, but with much larger size. \ It 
has 9,796 gold-plated tungsten sense wires and 29,682 gold-plated 
aluminum field wires, arranged in 47 layers. \ The wires are grouped 
in nearly square cells, with cell size 14~mm. \  Each cell contains
one sense wire surrounded by eight field wires. \ The voltage 
on the sense wires is $+2100$~V with respect to the grounded field 
wires. \ The gas in the DR is the same as the gas in the ZD, a mixture 
of 60\% helium and 40\% propane. \ Among the 47 layers
of wires, the first 16 are axial and the remainder are stereo layers 
that alternate angles about $3^{\circ}$ in groups of four to provide 
$z$ information throughout the volume of the detector. \ In order to 
determine the $z$ position of the particle at the stereo layer, the 
axial layers are used to predict the particle's $r-\phi$ position and 
to match the $r-\phi$ information from the stereo wires. \ The inner 
and outer radii of the drift chamber are covered with longitudinally 
and azimuthally segmented cathodes to provide precise measurements of 
the $z$ position of most tracks as they enter and emerge from the DR.

\subsubsection{Hadron Identification by Specific Ionization - $dE/dx$}

\begin{figure}[htbp]
\begin{center}
\includegraphics[width=.9\textwidth]{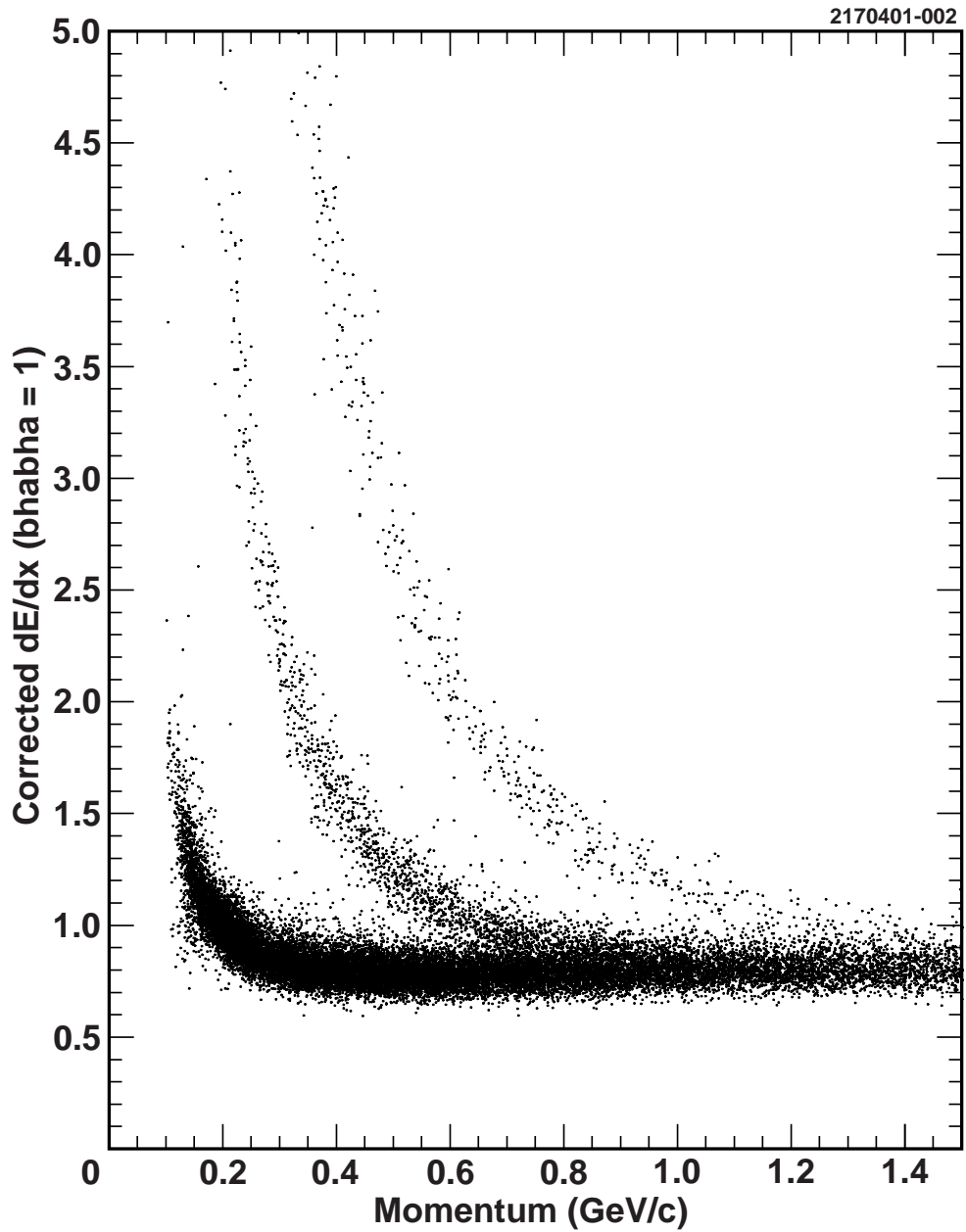}
\end{center}
\caption[Measured $dE/dx$ in the CLEO-c DR as a function of measured momentum]
{\label{fig:dedx} 
{Measured $dE/dx$ in the CLEO-c DR as a function of measured momentum. \ 
The bands represent particles of different mass: electrons, pions and 
kaons from left to right.}}
\end{figure}

Information about the rate of energy loss is provided by the DR for each  
charged particle, along with the the measured momentum. \ This can be 
used for particle identification. \ The energy lost per unit length 
depends upon a particle's velocity, as given by the Bethe-Bloch 
formula. \ A $\chi^{2}$ variable is formed for each particle
hypothesis of electron, muon, pion, kaon, or proton. \ The value of
$\chi^{2}$ is the sum of $\chi^{2}_{i}$ over a lot of hits. \  
$\chi_{i}$ is defined for hit $i$ as
\begin{equation}
\chi_{i} \equiv 
  \frac{dE/dx({\rm measured}) - dE/dx_{i}({\rm expected})}{\sigma_{i}} ,
\label{eq:dedx}
\end{equation}
where $\sigma_{i}$ is the uncertainty of the measurement, usually 
approximately 6\%. \ Figure~\ref{fig:dedx} shows the measured $dE/dx$ as 
a function of momentum for a large population of charged particles. \
At momenta below 500~MeV$/c$, pions and kaons are well separated. \ At 
higher momenta, the $dE/dx$ information is limited and additional 
information is needed for good particle identification.

\subsubsection{Track Reconstruction}
\label{sec:trackrecon}

Track reconstruction is the process to form tracks with hits, the 
electrical signals from the wires, using a pattern-recognition algorithm 
on three-dimensional ($r-\phi$ and $z$) position measurement from the ZD 
and DR. \ (The $r-\phi$ resolution varies over the cell and is of order 
$100~\mu{\rm m}$.) \ Once tracks are found they can be fit to 
determine interesting physics quantities like momenta, vertex positions 
and directions \cite{cbx96-20}. \ The CLEO-c fitter is an implementation 
of the Billoir or Kalman algorithm that optimizes the determination of the
momentum and trajectory, taking into account the expected energy loss.

The tracking system is within an axial magnetic field of 1~T. \ The 
charged particles follow helical paths within the constant magnetic 
field. \ As the magnetic field is parallel to the $z$ axis, the 
transverse momentum $P_{\perp}$, in the $r-\phi$ plane, can be related 
to the curvature of a particle's trajectory by
\begin{equation}
P_{\perp} = q B a,
\label{eq:trkrec}
\end{equation}
where $q$ is the magnitude of the particle's charge, $B$ is the magnitude
of the magnetic field, and $a$ is the radius of curvature. \ Therefore, the 
measurement of the full momentum vector can be obtained by the measurement 
of the track's curvature in $r \phi$, and its angular coordinates $\theta$
and $\phi$. \ At 1~GeV$/c$ the momentum resolution of the charged particle 
is about 0.6\%. \ The direction of the curvature within the magnetic field
indicates the sign of the charged particle.

\subsection{RICH Detector}
\label{sec:rich}

The Ring Image Cherenkov Detector (RICH) \cite{artuso-2005-554, cbx05-36} is 
located directly outside of the main drift chamber. \ It was the most 
significant improvement in the CLEO III upgrade from CLEO II.V and was 
not modified for CLEO-c. \ 
The measurement of Cherenkov radiation in the RICH provides the 
additional charged particle identification needed for the higher 
momenta where $dE/dx$ is inadequate.

When a charged particle passes through a transparent dielectric medium
at a speed greater than the speed of light in that medium, 
electromagnetic radiation, called Cherenkov radiation is produced. \ 
This effect was first observed by Cherenkov in 1934. \ 
As a charged particle travels through the medium, its electromagnetic 
field disturbs the local electromagnetic field in the medium, and 
displaces and polarizes the electrons in the atoms of the medium. \ 
Photons are emitted when the electrons of the medium return to their 
equilibrium state. \ When the
charged particle travels at a speed higher than the speed of light in
the medium, the photons constructively interfere to make the radiation
observable. \ The radiation is produced in a cone with its central 
axis along the trajectory of the traveling particle. \ The 
characteristic angle of the cone $\theta_{c}$, known as Cherenkov 
angle, is related to the velocity of the particle by
\begin{equation}
\cos\theta_{c}=\frac{1}{n\beta},
\label{eq:cherenkovangle}
\end{equation}
where $n$ is the index of refraction of the dielectric medium, and
$\beta$ is the ratio between the velocity of the charged particle
and the speed of light $c$. \ With the relations $\beta = p/E$ and 
$E^{2} = m^{2}+p^{2}$ the cosine of the Cherenkov angle can be 
expressed in terms of index of diffraction $n$, mass of the charged 
particle $m$, and momentum of the charged particle $p$ by
\begin{equation}
\cos\theta_{c} = \frac{1}{n}\sqrt{1+\frac{m^{2}}{p^{2}}}.
\label{eq:cherenkovmass}
\end{equation}
If the momentum of the charged particle can be measured independently, 
such as by fitting the trajectory in the tracking system, then the 
Cherenkov angle gives the mass and therefore the identity of the 
charged particle.

\begin{figure}[htbp]
\begin{center}
\includegraphics[width=.9\textwidth]{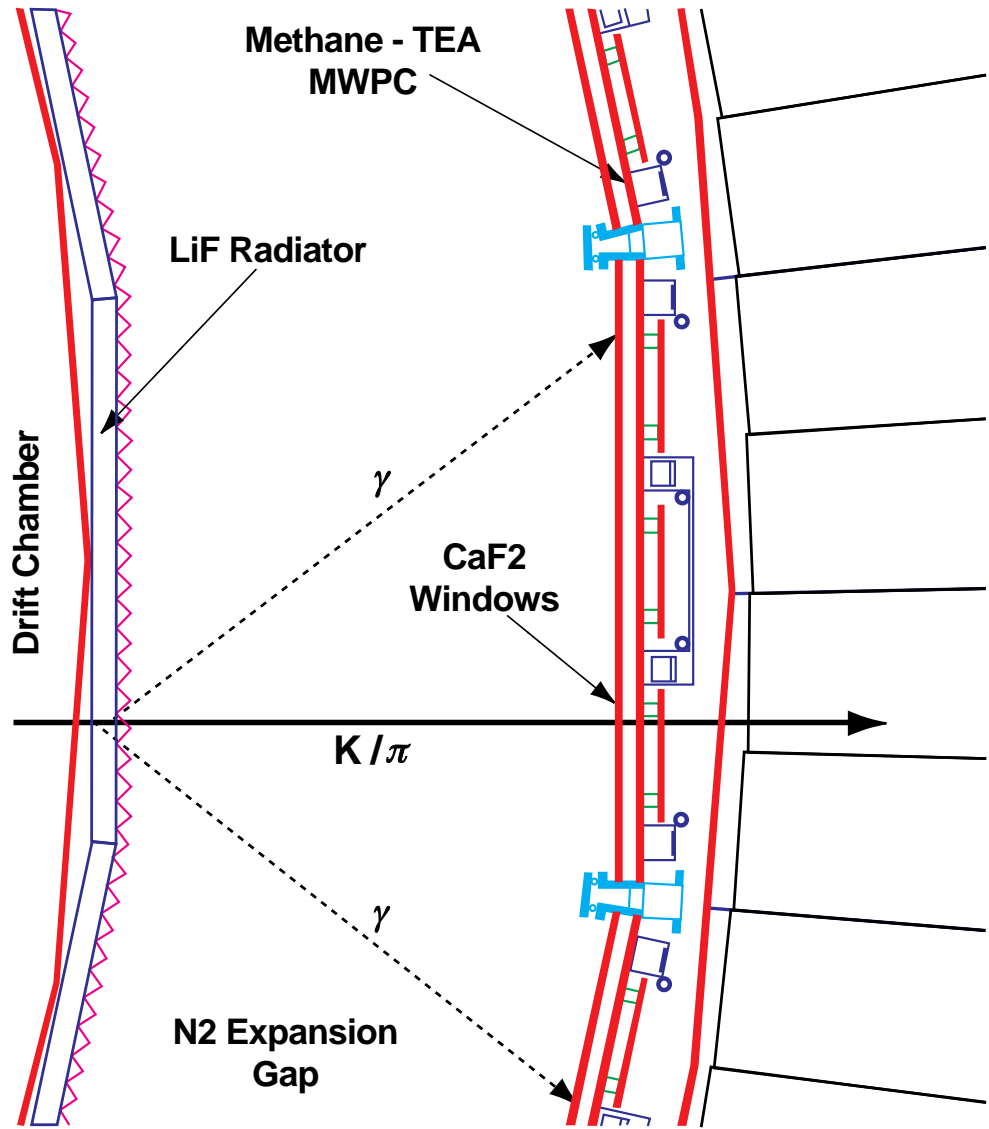}
\end{center}
\caption{\label{fig:rich} 
  {The $r-\phi$ cross-section view of the CLEO-c RICH detector.}}
\end{figure}

\begin{figure}[htbp]
\begin{center}
\subfigure{\includegraphics[width=.49\textwidth]{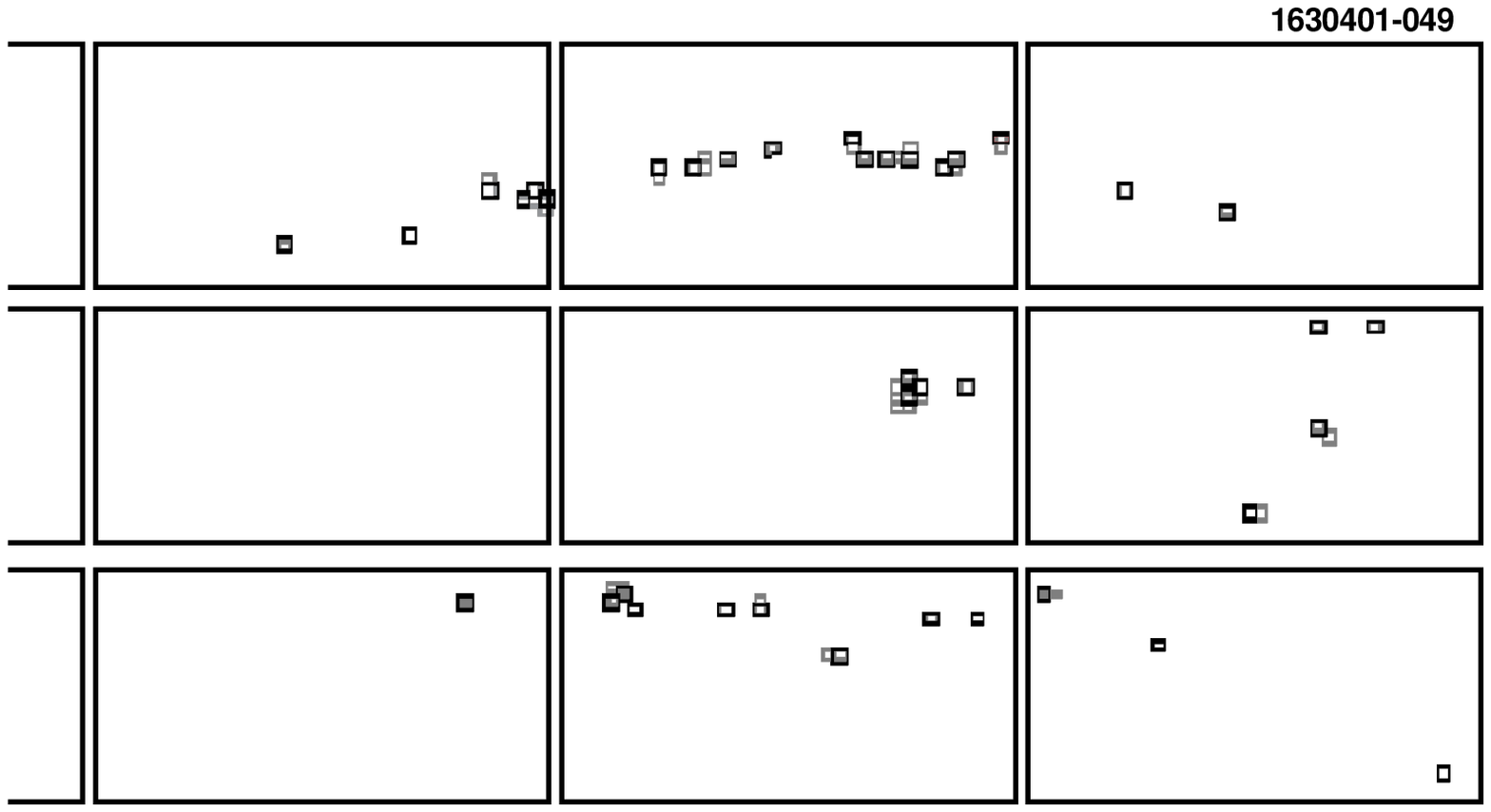}}
\subfigure{\includegraphics[width=.49\textwidth]{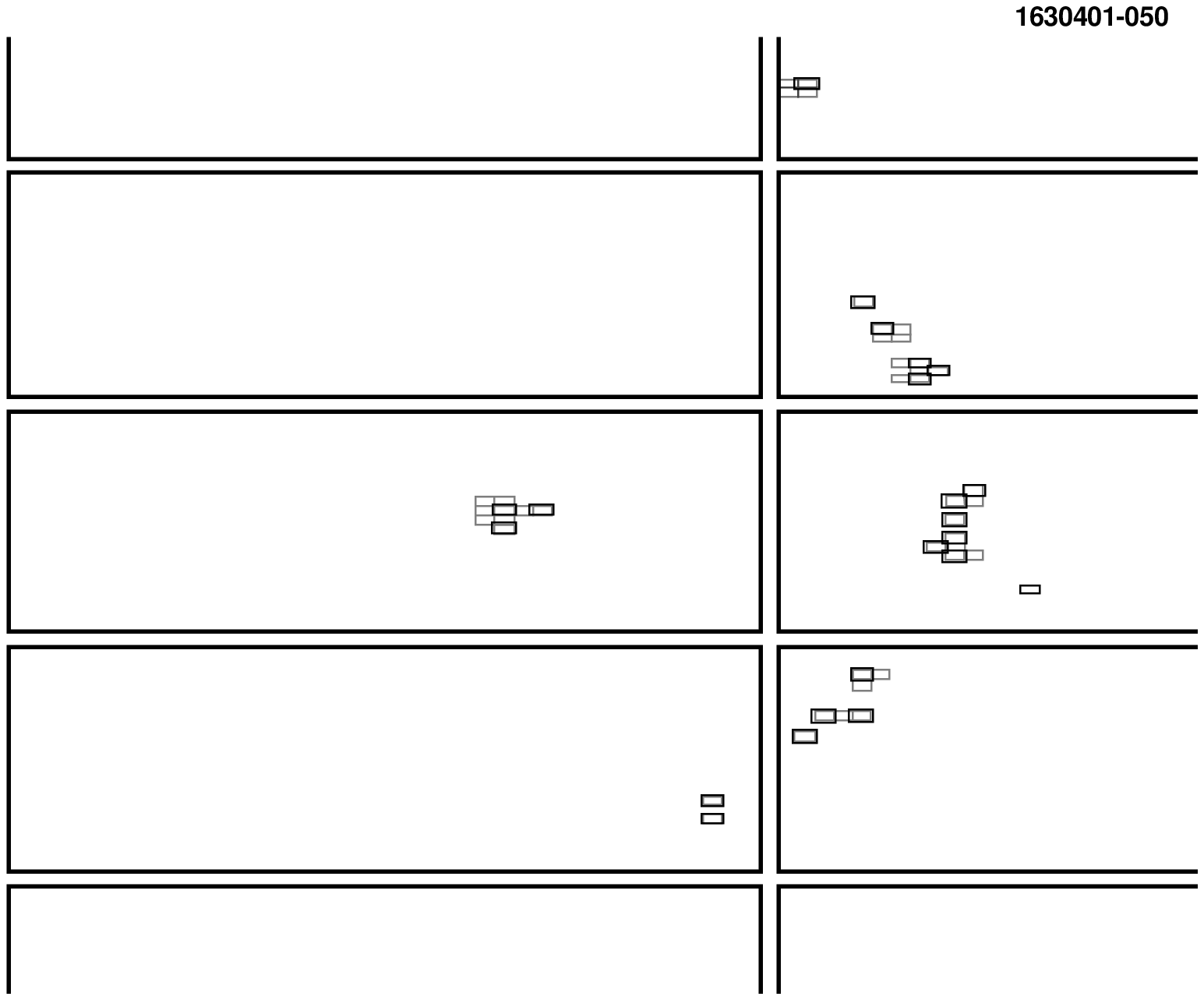}}
\end{center}
\caption[Cherenkov ring images produced by a charged track]
{\label{fig:richring} 
{Cherenkov ring images produced by a charged track crossing sawtooth 
(left) and flat (right) radiators. \ The rectangular grids are partial 
$24\times40$ pad arrays and the small squares represent the charge 
detected in pads. \ The hits at the center of the ring are due to the 
charged track crossing the wire chamber. \ The other hits are due to 
produced Cherenkov photons. \ The image of the flat radiator only shows 
half of the Cherenkov ring as the other half is trapped in the radiator 
by total internal reflection. The sawtooth radiator image is distorted 
by refraction.}}
\end{figure}

The RICH detector is shown in Figure~\ref{fig:rich} and  
covers 83\% of the $4\pi$ solid angle. \ The innermost parts of the RICH 
detector are the radiators, which are made from LiF and have an average 
thickness of 1.7~cm. \ For CLEO-c's magnetic field of 1~T, 
tracks with transverse momenta of 0.12~GeV/$c$ reach the 
RICH radiators. \ When a charged track passes through the radiators,
Cherenkov photons are produced. \ There are a total of 14 rows of 
radiators and the four central rows contain ``sawtooth'' radiators, with
triangular grooves on the surface, to overcome the total internal 
reflection for Cherenkov photons produced 
with nearly normal track incidence. \  The outer rows of radiators 
are flat. \ The Cherenkov photons produced in the radiators propagate 
outward through an expansion volume filled with pure nitrogen, which
is transparent to Cherenkov photons with the characteristic 
wavelength of about 150~nm. \ Propagation through the expansion volume 
allows the cone of radiation to be detected where its radius is large 
enough for accurate measurements. \ After expansion, the photons pass 
through CaF$_{2}$ windows and enter the multi-wire proportional 
chambers (MWPC). \ The 
chambers are filled with a gas mixture of methane (CH$_{4}$) and 
triethylamine (TEA, N(CH$_{2}$CH$_{3}$)$_{3}$) \cite{cbx01-20}, 
which converts Cherenkov photons in a narrow ultraviolet range of 
135-165~nm to photo-electrons. \ 
Gas multiplication of the primary ionization occurs as
described for the ZD and DR, although in the RICH the location of the photon
conversion is measured through charge induced on cathode pads. \ There 
are 230,400 pads in total. \ Sample Cherenkov ring images are shown in
Figure~\ref{fig:richring}.

\begin{figure}[htbp]
\begin{center}
\includegraphics[width=.9\textwidth]{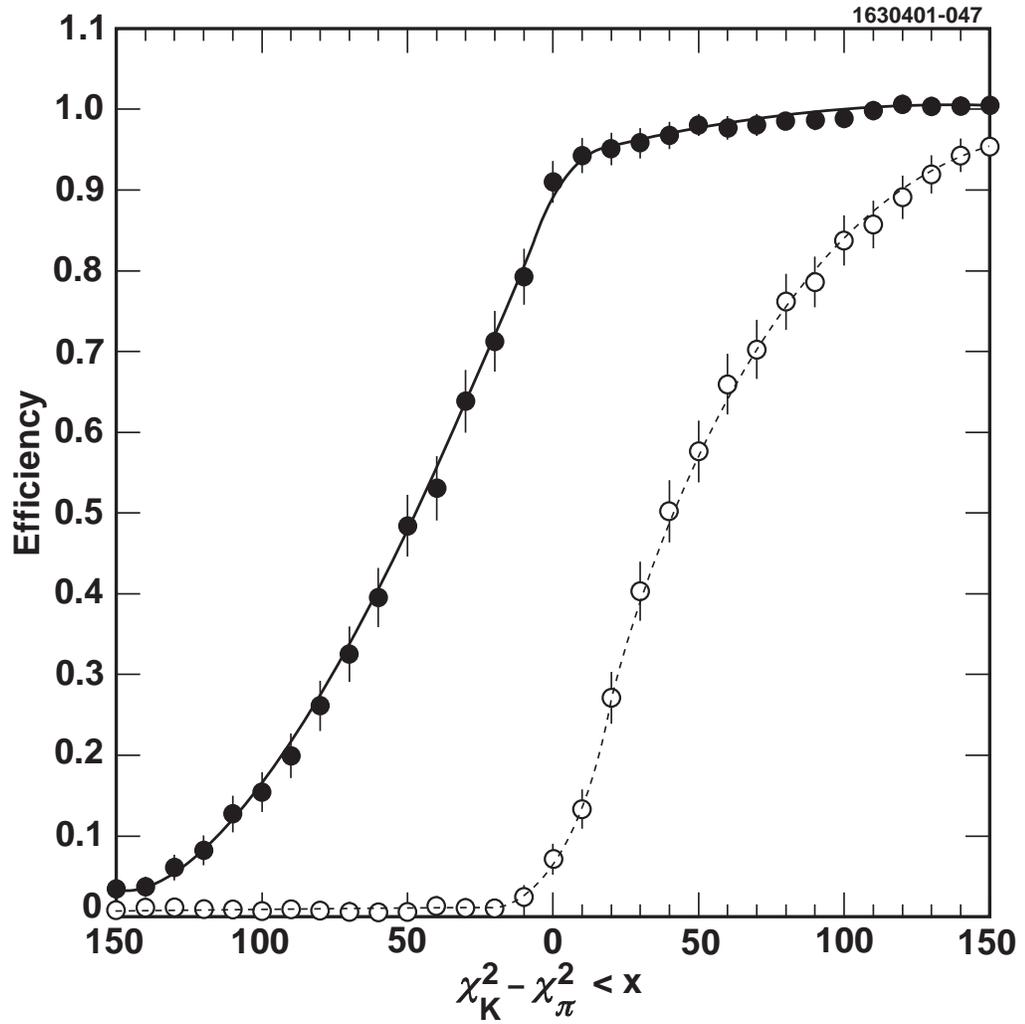}
\end{center}
\caption[Kaon efficiency and pion fake rate versus $\chi^{2}$ difference 
above kaon radiation threshold]
{\label{fig:richeffhimom}
{Kaon identification efficiency (filled circle) and pion fake rate (open 
circle) as a function of various cuts on $\chi^{2}$ difference between 
kaon and pion hypotheses. \ The momentum of tracks is between 0.7~GeV$/c$, 
just above kaon radiation threshold, and 2.7~GeV$/c$.}}
\end{figure}

\begin{figure}[htbp]
\begin{center}
\includegraphics[width=.9\textwidth]{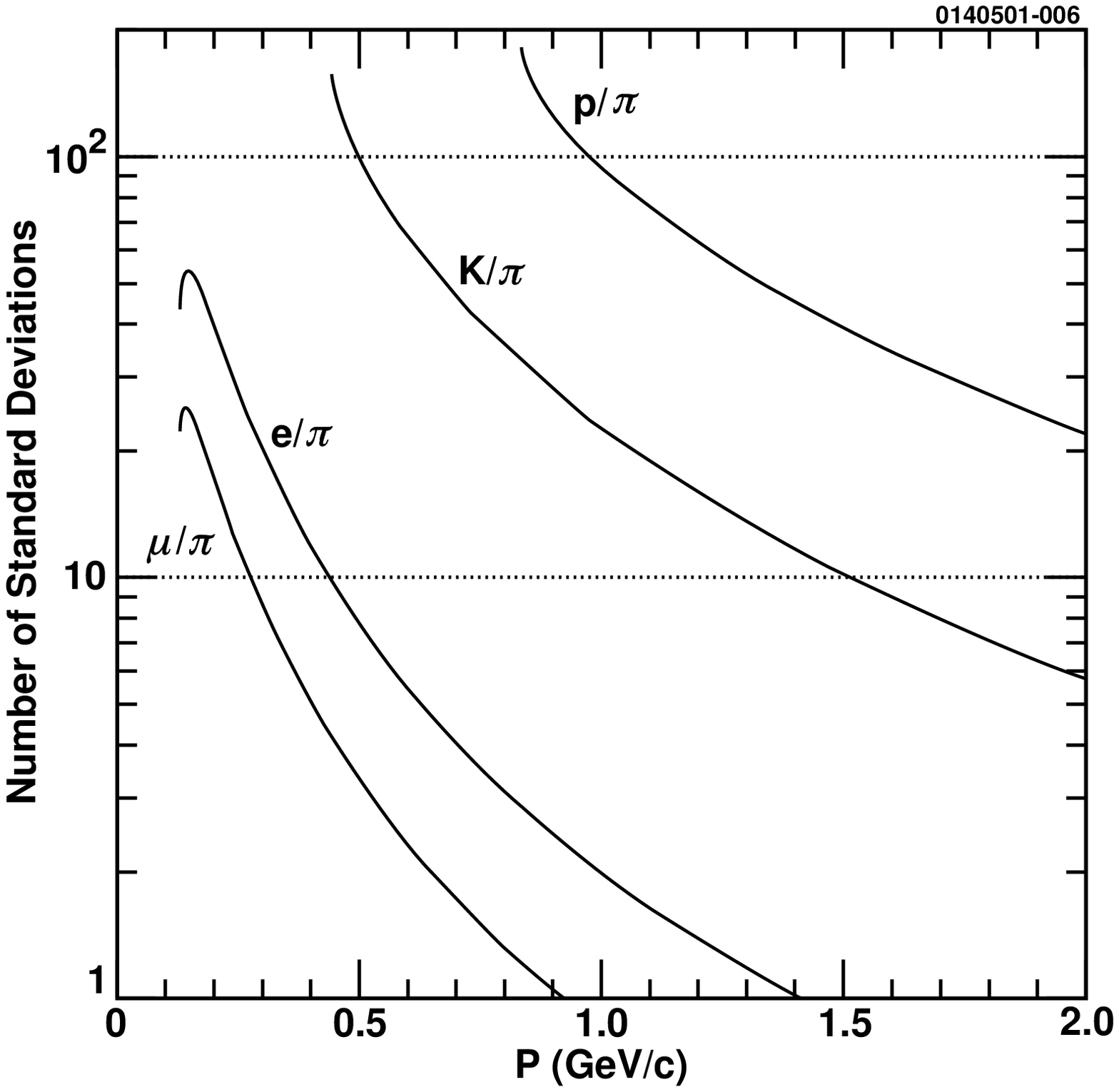}
\end{center}
\caption[Separation of particles in RICH]
{\label{fig:richsep}
{Separation of particles with different particle hypotheses in the RICH
detector as a function of momentum. \ All curves have cuts at minimum 
momentum because both particles are required to be above their 
radiation threshold, determined from the index of refraction of the
LiF radiator, $n=1.4$.}}
\end{figure}

To identify particles the radius of the 
Cherenkov cone is measured and combined with the expectations for a 
particular hypothesis to construct a 
likelihood ratio. \ For two different particle hypotheses, 
a $\chi^{2}$ difference variable for likelihood ratio is represented by
\begin{equation}
\chi_{i}^{2}-\chi_{j}^{2}=-2\ln L_{i}+2\ln L_{j},
\label{eq:richpid}
\end{equation}
where $i$, $j$ are different particle hypotheses and can be electrons,
muons, pions, kaons, or protons, and $L_{i}$, $L_{j}$ are their 
corresponding likelihoods. \ Figure~\ref{fig:richeffhimom} shows the 
measured fraction of kaons and pions as a function of the cut on 
$\chi_{K}^{2}-\chi_{\pi}^{2}$. \ When the cut is set at 
$\chi_{K}^{2}-\chi_{\pi}^{2}<0$, 92\% of kaons can be identified with 
a pion fake rate of 8\%. The measured momentum range of kaons or pions 
is from 700~MeV$/c$, which is just above kaon radiation threshold, to 
2.7~GeV$/c$. \ Figure~\ref{fig:richsep} shows the separation of the 
different particle hypotheses as a function of 
momentum. \ The curves are plotted with the momenta of particles
for both particle hypotheses to be above their respective thresholds, 
which is determined by the index of refraction of the LiF radiator, 
$n=1.4$.

\begin{figure}[htbp]
\begin{center}
\includegraphics[width=.9\textwidth]{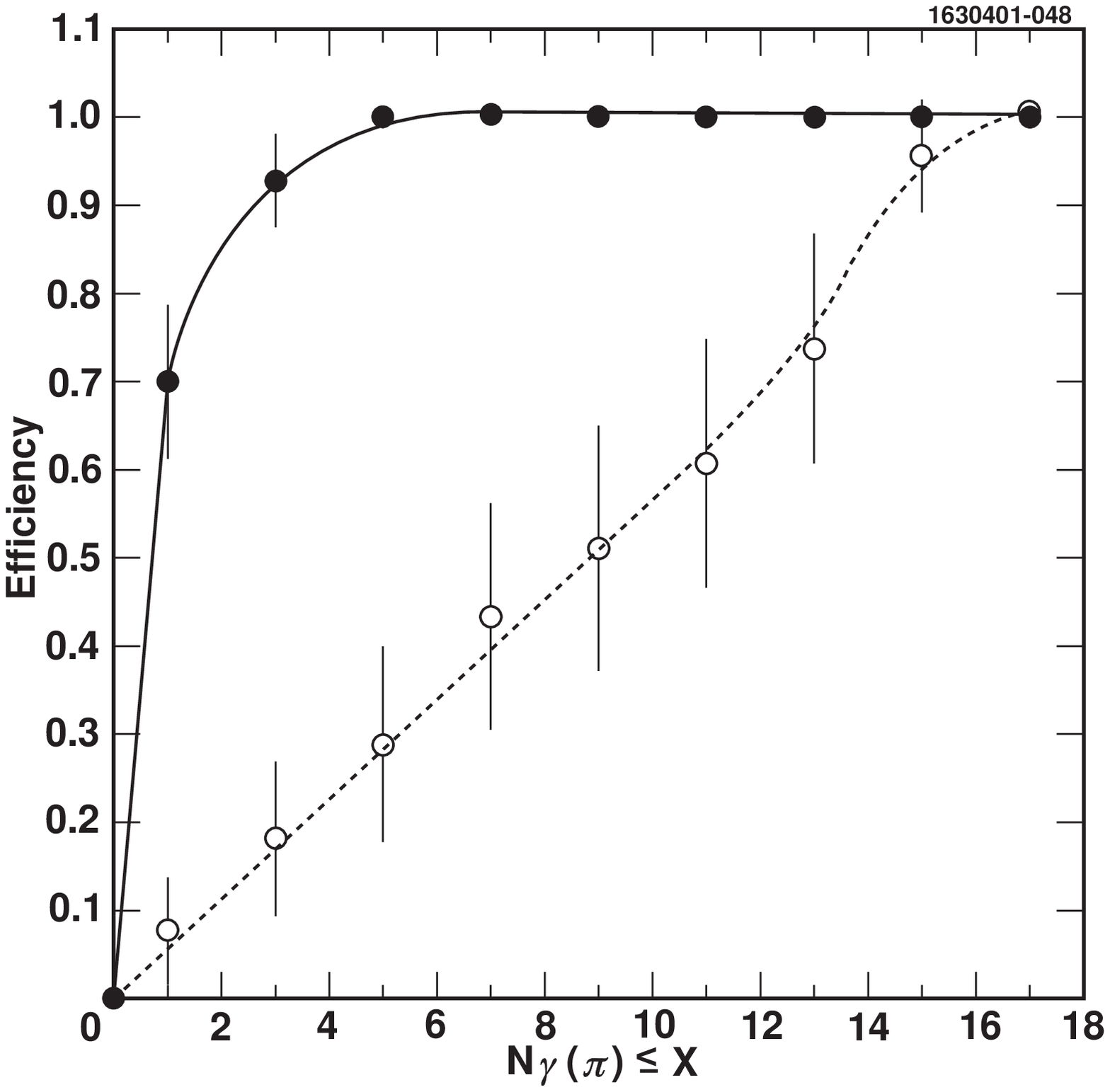}
\end{center}
\caption[Kaon efficiency and pion fake rate versus number of photons 
assigned to the pion hypothesis]
{\label{fig:richefflomom}
{Kaon identification efficiency (filled circles) and pion fake rate 
(open circles) as a function of the required number of photons 
assigned to the pion hypothesis. \ The momentum of tracks is less 
than 0.6~GeV$/c$. \  
The kaon radiation threshold is 0.44~GeV$/c$.}}
\end{figure}

For momenta below the threshold, RICH is used in a threshold mode and 
the number of photons assigned to the pion hypothesis is used instead 
of $\chi_{K}^{2}-\chi_{\pi}^{2}$ . \ The kaon efficiency and pion fake 
rate as a function of the number of photons assigned to the pion 
hypothesis are shown in Figure~\ref{fig:richefflomom}.

\subsection{Electromagnetic Calorimeter}
\label{sec:cc}

\begin{figure}[htbp]
\begin{center}
\includegraphics[width=.9\textwidth]{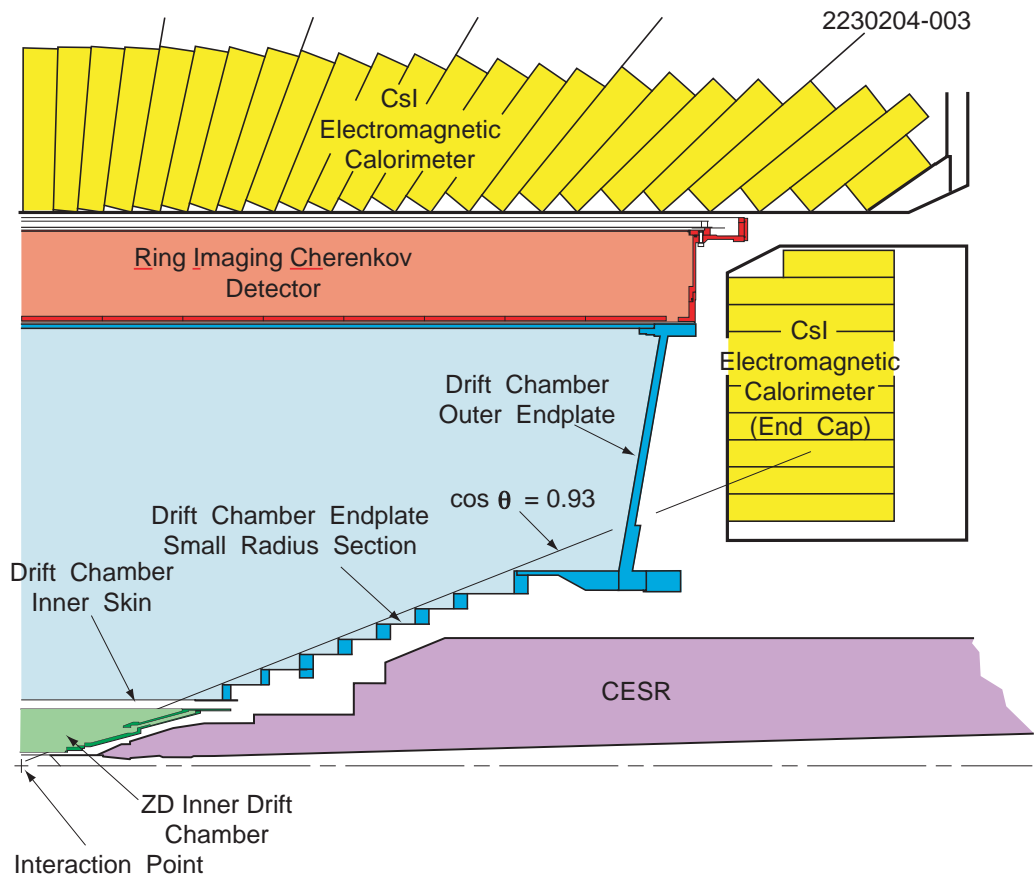}
\end{center}
\caption{\label{fig:ccxsec}{$r-z$ cross-section view of CLEO-c detector.}}
\end{figure}

\begin{figure}[htbp]
\begin{center}
\includegraphics[width=.9\textwidth]{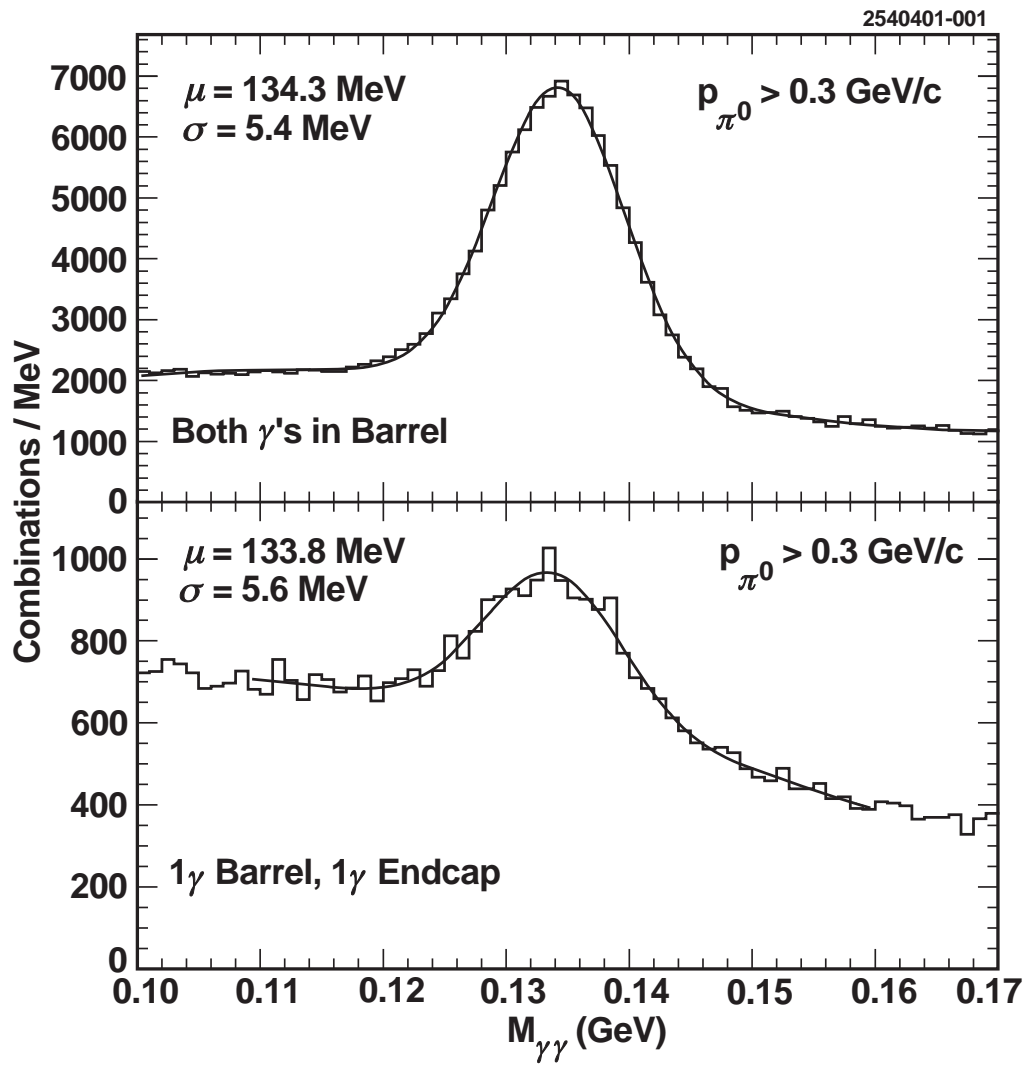}
\end{center}
\caption[$M_{\gamma\gamma}$ resolution for $\pi^{0}$ candidates]
{\label{fig:ccdiff}{$M_{\gamma\gamma}$ resolution for $\pi^{0}$ 
candidates in CLEO III $\Upsilon(4S)$ data.}}
\end{figure}

The electromagnetic crystal calorimeter (CC) is right outside of the RICH 
detector and inside of the superconducting solenoid, as shown in 
Figure~\ref{fig:cleoc} and \ref{fig:ccxsec}. \ It is used to measure 
the energies of electrons 
and photons, and covers approximately 93\% of the full $4 \pi$ 
solid angle. \ It consists of 7784 thallium-doped Cesium Iodide (CsI) 
scintillation crystals. \ There are 6144 crystals arranged in the barrel 
region, defined by $\left|\cos\theta\right|<0.80$, in a projective geometry 
pointed roughly toward the interaction point. \ The two endcap regions 
have 1640 crystals, covering $0.85<\left|\cos\theta\right|<0.93$. \ Each 
crystal measures approximately $5~{\rm cm} \times 5~{\rm cm}$ square by 
30~cm long. \ 
The energy resolution of the barrel and endcap regions are slightly 
different, as illustrated in Figure~\ref{fig:ccdiff}. \ The energy 
resolution of the transition region, between the barrel and endcap 
($0.85<\left|\cos\theta\right|<0.93$) is degraded by the substantially 
larger amount of material in front of the CC in this region. \ It is 
excluded in most analyses.

The material from which the crystals were made, CsI, has high density 
(4.53~g/cm$^3$) and short radiation length (8.93~g/cm$^3$), so it can stop 
photons and electrons effectively. \ 
In the calorimeter, electrons 
produce photons by Bremsstrahlung and photons undergo $e^+ e^-$ production 
near high-$Z$ nuclei. \ This process produces a cascade called an 
electromagnetic shower. \ 
The final low-energy electrons and positrons that result from this shower 
process excite atoms in the crystals which then ``scintillate,'' emitting 
visible light (560~nm) as they return to the ground state. \ The crystals 
are transparent for these visible light photons and they are read out 
using four 1~cm~$\times$~1~cm photodiodes mounted on the backs of the 
crystals. \ Muons and noninteracting hadrons are called minimum ionizing 
particles (MIPS) as they only deposit a small fraction of their energy by 
ionization. 

\begin{figure}[htbp]
\begin{center}
\includegraphics[width=.9\textwidth]{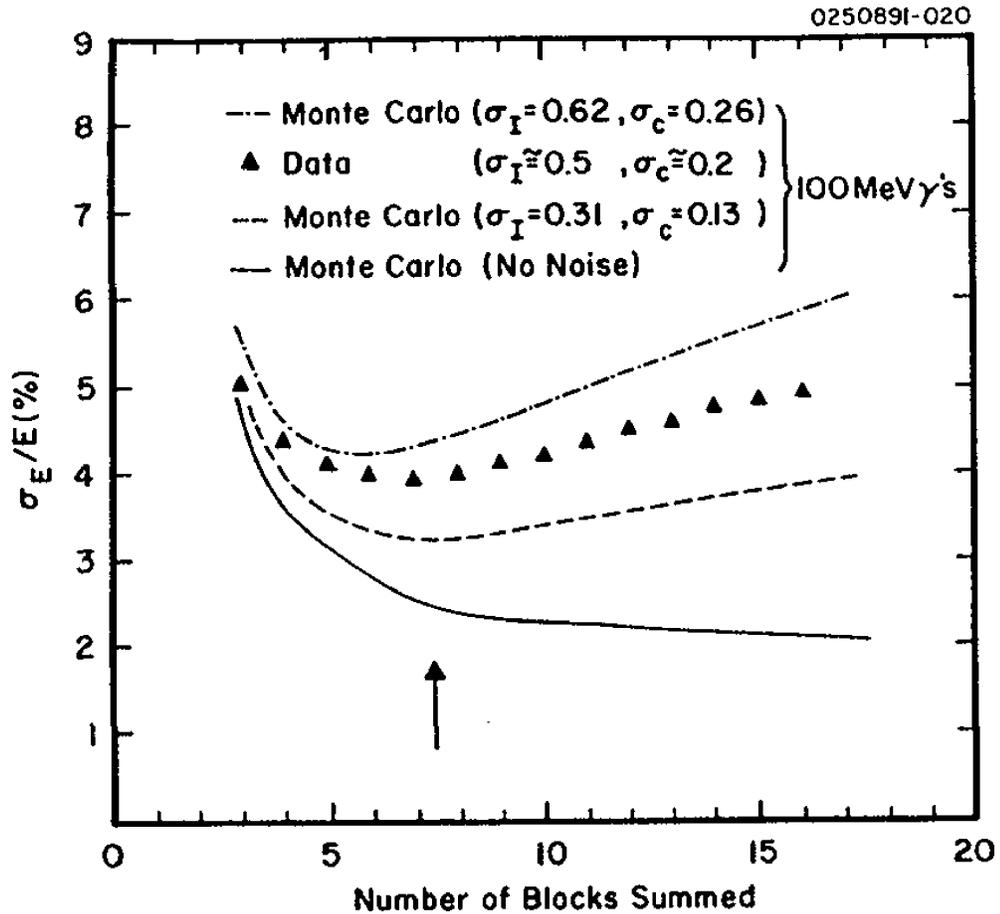}
\end{center}
\caption[Energy resolution versus number of crystals]
{\label{fig:ccnxtals}{Energy resolution as a function of the number of 
crystals (blocks) used for shower reconstruction in CLEO II. \ 
The same calorimeter has been used from CLEO II to CLEO-c. \ The 
smooth lines represent Monte Carlo simulations of 100~MeV photons with 
different noise levels. \ The points are the data from 100~MeV transition 
photons of the decay $\Upsilon(3S)\rightarrow\gamma\chi_{bJ}(2P)$, and the 
arrow indicates the number of crystals used for the 100~MeV photon 
reconstruction.}}
\end{figure}

The shower reconstruction process starts with calculating the energy 
deposited in a crystal from the amount of light detected. \ As the energy 
of a shower is generally deposited into multiple crystals, a cluster of 
several adjacent and near-adjacent crystals are considered together. \ 
Since including crystals with very small energy deposits can degrade the 
measurement by bringing in excessive noise, the number of blocks used has 
been optimized for the best precision. \ 
The uncertainty in the energy of the shower is calculated based on 
energy-weighted average of the blocks in the sum of the cluster, and an 
example optimization is shown in Figure~\ref{fig:ccnxtals}. \ The number 
of crystals used is logarithmic in energy and ranges from 4 at 25~MeV to 
17 at 4~GeV \cite{Kubota:1992ww}.

The photon energy resolution of CLEO-c calorimeter is about 4.0\% at 
0.1~GeV and 2.2\% at 1~GeV and the angular resolution is approximately 
10~mrad. \ The performance of the endcap region is not as good as that of 
barrel region because of the presence of the aluminum DR endplates and 
electronics in front of the crystals. \ The excellent energy and angular 
resolution of the calorimeter is critical for the reconstruction of the 
decay modes that include $\pi^0\to\gamma\gamma$ or $\eta\to\gamma\gamma$ 
decays, as well as the low-energy transition photons in $\psi(2S)$ 
radiative decays.

\subsection{Superconducting Magnet}

Right outside of the calorimeter is the superconducting solenoid, which has 
an inner diameter of 3~m and a length of 3.5~m. \ It provides a uniform 
(to $\pm2\%$) 1-T magnetic field parallel to the beam line over the full 
volume of the tracking system, the RICH detector and the calorimeter. \
The solenoid is cooled down to superconducting temperatures by liquid 
helium. \ Three layers of 36~cm thick iron flux return for the magnet 
also serve as part of the absorber for the muon identification system. \ 
The magnetic field used to be set at 1.5~T for CLEO III running at a 
center-of-mass energy of 10.5~GeV. \ In CLEO-c the center of mass energy 
is much lower and the momentum of charged particles is much smaller. \ 
From Equation \ref{eq:trkrec}, one can see that in order to keep the same 
radius of the curvature for the trajectory of particle, if the momentum of 
the particle decreases, the magnetic field has to be reduced to allow the 
tracks to reach the RICH detector and calorimeter.

\subsection{Muon Detector}

\begin{figure}[htbp]
\begin{center}
\includegraphics[width=.9\textwidth]{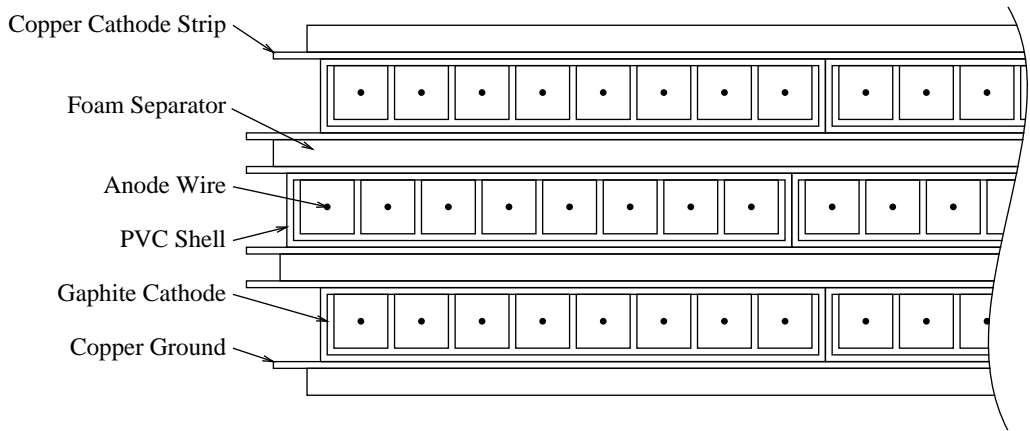}
\end{center}
\caption{\label{fig:mu}{A cross-section view of a muon chamber super-layer.}}
\end{figure}

Muons are distinguished from other charged particles by their low 
probability of interaction as they pass through material. \ The Muon
Detector (MU) is located at the outside of CLEO-c beyond a roughly 1-m 
steel absorber. \ The absorber absorbs other particles that pass through 
the calorimeter, so the hits in the MU chambers are almost exclusively 
from muons. \ The muon chambers (MU) operate in the similar way the 
tracking chambers do. \ A cross-section view of the muon super-layer 
is shown in Figure~\ref{fig:mu}.  
The muon detector uses gas-filled tracking chambers in between 36~cm iron 
absorbers surrounding the detector. \ It consists of three layers of 
proportional wire chambers of 4~m long, 8.3~cm wide, and 1.0~cm tall and
oriented with its long axis parallel to the z-axis.

The muon detector provides the information in terms of interaction length. \
The interaction lengths (also known as the absorption length) is defined
as the mean distance over which a particle travels before scattering
inelastically from a nucleus. \ The barrel super layers are located
at depths of 3, 5, and 7 interaction lengths, while the only super-layer
in each endcap is located at a depth of about 7 interaction lengths.

The MU is designed to detect muons with momentum of 0.8~Gev$/c$ or 
higher. \ It is not as useful for CLEO-c as it was for the previous 
experiment because of the much lower center of mass energies.

\subsection{Triggers and Data Acquisition System}
\label{sec:daq}

\begin{figure}[htbp]
\begin{center}
\includegraphics[width=.9\textwidth]{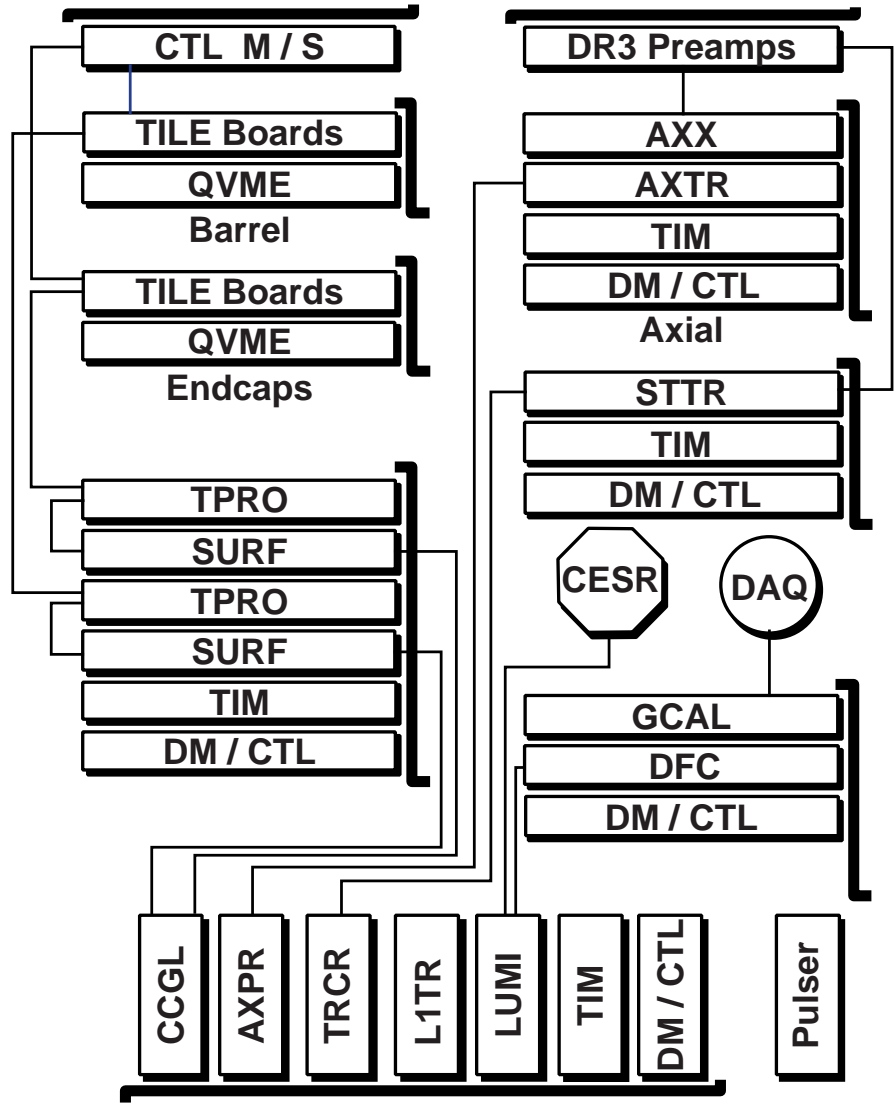}
\end{center}
\caption{\label{fig:trigger}{Overview of the trigger system.}}
\end{figure}

The trigger and data acquisition (DAQ) systems of CLEO-c are used to 
collect and process the data from all components of the detector, and save 
the data to mass storage devices. \ During CLEO-c runs, 
interactions happened at the frequency in the order of 1~MHz. \ The 
recording capability of the data acquisition system does not allow every 
collision to be recorded, and most of these interactions are of no 
interest. \ The actual rate of interesting physics events 
is only on the order of 1~Hz. \ The trigger system, as illustrated in 
Figure~\ref{fig:trigger}, is used to decide whether an event should be 
recorded based on the complicated information from many components of the 
detector.

\subsubsection{Trigger System}

The trigger system is operated in several levels
\cite{yellowbook, Hans:2001, Gollin:2001, Selen:2001}.
The first level of trigger, called level-zero (L0), is the hardware 
trigger. \ The tracking trigger, including axial and stereo tracking, 
and calorimeter trigger, including analog and digital calorimeter trigger, 
both belong to L0. \ Data from the drift chamber and calorimeter 
are received and processed in separate VME crates using the appropriate 
circuit boards to yield initial trigger parameters, such as track count 
and topology in the drift chamber, and shower count and topology in the 
electromagnetic calorimeter. \ The next level trigger is the global 
level-one (L1) trigger, which correlates the information from both tracking 
and calorimeter systems. \ The global trigger circuit generates an L1Pass 
strobe every time a valid trigger condition is satisfied. \ The generated 
L1Pass signals are conditionally transferred to the gating and calibration 
modules and then distributed to the data acquisition system. \ The L1 Decision 
and Data Flow Control system makes a trigger decision every 42~ns based on 
the information from the tracking system and calorimeter. \ Programmable 
trigger decision boards (L1TR), which can be configured for a wide range of 
topologies, are used to look at this information and respond to various 
trigger conditions. \ It takes the tracking system about 2~$\mu$s and 
calorimetry over 2.5~$\mu$s to deliver information for the decision. \ 
There are currently about eight trigger lines, which definitions and 
relative rates are shown in Table \ref{tab:triggerlines}. \ The total 
trigger rate at a luminosity of 
$\mathcal{L} = 1 \times 10^{33} ~ {\rm cm}^{-2}{\rm s}^{-1}$ 
is between 80 and 90~Hz \cite{yellowbook}. 

\begin{table}[tbph]
\begin{center}
\begin{tabular}{l|l|c}
\hline
Name             & Definition              & Relative Rate   \\ \hline
Hadronic         & number of axial track $>1$ \&              & 0.41 \\
                 & number of showers below a min. energy $>0$ & \\
$\mu$-pair       & back-to-back stereo tracks                 & 1.40 \\
Barrel Bhabha    & back-to-back high showers in CB            & 1.0  \\
Endcap Bhabha    & back-to-back high showers in CE            & 0.23 \\
electron+track   & number of axial track $>1$ \&              & 1.48 \\
                 & number of showers above a min. energy $>0$ & \\
tau/radiative    & number of stereo track $>1$ \&             & 2    \\
                 & number of showers below a min. energy $>0$ & \\
Two Track        & number of axial track $>1$                 & 0.69 \\
Random           & random 1 kHz source                        & 1    \\ 
\hline
\end{tabular}
\end{center}
\caption[Definition of CLEO III trigger lines.]
{\label{tab:triggerlines}Definition of CLEO III trigger lines. \ The rate is 
relative to barrel Bhabhas line.}
\end{table}

The final trigger stage is called level-three (L3). \ It is a software
trigger, which is implemented on a fast workstation. \ It 
takes an event passed by the lower level triggers and either rejects or 
accepts it based on more sophisticated reconstruction algorithms.\ 
Events such as those from cosmic rays or interactions of the beam with 
residual gas molecules in the beam pipe are rejected by L3 before they are 
permanently recorded.

\subsubsection{Data Acquisition System}

The data acquisition system consists of two equally important parts 
\cite{yellowbook}. \ Data quality and the detector performance are 
monitored by the slow control system, and the data collection system is 
responsible for the data transfer from the front end electronics to the 
mass storage device. \ The slow control and monitoring system is required 
to provide sufficient flexibility for online data quality and detector 
performance monitoring. \ It also includes run control, alarm and message 
handling, and calibration constants management. \ For the data collection 
system the knowledge of cross sections of the interesting physics 
processes and expected CESR luminosity are needed to establish the
performance requirement. \ The key parameters for the DAQ system are the
expected trigger rate, the acceptable detector dead-time and the average
event size. \ In the current configuration the data acquisition system 
can operate at trigger rates up to 150~Hz, significantly below the design 
rate of 1000~Hz. \ A read-out time, defined as the time between the 
trigger signal and the end of the digitization process in the front-end 
electronic modules, causes dead-time and therefore should be kept minimal. \
The detector dead-time is less than 3\% when the average read-out time is
20-30~$\mu$s at a trigger rate of 1000~Hz.

\begin{figure}[htbp]
\begin{center}
\includegraphics[width=.9\textwidth]{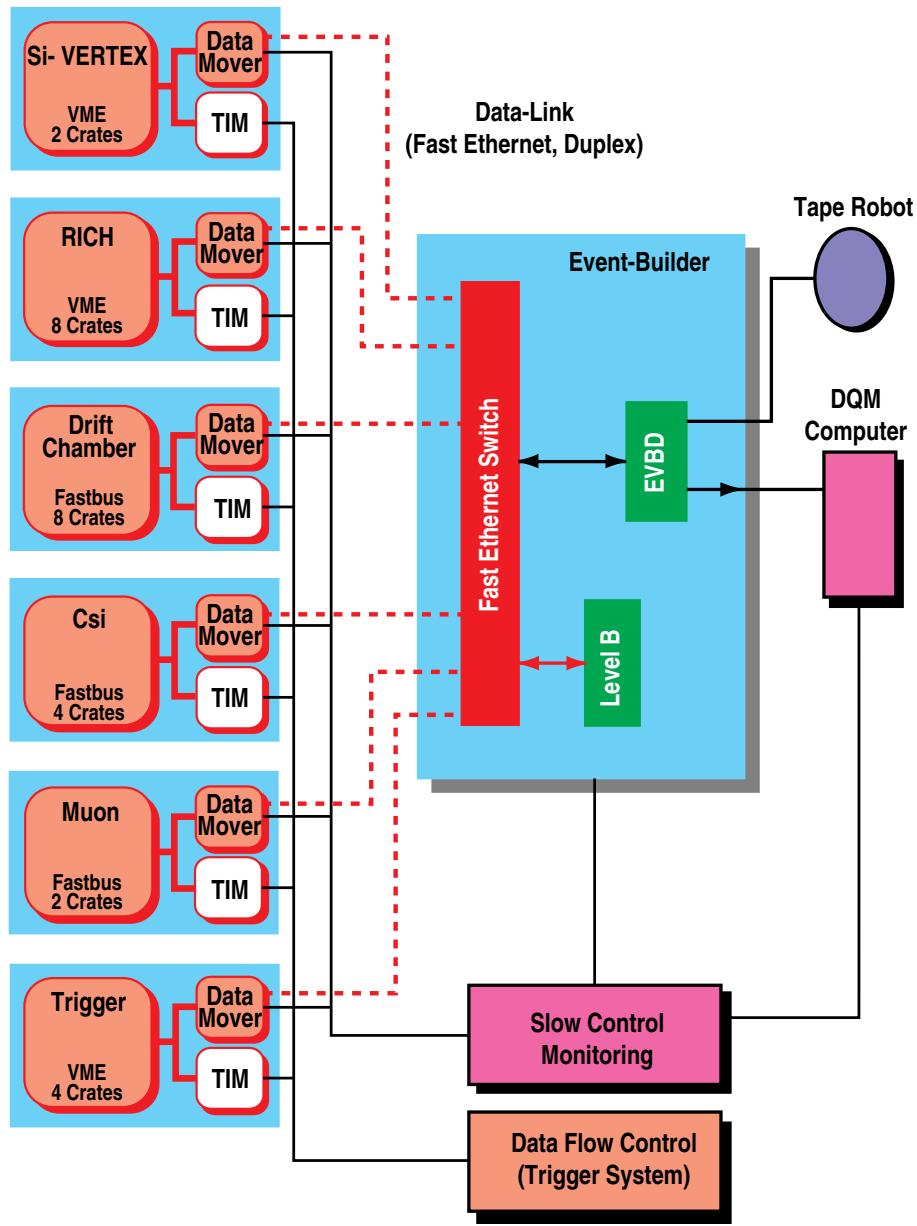}
\end{center}
\caption[Data Acquisition System.]
{\label{fig:daq}{Architecture of the CLEO III Data Acquisition System. \ 
The only change from CLEO III to CLEO-c in this diagram is to replace 
the silicon vertex detector by the ZD.}}
\end{figure}

The structure of the data acquisition system is shown in 
Figure~\ref{fig:daq}. \ 
For each event that is accepted by the trigger, all detector
channels need to be digitized. \  Front-end data conversion is performed 
in parallel and local buffers on each data-board hold the data for later 
asynchronous readout by the data acquisition system. \ Data sparsification 
(the suppression of channels without valid hits) 
is performed directly on the data-boards. The Data-Mover, a dedicated 
module in each front-end crate, assures transfer times below 500~$\mu$s 
and provides a second buffer level. \ 
About 30 front-end crates are needed for the detector. \ 
The data is sent to L3 and the L3 decision is sent back to the front-end 
crates. \ The fragments of the events that are accepted by L3 are 
transmitted from the crates to the Event-Builder by the Data-Mover. \ 
Completely reconstructed events are transferred to the mass storage device and 
a fraction of the data is analyzed online by a monitor program to quickly 
discover problems of the detector and to ensure the quality of the stored 
data.

\section{Event Reconstruction}
\label{sec:reconstruction}

Before the data can be used for physics analyses, the raw data collected 
needs to be processed by software packages to reconstruct tracks and 
showers. \ Raw data collected from different components of the detector 
are processed through {\tt Pass1} and {\tt Pass2}, which consist 
of collections of processors and routines for event reconstruction. \ 
The detailed processes of the track and shower reconstruction have been 
described in Sections~\ref{sec:trackrecon} and \ref{sec:cc}.

The function of {\tt Pass1} is to perform fast classification of the raw
data events. It runs online to provide real-time feedback about the 
performance of the detector while taking data for data quality control. \ 
Only one-tenth of the collected events are processed with fast 
reconstruction. \ {\tt Pass2} processes all events offline and it provides 
more detailed and accurate event reconstruction. \ After {\tt Pass2} the 
data are ready for the physics analyses and one can access the data with 
personal computer program written with a computer language like C++.

\section{Monte Carlo Simulation}
\label{sec:cleog}

Monte Carlo (MC) techniques allow the simulation of the data based on
the best knowledge of physics properties, detector configuration and 
detector performance. \ The simulated data are used to 
determine detection efficiencies, estimate backgrounds, or determine  
background and signal shapes for specific physical processes studied, 
including in this dissertation.

The CLEO-c MC production \cite{Ryd:evtgen, Lange:2001} starts with the 
generation of a list of particles. \ In this stage, for example, $\psi(2S)$ 
events are generated with the software package EvtGen \cite{Ryd:evtgen} 
and final state radiation (FSR) is simulated with \texttt{PHOTOS} 
\cite{Barberio:2001}. \ EvtGen produces Monte Carlo events with a list of 
particles (and their four-momenta) created in an $e^+ e^-$ interaction 
at a desired energy. \ In addition to specific states like  $\psi(2S)$, 
it simulates events for $e^+ e^-\to q\bar{q}$ where $q$ can be an up, 
down, strange, or charm quark. \ The quark fragmentation simulation uses 
an interface to the JetSet software package 
\cite{Sjostrand:1985ys, Sjostrand:1986hx} that models 
quark and gluon hadronization. \ The subsequent decays of unstable 
particles are simulated based on the amplitudes, which are used to 
calculate the probability for any decay tree from a decay table that is 
updated periodically to reflect our best knowledge.  

The CLEOg is the simulation program based on the GEANT3 \cite{geant}
package developed at CERN. \ CLEOg simulates the passage of particles 
through the CLEO-c detector material and detector elements. \ At each 
step, a random number is generated to determine 
if the particle interacts with any material. \ If it does, more random 
numbers are generated to determine its effect on the particle and still 
more to calculate the signal that would result in the detector element. \ 
Using calibration data or samples of data events taken with a random trigger, 
the detector noise and the beam-related backgrounds are also simulated 
when no colliding physics events were present. \ Finally, the simulated 
events are saved in the same format as raw data so they are processed 
with the same reconstruction program {\tt Pass2}.

\chapter{Measurement of $\psi(2S) \to \gamma \eta_{c}(2S)$}
\label{chap:analysis_etac2s}

In this chapter, the analysis of the process of 
$\psi(2S) \to \gamma \eta_{c}(2S)$ is described in detail. \ 
As the channel had never
been accurately measured before, I start by introducing the
strategy and procedures of this analysis, followed by describing the 
samples used and the event selection criteria. \ 
The study of $\psi(2S) \to \gamma \chi_{c2}$ decays is presented 
to justify the analysis procedures 
of $\psi(2S) \to \gamma \eta_{c}(2S)$ decays in Section~\ref{sec:chic2}. \
Finally the results of $\psi(2S) \to \gamma \eta_{c}(2S)$ decays and 
systematic uncertainty studies are presented.

\section{Strategy and Procedures}

The analysis was initiated from the CLEO exclusive study of 
$\psi(2S) \to \gamma \eta_{c}(1S)$, which was included in the 
search for $\psi(2S) \to \pi^{0} h_{c}$ \cite{cbx05-3, rosner:102003}. \
We reconstruct $\psi(2S) \to \gamma \eta_{c}(2S)$ candidates and extract 
information from the energy of the transition photons.

Various modes have been searched. \ We constructed an extensive list 
of possible decay modes following the model of 
a CLEO-c 
analysis of $J/\psi \to \gamma \eta_{c}(1S)$ \cite{R.Mitchell:1870}. \ 
We considered  modes with relatively high $\eta_{c}(1S)$ yields, and 
selected the following final states: 
$4\pi$, $6\pi$, $KK\pi\pi$, $KK\pi^{0}$, $K_{S}K\pi$, 
$\pi\pi\eta$ with $\eta\to\gamma\gamma$ or $\pi\pi\pi^{0}$,
$\pi\pi\eta^{\prime}$ with $\eta^{\prime}\to\pi\pi\eta(\gamma\gamma)$,
$KK\eta$ with $\eta\to\gamma\gamma$ or $\pi\pi\pi^{0}$, 
$KK\pi\pi\pi^{0}$, $KK4\pi$, and $K_{S}K3\pi$.

The major challenge of this analysis is to 
detect the low energy transition photon while suppressing 
backgrounds. \ We reconstruct the events exclusively by considering 
all combinations of charged tracks and photons, and remove the possible 
candidates of other decays by applying various mode dependent
cuts, as described in Section \ref{sec:eventselection}. \ 
Signal-squared over signal-plus-background [$S^2/(S+B)$] studies  
were used to optimize selection criteria for the 4-C kinematic fit 
$\chi^2/{\rm dof}$ and the distance to the nearest track for the transition photon 
and the angle between the transition photon and the closest track. \

Some other modes have been considered but were not included 
in the final analysis.  The modes with large $\psi(2S)$ decay branching 
fractions are expected to have large backgrounds. \ These modes include 
$4\pi2\pi^0$ and $K_{S}K_{S}\pi\pi$. \ In addition to these two modes, 
the following modes are not considered because of high background level 
based on studies of 
generic $\psi(2S)$ Monte Carlo (MC):
$\pi\pi\eta^{\prime}, \eta^{\prime}\to\gamma\rho^{0}(\pi\pi)$,
$2\pi2\pi^0$. 
The modes $\pi\pi\eta^{\prime}, 
\eta^{\prime}\to\pi\pi\eta(\pi\pi\pi^{0})$, $4K$, $K_{S}K\pi\pi^0$
are not considered due to 
poor signal-to-background ratio based on 
small expected yields from extrapolating the partial widths from the 
$\eta_{c}(1S)$ to the $\eta_{c}(2S)$. \ 
Modes $KK4\pi\pi^{0}$ and $K_{S}K3\pi\pi^{0}$ are excluded 
because signal MC studies show they have low efficiency. \
When optimizing the cuts for the considered modes, the branching fraction 
of each $\eta_{c}(2S)$ mode was arbitrarily assumed to be 1\% because there are 
no previous measurements that can provide reliable estimates.

\section{Samples}

\subsection{Data}

In order to search for $\psi(2S) \to \gamma \eta_{c}(2S)$, CLEO-c 
Datasets 32 and 42 were used. \ 
Dataset 32 has 1.44 million $\psi(2S)$ decays and Dataset 42 has 24.45
million, giving a total for the CLEO-c data sample of 25.89 million 
$\psi(2S)$ decays \cite{cbx07-4}. \ 
For our final processing we selected a recent update of the CLEO-c software
that allowed us to use updated endcap calorimeter (CC) calibrations and a
new CC energy-correction procedure (CCFIX) \cite{ccfix}.

\subsection{Generic Monte Carlo}

The 5 times luminosity generic $\psi(2S)$ 
and the 5 times luminosity continuum MC samples, 
both of which were generated by the Minnesota MC farm, were used
for cut optimization and background suppression. \ The full 10 times generic 
$\psi(2S)$ and continuum MC samples were used for fitting the measured 
photon energy 
distribution in the $\eta_c(2S)$ signal region. \ The new 5 times luminosity 
generic $\psi(2S)$ sample is found to be consistent with that of the original 
5 times luminosity sample for this study.

\subsection{Signal Monte Carlo}
\label{subsec:signalmc}

The signal MC sample consists of one million 
events, distributed over 15 modes, including the 13 modes that were 
used in this study. (We count the modes with an $\eta$ decay as two separate 
modes, although they are combined for final measurements.) \ 
The sample was generated to replicate the full CLEO-c 
$\psi(2S)$ data sample. \ The $\eta_c(2S)$ mass ($3638\pm4~{\rm MeV}$) 
and width ($14\pm7~{\rm MeV}$) were taken from the PDG \cite{PDBook2006}. \ 
This MC sample 
was generated using the appropriate angular distribution for a vector to 
vector-pseudoscalar decay, i.e., $P(\cos\theta) = 1 + \cos^2\theta$, 
where $\theta$ is the angle between the 
M1 transition photon and the positron beam. \ The angular distributions for the 
$\eta_c(2S)$ decays were thrown according to phase space. \ The minimum 
hadronic mass of the $\eta_c(2S)$ 
was decreased from its 
default value of 20~MeV below the nominal mass of $\eta_c(2S)$ (3638~MeV) to 
210~MeV below in order to appropriately model the tails of the 
$\eta_c(2S)$ decay.  

A signal MC sample to study the $\chi_{c2}$ decaying to the same hadronic 
final states 
was also generated. \ The study of $\chi_{c2}$ decays was used to validate the 
analysis procedure.  
A one million event sample, distributed over the same modes, was generated. \  
This $\chi_{c2}$  MC sample was generated using the appropriate angular 
distribution 
for a vector to axial vector-pseudoscalar decay, i.e., 
$P(\cos\theta) = 1 + (1/13) \cdot \cos^2\theta$ \cite{PhysRevD.13.1195}. \ 
As with the $\eta_c(2S)$ signal MC sample, 
this sample was generated to replicate the full CLEO-c $\psi(2S)$ data 
sample. \ 
The $\chi_{c2}$ mass ($3556.20~{\rm MeV}$) and width ($2.05~{\rm MeV}$) \cite{PDBook2006} 
were used. \ 
The minimum hadronic mass of the $\chi_{c2}$ 
was decreased from its default value of 6.0~MeV below the nominal mass
of $\chi_{c2}$ ($3556.20~{\rm MeV}$) to 41.6~MeV below in order to 
appropriately model the tails of the $\chi_{c2}$ decay.

The numbers of generated events for both $\eta_c(2S)$ and $\chi_{c2}$ 
decays are listed in Table~\ref{table:etac2ssignalmc}.
\begin{table}[htbp]
\caption[Number of events generated events in signal MC samples]
{\label{table:etac2ssignalmc}Number of generated signal MC events. The 
resonance parameters used to generate these samples are described in the text.}
\begin{center}
\begin{tabular}{|l|c|c|}
  \hline
  Mode & $N_{\eta_{c}(2S)}$ & $N_{\chi_{c2}}$ \\ \hline
  $4\pi$ & 62616 & 71388 \\ \hline
  $6\pi$ & 63103 & 71444 \\ \hline
  $KK\pi\pi$ & 62649 & 71659 \\ \hline
  $KK\pi^{0}$ & 71112 & 71248 \\ \hline
  $K_{S}K\pi$ & 73383 & 70853 \\ \hline
  $\pi\pi\eta(\gamma\gamma)$ & 71052 & 71904 \\ \hline
  $\pi\pi\eta(\pi\pi\pi^{0})$ & 70928 & 71307 \\ \hline
  $\pi\pi\eta^{\prime}, \eta^{\prime}\to\pi\pi\eta(\gamma\gamma)$
    & 70986 & 49970 \\ \hline
  $KK\eta(\gamma\gamma)$ & 70879 & 71145 \\ \hline
  $KK\eta(\pi\pi\pi^{0})$ & 71261 & 71295 \\ \hline
  $KK\pi\pi\pi^{0}$ & 62972 & 71742 \\ \hline
  $KK4\pi$ & 62426 & 70908 \\ \hline
  $K_{S}K3\pi$ & 64556 & 71063 \\ \hline
\end{tabular}
\end{center}
\end{table}

\section{Event Selection}
\label{sec:eventselection}

It was required that the net charge of the event equal zero and the 
number of good tracks match the number of charged tracks in the decay 
mode.

\subsection{Track Selection}

We used standard CLEO-c selection criteria for track quality and 
particle identification (PID). \ These have been developed for use 
in high-precision measurements of $D$ decays by the CLEO-c ``DTag''
subgroup \cite{cbx07-30}. \ They are applied in selecting
$K^{\pm}$'s and $\pi^{\pm}$'s.

\subsubsection{Track Quality}

For track quality, it was required that the track have a successful 
Kalman fit. \ Also the projection of the distance from the interaction 
point of the $e^+e^-$ annihilation (IP) to the origin of the track on 
the $r-\phi$ plane 
$|D_{0}|< 5~{\rm mm}$ and the projection of the distance between the 
IP and origin of the track along the $z$-axis 
$|Z_{0}|< 5~{\rm cm}$ were required for each qualified track, with 
exception of charged tracks from $K_{S} \to \pi^+\pi^-$ decays. \ 
(If charged tracks are daughters of a $K_{S}$, the $D_{0}$ and $Z_{0}$ of the 
track 
tend to be larger as the charged particles are created at where the 
$K_{S}$ decays rather than the IP.)

\subsubsection{Particle ID}

The Particle ID criteria developed by the DTag group were used for 
charged pions and kaons. \ Those criteria are
\begin{itemize}
\item $|\cos \theta | < 0.93$, where $\theta$ is the angle between the 
initial momentum of the track and the direction of the beam, the $z$ axis
\item RICH information is valid
\item Both $\pi^{\pm}$, $K^{\pm}$ hypotheses analyzed
\item $3 \sigma$ $dE/dx$ consistency with the $\pi^{\pm}$ or $K^{\pm}$ 
hypothesis, which means if the $dE/dx$ distribution for the $\pi^{\pm}$ 
or $K^{\pm}$ hypothesis is Gaussian, then 
\[
\frac{|(dE/dx)_{\rm measured}-(dE/dx)_{\pi^{\pm}~{\rm or}~K^{\pm}}|}
{\sigma_{dE/dx}}\le 3,
\] 
where $(dE/dx)_{\pi^{\pm}~{\rm or}~K^{\pm}}$ and $\sigma_{dE/dx}$ are the 
mean and 
standard deviation of the the $dE/dx$ distribution for the $\pi^{\pm}$ or 
$K^{\pm}$ 
hypothesis
\item If the momentum of the particle is greater than or equal to 
$700~{\rm MeV}$:
  \begin{itemize}
  \item More than two photons on the Cherenkov ring (See 
Section~\ref{sec:rich}) are associated to the pion or kaon
  \item Combined log-likelihood 
    $L = L_{\pi^{\pm}}-L_{K} + \sigma_{\pi^{\pm}}^{2} - 
    \sigma_{K^{\pm}}^{2} > 0$ for $\pi^{\pm}$, $< 0$ for $K^{\pm}$, 
    where $L_{\pi^{\pm}}$ and $L_{K^{\pm}}$ are the log-likelihoods for 
    the $\pi^{\pm}$ and $K^{\pm}$ hypothesis, respectively, 
    and $\sigma_{\pi^{\pm}}$ and $\sigma_{K^{\pm}}$ are the standard 
    deviation of the $dE/dx$ distribution for the $\pi^{\pm}$ and 
    $K^{\pm}$ hypothesis as defined above as $\sigma_{dE/dx}$
  \end{itemize}
\item If the momentum of the particle is smaller than $700~{\rm MeV}$, or
if RICH information is not available: 
  $\sigma_{\pi^{\pm}}^{2} - \sigma_{K^{\pm}}^{2} > 0$ for $\pi^{\pm}$, 
  $< 0$ for $K^{\pm}$

\end{itemize}

For any mode including a $K^+K^-$ pair, only one kaon is required to pass
the PID to increase the efficiency.

\subsection{Photon Selection}

General requirements for selecting photons, both from radiative 
transitions and in 
$\pi^{0}, \eta \to \gamma \gamma$ reconstruction, were applied. \ 
We required that a shower does not contain any ``hot'' crystals, is not 
matched to the projection of a charged track into the CC, and has E9/E25 OK.
\ E9/E25 is the ratio between the total energy from 9 central crystals 
around the highest energy crystal and the total energy from the 25 
crystals centering at the highest energy crystal in the calorimeter. \ 
E9/E25 OK means that for a shower with a 
certain energy, the ratio is within 99\% of the E9/E25 photon like 
distribution at this energy.

In selecting candidate transition photons for the decay 
$\psi(2S)\to\gamma\eta_{c}(2S)$, 
we required the energy to be in the range
$E_{\gamma} = [30-110]~{\rm MeV}$. \ Because of the degraded photon 
resolution and efficiency in the endcap, we selected showers in the good 
barrel only, $|\cos \theta| < 0.81$.

\subsection{$K^{0}_{S}$ Selection}

We required the $K^{0}_{S}$ flight significance to be greater than 3. \
$K^{0}_{S}$ flight significance is the ratio of the distance between the 
position of $K^{0}_{S}$ where it decays to two charged tracks and the 
beam spot to the measurement error of the position determined by the
error matrix. \ In another word, the intersection of the two charged 
tracks that are $K^{0}_{S}$ daughters was required to be far enough from
the interaction point.

The mass of a $K^{0}_{S}$ candidates was required to be within 
$10~{\rm MeV}$ of the $K^{0}_{S}$ nominal mass, i.e., 
$|M_{K^{0}_{S}} -  0.498~{\rm GeV}| < 10~{\rm MeV}$.

\subsection{Neutral Pion Selection}

For selecting photons used to compose a $\pi^{0}$ or $\eta$, CLEO-c
software 
PhotonDecaysProd was used with the following defaults for 
the CLEO-c software packages NavPi0ToGG and 
NavEtaToGG: E9/E25 OK, unmatched shower in good barrel or good endcap,
and $E_{\gamma}> 30~{\rm MeV}$.

$\pi^{0}$ candidates were obtained from the table of $\pi^{0}$ candidates 
that was produced by NavPi0toGG. \ 
The default photon energy threshold of NavPi0ToGG table is 30 MeV in the
good barrel and 50 MeV elsewhere. \ No hot crystals and E9/E25 OK were 
also required. \ The only additional requirement for selecting 
$\pi^{0}$'s was to require the pull mass of the $\pi^{0}$ to be within 
$\pm 3 \sigma$. \ The pull mass is defined as the ratio of the difference 
between the mass of the $\pi^{0}$ candidate and the $\pi^{0}$ nominal
mass divided by the standard deviation of the $\pi^{0}$ mass 
distribution. \ This
means that the mass of the $\pi^{0}$ candidate was required to be close
enough to the $\pi^{0}$ nominal mass.

\subsection{$\eta$ Selection}

The $\eta$ candidates were reconstructed in two ways, 
$\eta\to\gamma\gamma$ and $\eta\to\pi^{+}\pi^{-}\pi^{0}$.

The $\eta\to\gamma\gamma$ candidates were obtained from the NavEtaToGG
(CLEO-c software package) table similar to selecting the $\pi^{0}$ 
candidates. \ The default
requirements of photons in NavEtaToGG are the same as NavPi0ToGG. \
Similarly, the only additional requirement for selecting 
$\eta\to\gamma\gamma$ decays was to require the pull mass of the
$\eta$ to be within $\pm 3 \sigma$.  

\begin{figure}[htbp]
  \centering
  \subfigure
    {\includegraphics[height=.40\textheight]{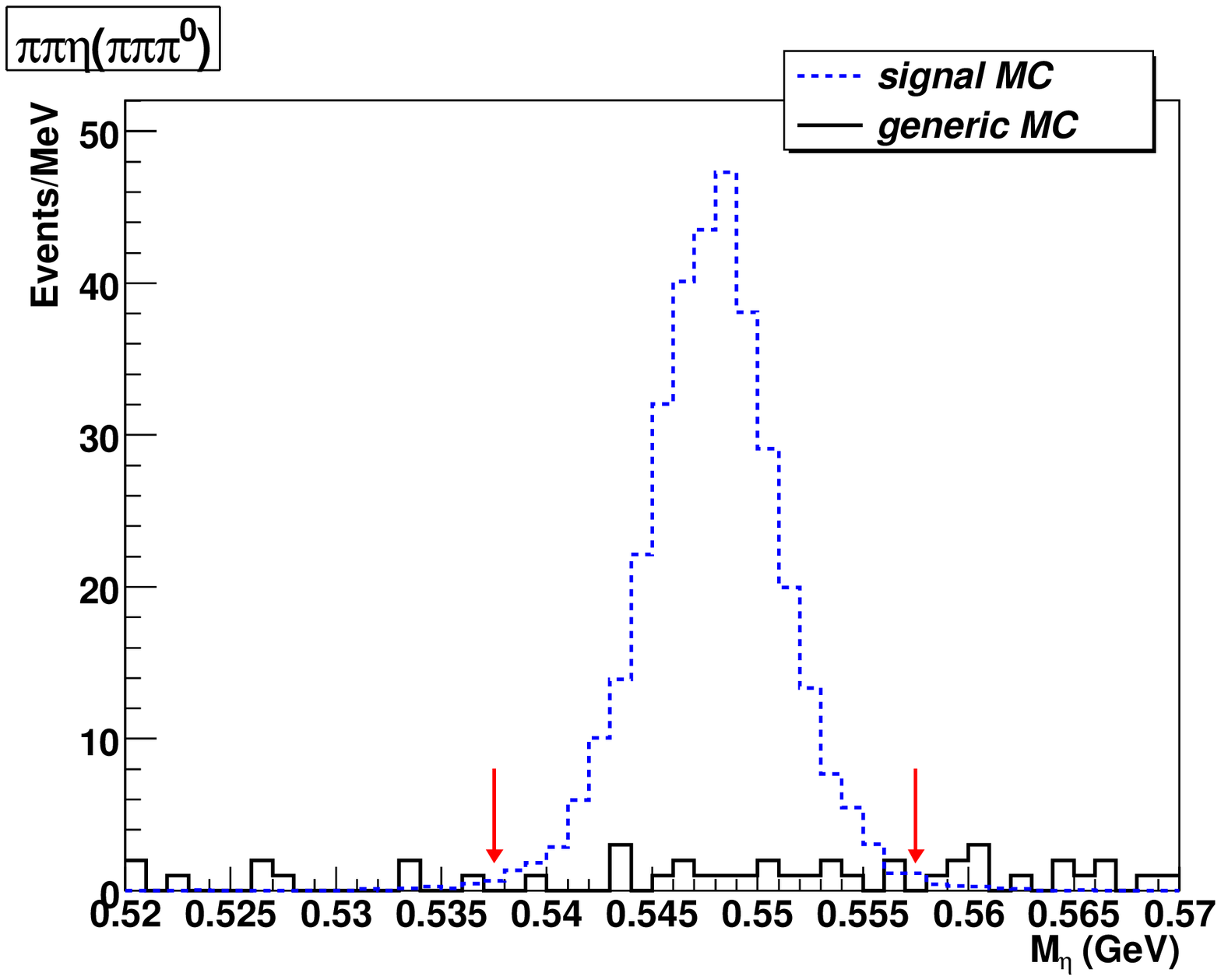}}
  \subfigure
    {\includegraphics[height=.40\textheight]{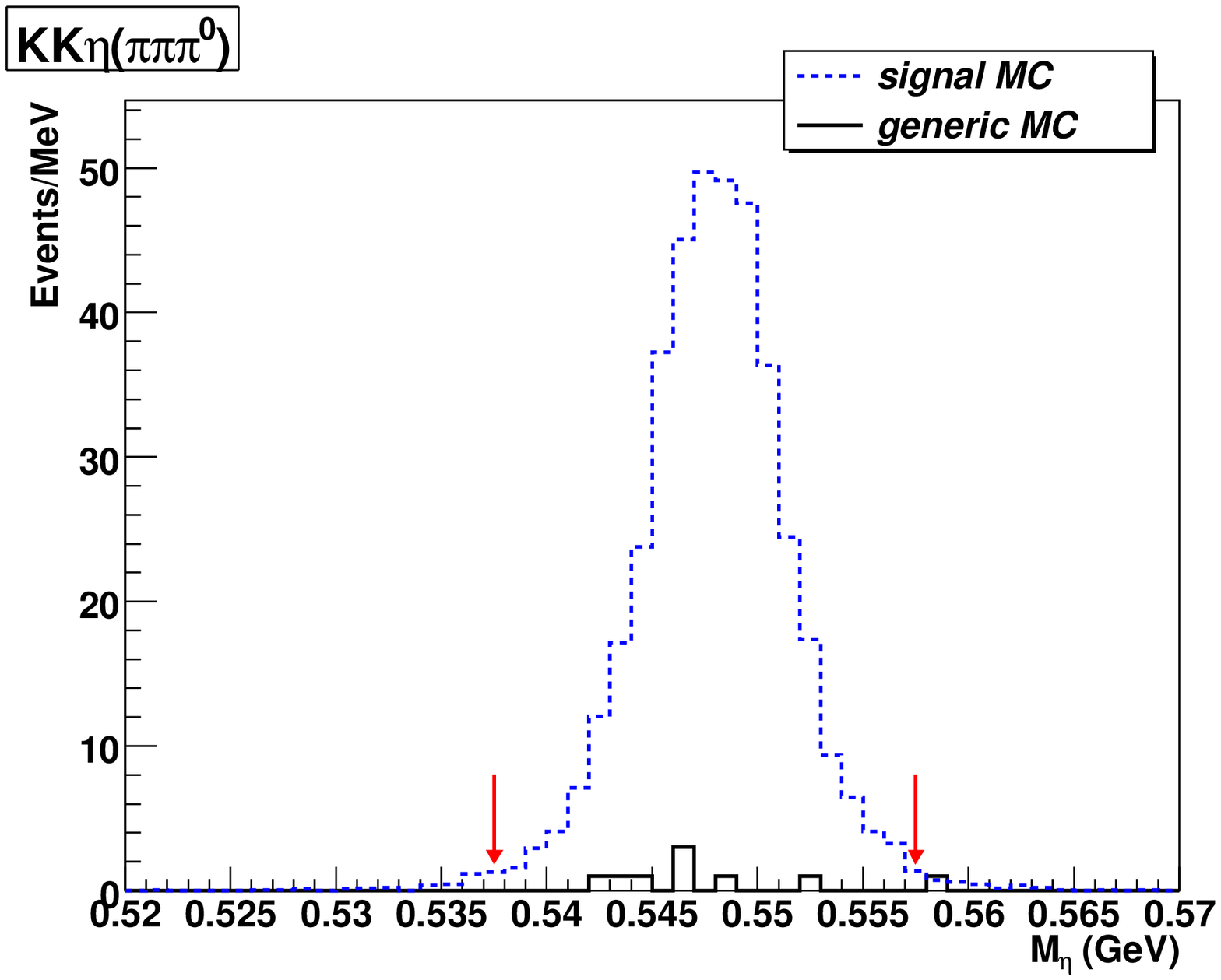}}
  \caption[MC simulations of $\eta$ candidate masses]
{MC simulations of $\eta$ candidate masses in $\psi(2S)$ events 
with the final states with $\pi\pi\eta(\pi\pi\pi^{0})$ (top) and 
$KK\eta(\pi\pi\pi^{0})$ (bottom). 
     \ The solid histogram is the 5 times luminosity background MC samples, 
     while the dashed histogram is signal MC, arbitrarily scaled for clarity. 
     \ The arrows show the selection cuts that were applied.
     \ All other event selection criteria have been applied.}
  \label{fig:cut_EtaMass}
\end{figure}

For $\eta\to\pi^{+}\pi^{-}\pi^{0}$, we required the mass of 
$\pi^{+}\pi^{-}\pi^{0}$ combination to be within $10~{\rm MeV}$ of the 
nominal $\eta$ 
mass, i.e., $|M_{\pi^{+}\pi^{-}\pi^{0}} - M_{\eta}| < 10~{\rm MeV}$. \
Figure~\ref{fig:cut_EtaMass} shows simulated $\eta$ mass distributions 
before the cut was applied in the two $\eta_{c}(2S)$ modes with 
$\eta\to\pi\pi\pi^{0}$.  

\subsection{$\eta^{\prime}$ selection}

\begin{figure}[htbp]
  \centering
  \subfigure
    {\includegraphics[width=.95\textwidth]{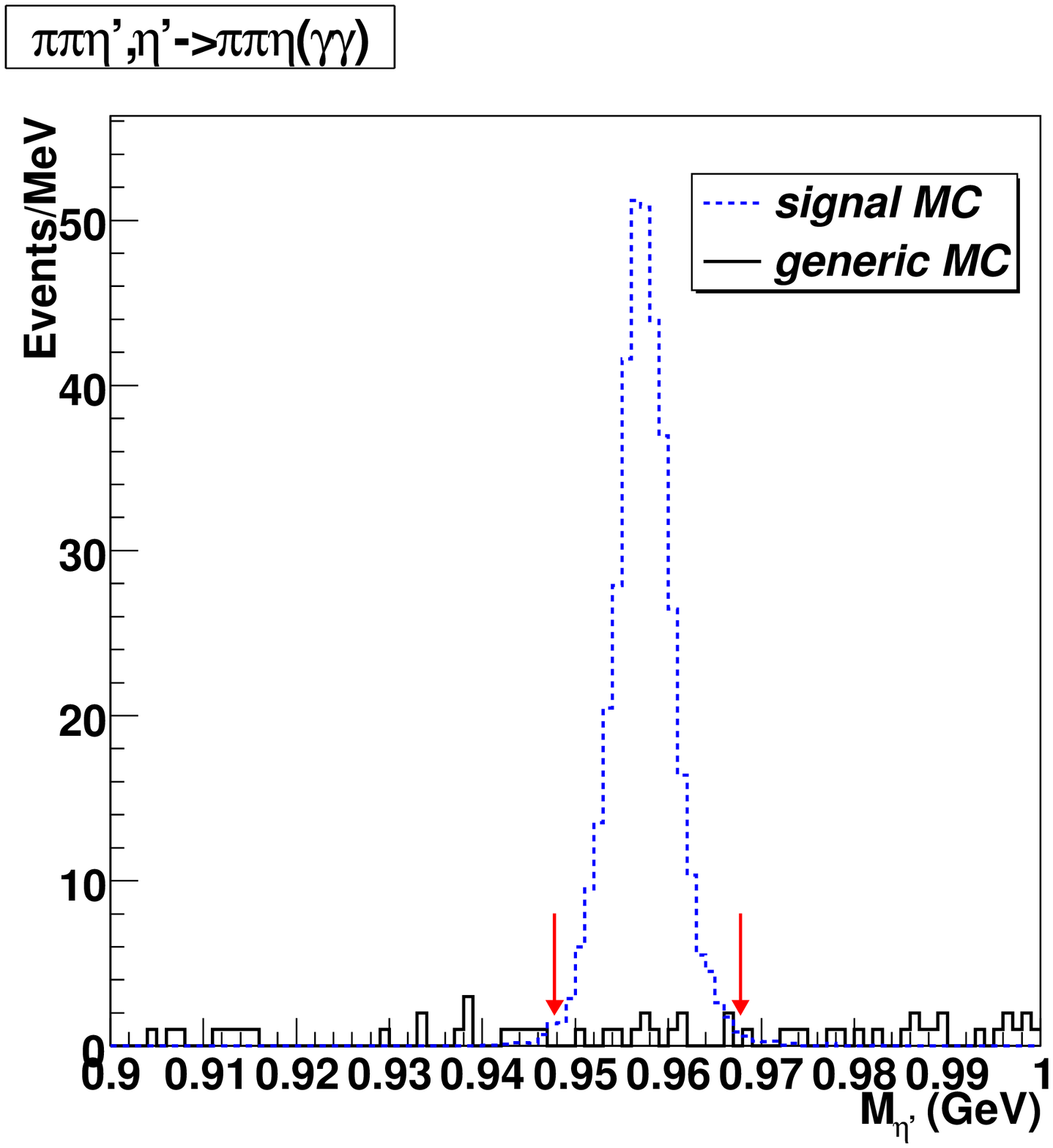}}
  \caption[MC simulations of $\eta^{\prime}$ candidate masses]
{MC simulations of $\eta^{\prime}$ candidate masses in $\psi(2S)$ 
events with the final state $\pi\pi\eta^{\prime}$, 
$\eta^{\prime}\to\pi\pi\eta(\gamma\gamma)$. 
     \ The solid histogram is the 5 times luminosity background MC samples, 
     while the dashed histogram is signal MC, arbitrarily scaled for clarity. 
     \ The arrows show the selection cuts that were applied.
     \ All other event selection criteria have been applied.}
  \label{fig:cut_EtaPMass}
\end{figure}

Among the seven modes we considered, only one mode has an  
$\eta^{\prime}$ decay, $\eta_{c}(2S)\to\pi\pi\eta^{\prime}$, 
$\eta^{\prime}\to\pi\pi\eta(\gamma\gamma)$. \  
For this mode we required 
the invariant mass of the hadrons that compose the $\eta^{\prime}$ to 
be within $10~{\rm MeV}$ of the expected $\eta^{\prime}$ mass,  
$|M(\pi \pi \eta) - M(\eta')| < 10~{\rm MeV}$. \
Figure~\ref{fig:cut_EtaPMass} shows simulated $\eta^{\prime}$ mass 
distributions for this mode before the cut was applied.

\subsection{Suppression of Background Charmonium States}

In following section, we describe selection criteria used to suppress production 
of other charmonium states observed in the generic $\psi(2S)$ MC samples.  

\subsubsection{Hadronic Invariant Mass Difference}

To suppress decays of direct $\psi(2S)$ and $\chi_{c2}$, produced via 
$\psi(2S) \to \gamma \chi_{c2}$, to the hadronic final states 
under investigation, we require the hadronic invariant mass difference 
$\Delta M \equiv M(\psi(2S)) - M(X)$ to be in the range [0,100]~MeV,
where $M(\psi(2S))=3686.1~{\rm MeV}$ is the PDG value of the $\psi(2S)$
mass \cite{PDBook2006}, 
and $M(X)$ is the 
reconstructed hadronic invariant mass.

\subsubsection{$\eta$ Recoil Mass}

In selecting $\eta_c(2S)$ decays to final states with an $\eta$ decay, large backgrounds 
arise from $\psi(2S) \to J/\psi(1S) \eta$ decays. \ The branching 
fraction of this decay mode is $(3.09\pm0.08)\%$ \cite{PDBook2006}. \   
Therefore we applied $\eta$ recoil mass cuts to the modes 
$\eta_c(2S) \to \pi\pi\eta(\gamma\gamma)$ and $\pi\pi\eta(\pi\pi\pi^{0})$. \ 
In the mode $\pi\pi\eta^{\prime}$, $\eta^{\prime}\to\pi\pi\eta(\gamma\gamma)$, 
although an $\eta$ is in the decay, we did not apply the $\eta$ recoil mass 
cut because the selection of $\eta^{\prime}$ has already eliminated 
$\psi(2S) \to J/\psi(1S) \eta$.

\begin{figure}[htbp]
  \centering
  \subfigure
    {\includegraphics[height=.40\textheight]{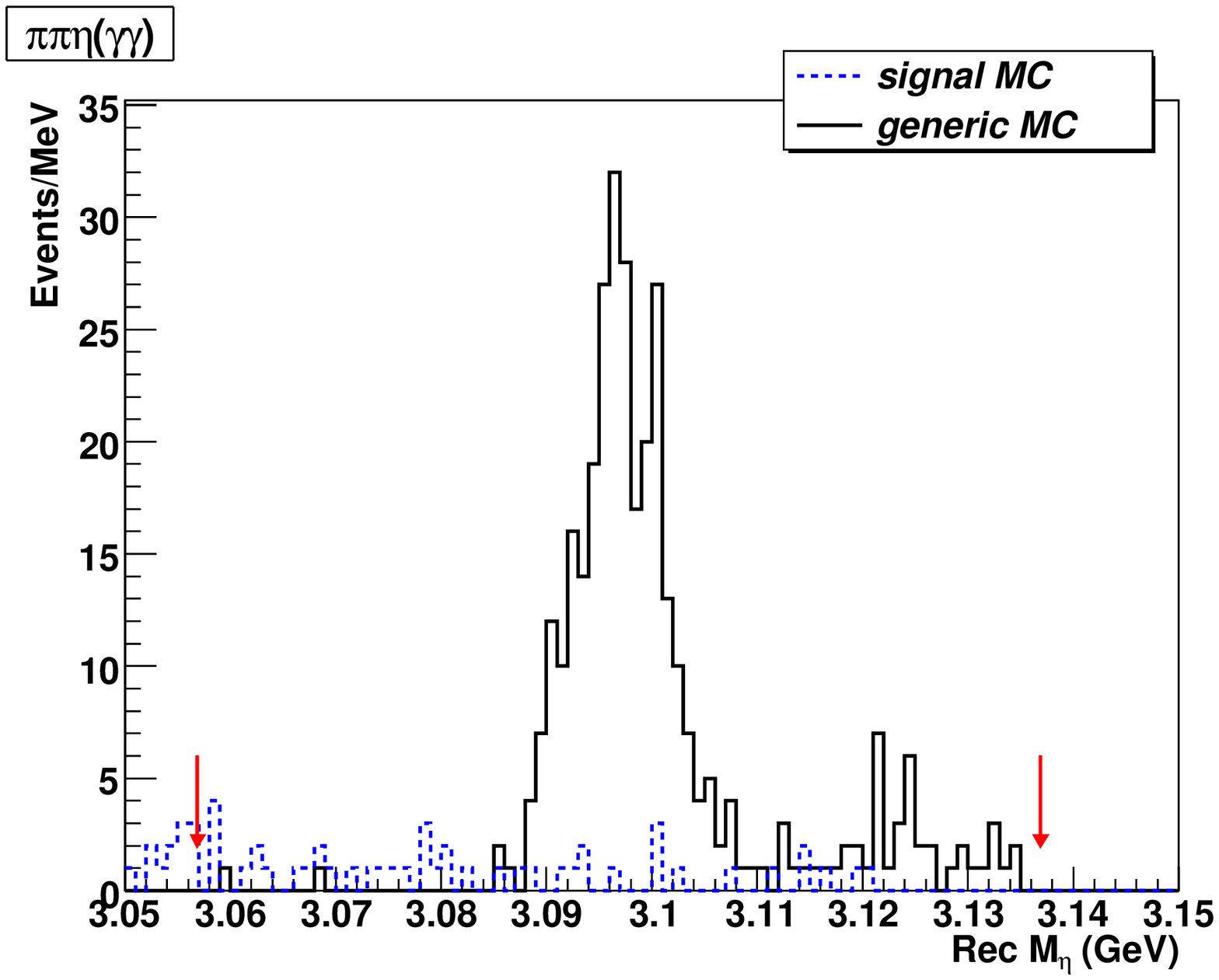}}
  \subfigure
    {\includegraphics[height=.40\textheight]{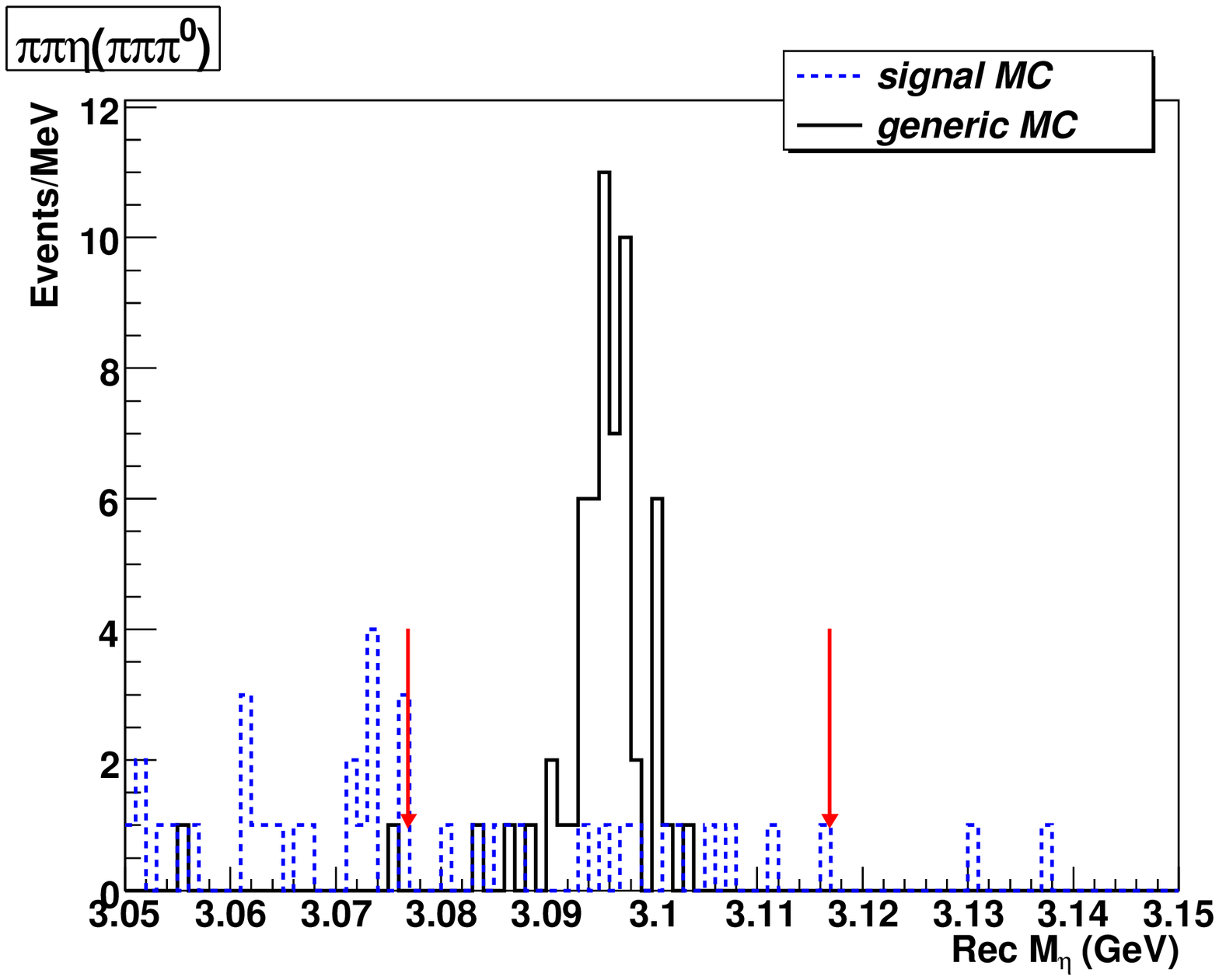}}
  \caption[MC simulations of distributions of $\eta$ recoil mass]
    {MC simulations of the $\eta$ recoil mass distributions 
     for $\psi(2S)$ events with final states $\pi\pi\eta(\gamma\gamma)$ (top)
     and $\pi\pi\eta(\pi\pi\pi^{0})$ (bottom).
     \ The solid histogram is the 5 times luminosity background MC samples, 
     while the dashed histogram is signal MC, arbitrarily scaled for clarity. 
     \ The arrows show the selection cuts that were applied.
     \ All other event selection criteria have been applied.}
  \label{fig:cut_EtaRecMass}
\end{figure}

The $\eta$ recoil mass cut was mode dependent. \ For mode 
$\eta_{c}(2S) \to \pi\pi\eta(\gamma\gamma)$, we rejected events with the 
$\eta$ recoil mass within $40~{\rm MeV}$ of the $J/\psi$ mass by requiring 
$|{\rm Rec} M(\eta) - M(J/\psi)| \geq 40~{\rm MeV}$. \ For mode 
$\eta_{c}(2S) \to \pi\pi\eta(\pi\pi\pi^{0})$, we required 
$|{\rm Rec} M(\eta) - M(J/\psi)| \geq 20~{\rm MeV}$. \ Simulated $\eta$ 
recoil mass distributions before applying this cut are shown in 
Figure~\ref{fig:cut_EtaRecMass}. \ 
These cuts were found to be unnecessary for the 
$\eta_c(2S) \to KK\eta(\gamma\gamma)$ and $KK\eta(\pi\pi\pi^{0})$ modes.

\subsubsection{Photon Recoil Mass}

In modes that contain $\eta \to \gamma\gamma$ and 
$\pi^{0} \to \gamma\gamma$ decays, a large amount of background comes from
using the photon from the decay $\psi(2S)\to\gamma\chi_{cJ}$ with another 
photon.  \ 
By applying mode dependent recoil mass cuts, we effectively remove a 
large amount of this background.

For modes
$\eta_{c}(2S) \to \pi\pi\eta(\gamma\gamma)$ and 
$\eta_{c}(2S) \to KK\eta(\gamma\gamma)$, since 
$M(\eta) = (547.51\pm0.18)$~MeV \cite{PDBook2006},
a relatively large mass compared to the energy of the transition photons in the 
$\psi(2S) \to \gamma \chi_{cJ}$ decays (100-300~MeV), the transition 
photon 
is sometimes used as the low energy photon in the $\eta$ decay. \ 
Therefore we examined the photon recoil mass for the lower energy photon in 
$\eta\to\gamma\gamma$ decays. \ For these two modes, we rejected events by
requiring 
\begin{itemize}
\item $|{\rm Rec} M(\eta~{\rm low~energy}~\gamma) - M(\chi_{c2})| \geq 20~{\rm MeV}$
\item $|{\rm Rec} M(\eta~{\rm low~energy}~\gamma) - M(\chi_{c1})| \geq 20~{\rm MeV}$
\item $|{\rm Rec} M(\eta~{\rm low~energy}~\gamma) - M(\chi_{c0})| \geq 30~{\rm MeV}$
\end{itemize}

For the mode
$\eta_{c}(2S) \to KK\pi^{0}$, since $M(\pi^{0}) = (134.9766\pm0.0006)$~MeV
\cite{PDBook2006}, 
a relatively small mass compared to the energy of the transition photon 
(100-300~MeV), 
the high energy photon is more likely swapped with the 
transition photon. \ Therefore we examined the photon recoil mass for the 
higher energy photon in 
$\pi^0\to\gamma\gamma$ decays. \ For this mode, we rejected events by requiring

\begin{itemize}
\item $|{\rm Rec} M(\pi^{0}~{\rm high~energy}~\gamma)-M(\chi_{c2})| \geq 20~{\rm MeV}$
\item $|{\rm Rec} M(\pi^{0}~{\rm high~energy}~\gamma)-M(\chi_{c1})| \geq 20~{\rm MeV}$
\item $|{\rm Rec} M(\pi^{0}~{\rm high~energy}~\gamma)-M(\chi_{c0})| \geq 30~{\rm MeV}$
\end{itemize}

Similar to the case of the $\eta$ recoil mass cut, modes 
$\eta_c(2S)\to\pi\pi\eta(\pi\pi\pi^{0})$ and $KK\eta(\pi\pi\pi^{0})$, even 
though
they contain $\pi^{0}$ decays, the photons used to form the $\pi^0$ candidate 
did not contain 
contamination from $\psi(2S)\to\gamma\chi_{cJ}$ decays. \ 
Therefore no photon recoil mass cut was applied to these two modes.

\begin{figure}[htbp]
  \centering
  \subfigure
    {\includegraphics[height=.40\textheight]{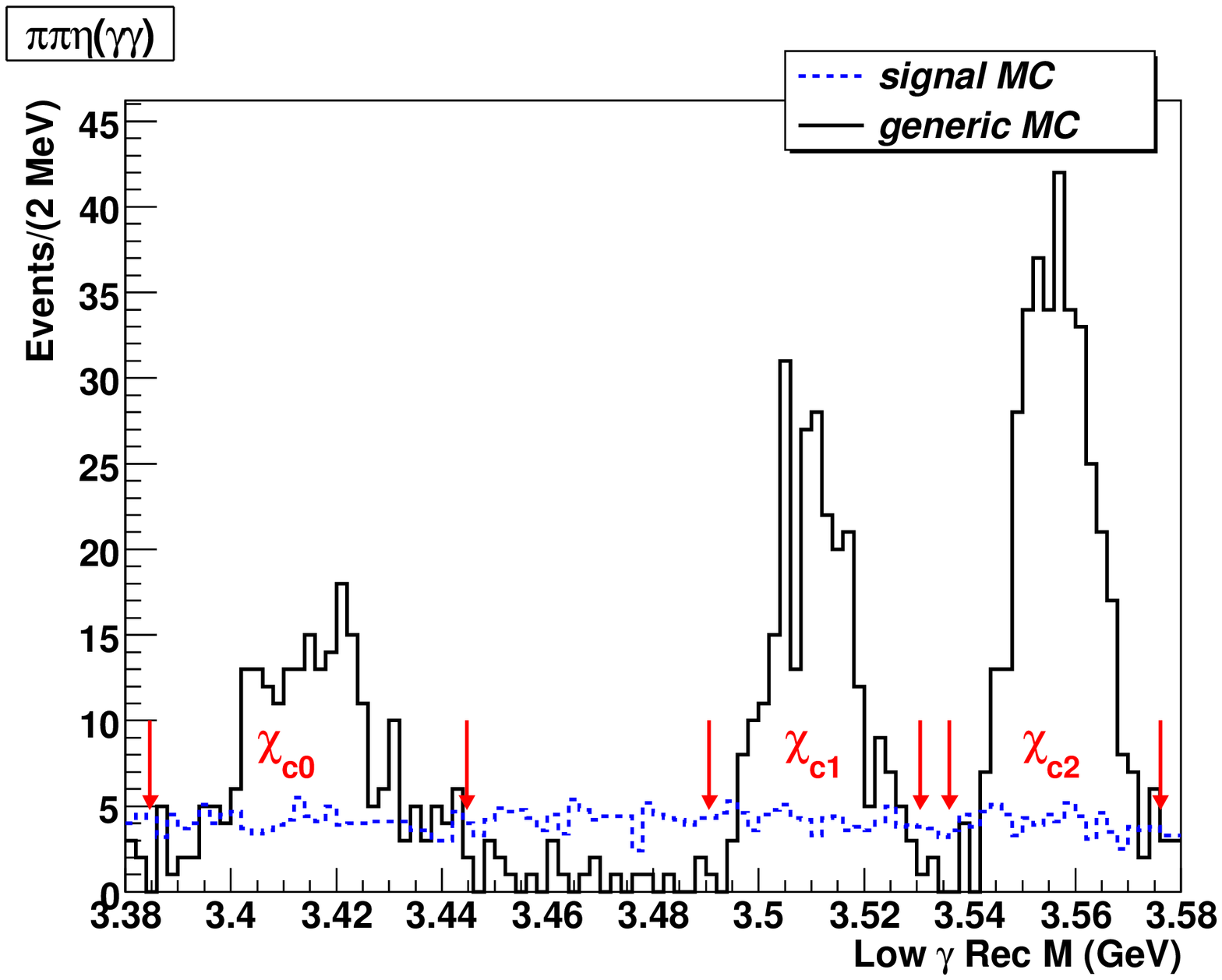}}
  \subfigure
    {\includegraphics[height=.40\textheight]{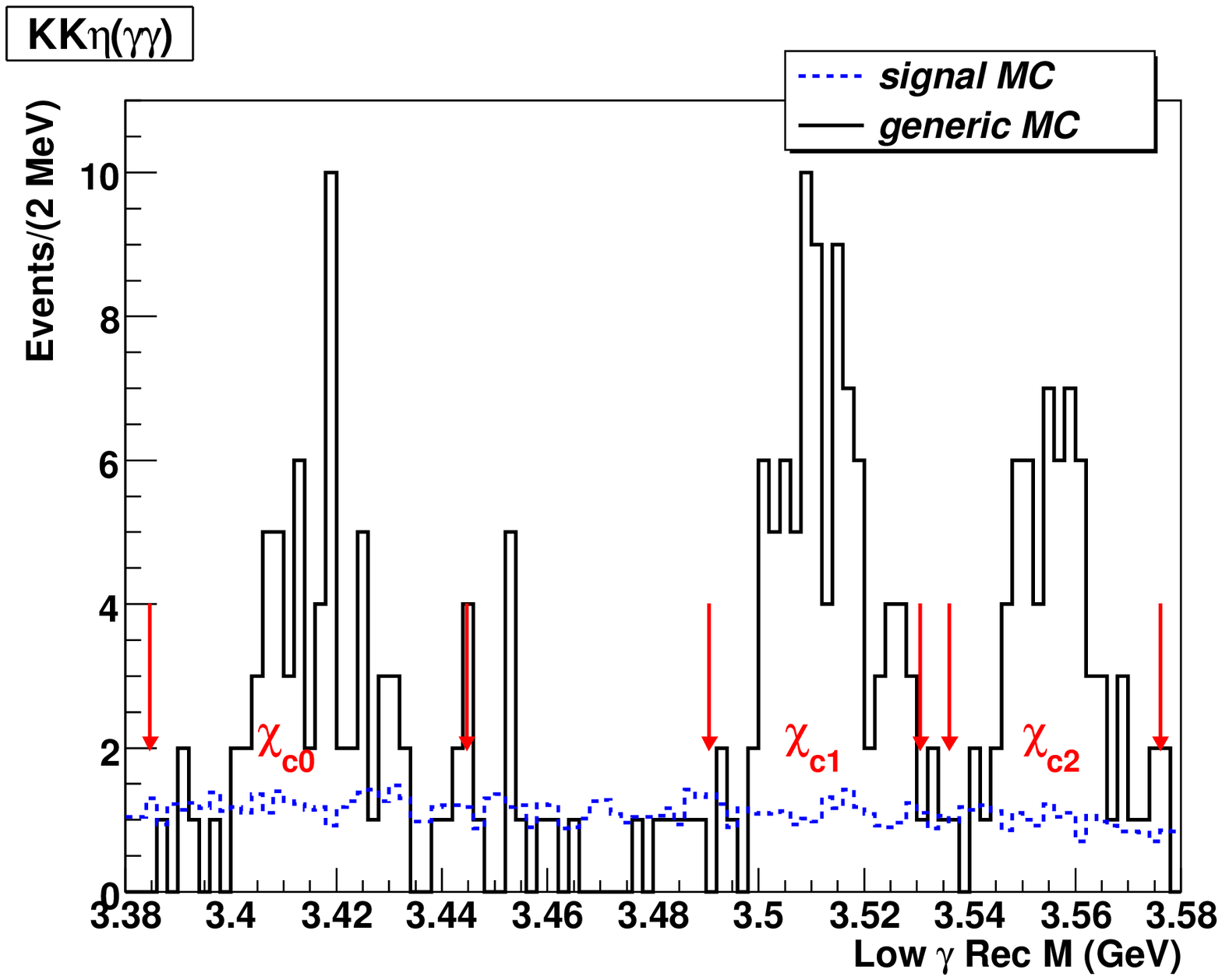}}
  \caption[MC simulations of distributions of low energy photon recoil mass]
    {MC simulations of distributions of low energy photon 
     recoil mass for $\psi(2S)$ events with final states
     $\pi\pi\eta(\gamma\gamma)$ (top) and $KK\eta(\gamma\gamma)$ (bottom). 
     \ The solid histogram is the 5 times luminosity background MC samples, 
     while the dashed histogram is signal MC, arbitrarily scaled for clarity. 
     \ The arrows show the selection cuts that were applied.
     \ All other event selection criteria have been applied.}
  \label{fig:cut_PhRecMass_low}
\end{figure}

\begin{figure}[htbp]
  \centering
  \subfigure
    {\includegraphics[width=.95\textwidth]{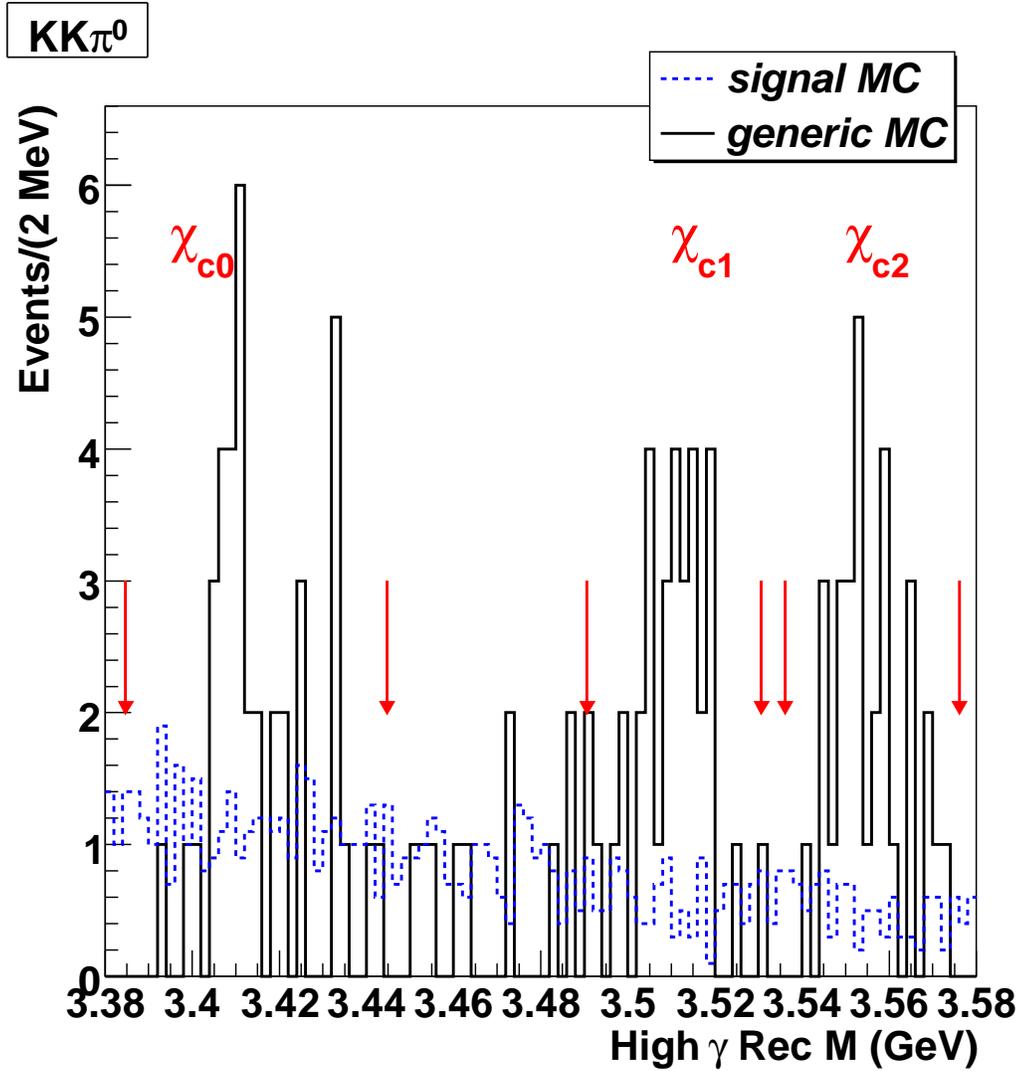}}
  \caption[MC simulations of distributions of high energy photon recoil mass]
    {MC simulations of distributions of low energy photon
     recoil mass for $\psi(2S)$ events with the final state $KK\pi^{0}$.
     \ The solid histogram is the 5 times luminosity background MC samples, 
     while the dashed histogram is signal MC, arbitrarily scaled for clarity. 
     \ The arrows show the selection cuts that were applied.
     \ All other event selection criteria have been applied.}
  \label{fig:cut_PhRecMass_high}
\end{figure}

Figures~\ref{fig:cut_PhRecMass_low} and \ref{fig:cut_PhRecMass_high} show 
the simulated distributions of photon recoil mass for the decay modes 
$\pi\pi\eta(\gamma\gamma)$, $KK\eta(\gamma\gamma)$, and $KK\pi^{0}$ before
applying these cuts.

\subsubsection{$J/\psi$ Rejection - $\pi\pi$ Recoil Mass}

For the modes that include charged pion pairs,  
we rejected events which contained decays of
$\psi(2S) \to \pi^{+}\pi^{-}J/\psi$. \ Specifically, we required
$|{\rm Rec} M_{\pi^{+}\pi^{-}} - M_{J/\psi}| > 20~{\rm MeV}$ for modes 
$6\pi$, $KK\pi\pi$, $KK\pi\pi\pi^{0}$, $KK4\pi$, and $K_{S}K3\pi$, and
${\rm Rec} M_{\pi^{+}\pi^{-}} - M_{J/\psi} < -30~{\rm MeV}$ for mode 
$4\pi$. \ With these requirements, we can suppress background from the 
decay $\psi(2S)\to\pi^{+}\pi^{-}J/\psi$, which is a dominant decay mode
of $\psi(2S)$ (${\cal B} \approx 31.8\%$).

\begin{figure}[htbp]
  \centering
  \subfigure
    {\includegraphics[height=.40\textheight]{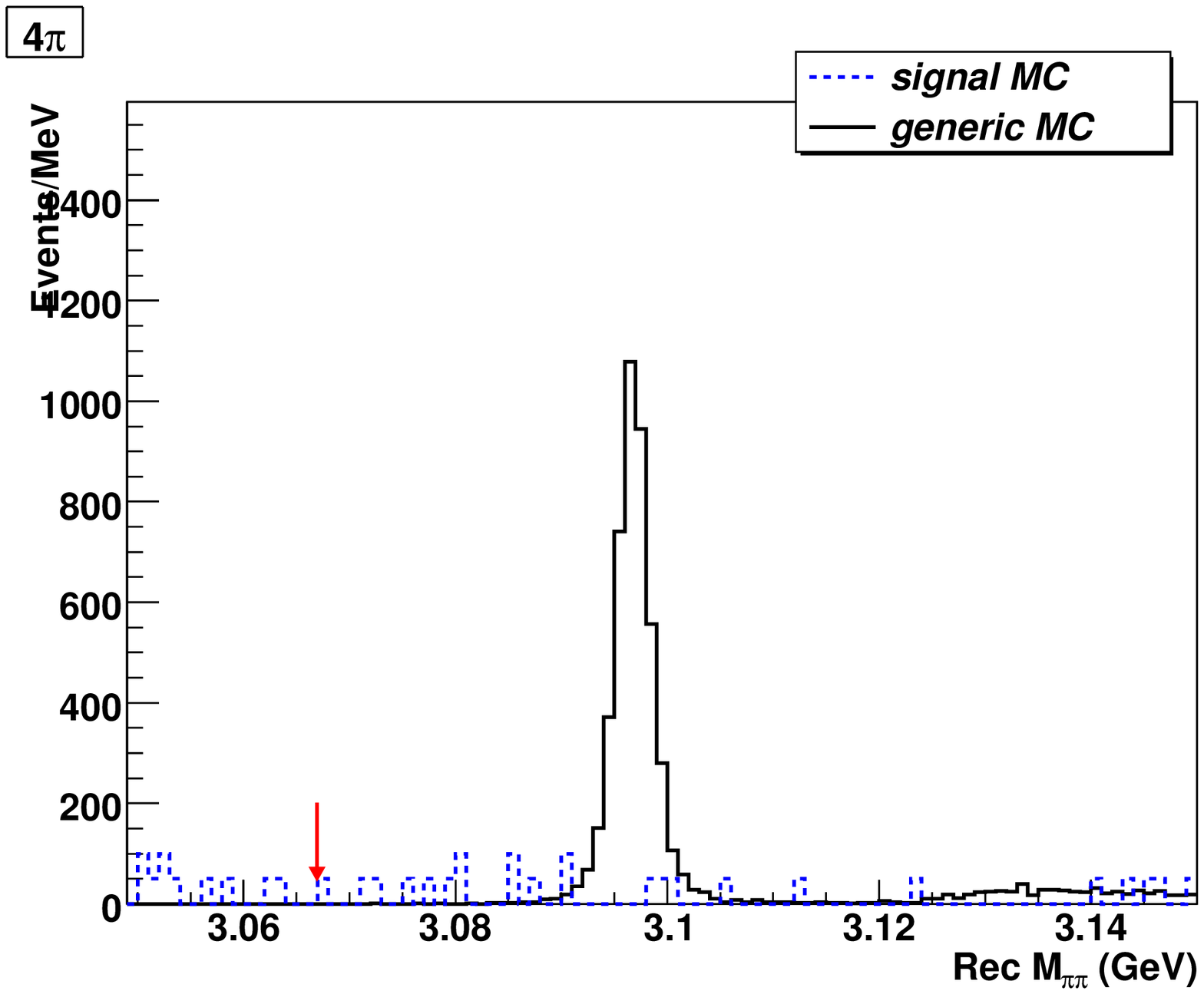}}
  \subfigure
    {\includegraphics[height=.40\textheight]{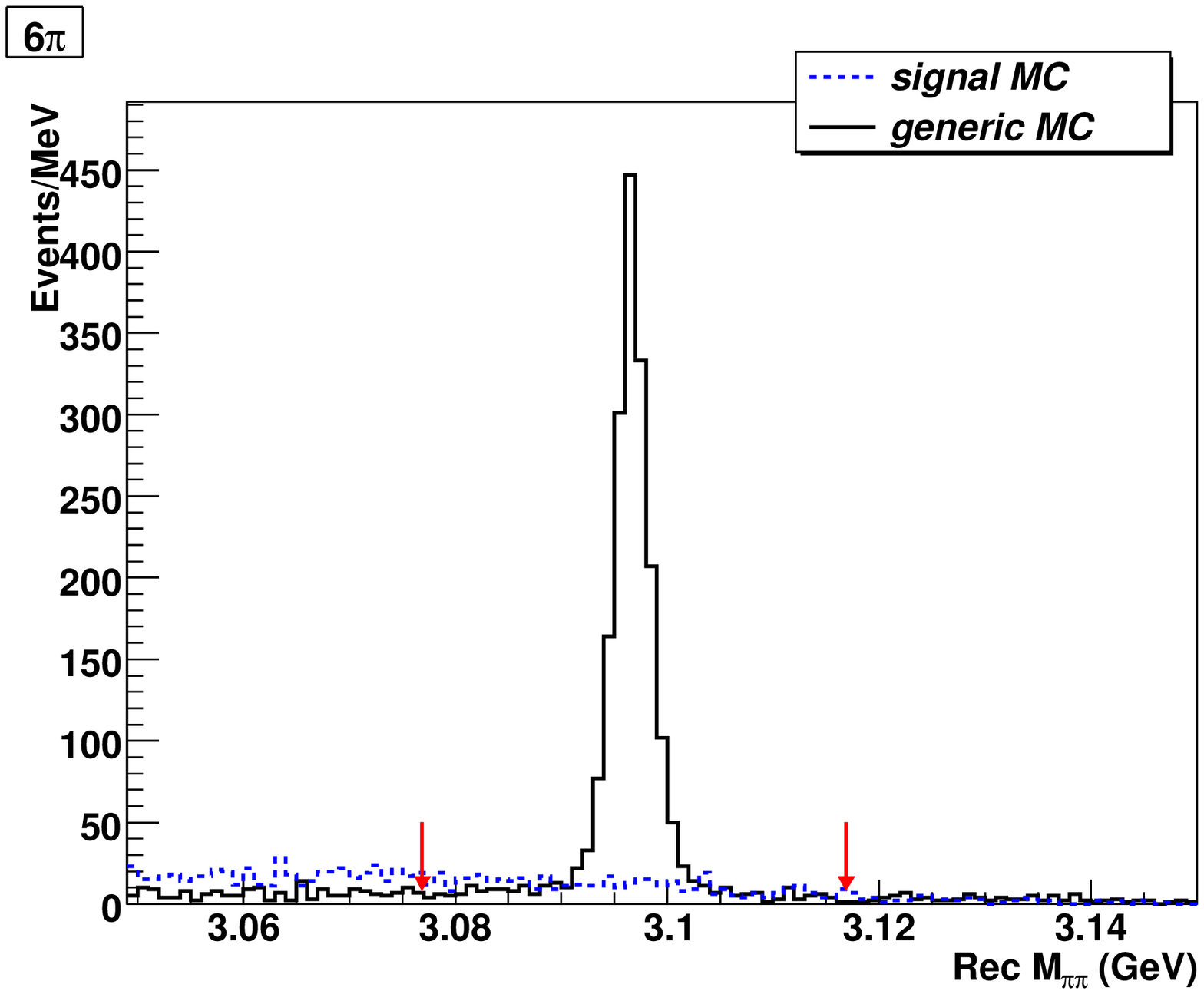}}
  \caption[MC simulations of distributions of $\pi\pi$ recoil mass (1)]
    {MC simulations of distributions of $\pi\pi$ recoil mass 
     for $\psi(2S)$ events with final states $4\pi$ (top) and $6\pi$ (bottom).
     \ The solid histogram is the 5 times luminosity background MC samples, 
     while the dashed histogram is signal MC, arbitrarily scaled for clarity. 
     \ The arrows show the selection cuts that were applied.
     \ All other event selection criteria have been applied.}
  \label{fig:cut_ppRecMass1}
\end{figure}

\begin{figure}[htbp]
  \centering
  \subfigure
    {\includegraphics[height=.40\textheight]{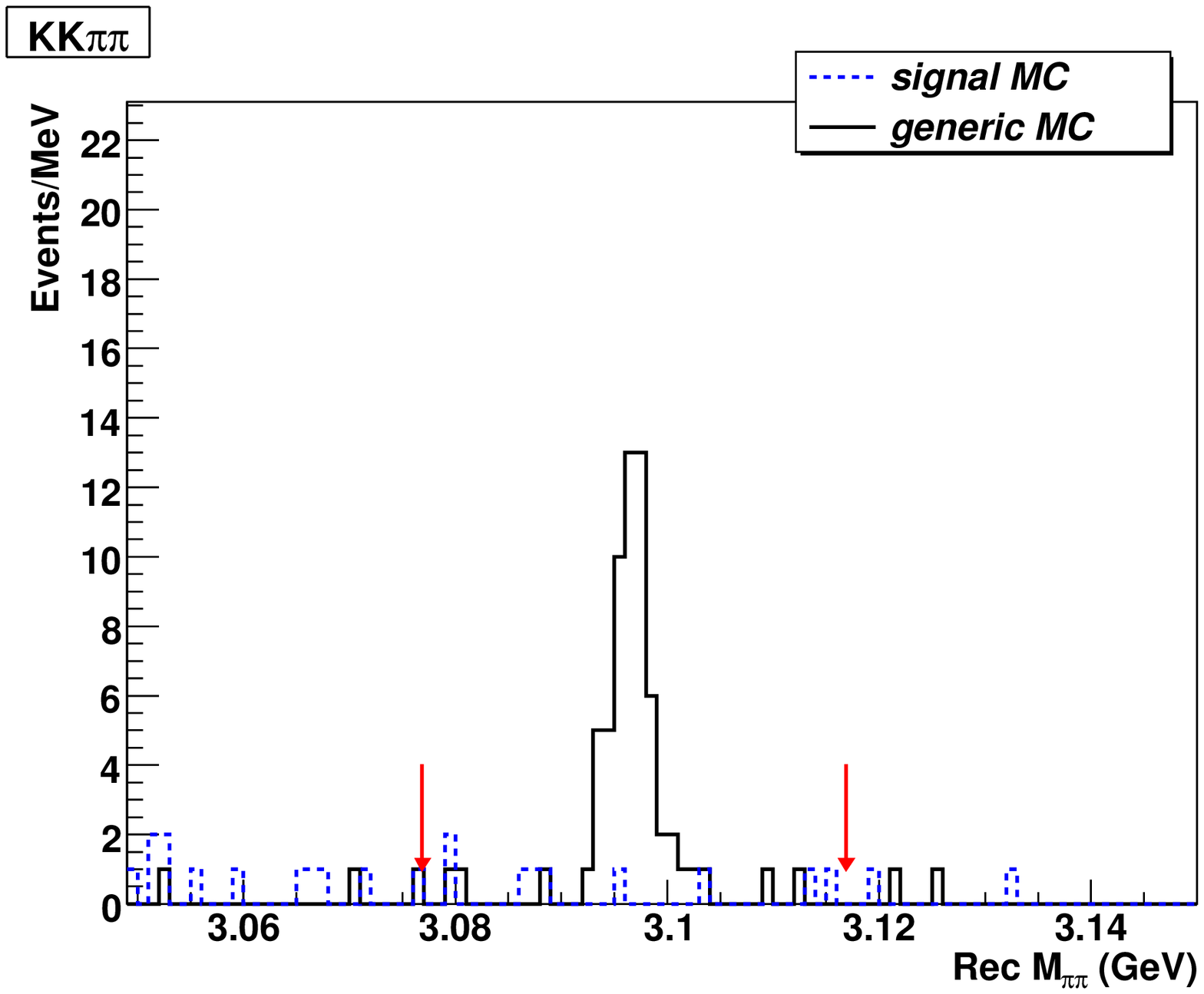}}
  \subfigure
    {\includegraphics[height=.40\textheight]{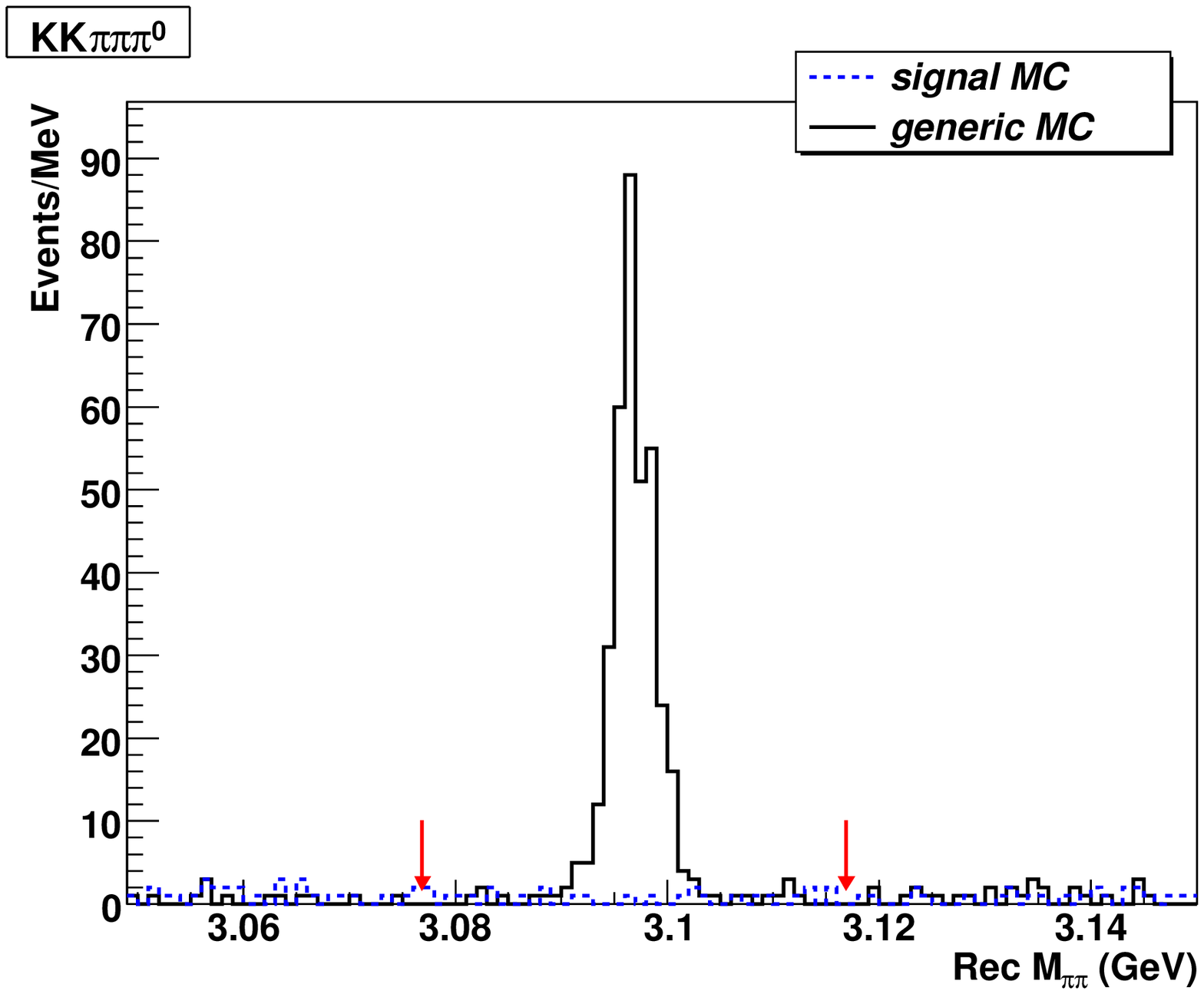}}
  \caption[MC simulations of distributions of $\pi\pi$ recoil mass (2)]
    {MC simulations of distributions of $\pi\pi$ recoil mass 
     for $\psi(2S)$ events with final states $KK\pi\pi$ (top) and 
     $KK\pi\pi\pi^{0}$ (bottom).
     \ The solid histogram is the 5 times luminosity background MC samples, 
     while the dashed histogram is signal MC, arbitrarily scaled for clarity. 
     \ The arrows show the selection cuts that were applied.
     \ All other event selection criteria have been applied.}
  \label{fig:cut_ppRecMass2}
\end{figure}

\begin{figure}[htbp]
  \centering
  \subfigure
    {\includegraphics[height=.40\textheight]{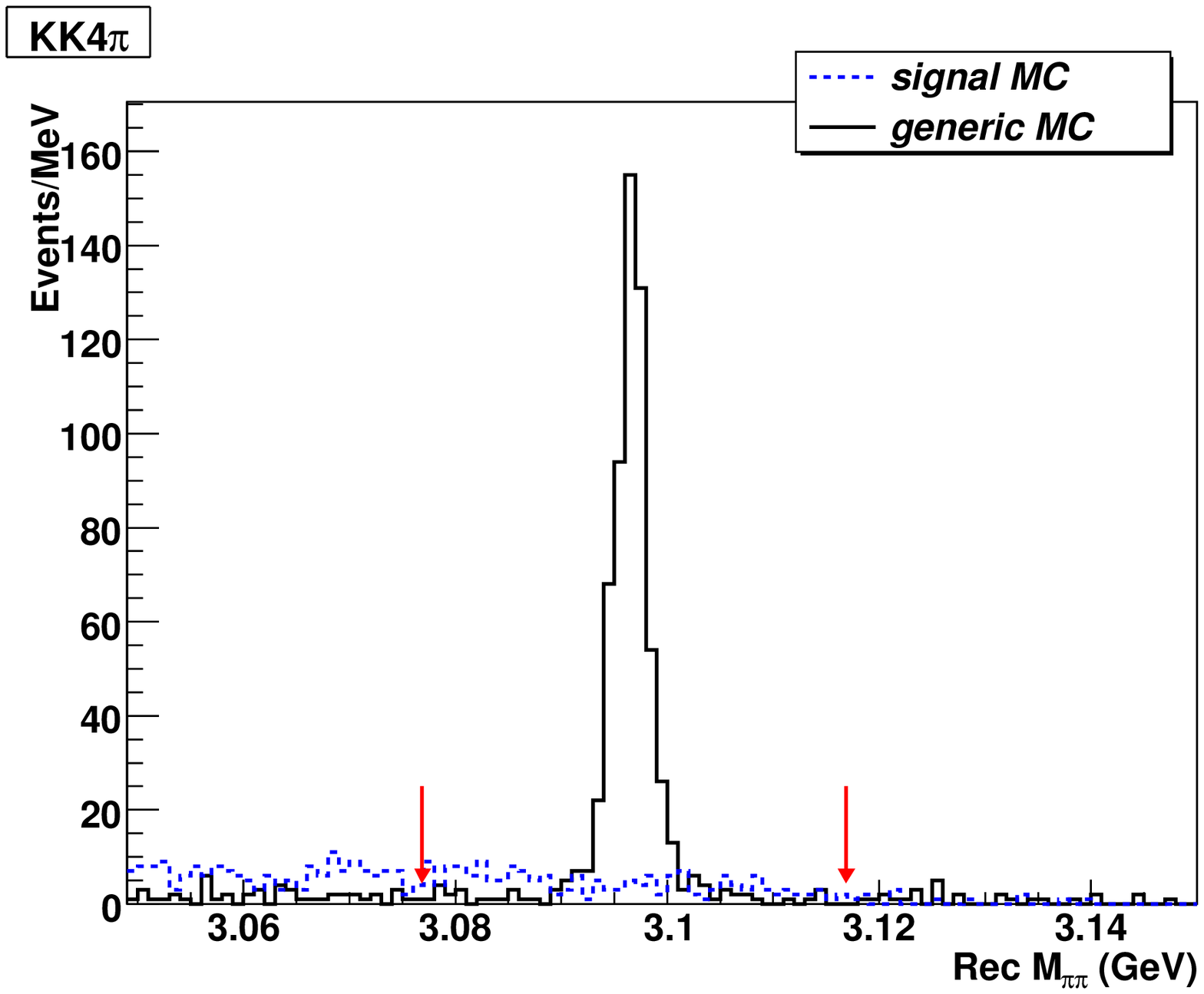}}
  \subfigure
    {\includegraphics[height=.40\textheight]{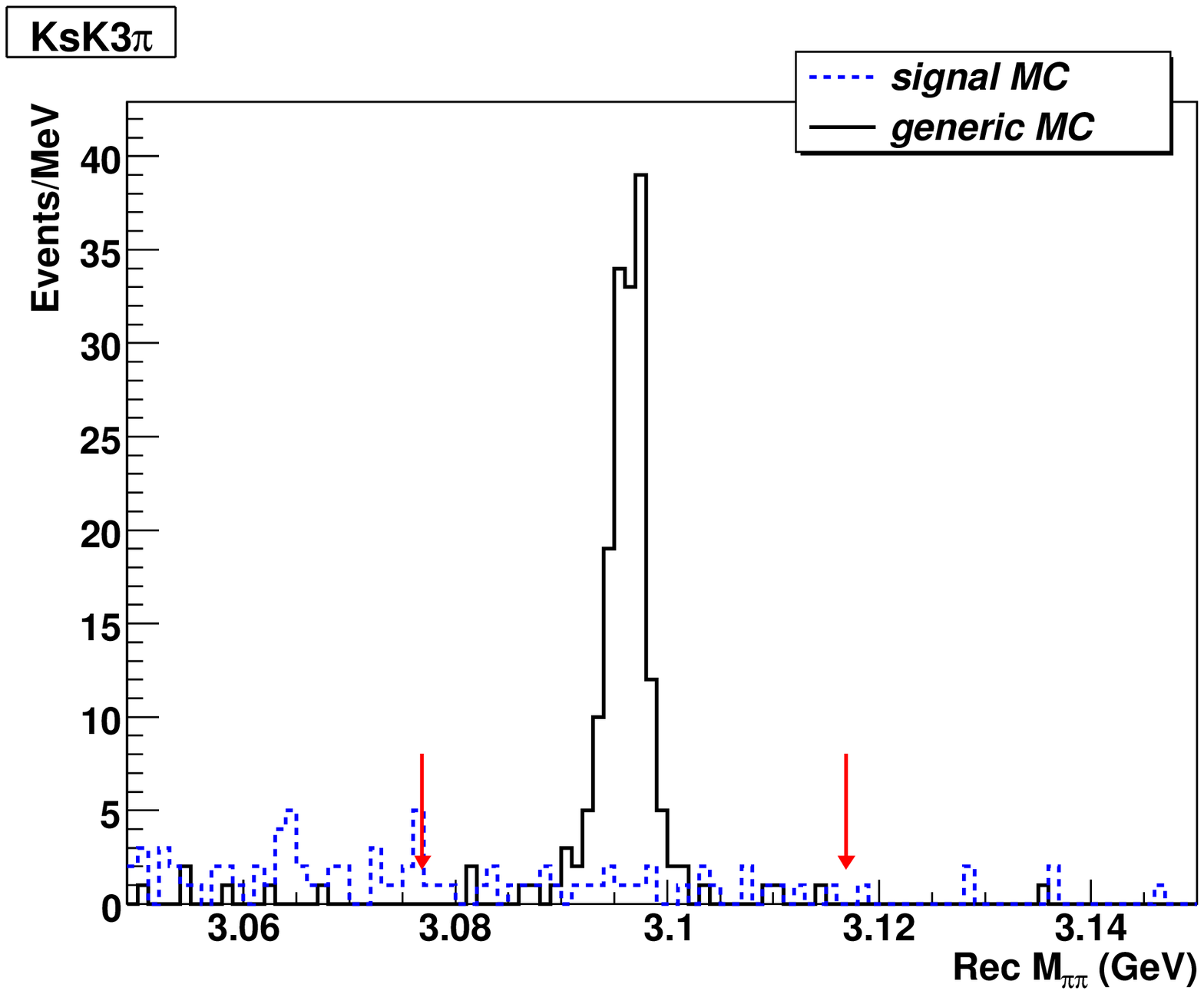}}
  \caption[MC simulations of distributions of $\pi\pi$ recoil mass (3)]
    {MC simulations of distributions of $\pi\pi$ recoil mass \
     for $\psi(2S)$ events with final states $KK4\pi$ (top) and 
     $K_{S}K3\pi$ (bottom).
     \ The solid histogram is the 5 times luminosity background MC samples, 
     while the dashed histogram is signal MC, arbitrarily scaled for clarity. 
     \ The arrows show the selection cuts that were applied.
     \ All other event selection criteria have been applied.}
  \label{fig:cut_ppRecMass3}
\end{figure}

The distributions of $\pi\pi$ recoil mass for modes $4\pi$, $6\pi$, 
$KK\pi\pi$, $KK\pi\pi\pi^{0}$, $KK4\pi$, $K_{S}K3\pi$ before applying this 
cut are shown in Figures~\ref{fig:cut_ppRecMass1} through
\ref{fig:cut_ppRecMass3}.

\subsubsection{$J/\psi$ Rejection - Hadron Invariant Mass}

Even after we rejected events with the $\pi\pi$ recoil mass cuts, 
$J/\psi\to(X~-~2\pi)$ 
decays, where $X$ is a particular final state under study, were 
still observed in the generic $\psi(2S)$ MC sample. \ This 
background can be further 
suppressed by requiring the invariant mass of 
hadrons (excluding the $\pi\pi$ pair) to be displaced from the $J/\psi$ mass. \ 
More specifically, for the modes $X$ = $6\pi$, $KK\pi\pi$, $KK\pi\pi^0$, 
$KK4\pi$, $KsK3\pi$, 
we reject events by requiring $|M(X~-~2\pi) - M(J/\psi)| > 30~{\rm MeV}$. 

\begin{figure}[htbp]
  \centering
  \includegraphics[width=.95\textwidth]{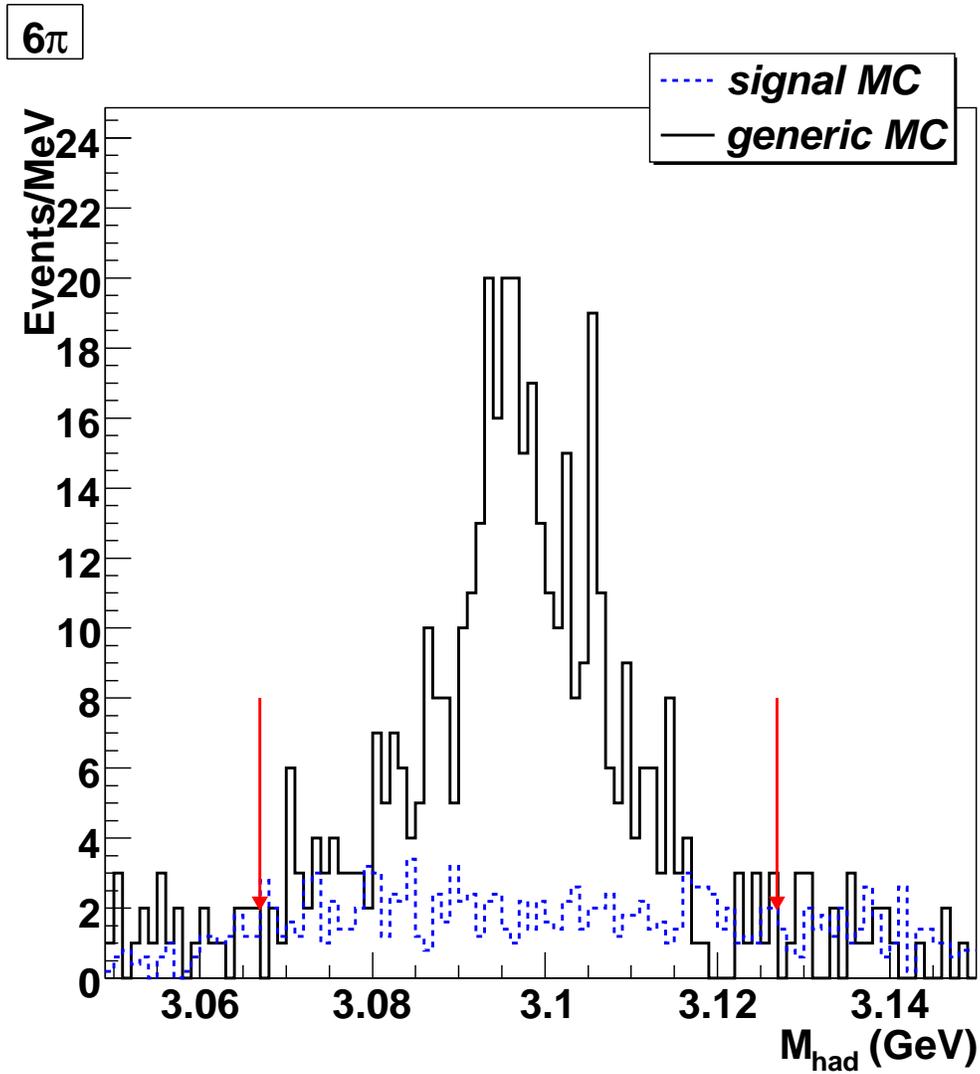}
  \caption[MC simulations of distributions of hadron invariant mass (1)]
    {MC simulations of distributions of hadron invariant mass 
     $M(X~-~2\pi)$ for $\psi(2S)$ events with the final state $6\pi$. \ 
     \ The solid histogram is the 5 times luminosity background MC samples, 
     while the dashed histogram is signal MC, arbitrarily scaled for clarity. 
     \ The arrows show the selection cuts that were applied.
     \ All other event selection criteria have been applied.}
  \label{fig:cut_hadinvMass1}
\end{figure}

\begin{figure}[htbp]
  \centering
  \subfigure
    {\includegraphics[width=.49\textwidth]{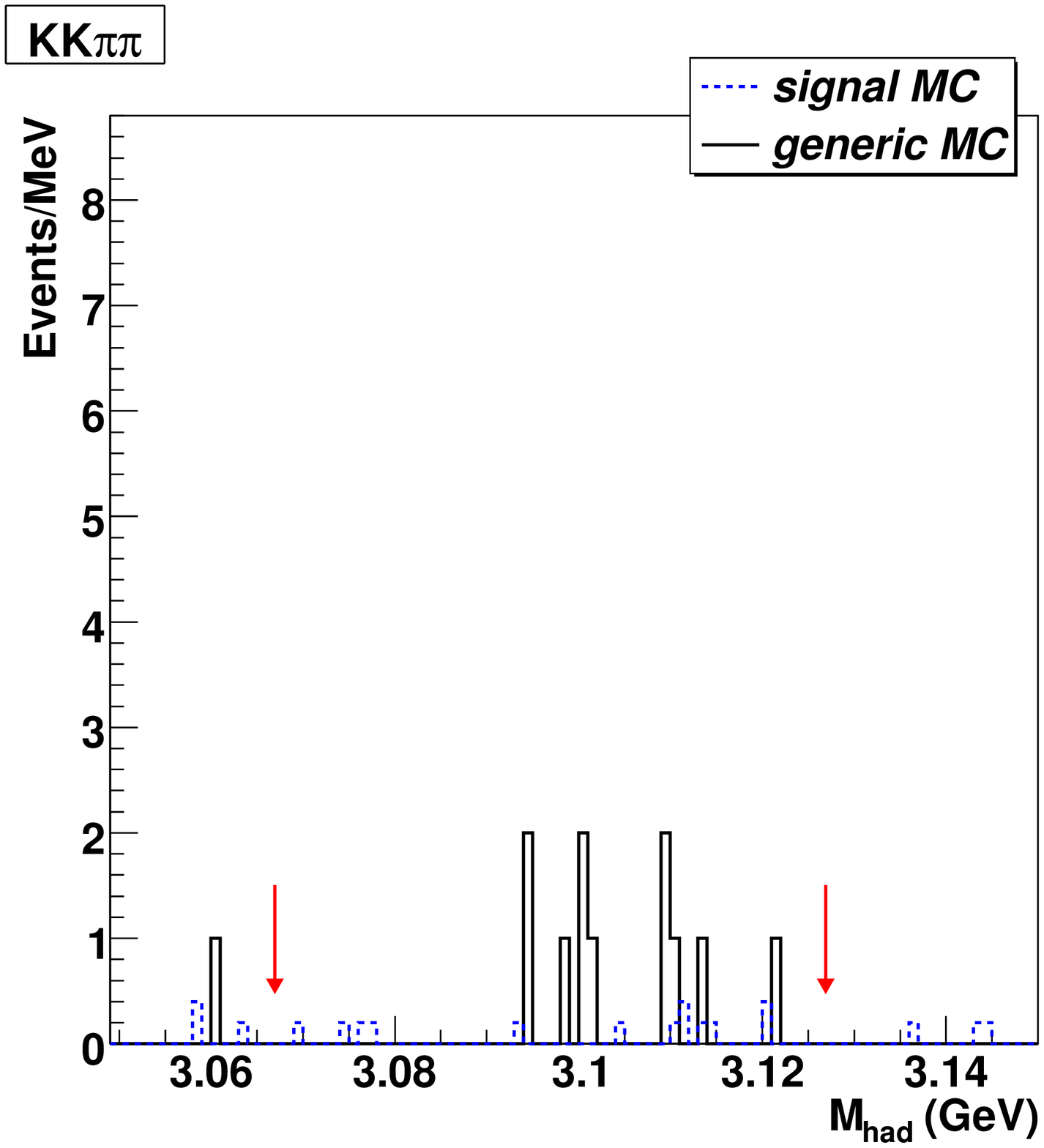}}
  \subfigure
    {\includegraphics[width=.49\textwidth]{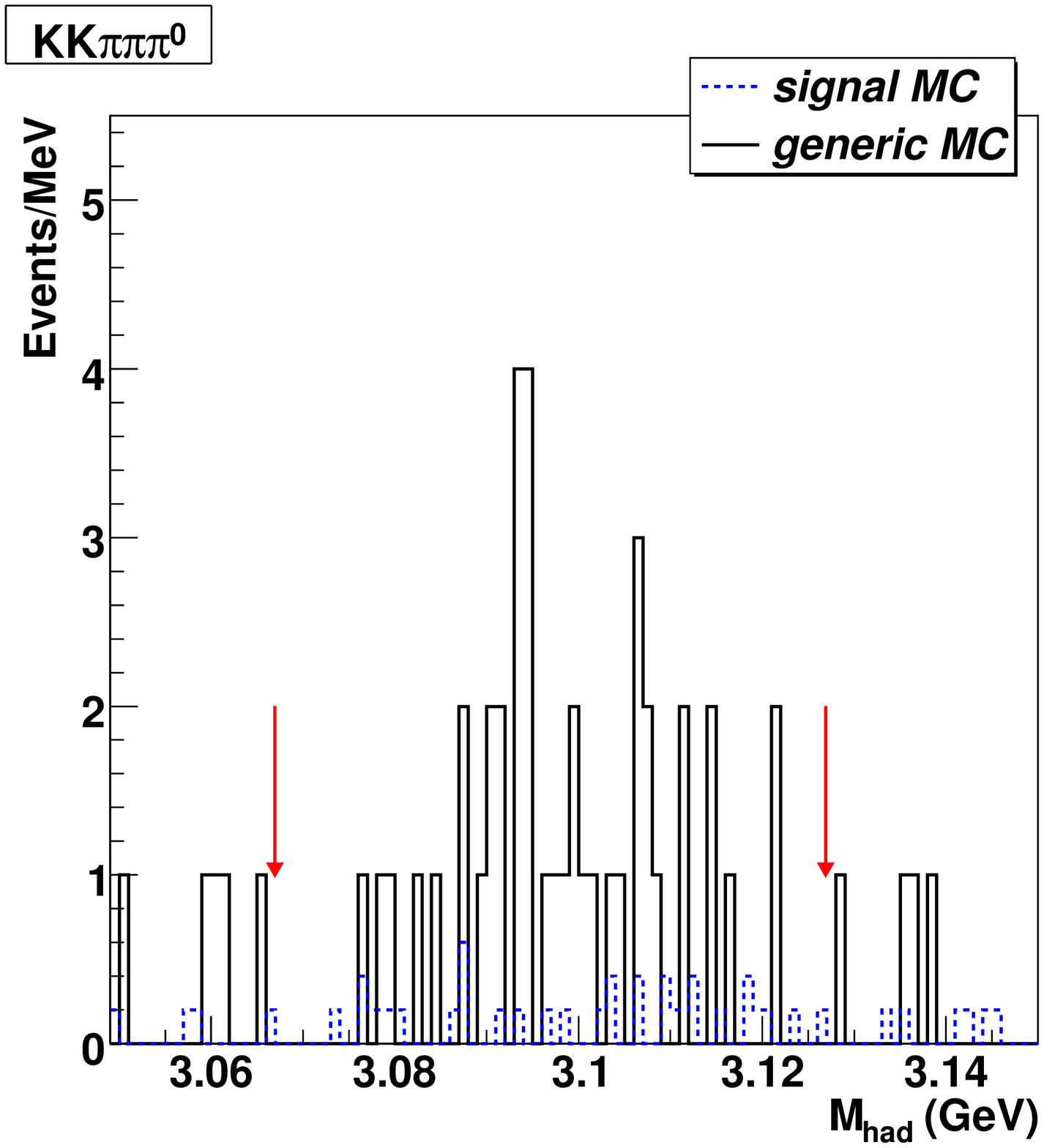}}

  \subfigure
    {\includegraphics[width=.49\textwidth]{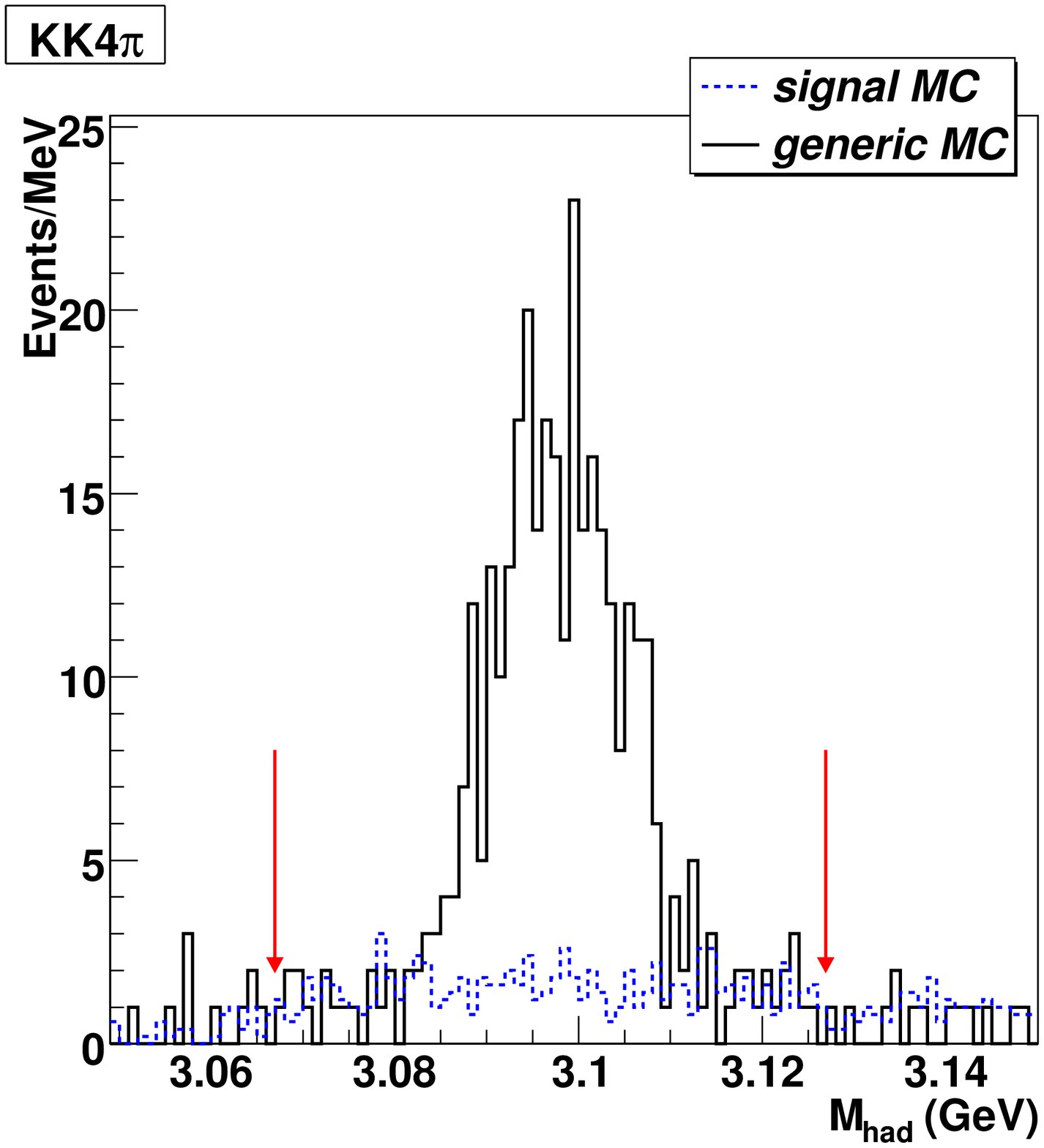}}
  \subfigure
    {\includegraphics[width=.49\textwidth]{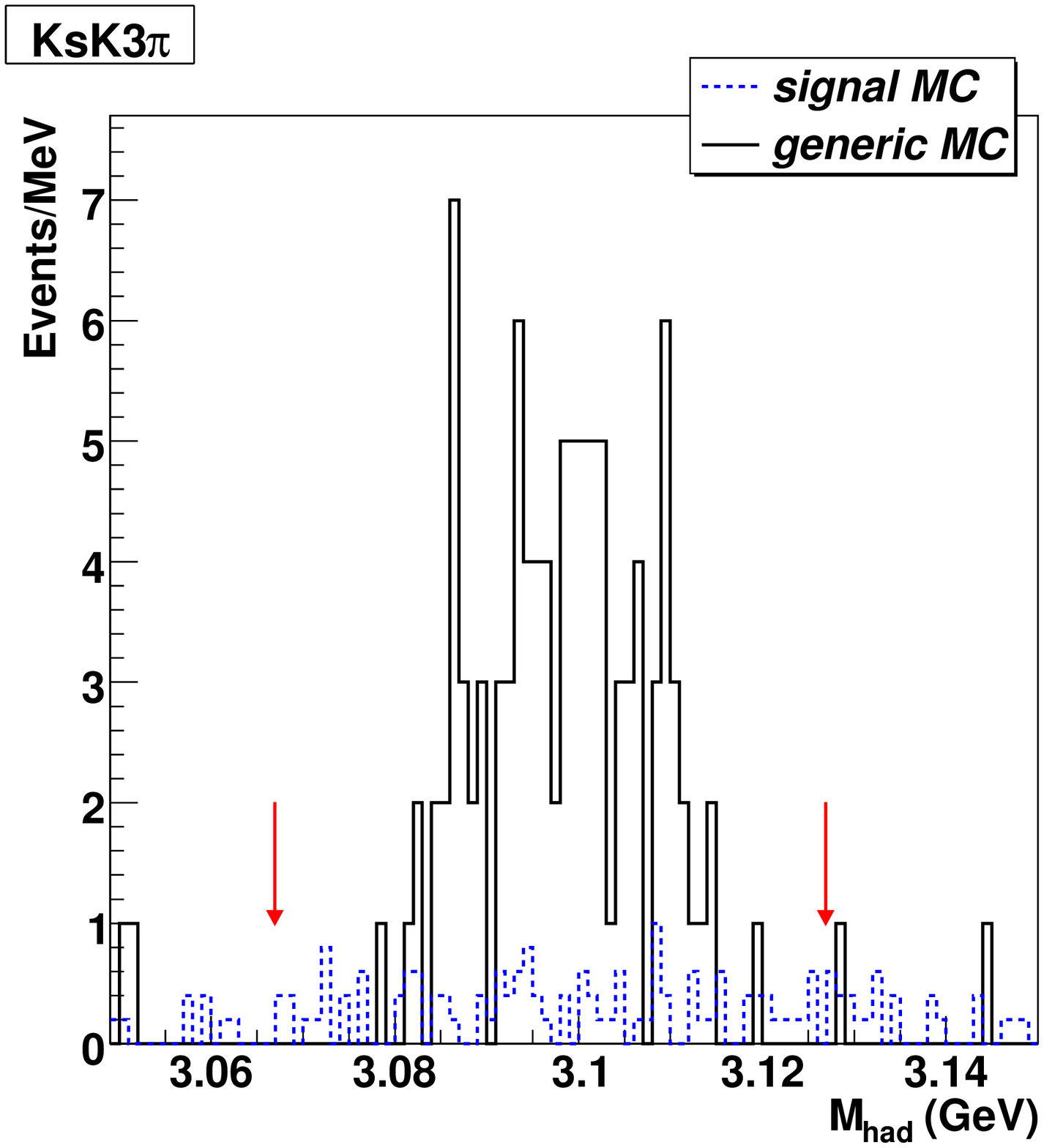}}
  \caption[MC simulations of distributions of hadron invariant mass (2)]
    {MC simulations of distributions of hadron invariant mass 
     $M(X~-~2\pi)$ for $\psi(2S)$ events with final states $KK\pi\pi$ 
     (top left), $KK\pi\pi\pi^{0}$ (top right), $KK4\pi$ (bottom left) and 
     $K_{S}K3\pi$ (bottom right). 
     \ The solid histogram is the 5 times luminosity background MC samples, 
     while the dashed histogram is signal MC, arbitrarily scaled for clarity. 
     \ The arrows show the selection cuts that were applied.
     \ All other event selection criteria have been applied.}
  \label{fig:cut_hadinvMass2}
\end{figure}

Figure~\ref{fig:cut_hadinvMass1} shows the distributions of hadron invariant 
mass for the $6\pi$ mode before applying the cut. \ The distributions for 
other four modes ($KK\pi\pi$, $KK\pi\pi\pi^{0}$, $KK4\pi$, and $K_{S}K3\pi$) 
before applying the cut are shown in Figure~\ref{fig:cut_hadinvMass2}.

\subsection{Photon Background Suppression}
\label{subsubsec:PhDNrTrk}

We investigated background suppression strategies related to the relationship 
between the candidate transition photon and the charged tracks in the event.  

\begin{figure}[htbp]
  \centering
    \includegraphics[width=.95\textwidth]{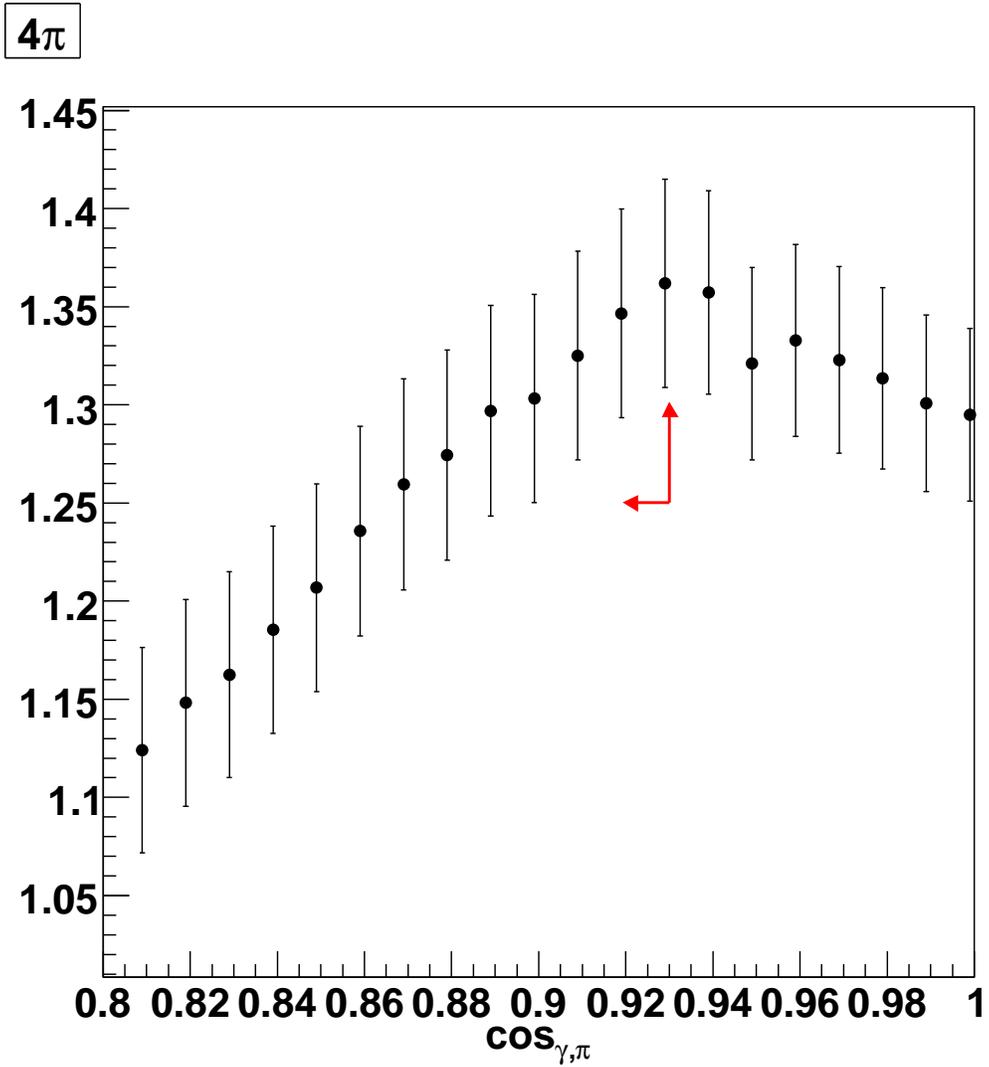}
  \caption[$d_{\gamma, trk}$ or $\cos \theta_{\gamma,\pi}$ cut (1)]
    {MC study of $S^2/(S+B)$ for different cuts on the angle between 
     the transition photon candidate and 
     the closest pion track for the $4\pi$ mode.
     \ The arrow shows the cut value that was selected.}
  \label{fig:cut_PhDNrTrk1}
\end{figure}

\begin{figure}[htbp]
  \centering
  \subfigure
    {\includegraphics[width=.49\textwidth]{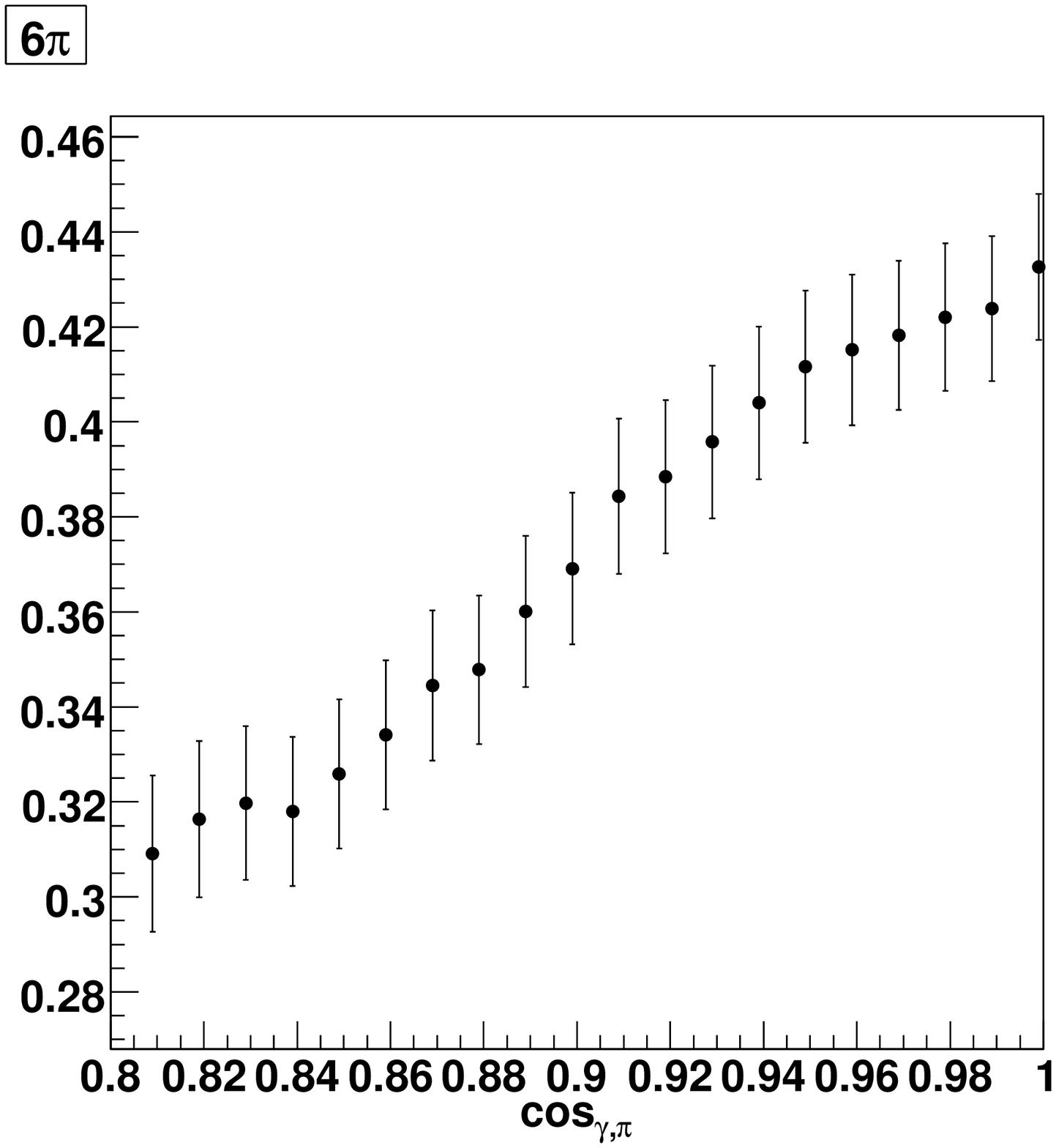}}
  \subfigure
    {\includegraphics[width=.49\textwidth]{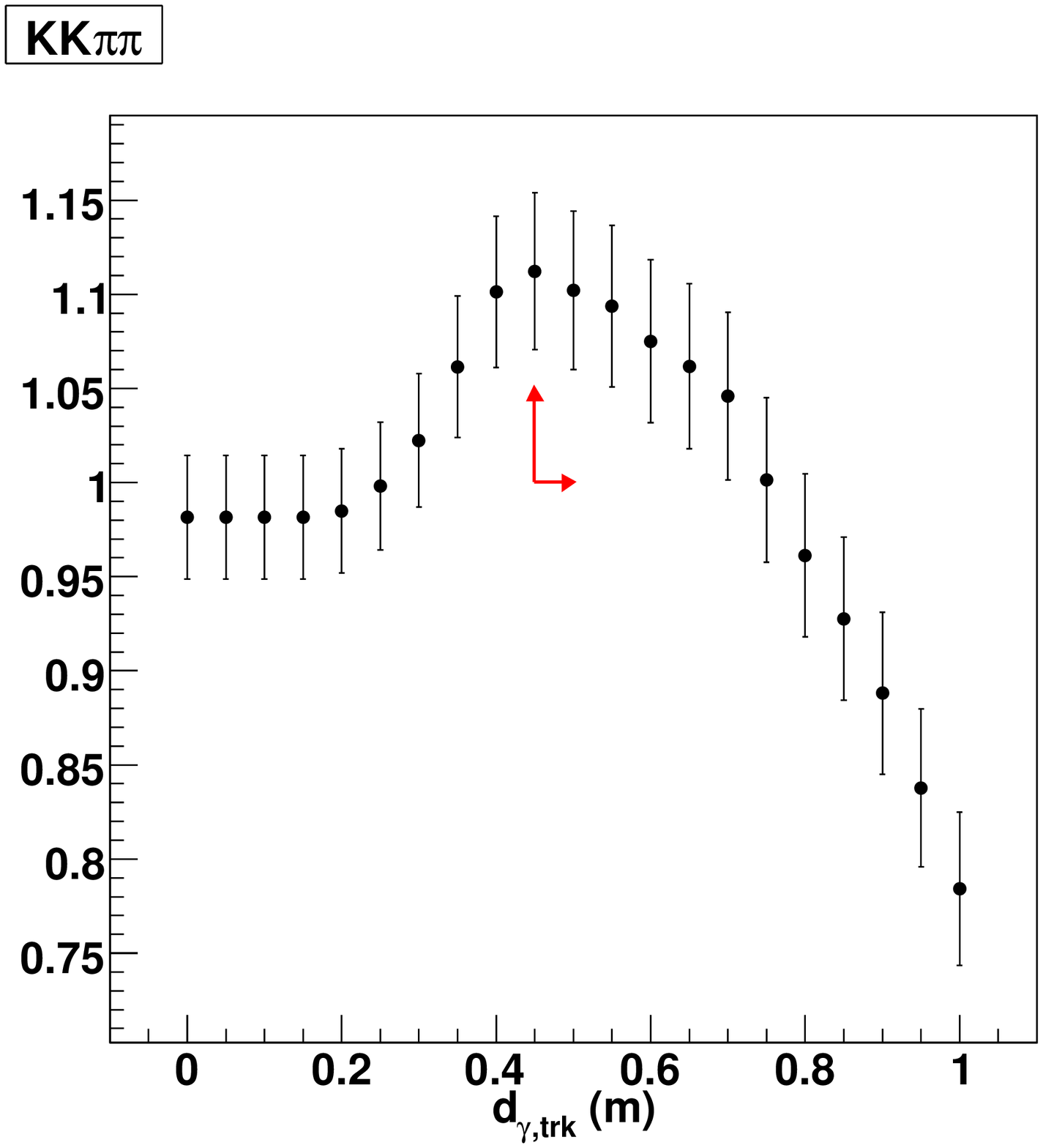}}
  \subfigure
    {\includegraphics[width=.49\textwidth]{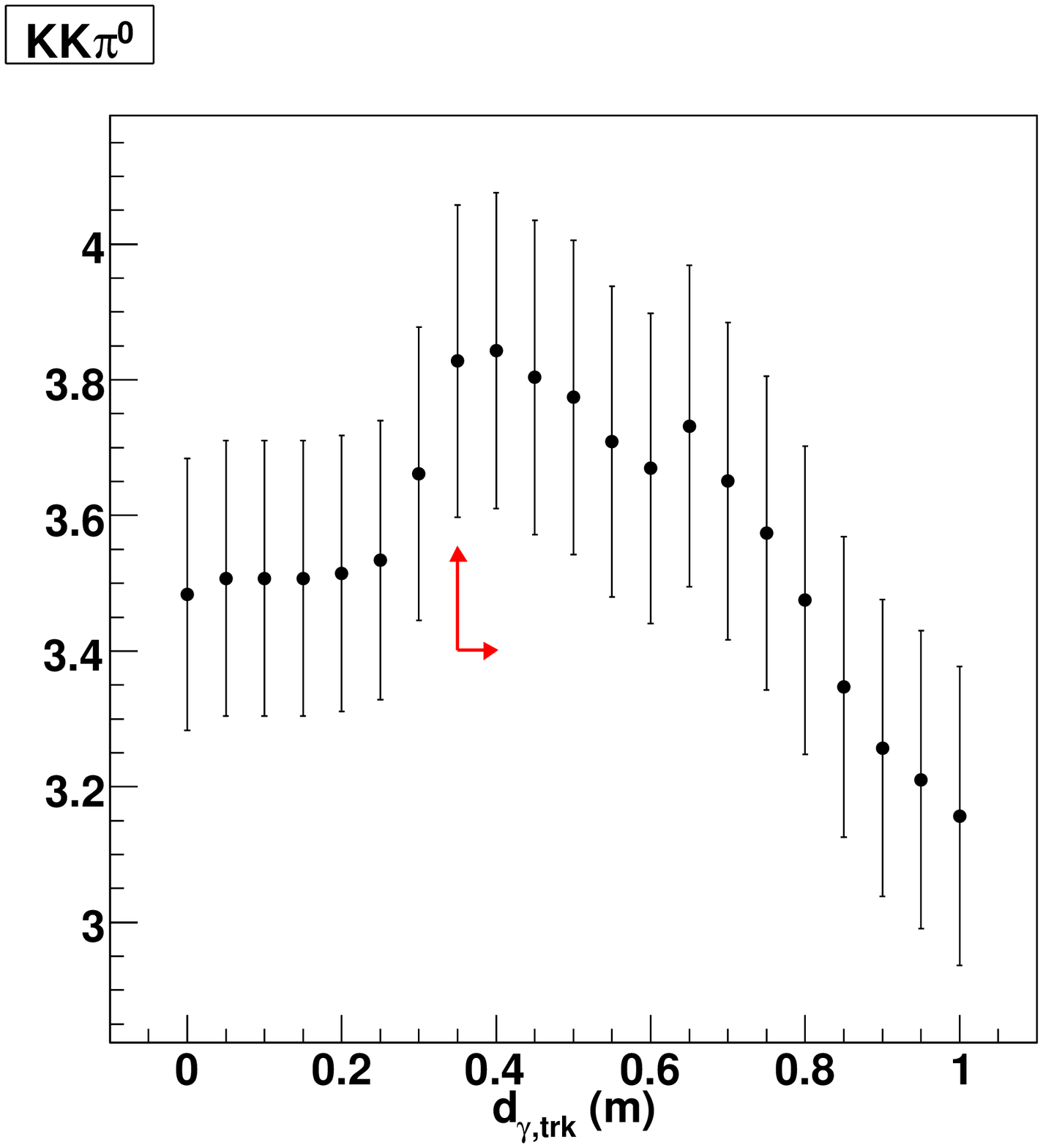}}
  \subfigure
    {\includegraphics[width=.49\textwidth]{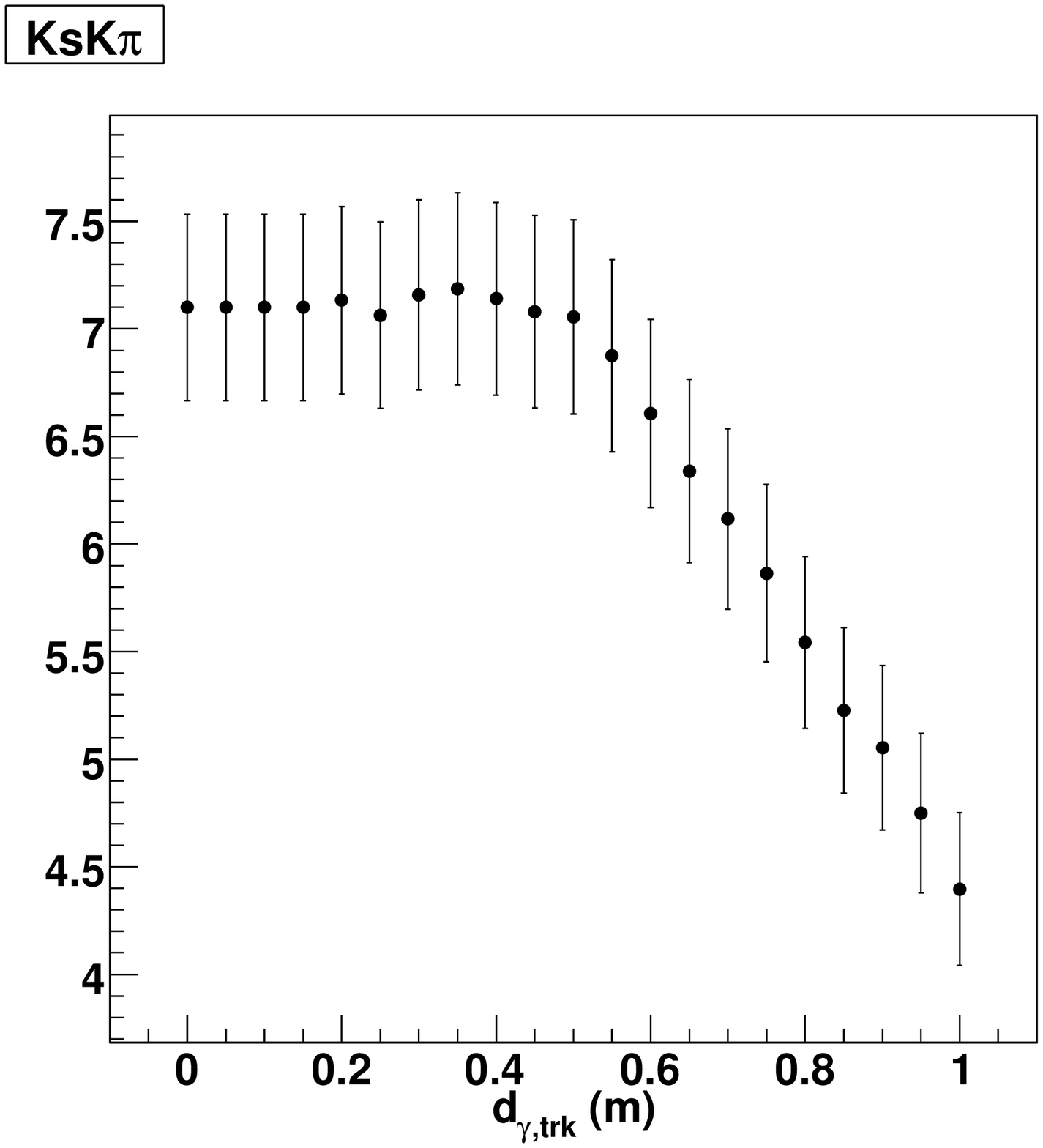}}
  \caption[$d_{\gamma, trk}$ or $\cos \theta_{\gamma,\pi}$ cut (2)]
    {MC study of $S^2/(S+B)$ for different cuts on the distance between a
     candidate transition photon shower and the nearest track
     in the CC or the angle between the transition photon candidate and
     the closest pion track 
     for modes $6\pi$ (top left), $KK\pi\pi$ (top right), $KK\pi^{0}$ (bottom left), 
     and $K_{S}K\pi$ (bottom right).
     \ The arrows show the cut values that were selected.}
  \label{fig:cut_PhDNrTrk2}
\end{figure}

\begin{figure}[htbp]
  \centering
  \subfigure
    {\includegraphics[width=.49\textwidth]{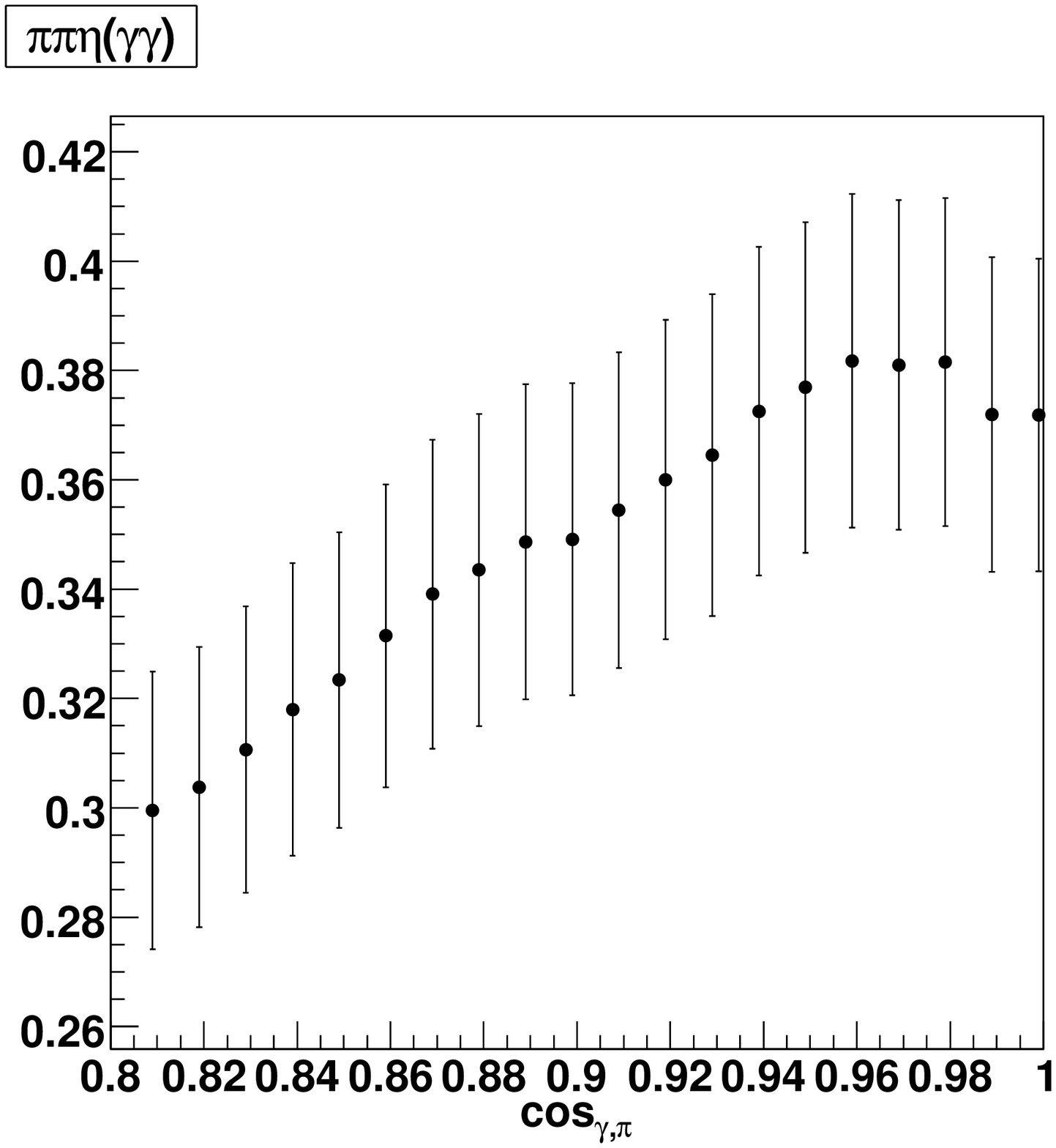}}
  \subfigure
    {\includegraphics[width=.49\textwidth]{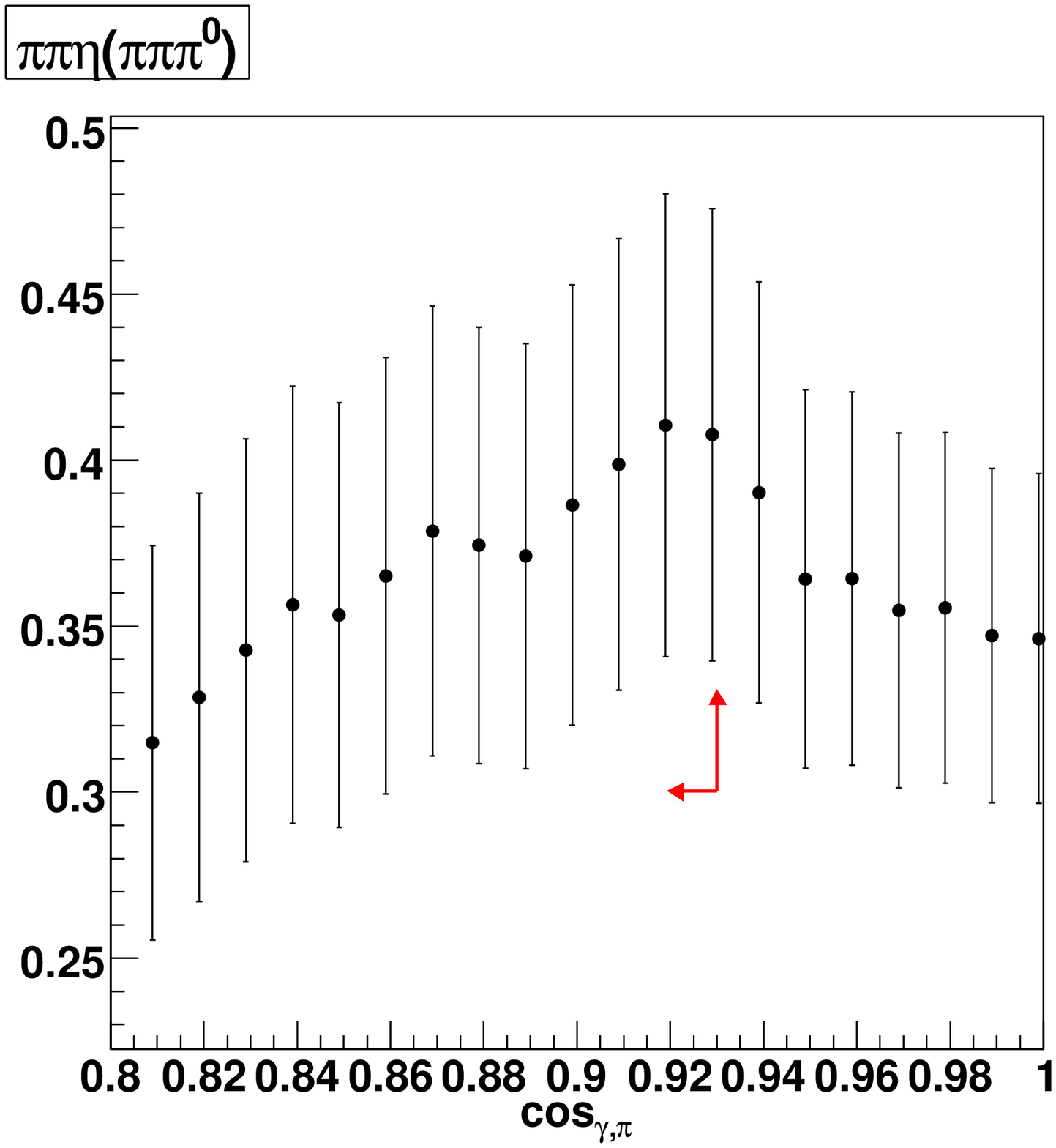}}
  \subfigure
    {\includegraphics[width=.49\textwidth]{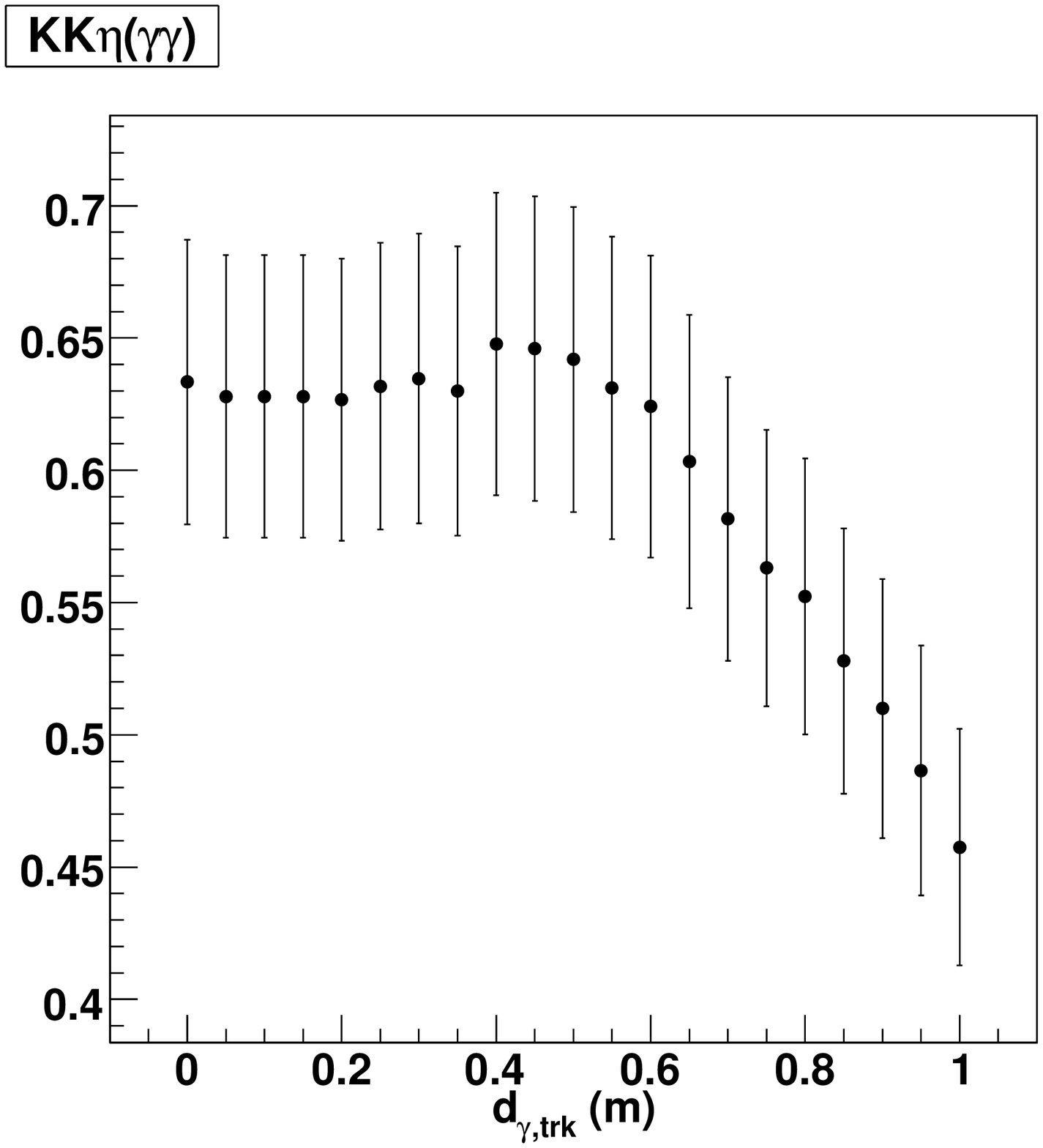}}
  \subfigure
    {\includegraphics[width=.49\textwidth]{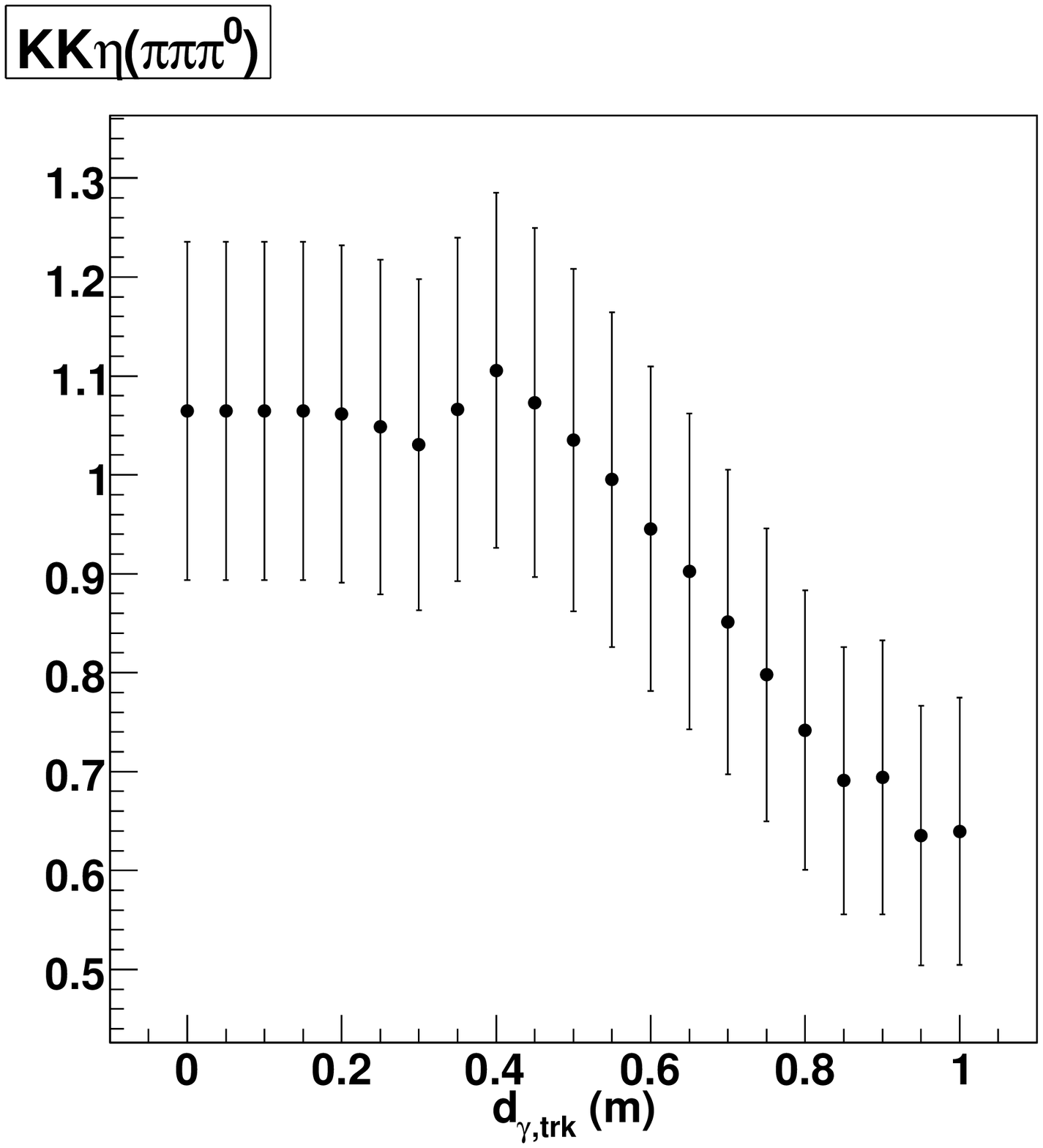}}
  \caption[$d_{\gamma, trk}$ or $\cos \theta_{\gamma,\pi}$ cut (3)]
    {MC study of $S^2/(S+B)$ for different cuts on the distance between a
     candidate transition photon shower and the nearest track projection
     into the CC or the angle between the transition photon candidate and
     the closest pion track 
     for modes $\pi\pi\eta(\gamma\gamma)$ (top left), $\pi\pi\eta(\pi\pi\pi^{0})$ (top right), 
     $KK\eta(\gamma\gamma)$ (bottom left), and $KK\eta(\pi\pi\pi^{0})$ (bottom right).
     \ The arrow shows the cut value that was selected.}
  \label{fig:cut_PhDNrTrk3}
\end{figure}

\begin{figure}[htbp]
  \centering
  \subfigure
    {\includegraphics[width=.49\textwidth]{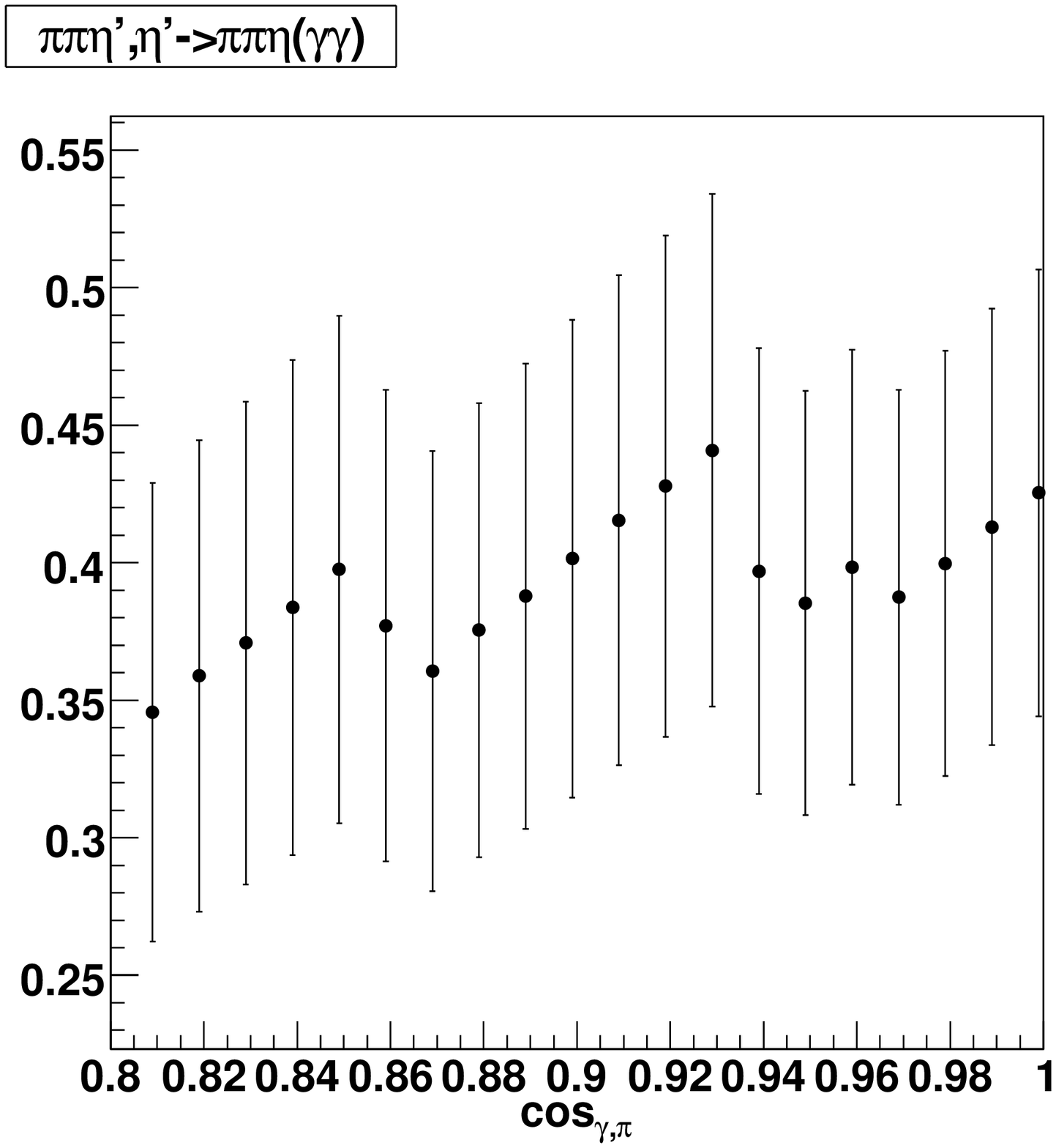}}
  \subfigure
    {\includegraphics[width=.49\textwidth]{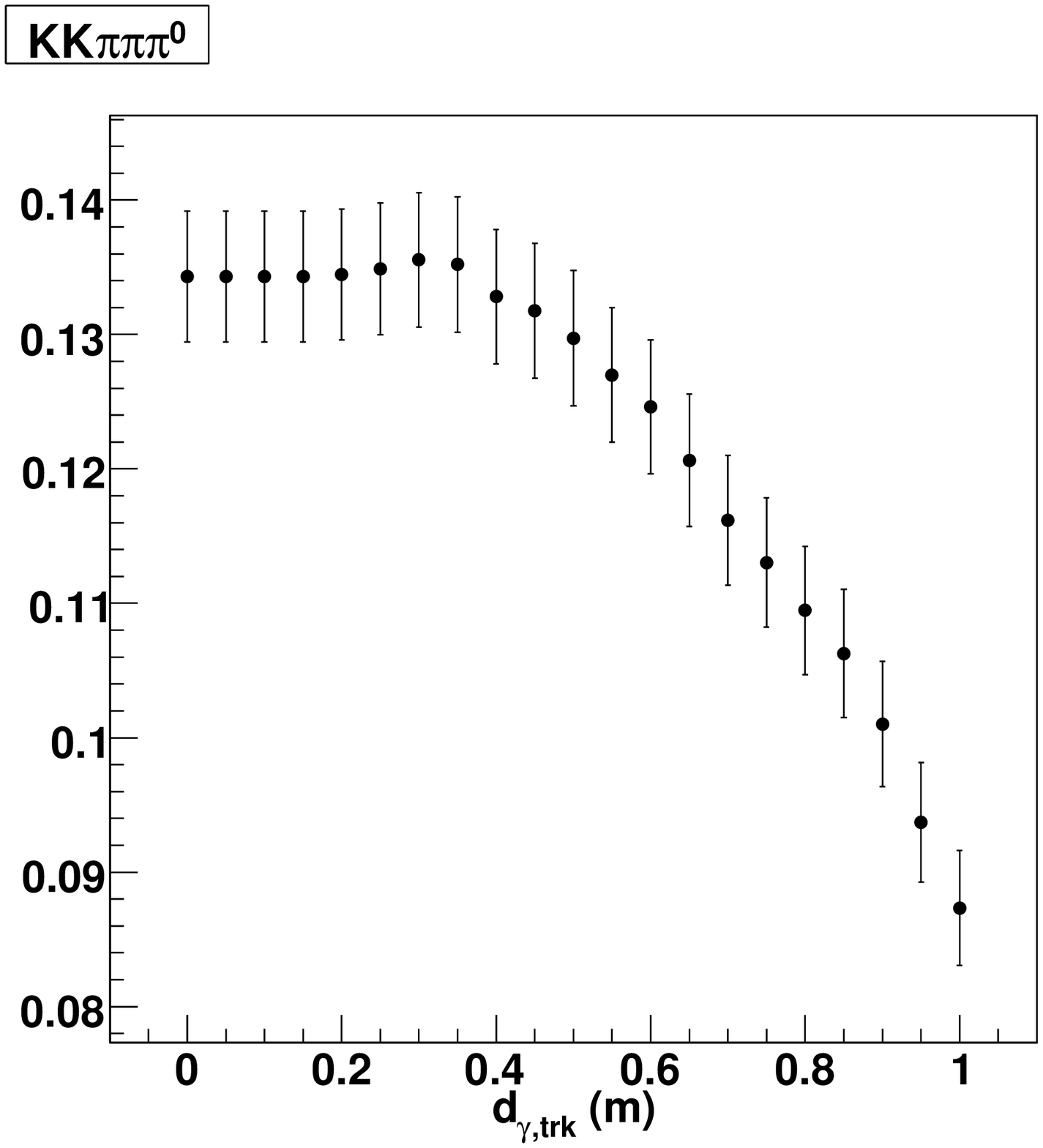}}
  \subfigure
    {\includegraphics[width=.49\textwidth]{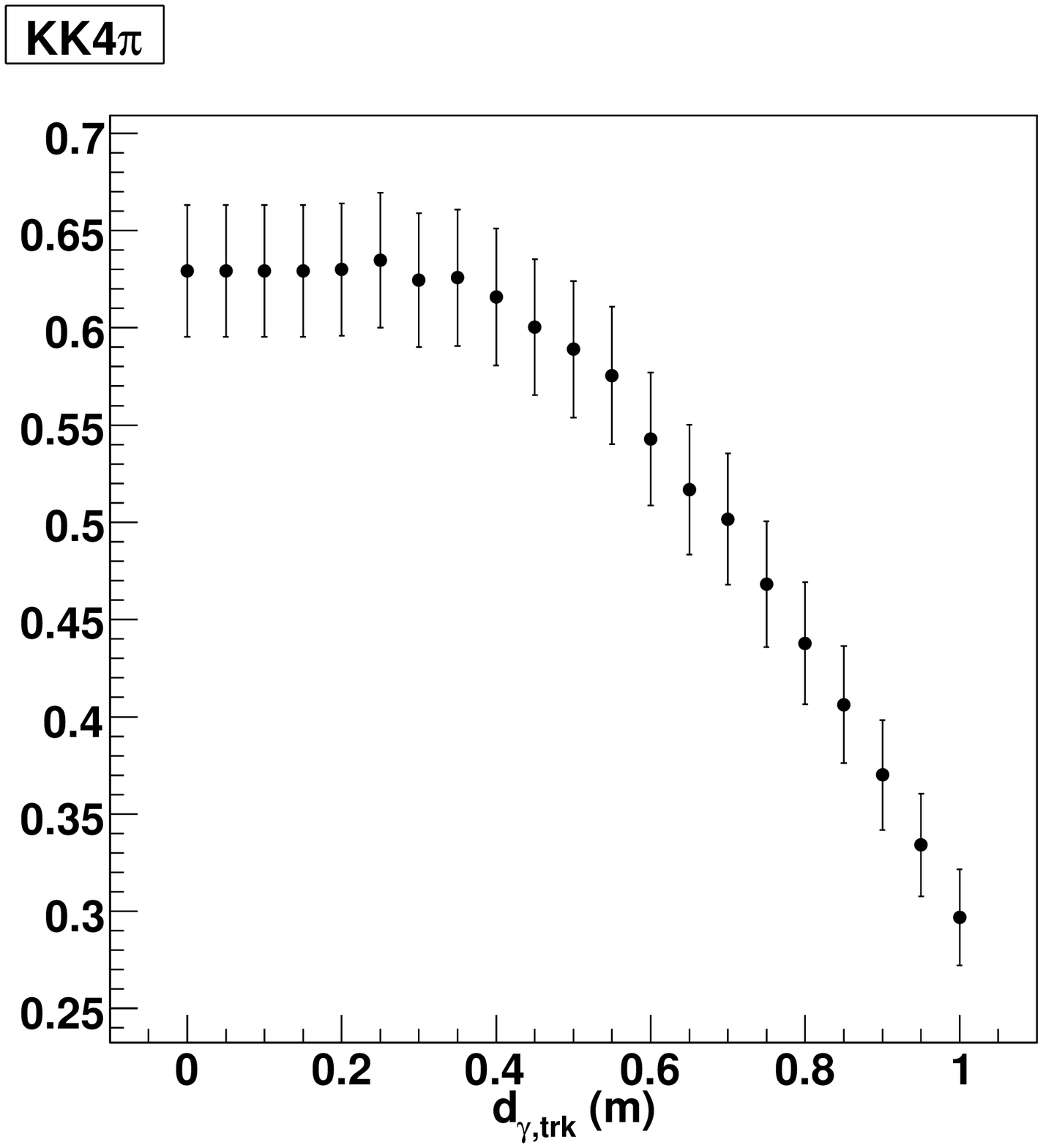}}
  \subfigure
    {\includegraphics[width=.49\textwidth]{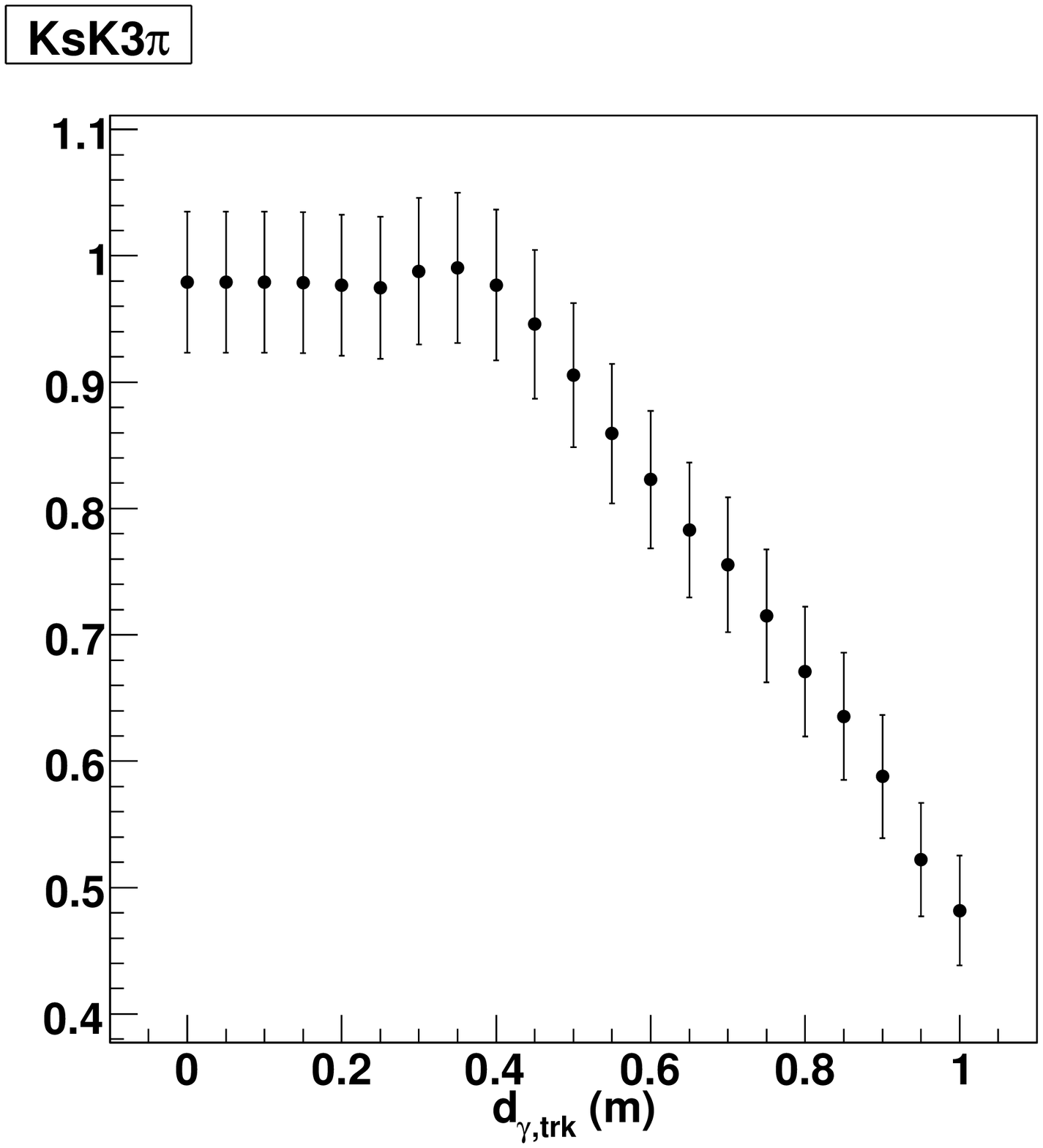}}
  \caption[$d_{\gamma, trk}$ or $\cos \theta_{\gamma,\pi}$ cut (4)]
    {MC study of $S^2/(S+B)$ for different cuts on the distance between a
     candidate transition photon shower and the nearest track projection
     into the CC or the angle between the transition photon candidate and
     the closest pion track
     for modes $\pi\pi\eta^{\prime}$ with $\eta^{\prime}\to\pi\pi\eta(\gamma\gamma)$ (top left),
     $KK\pi\pi\pi^{0}$ (top right), $
     KK4\pi$ (bottom left), and $K_{S}K3\pi$ (bottom right).}
  \label{fig:cut_PhDNrTrk4}
\end{figure}

Photons which are emitted from charged particles are typically in the same 
direction as the initial momentum vector of the charged particle at the IP. \ 
These photons are called final state radiation (FSR). \ 
If a shower is close to a nearby track in the CC, it may be a splitoff shower 
(a shower that split off from a hadronic interaction of a nearby track in 
the CC). \ To identify and remove these backgrounds,
the angle between the transition photon and the initial momentum of
the closest pion ($\cos \theta_{\gamma,\pi}$) and the distance between
the transition photon and the  nearest track ($d_{\gamma, trk}$) were 
studied. \ Although either of these can help remove FSR and 
splitoff showers, a cut on the angle between the 
transition photon and the initial momentum of the closest pion is 
more efficient in suppressing final state radiation and a cut on the 
distance between the transition photon and the nearest track is more 
efficient in removing splitoff showers. \ In this analysis, the 
following rules were applied while making the decision on which of these two 
cuts would be applied to a mode:
\begin{itemize}
\item
 If the hadronic decay only contained charged pions, cut on the angle
 between the initial pion momentum and the candidate transition photon.
\item
 If the hadronic decay only contained charged kaons, cut on the distance
 of the closest track to the candidate transition photon.
\item
 If the hadronic decay contained both charged pions and kaons, cut on the
 distance of the closest track to the candidate transition photon, then we
 examined if a cut on the angle between initial pion momentum and the
 candidate transition photon improves the $S^{2}/(S+B)$.  
 (We did not apply both types of cuts to the same mode.)
\end{itemize}
We optimized the selection criteria by evaluating the figure of merit 
$S^2/(S+B)$ at different cut values. \ The value of $S$ was from the signal 
MC sample with the assumption that 
${\cal B}(\psi(2S)\to\gamma\eta_c(2S)) \times 
{\cal B}(\eta_c(2S) \to X)  = 2.6\times10^{-6}$ 
for all $\eta_{c}(2S)$ decay modes. \ The value of $B$ was from the sum of the 
5 times generic $\psi(2S)$ and continuum MC samples scaled to the true
luminosity of data. \ The plot of $S^2/(S+B)$ versus cut value was 
examined to determine the optimal cut value. \ An example of the 
$S^2/(S+B)$ is shown in Figure~\ref{fig:cut_PhDNrTrk1} for the $4\pi$ 
mode. \ The $S^2/(S+B)$ for the other modes are shown in 
Figures~\ref{fig:cut_PhDNrTrk2} through \ref{fig:cut_PhDNrTrk4}. \ 
The optimized cut for each mode is listed in Table~\ref{table:cut_kinchi2}.

\subsection{Global Event Selection}

\subsubsection{4-Constraint Kinematic Fit $\chi^2/$dof}

\begin{figure}[htbp]
  \centering
    \includegraphics[width=.95\textwidth]{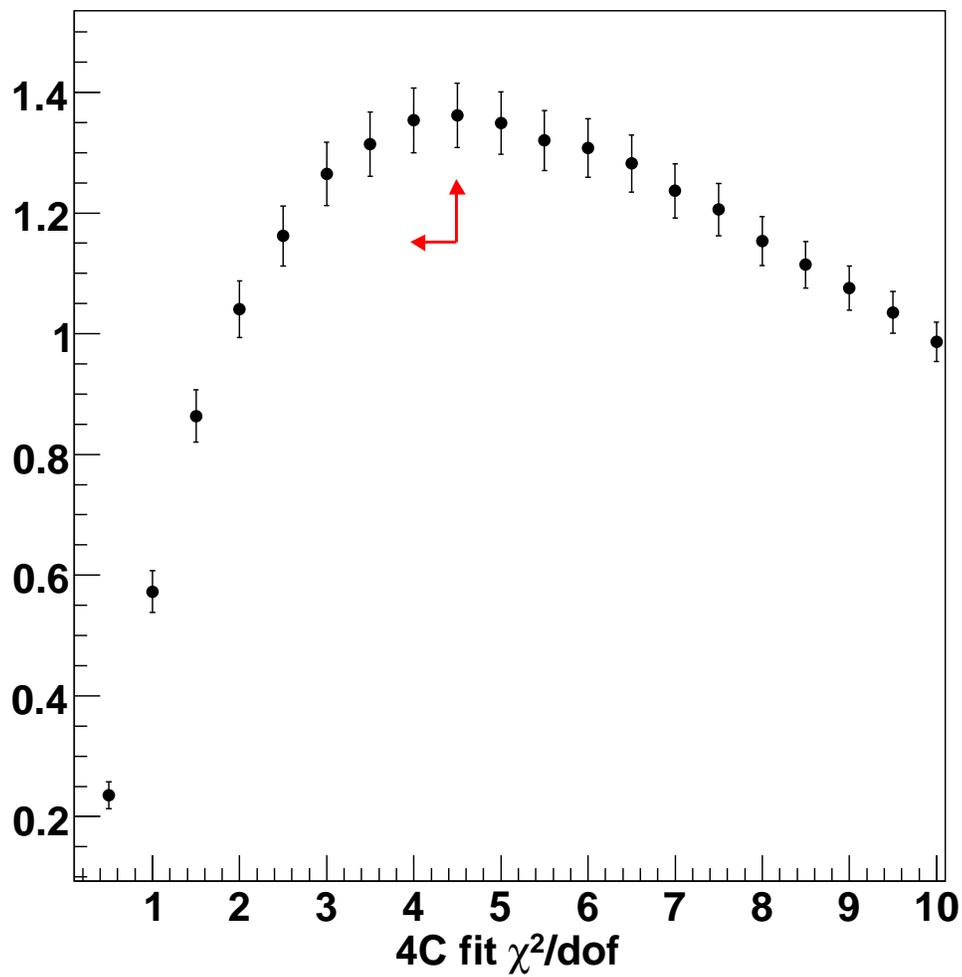}
  \caption[Four momentum kinematic fit $\chi^2/$dof cut (1)]
    {MC study of $S^2/(S+B)$ for different cuts on the 4-momentum 
     kinematic fit $\chi^2/$dof for the $4\pi$ mode.
     \ The arrow shows the cut value that was selected.}
  \label{fig:cut_KinPFRedChi2Fit1}
\end{figure}

\begin{figure}[htbp]
  \centering
  \subfigure
    {\includegraphics[width=.49\textwidth]{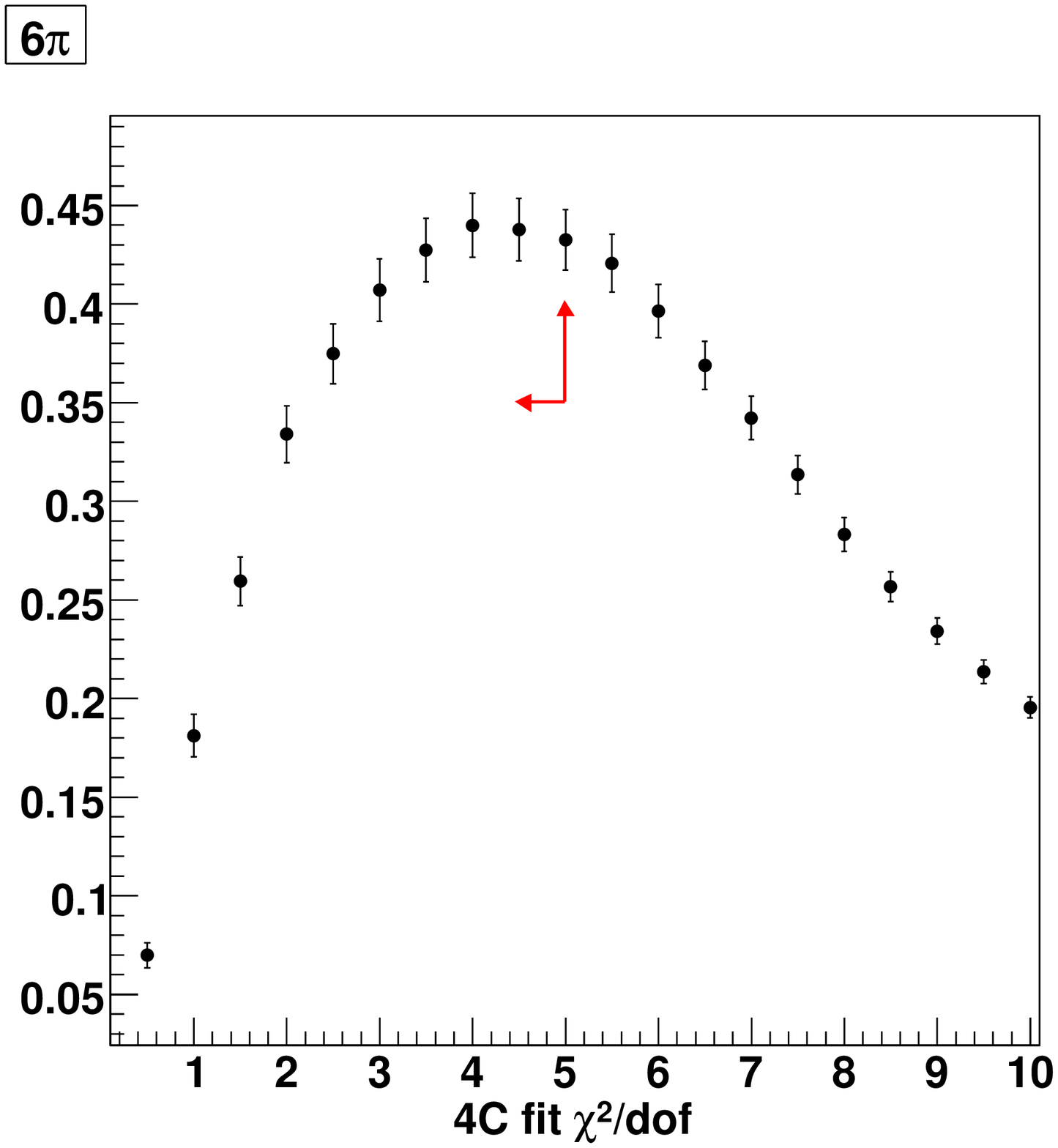}}
  \subfigure
    {\includegraphics[width=.49\textwidth]{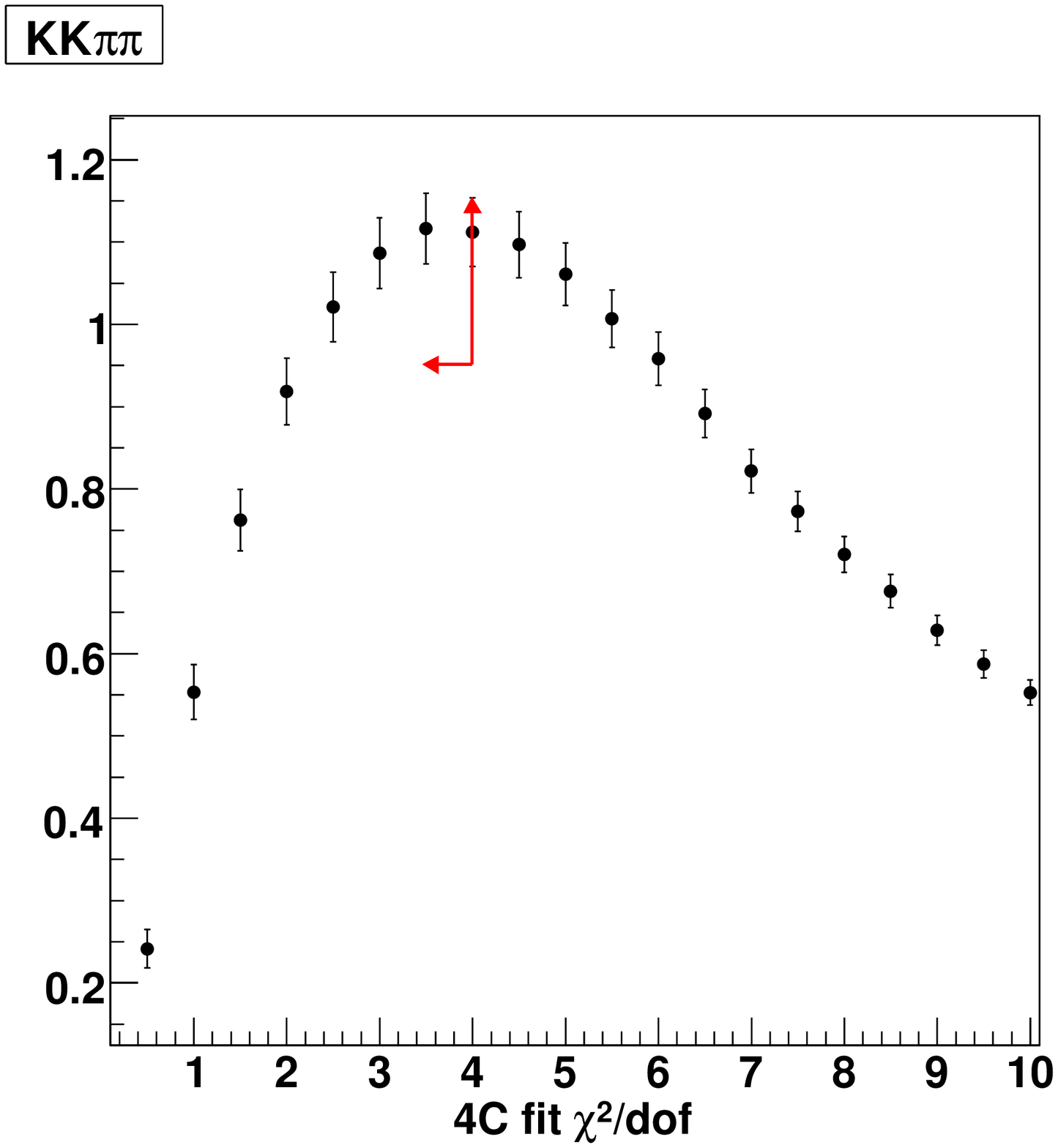}}
  \subfigure
    {\includegraphics[width=.49\textwidth]{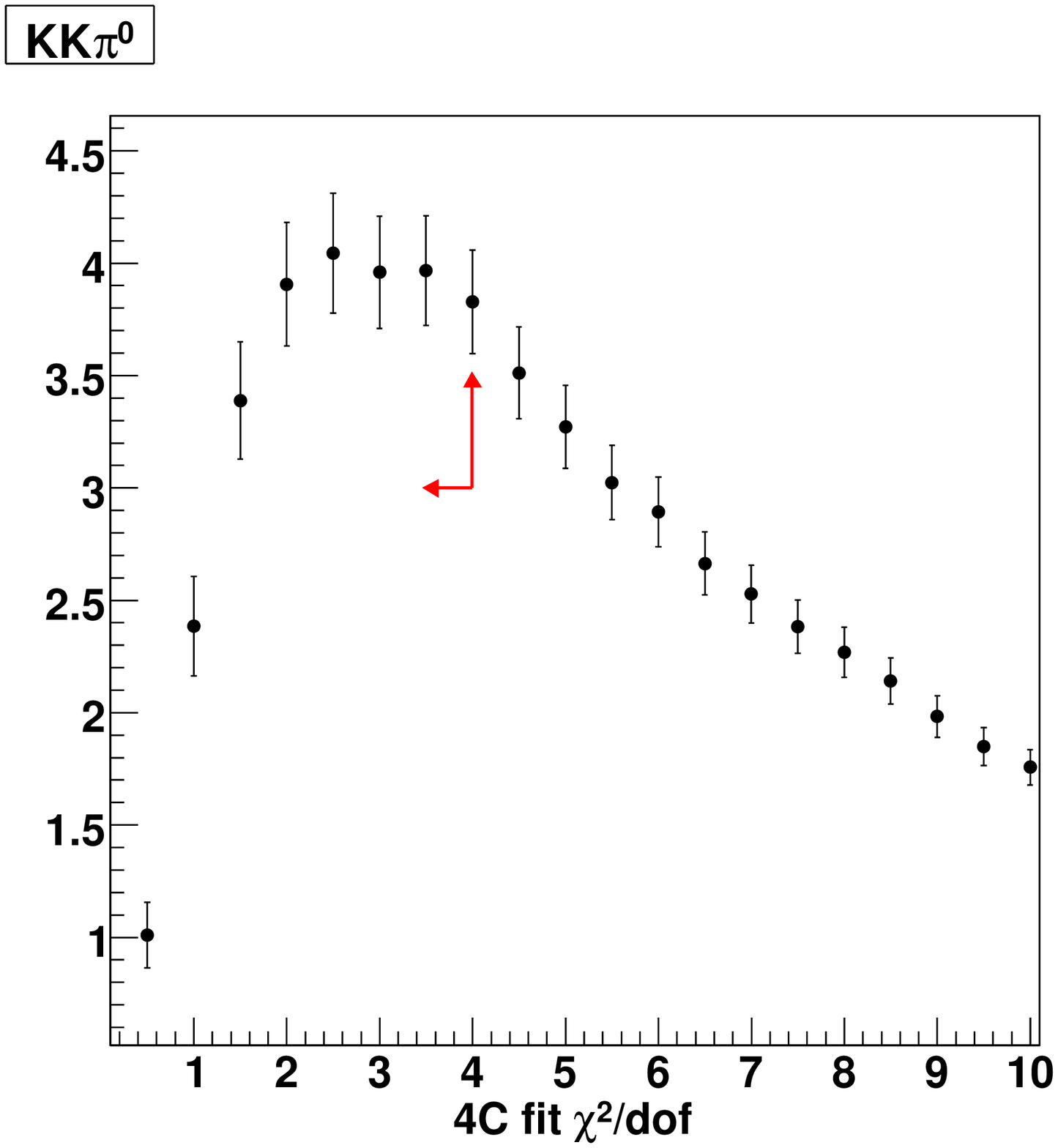}}
  \subfigure
    {\includegraphics[width=.49\textwidth]{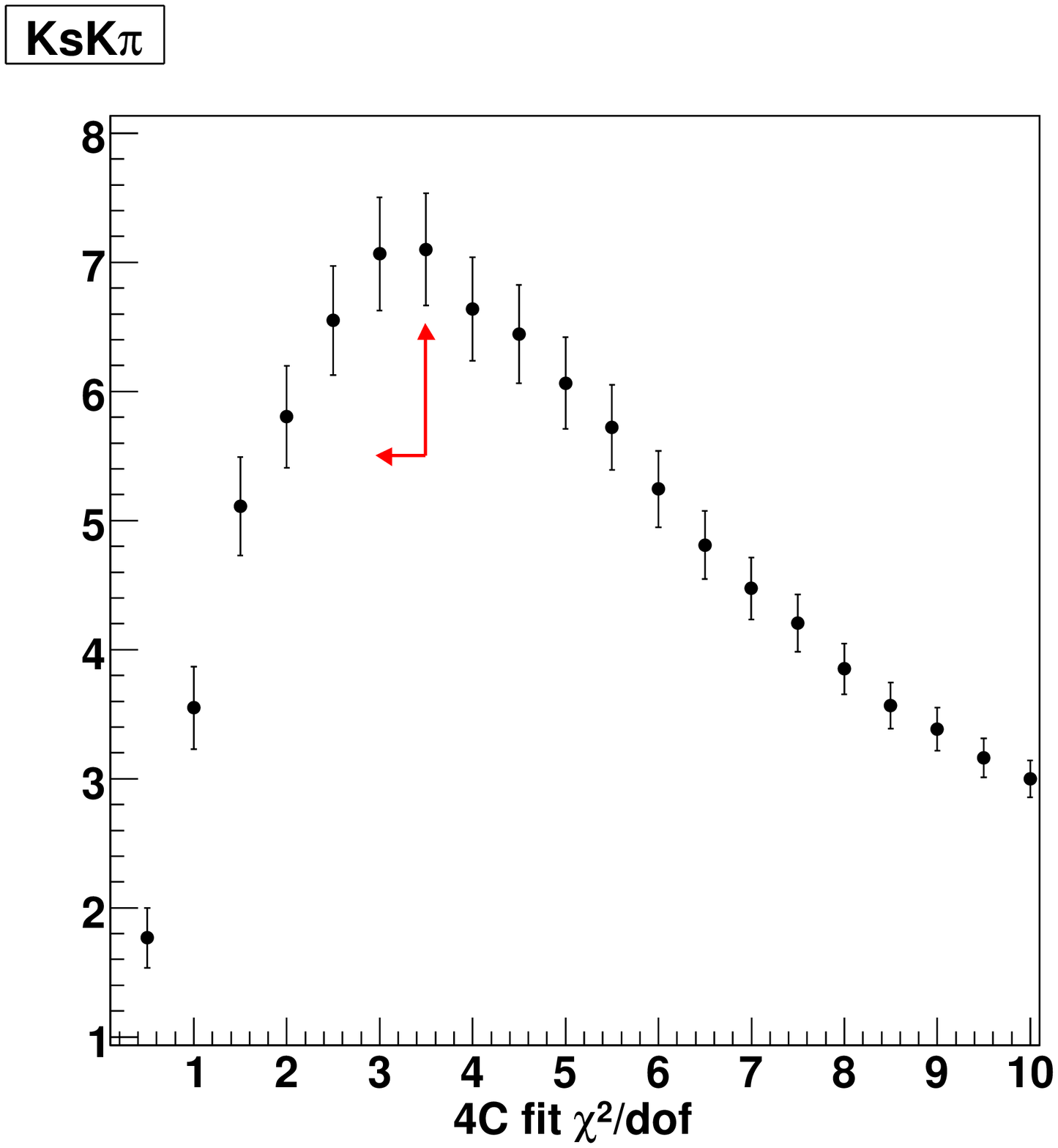}}
  \caption[Four momentum kinematic fit $\chi^2/$dof cut (2)]
    {MC study of $S^2/(S+B)$ for different cuts on the 4-momentum
     kinematic fit $\chi^2/$dof for modes $6\pi$ (top left), $KK\pi\pi$ (top right), 
     $KK\pi^{0}$ (bottom left), and $K_{S}K\pi$ (bottom right).
     \ The arrows show the cut values that were selected.}
  \label{fig:cut_KinPFRedChi2Fit2}
\end{figure}

\begin{figure}[htbp]
  \centering
  \subfigure
    {\includegraphics[width=.49\textwidth]{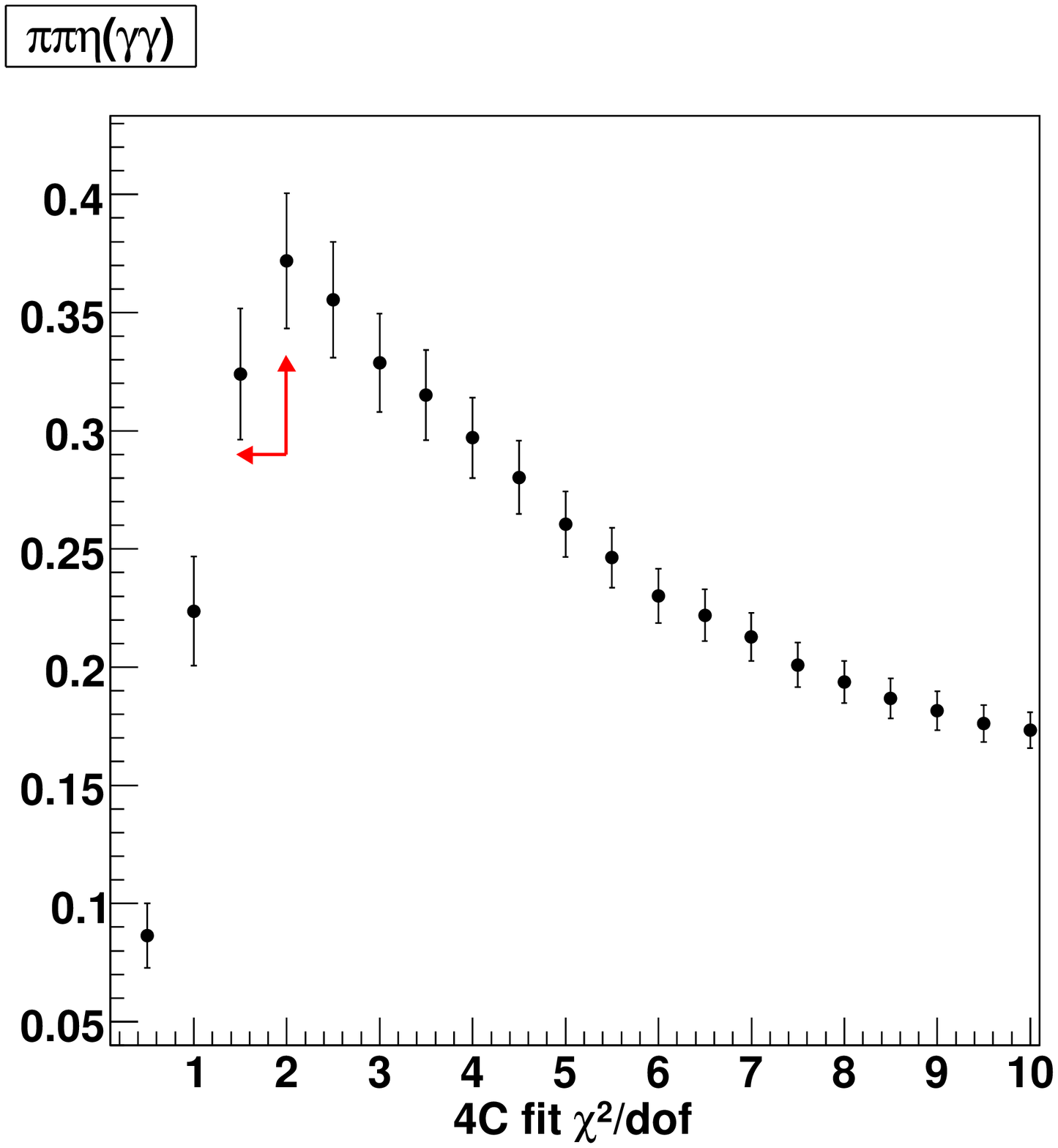}}
  \subfigure
    {\includegraphics[width=.49\textwidth]{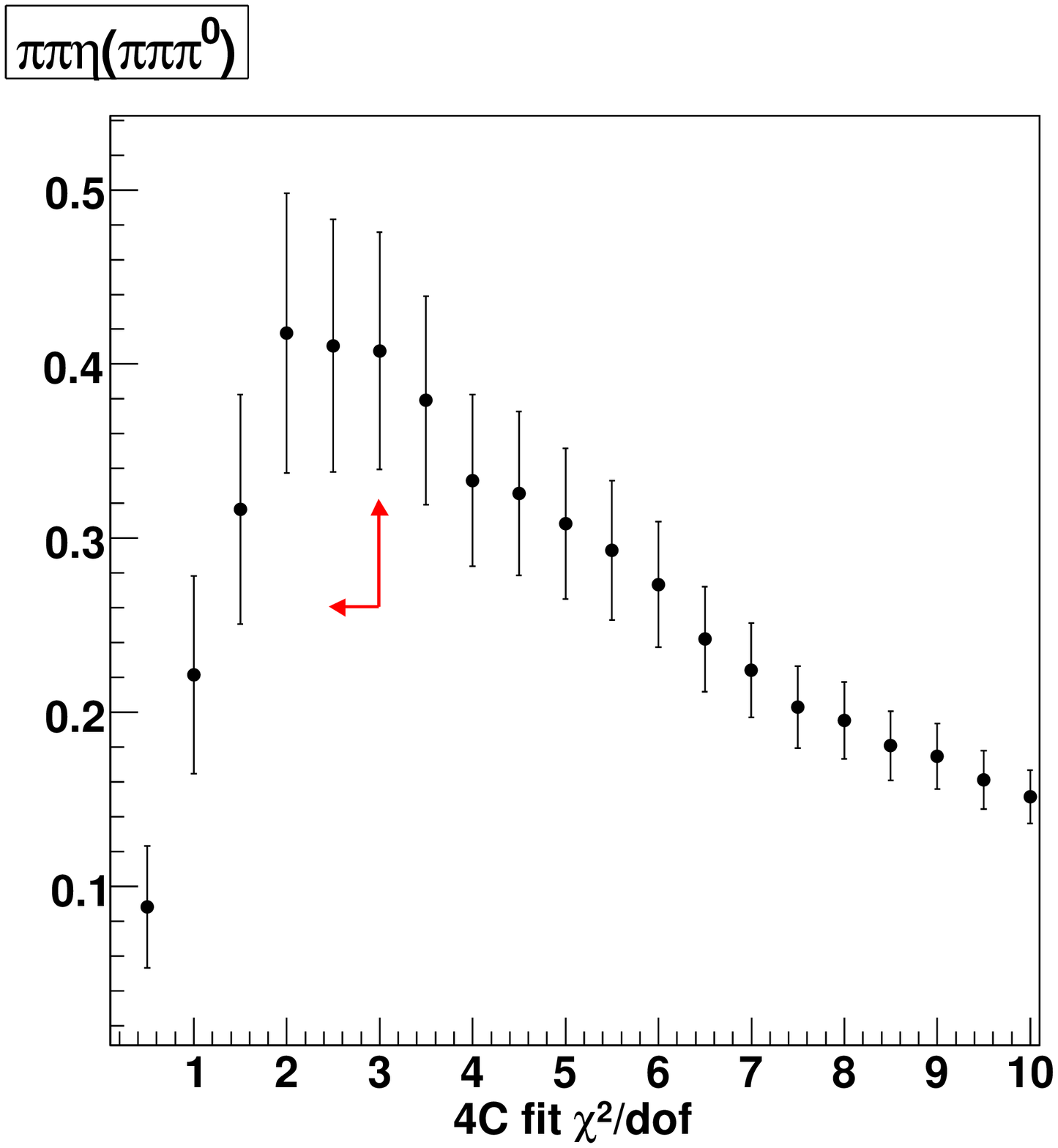}}
  \subfigure
    {\includegraphics[width=.49\textwidth]{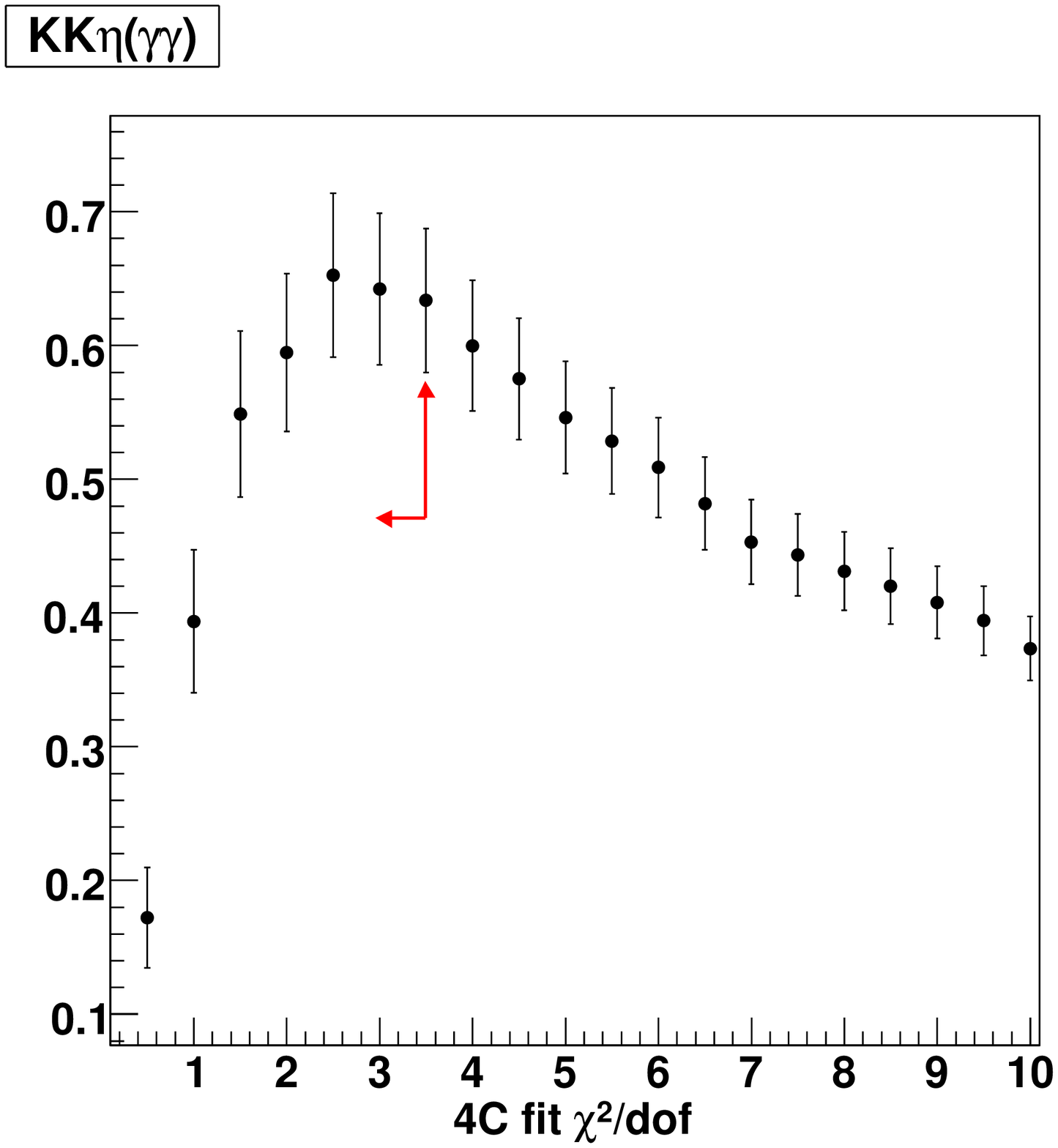}}
  \subfigure
    {\includegraphics[width=.49\textwidth]{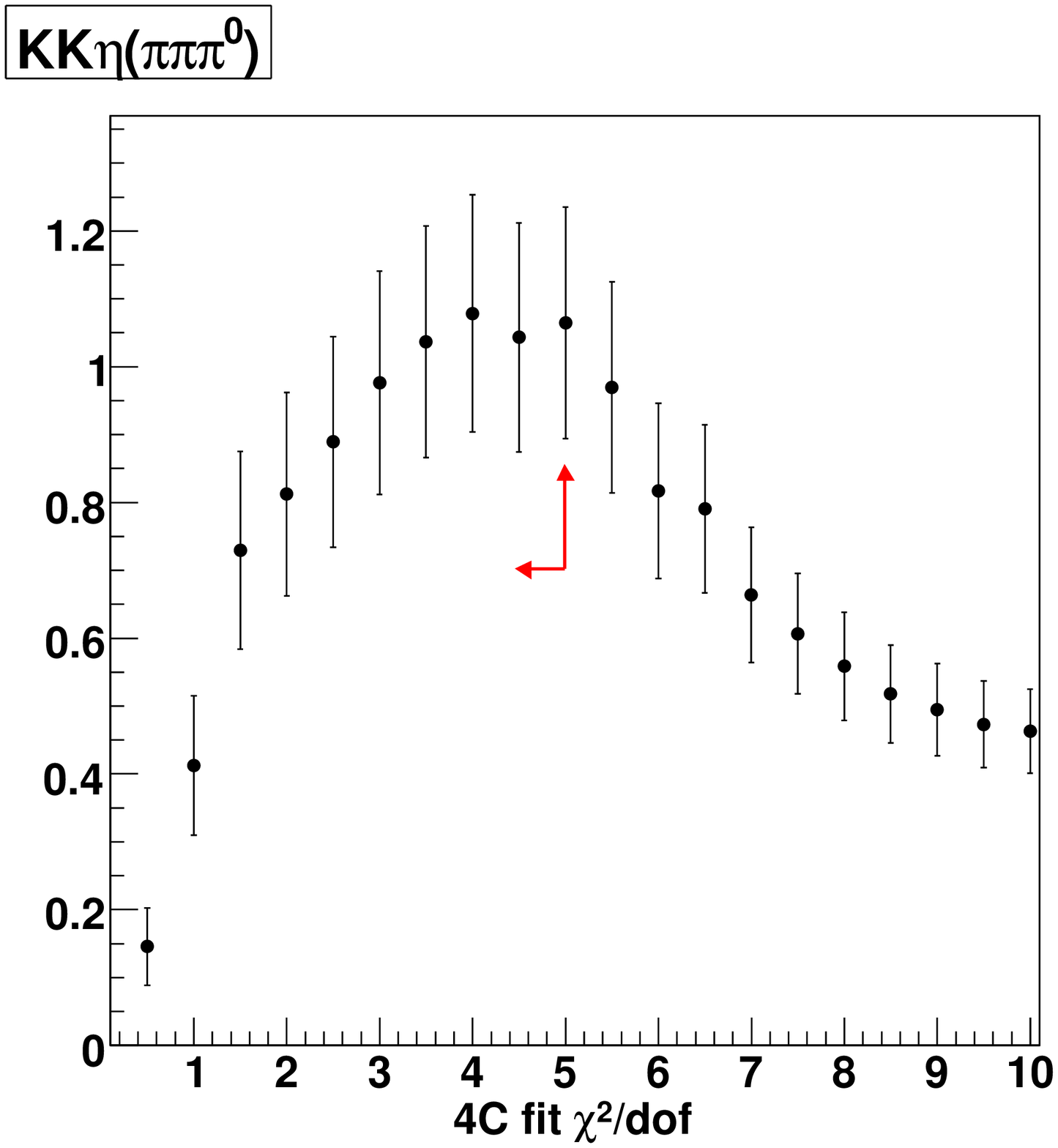}}
  \caption[Four momentum kinematic fit $\chi^2/$dof cut (3)]
    {MC study of $S^2/(S+B)$ for different cuts on the 4-momentum
     kinematic fit $\chi^2/$dof 
     for modes $\pi\pi\eta(\gamma\gamma)$ (top left), $\pi\pi\eta(\pi\pi\pi^{0})$ (top right),
     $KK\eta(\gamma\gamma)$ (bottom left), and $KK\eta(\pi\pi\pi^{0})$ (bottom right).
     \ The arrows show the cut values that were selected.}
  \label{fig:cut_KinPFRedChi2Fit3}
\end{figure}

\begin{figure}[htbp]
  \centering
  \subfigure
    {\includegraphics[width=.49\textwidth]{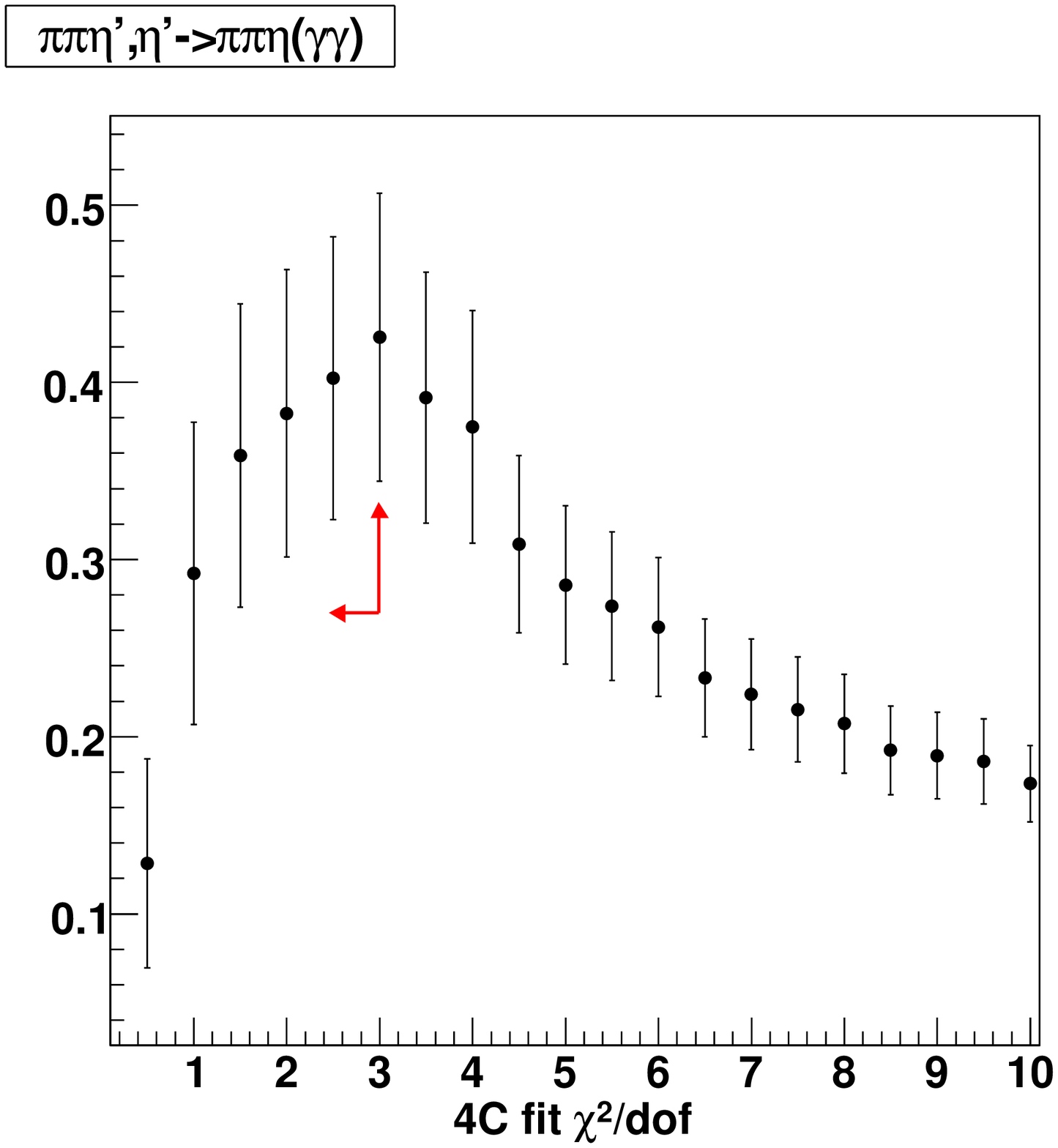}}
  \subfigure
    {\includegraphics[width=.49\textwidth]{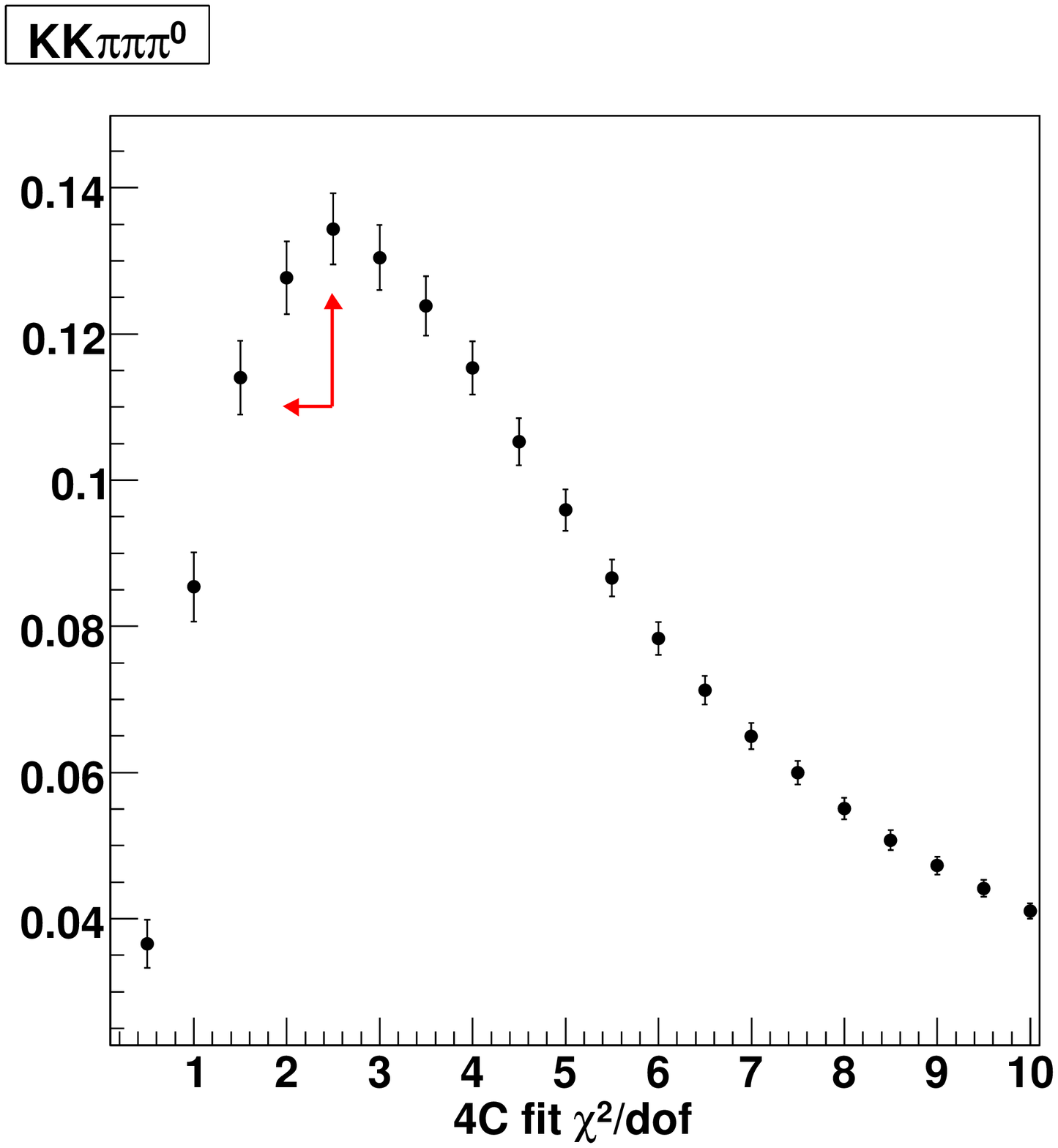}}
  \subfigure
    {\includegraphics[width=.49\textwidth]{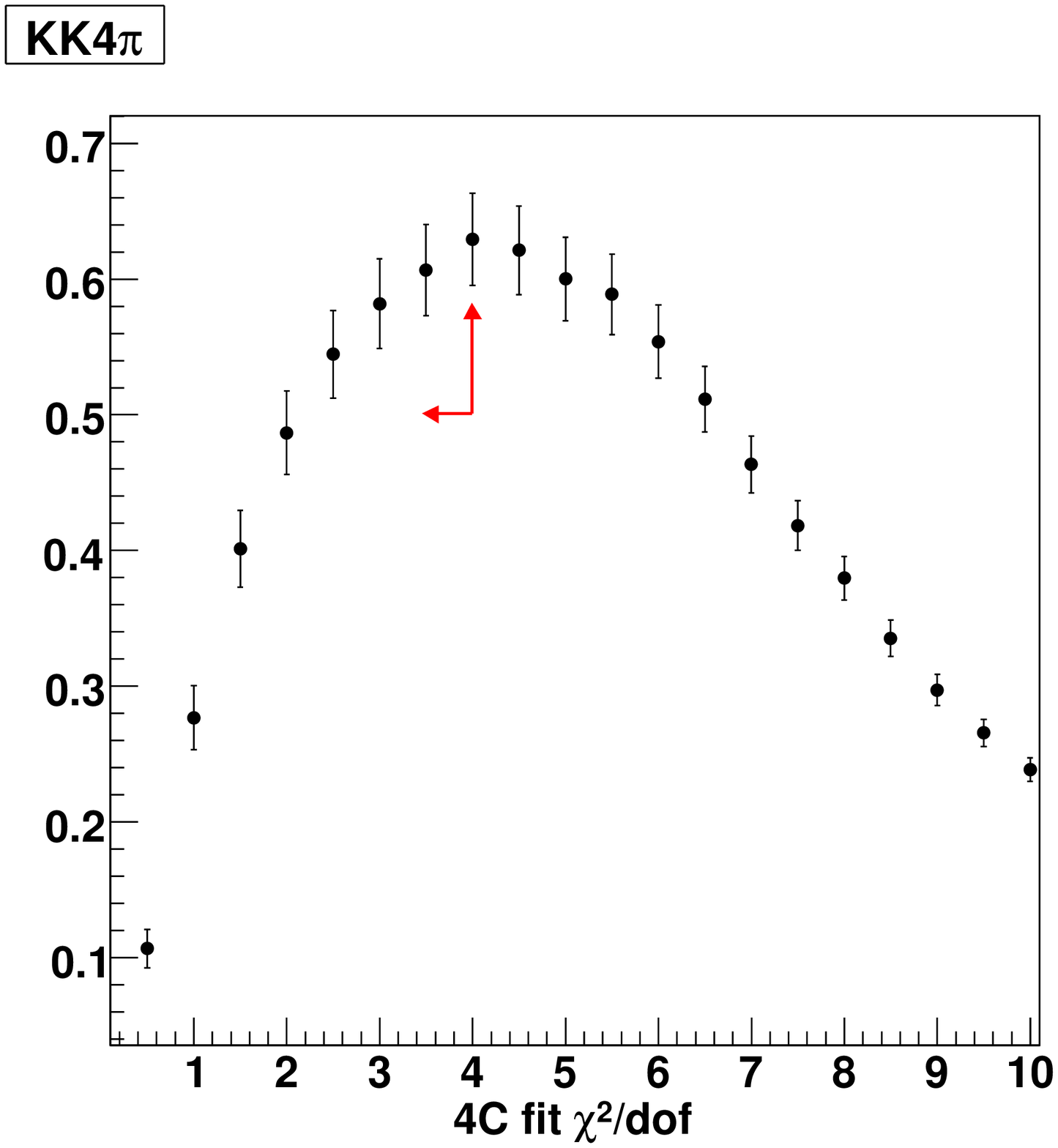}}
  \subfigure
    {\includegraphics[width=.49\textwidth]{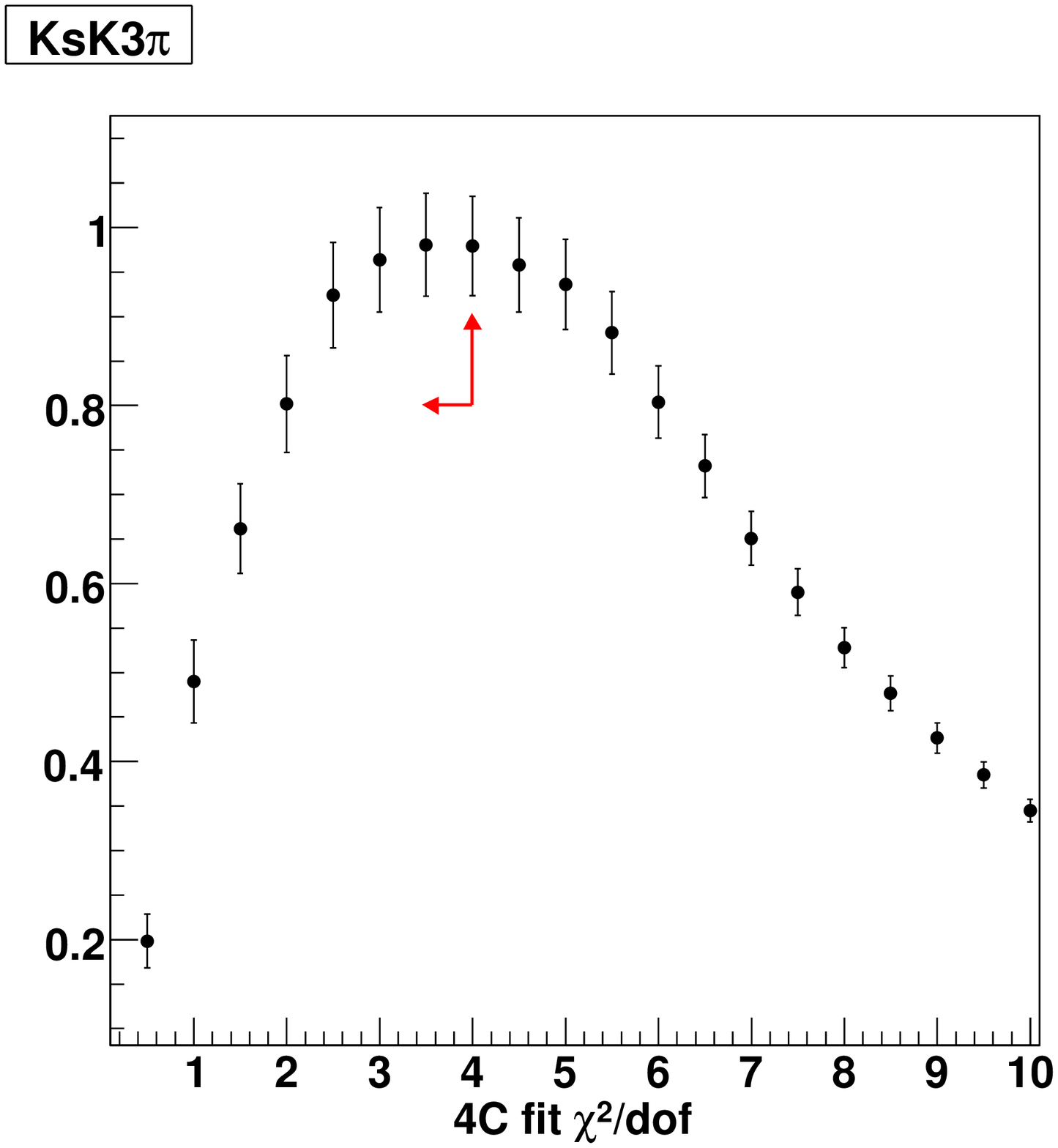}}
  \caption[Four momentum kinematic fit $\chi^2/$dof cut (4)]
    {MC study of $S^2/(S+B)$ for different cuts on the 4-momentum
     kinematic fit $\chi^2/$dof 
     for modes $\pi\pi\eta^{\prime}$ with $\eta^{\prime}\to\pi\pi\eta(\gamma\gamma)$ (top left),
     $KK\pi\pi\pi^{0}$ (top right), 
     $KK4\pi$ (bottom left), and $K_{S}K3\pi$ (bottom right).
     \ The arrows show the cut values that were selected.}
  \label{fig:cut_KinPFRedChi2Fit4}
\end{figure}

To select events consistent with full detection of 
$\psi(2S) \to \gamma \eta_{c}(2S), \eta_{c}(2S) \to X$ we compute the 
4-constraint kinematic 
fit $\chi^2/{\rm dof}$ using the package FitEvt 
\cite{cbx06-28}
. \ Both 
momentum and vertex constraints were imposed. \ For each mode, a vertex fit 
of the charged tracks from the IP
was performed first and the fit result was used as the origin of the photon(s) 
in the kinematic fit. \ In the fit we imposed the constraint that the total 
4-momentum of all tracks and neutral particles be equal to the total 
4-momentum of the $\psi(2S)$. \ We optimized a cut on the resulting overall
$\chi^2/{\rm dof}$ in the same manner as the optimization of the cuts on 
nearby tracks (see Section \ref{subsubsec:PhDNrTrk}), using $S^2/(S+B)$ 
as the figure of merit. \ An example of the $S^2/(S+B)$ optimization for the
4-momentum kinematic fit $\chi^2/{\rm dof}$ is shown in 
Figure~\ref{fig:cut_KinPFRedChi2Fit1} for the $4\pi$ mode. \ The $S^2/(S+B)$ 
for the other modes are shown in Figures~\ref{fig:cut_KinPFRedChi2Fit2} 
through \ref{fig:cut_KinPFRedChi2Fit4}. 
The optimized cuts are listed in Table~\ref{table:cut_kinchi2}.
\begin{table}[htbp]
\caption[Cut optimization]{\label{table:cut_kinchi2}Optimization of 4-C 
kinematic fit $\chi^2/$dof, along with the transition photon distance to the 
nearest track ($d_{\gamma, trk}$) or the angle between the transition photon 
and the closest pion ($\cos \theta_{\gamma,\pi}$).}
\begin{center}
\begin{tabular}{|l|c|c|}
  \hline
  Mode & $d_{\gamma, trk}$~(cm) or $\cos \theta_{\gamma,\pi}$ & $\chi^2/{\rm dof}$ \\ \hline
  $4\pi$ 
    & $\cos \theta_{\gamma,\pi} < 0.93$ & $< 4.5$ \\ \hline
  $6\pi$ 
    & - & $< 5.0$ \\ \hline
  $KK\pi\pi$ 
    & $d_{\gamma, trk} \ge 45$ & $< 4.0$ \\ \hline
  $KK\pi^{0}$
    & $d_{\gamma, trk} \ge 35$ & $< 4.0$ \\ \hline
  $K_{S}K\pi$ 
    & - & $< 3.5$ \\ \hline
  $\pi\pi\eta(\gamma\gamma)$
    & - & $< 2.0$ \\ \hline
  $\pi\pi\eta(\pi\pi\pi^{0})$
    & $\cos \theta_{\gamma,\pi} < 0.93$ & $< 3.0$ \\ \hline
  $\pi\pi\eta^{\prime}, \eta^{\prime}\to\pi\pi\eta(\gamma\gamma)$
    & - & $< 3.0$ \\ \hline
  $KK\eta(\gamma\gamma)$
    & - & $< 3.5$ \\ \hline
  $KK\eta(\pi\pi\pi^{0})$
    & - & $< 5.0$ \\ \hline
  $KK\pi\pi\pi^{0}$ 
    & - & $< 2.5$ \\ \hline
  $KK4\pi$ 
    & - & $< 4.0$ \\ \hline
  $K_{S}K3\pi$ 
    & - & $< 4.0$ \\ \hline
\end{tabular}
\end{center}
\end{table}


\subsubsection{4-Constraint Kinematic Fit $\chi^2/$dof of Hadrons Only}

Similar to the 4-constraint kinematic fit for the whole event, we used the 
FitEvt package \cite{cbx06-28}
to perform the fit only for the hadrons of 
$\psi(2S) \to X$ candidates, both momentum and vertex. \ 
In the fit we impose the constraint that the total 4-momentum of all 
hadrons be equal to the total 4-momentum of the $\psi(2S)$. \ If a candidate is 
a $\psi(2S) \to \gamma \eta_{c}(2S)$ event, then the 4-C kinematic fit 
$\chi^2/{\rm dof}$ of the whole event should peak at zero, and that of 
just the hadrons should be distributed away from zero. \ This cut was 
evaluated, but was found not to suppress the background as efficiently as 
hoped. \ Therefore it was excluded from the final set of selection criteria.

\section{Signal Extraction Procedures}

\begin{figure}[htbp]
  \centering
  \subfigure
    {\includegraphics[width=.79\textwidth]{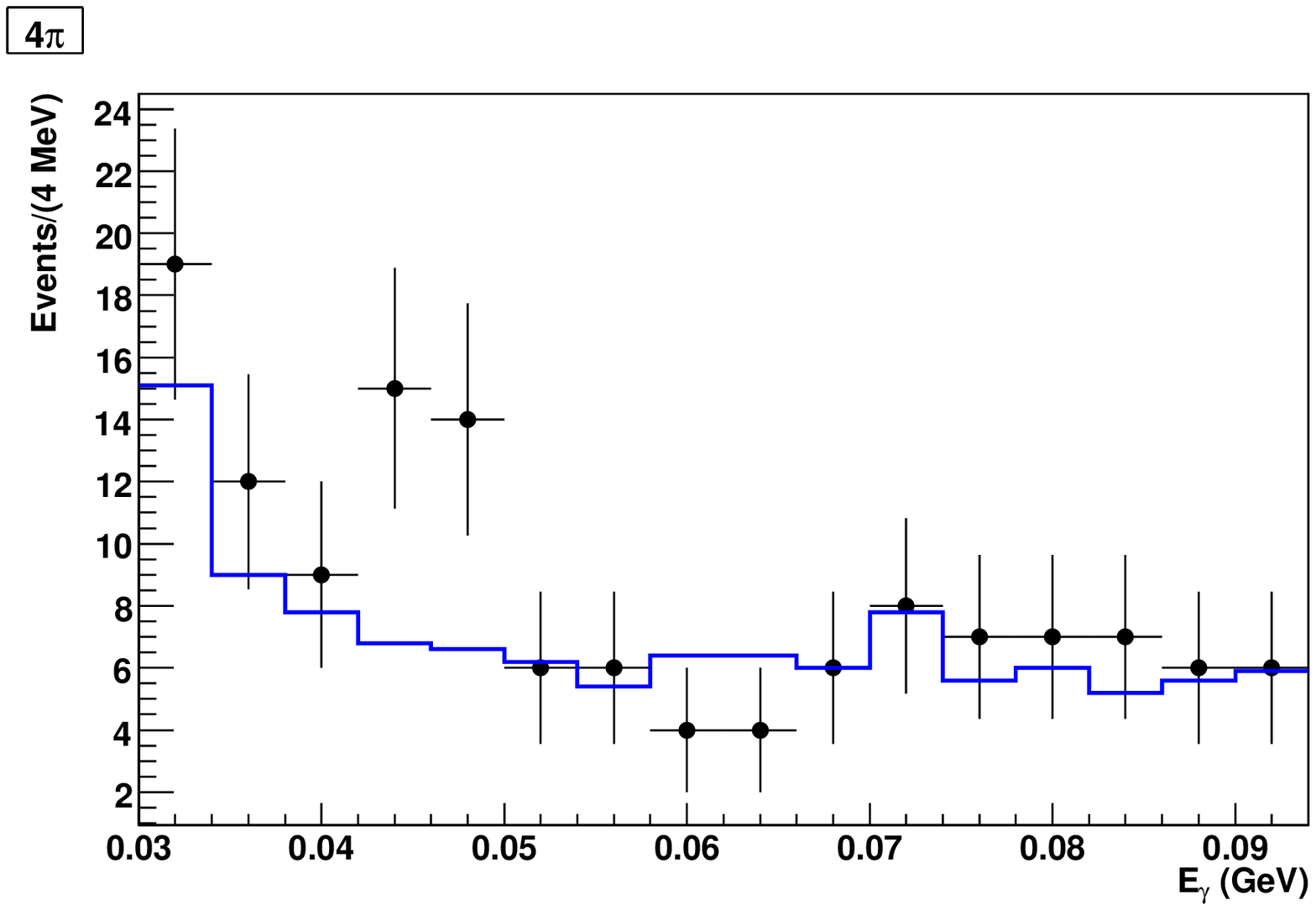}}
  \subfigure
    {\includegraphics[width=.79\textwidth]{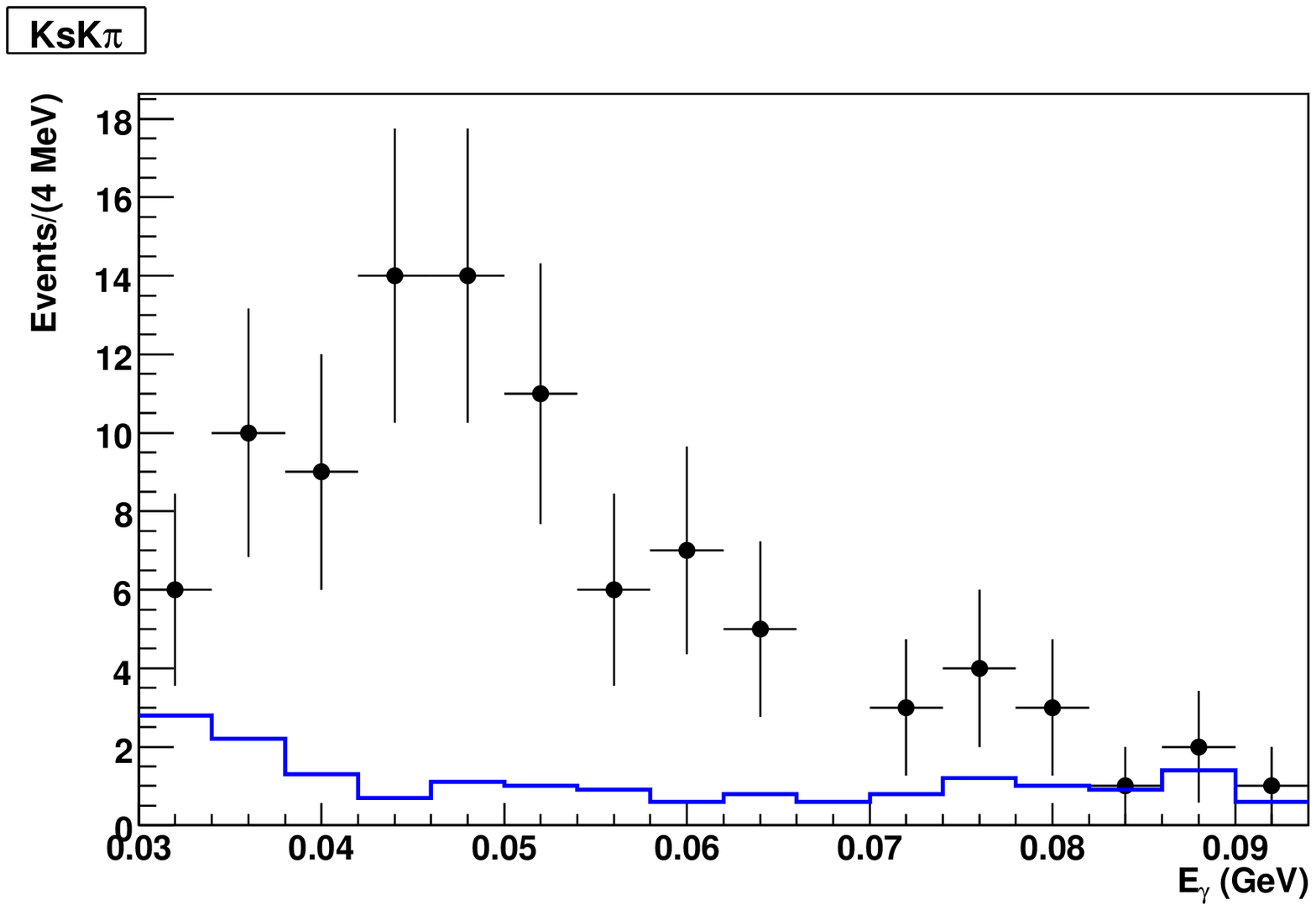}}
  \caption[Expected signal distributions]
    {The expected signal distributions for modes $4\pi$ (top) and $K_{S}K\pi$ 
     (bottom) for a simulated data-sized sample based on the assumed 
     product branching fractions listed in Table~\ref{table:numofproduced}. \ 
     The solid histogram is the sum of the 10 times generic $\psi(2S)$ and 
     5 times continuum MC samples scaled to one time
     luminosity of data.  The points are randomly thrown signal MC events, 
     normalized to the product branching fraction expectation listed in 
     Table~\ref{table:numofproduced}. \ The signal MC was generated with 
     the expected signal mean of 48~MeV and width of 14~MeV.}
  \label{fig:expsigdist}
\end{figure}

In this section, we describe the procedure we developed for extracting 
the signal yields and resonance parameters from the 
25.9~M $\psi(2S)$ decays. \ We fit the measured, or unconstrained, 
photon energy distribution in the range $E_{\gamma} = [30,94]~{\rm MeV}$. \ 
Figure~\ref{fig:expsigdist} shows the expected signal distribution for the 
modes $4\pi$ and $K_{S}K\pi$ for a mock data sample of 25.9~M $\psi(2S)$ 
decays generated under the assumption that the partial widths for 
$\eta_{c}(2S) \to X$ are the same as for $\eta_{c}(1S) \to X$, and 
${\mathcal B}(\psi(2S) \to \gamma \eta_{c}) = 2.6 \times 10^{-4}$.

\subsection{Detector Resolution}
\label{sec:resfunct}

Since we extract the signal from a measured photon distribution, a Crystal 
Ball function \cite{Skwarnicki:1986xj, Gaiser:1982yw} is used to fit the 
detector resolution distribution from the signal MC samples described in 
Section~\ref{subsec:signalmc}. \ 
The Crystal Ball function (named for the Crystal Ball experiment) is designed 
to provide a good representation of the photon energy distribution as 
measured with a high-precision crystal calorimeter like CLEO-c's. \
The resolution is defined as the 
standard deviation of the difference between the measured and generator-level 
photon energy. \ The RooFit software package \cite{roofit} was used to perform 
these $\chi^{2}$ fits. \ 

The Crystal Ball function is defined as
\begin{equation}
f(x;\alpha,n,\bar{x},\sigma) = N \cdot \left\{
\begin{array}{l@{\quad\text{for}\quad}l}
\exp\left(-\frac{(x - \bar{x})^{2}}{2\sigma^{2}}\right), 
& \frac{x - \bar{x}}{\sigma} > - \alpha \\
\left(\frac{n}{|\alpha|}\right)^{n} e^{-\frac{|\alpha|^{2}}{2}}
\left(\frac{n}{|\alpha|} - |\alpha| - \frac{x - \bar{x}}{\sigma}\right)^{-n}, 
& \frac{x - \bar{x}}{\sigma} \leq - \alpha
\end{array}
\right. ,
\end{equation}
where $N$ is the normalization factor and $\alpha$, $n$, $\bar{x}$, and 
$\sigma$ are parameters that govern the shape. \ The parameter $\bar{x}$ 
represents the mean of the resolution function and $\sigma$ is the 
resolution in the photon energy. \ The Crystal Ball function features
a Gaussian distribution with a power-law tail (index $n$) at the low end 
below a threshold (given by the parameter $\alpha$) to account for 
energy leakage within the calorimeter.

To obtain successful fits for the resolution function, the parameter $n$ 
was fixed at 140 for all modes. \ It was observed that when $n$ varied 
over the range from 40 to 140, the shape of the Crystal Ball function 
and the fit $\chi^{2}$ showed very little variation. 

\begin{figure}[htbp]
\begin{center}
\includegraphics[width=.95\textwidth]{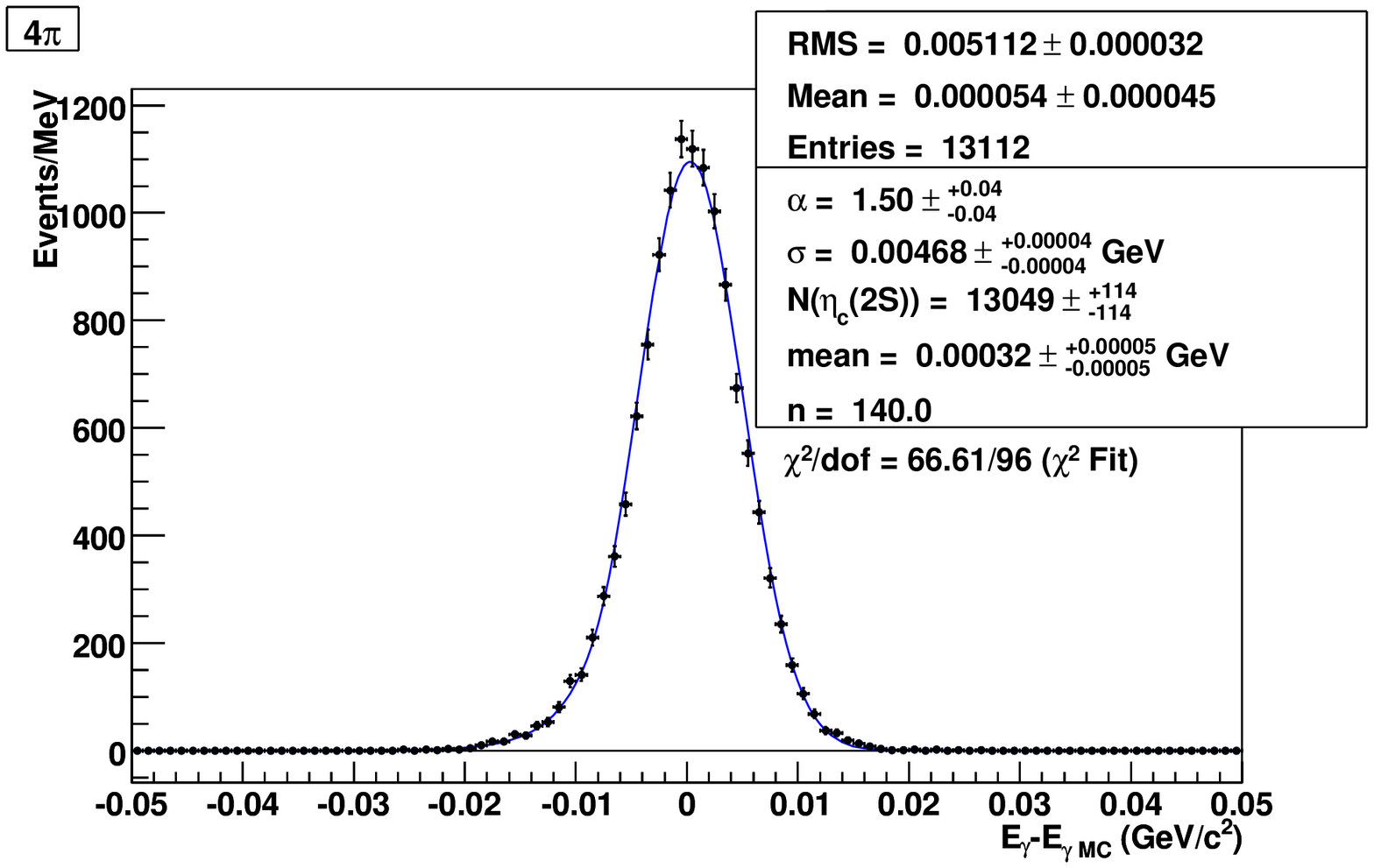}
\end{center}
\caption[Resolution fit of $\eta_{c}(2S)$ decay $4 \pi$ mode]
{\label{fig:etac2s_res_4Pi}
{Resolution fit of decay $\psi(2S)\to\gamma\eta_{c}(2S)$, 
$\eta_{c}(2S)\to 4 \pi$.}}
\end{figure}

As an example Figure~\ref{fig:etac2s_res_4Pi} shows the resolution fit of 
the $4\pi$ mode. \ Resolution fits of other modes
are shown in Appendix ~\ref{appendix:resfnctfits}.

\subsection{Background Parametrization}

A necessary ingredient for obtaining reliable results was 
the use of an appropriate function to describe the background shape. \ 
For this purpose we can either use a polynomial or other function to 
parametrize the background or rely on a MC simulation that at 
least approximates the underlying physics. \ Clearly the latter is 
potentially more reliable, but it is first necessary
to study the reliability of our MC samples and the composition 
of the background. \ An independent data sample is required for such a
study, and we used $\psi(2S) \to \pi^{+} \pi^{-} J/\psi$ events 
with four specific $J/\psi$ decay modes. \ 
These modes are 
$J/\psi \to 4\pi$, $KK\pi\pi$, $KK\pi^{0}$, and $K_{S}K\pi$. \ 
Note that these modes gave us a sample of decays modes with 
$\pi, K, \pi^0,$ and $K_{S}$. \  When selecting $J/\psi$ events, 
the $\pi\pi$ recoil mass was 
required to be within 20 MeV of the mass of $J/\psi$. \ All other event 
selection criteria for $\psi(2S)\to\gamma\eta_{c}(2S)$ decays were 
applied, except that the $\pi\pi$ recoil mass and 
$J/\psi \to (X~-~2\pi)$ rejection cuts were removed and the hadronic 
invariant mass cut was adjusted for the $J/\psi$ mass
instead of the mass of $\psi(2S)$.

\begin{figure}[htbp]
  \centering
  \subfigure
    {\includegraphics[width=.49\textwidth]{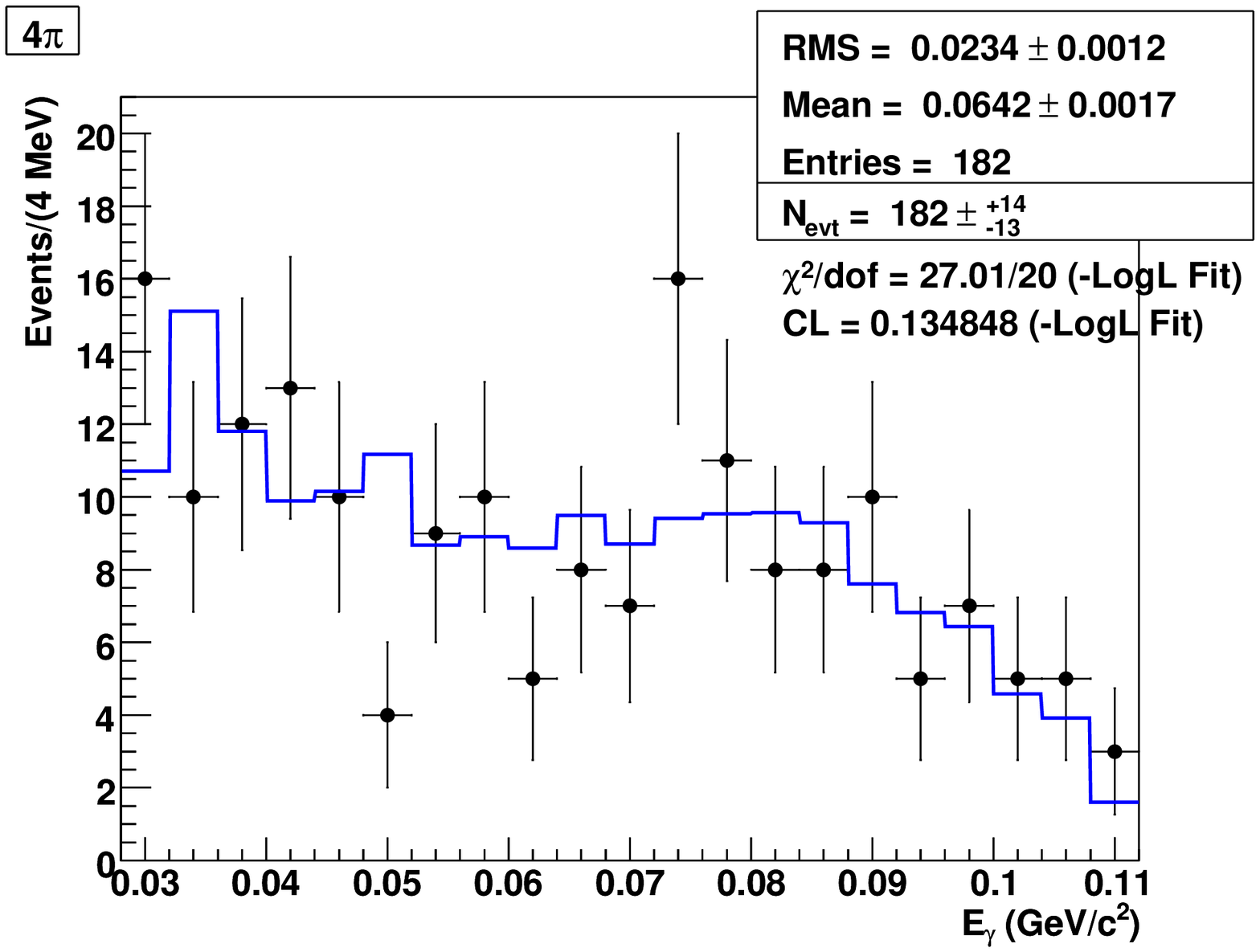}}
  \subfigure
    {\includegraphics[width=.49\textwidth]{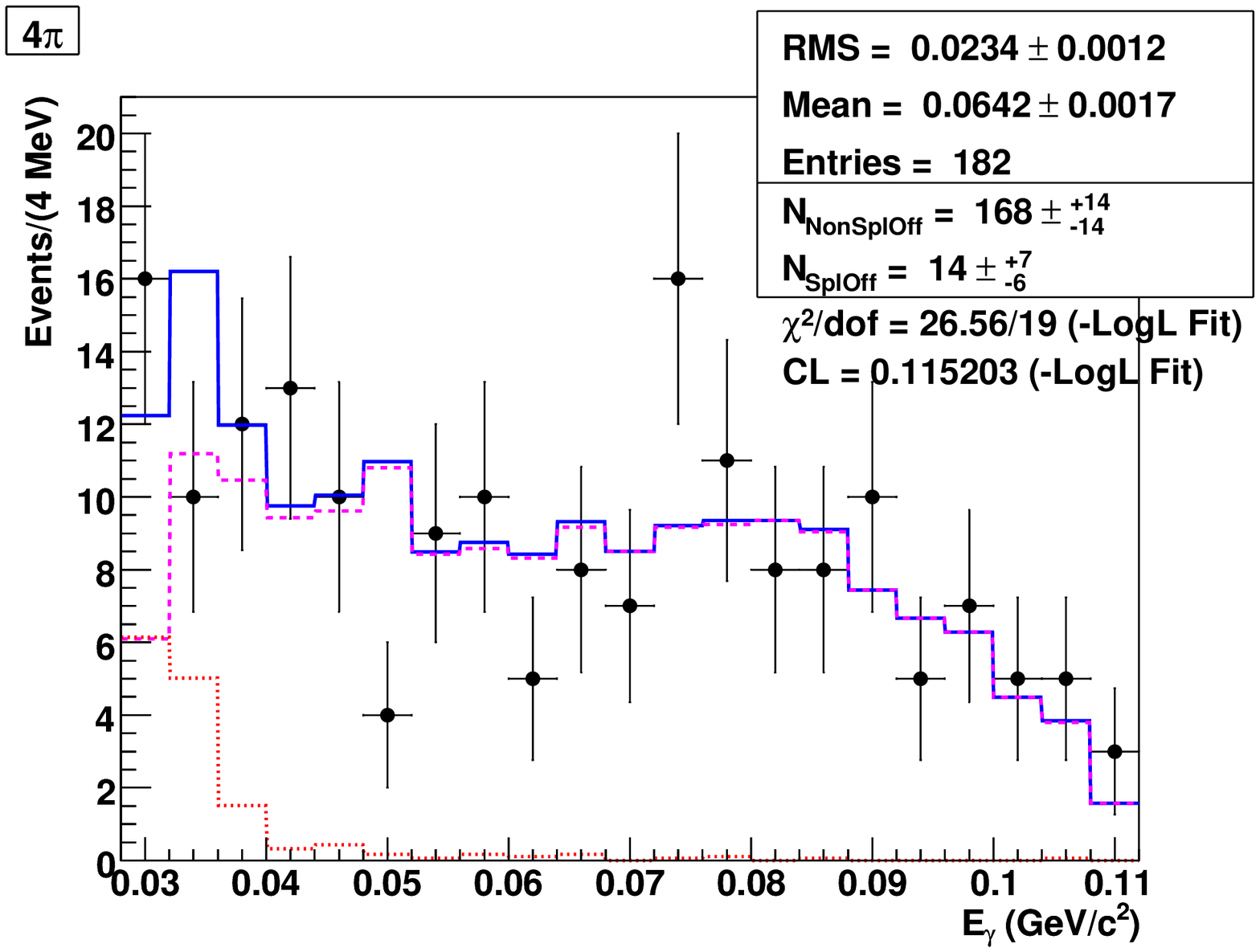}}
  \subfigure
    {\includegraphics[width=.49\textwidth]{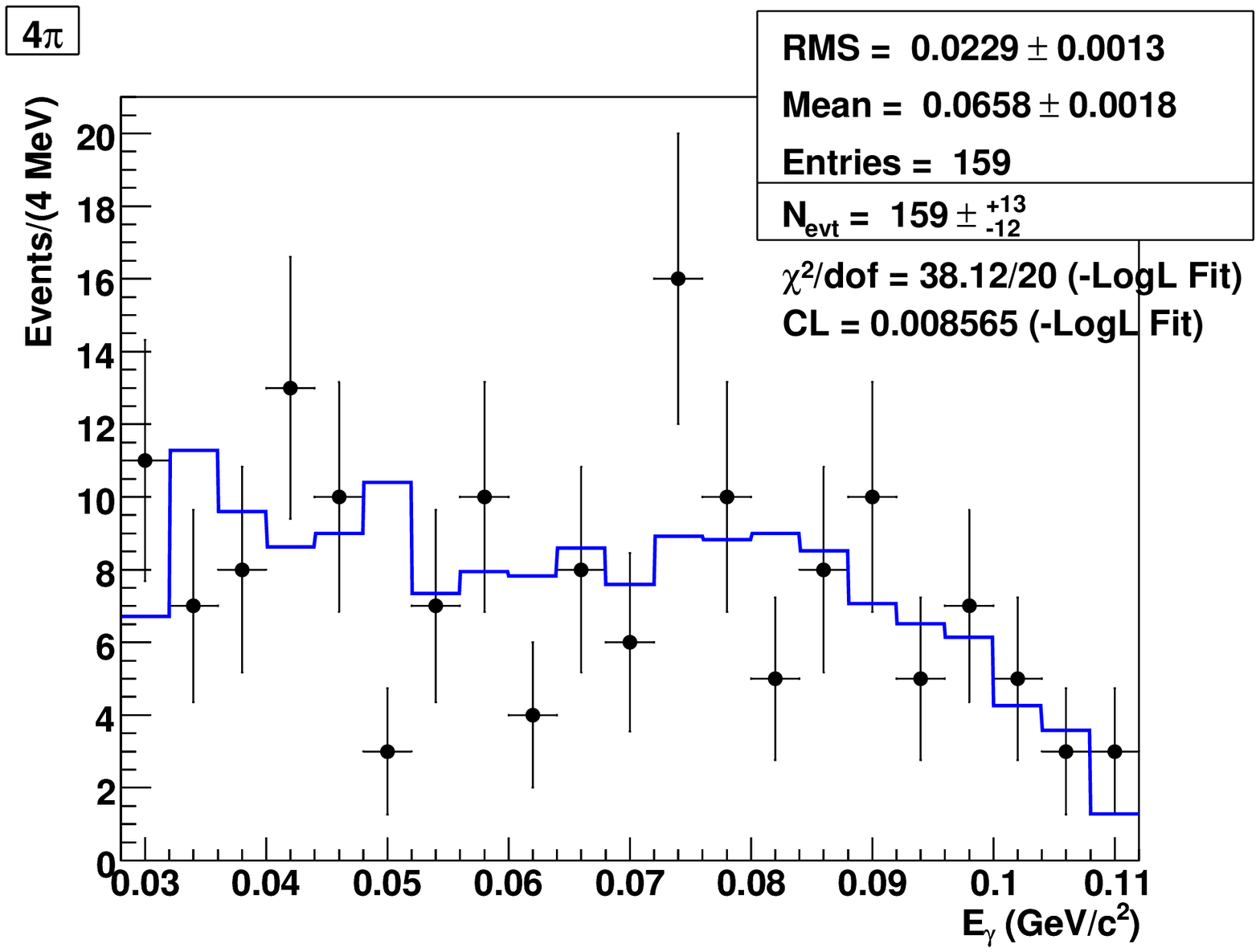}}
 \subfigure
    {\includegraphics[width=.49\textwidth]{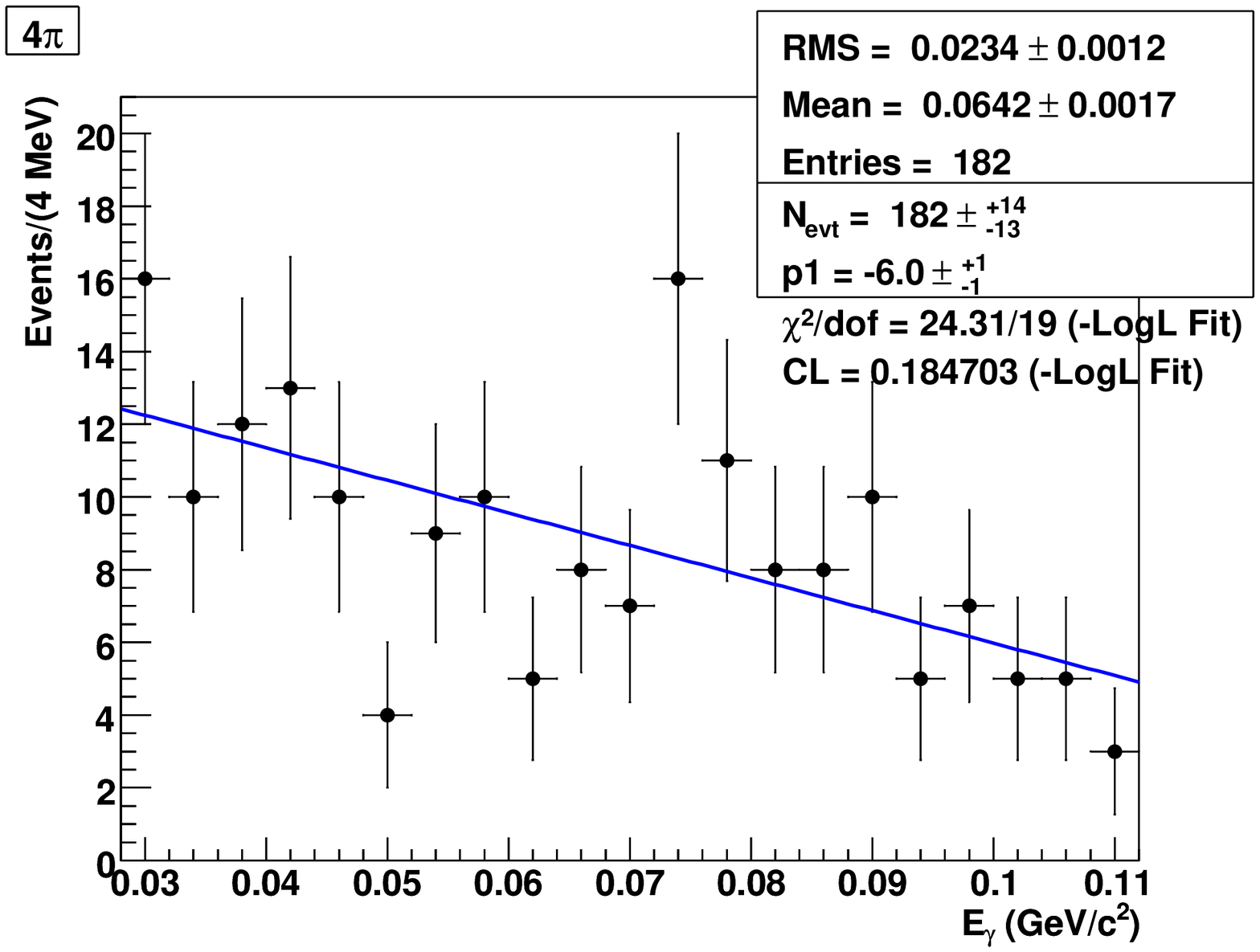}}
  \caption[Background fits of decay $J/\psi \to 4\pi$]
    {Background fits of the measured photon-energy distribution for
    $\psi(2S) \to \pi^{+} \pi^{-} J/\psi$ events with $J/\psi \to 4\pi$
    and a low energy shower. \ The upper left plot is the fit of 
    data to a single MC histogram. \ The upper right plot is the fit of data 
    to two MC histograms with independent parameters, one (dashed) histogram 
    excluding splitoff showers and the other (dotted) including only splitoff 
    showers. \ The lower left plot is the fit of data to a single MC 
    histogram with a tighter cut of full event 4-C fit $\chi^{2}/{\rm dof}<3.0$. \ 
    The lower right plot is the fit of data with a linear function.}
  \label{fig:jpsi_6Pi}
\end{figure}

\begin{figure}[htbp]
  \centering
  \subfigure
    {\includegraphics[width=.49\textwidth]{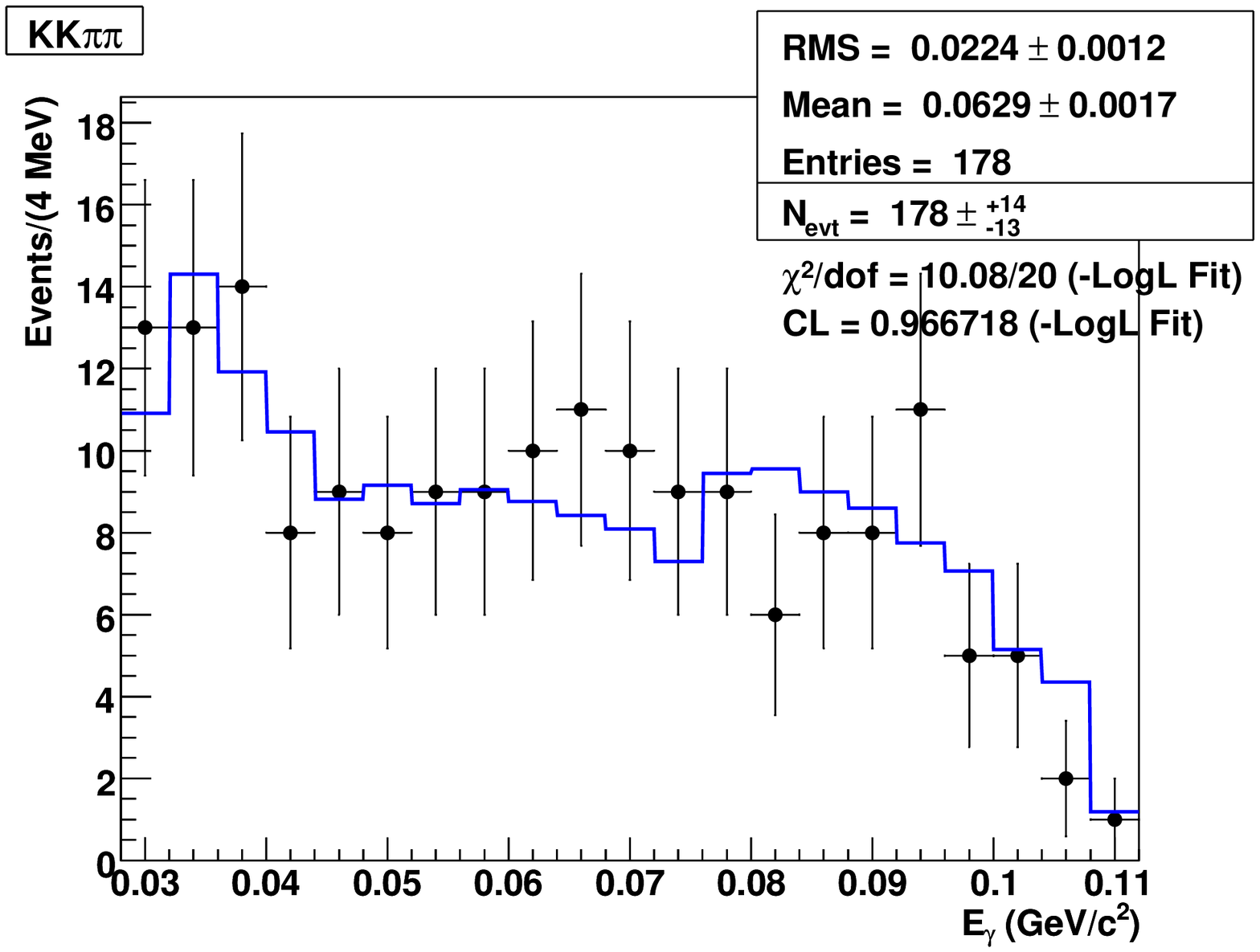}}
  \subfigure
    {\includegraphics[width=.49\textwidth]{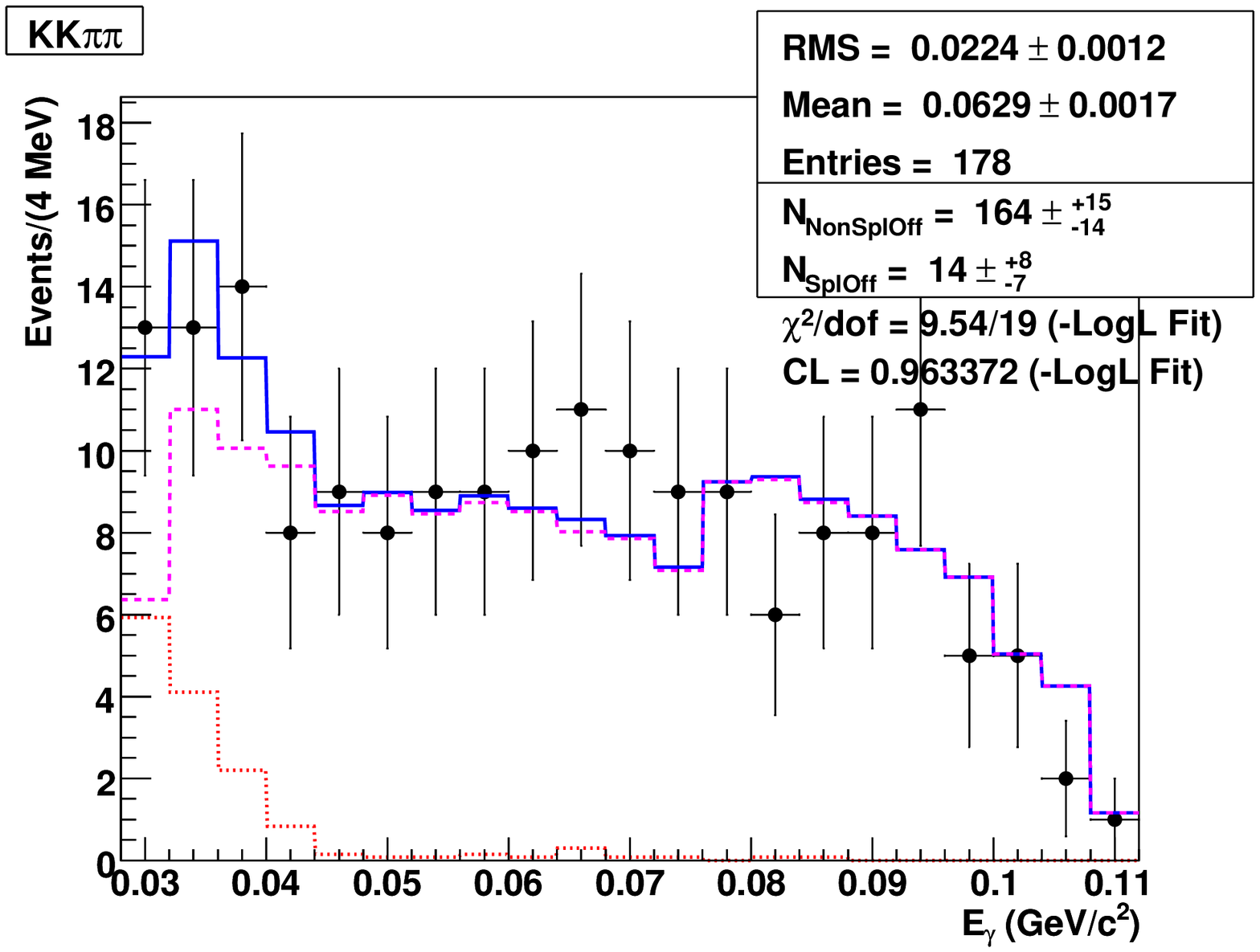}}
  \subfigure
    {\includegraphics[width=.49\textwidth]{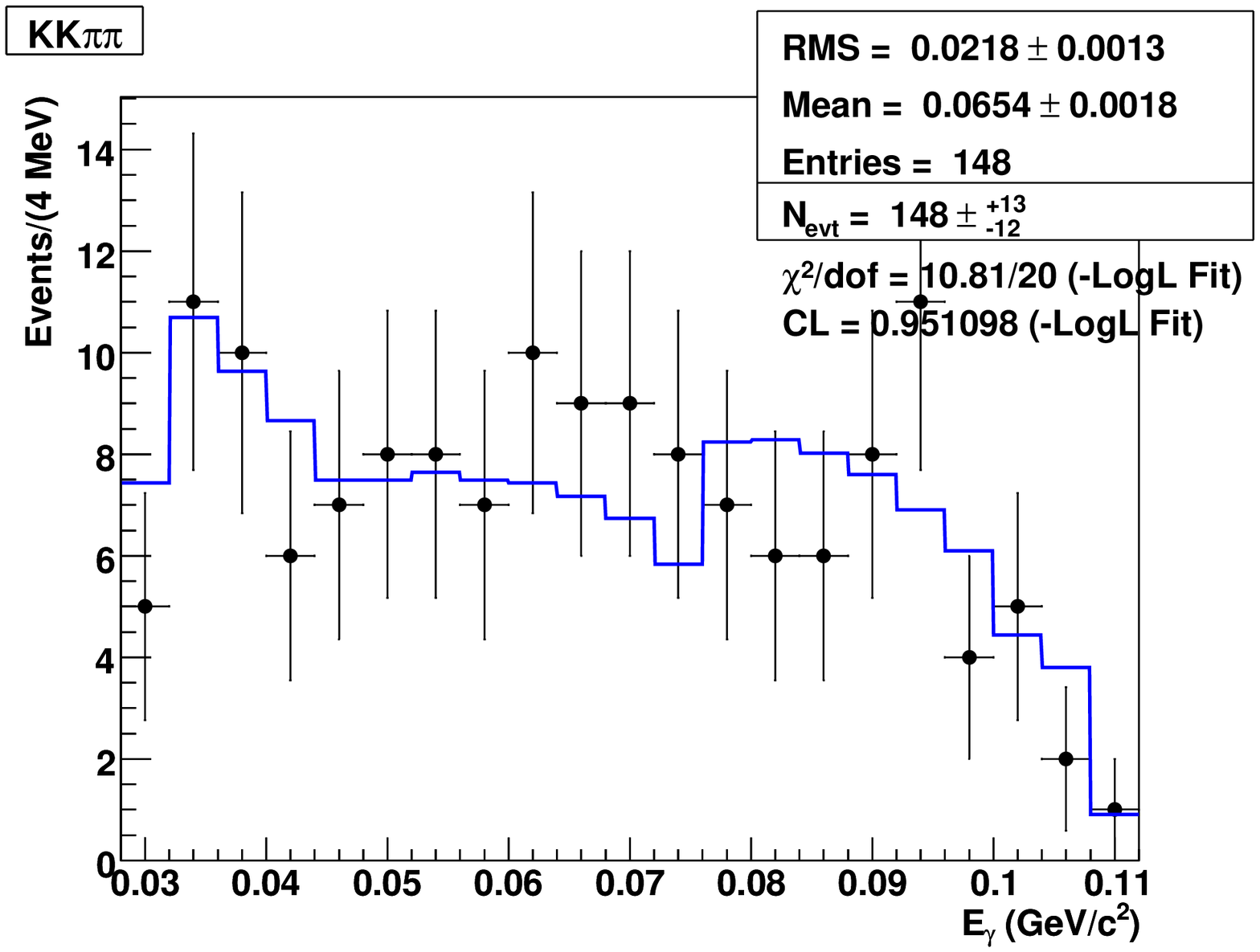}}
  \subfigure
    {\includegraphics[width=.49\textwidth]{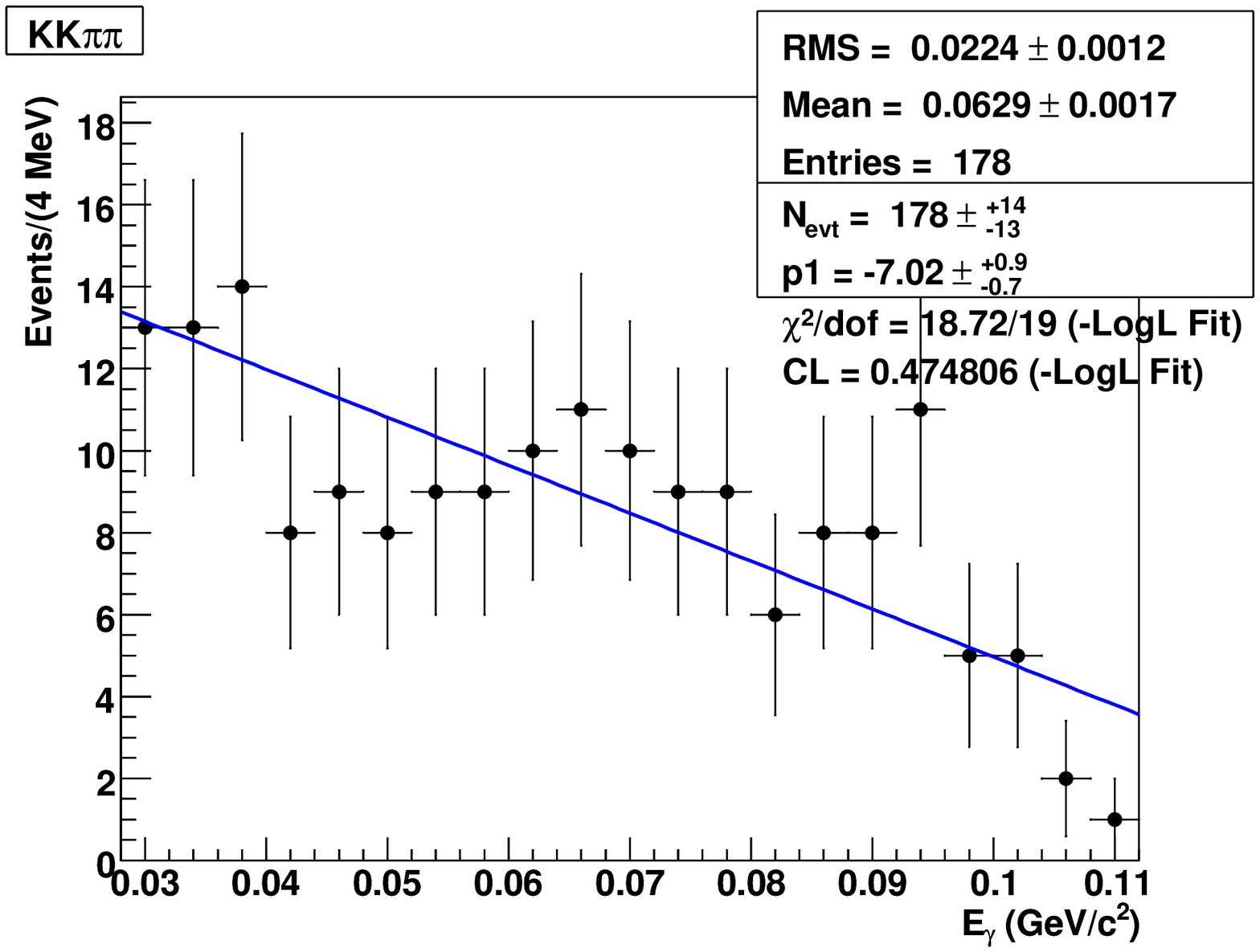}}
  \caption[Background fits of decay $J/\psi \to KK\pi\pi$]
    {Background fits of the measured photon-energy distribution for
    $\psi(2S) \to \pi^{+} \pi^{-} J/\psi$ events with $J/\psi \to KK\pi\pi$ 
    and a low energy shower. \ The upper left plot is the fit of 
    data to a single MC histogram. \ The upper right plot is the fit of data 
    to two MC histograms with independent parameters, one (dashed) histogram 
    excluding splitoff showers and the other (dotted) including only splitoff 
    showers. \ The lower left plot is the fit of data to a single MC 
    histogram with a tighter cut of full event 4-C fit $\chi^{2}/{\rm dof} <3.0$. \ 
    The lower right plot is the fit of data with a linear function.}
  \label{fig:jpsi_KK4Pi}
\end{figure}

\begin{figure}[htbp]
  \centering
  \subfigure
    {\includegraphics[width=.49\textwidth]{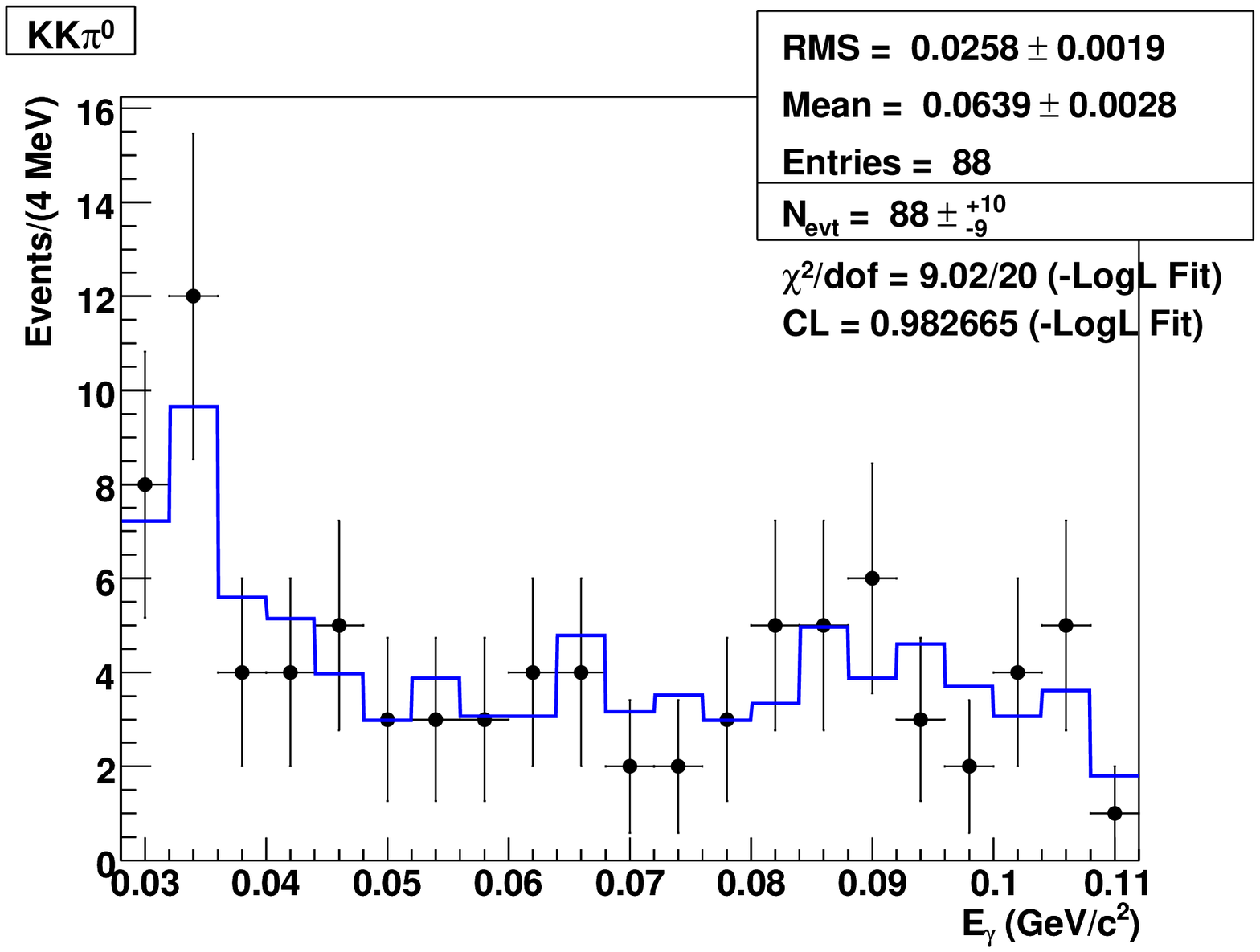}}
  \subfigure
    {\includegraphics[width=.49\textwidth]{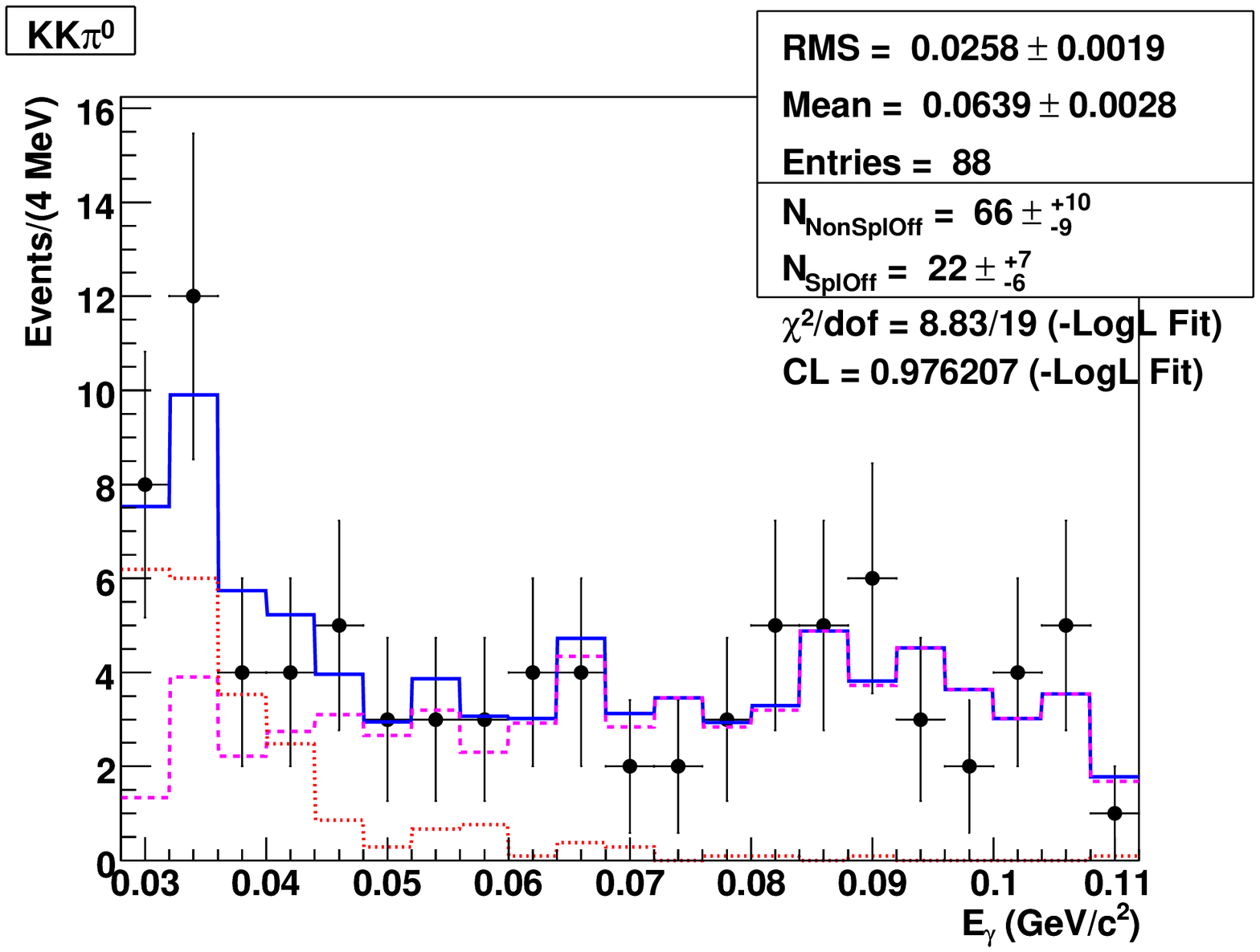}}
  \subfigure
    {\includegraphics[width=.49\textwidth]{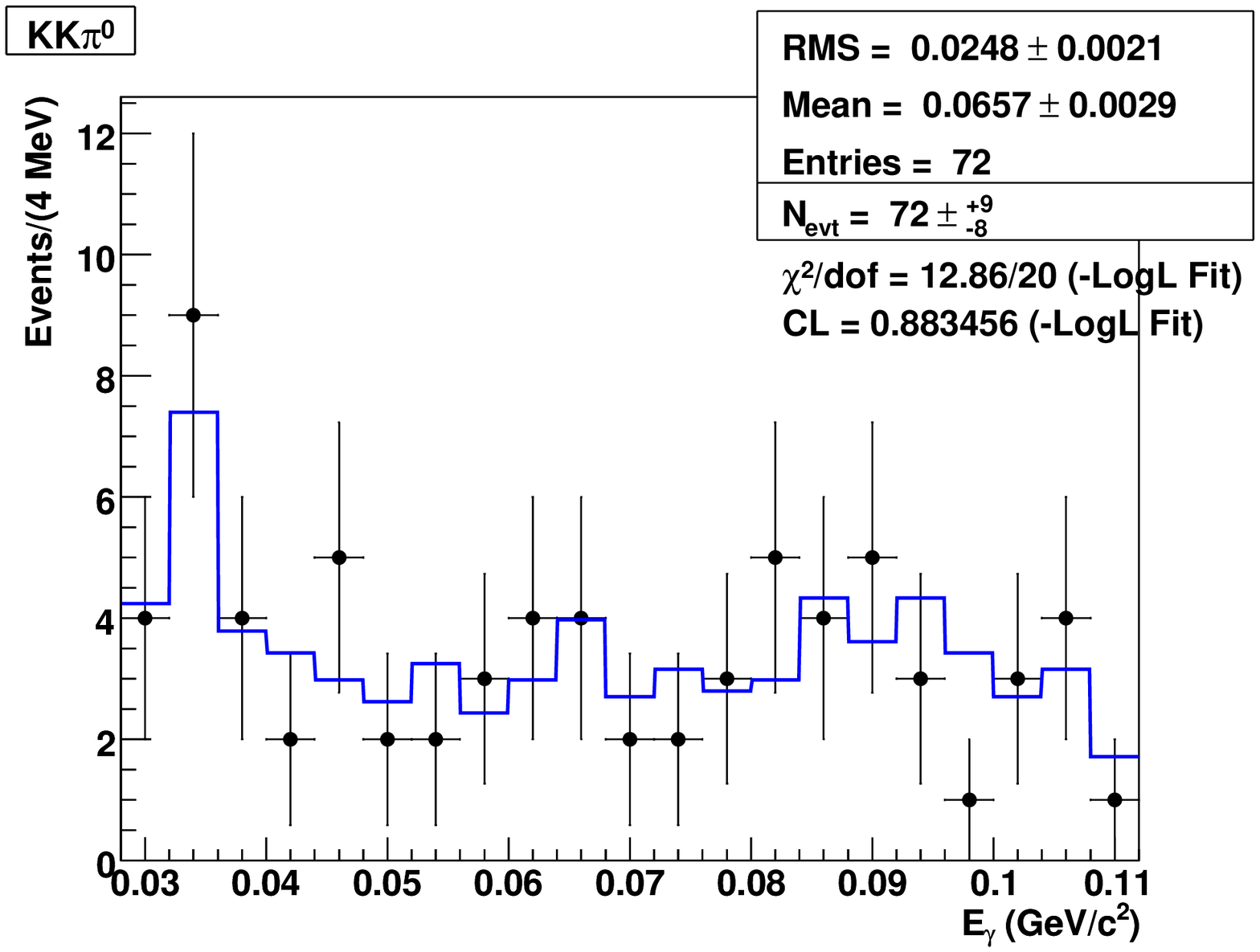}}
  \subfigure
    {\includegraphics[width=.49\textwidth]{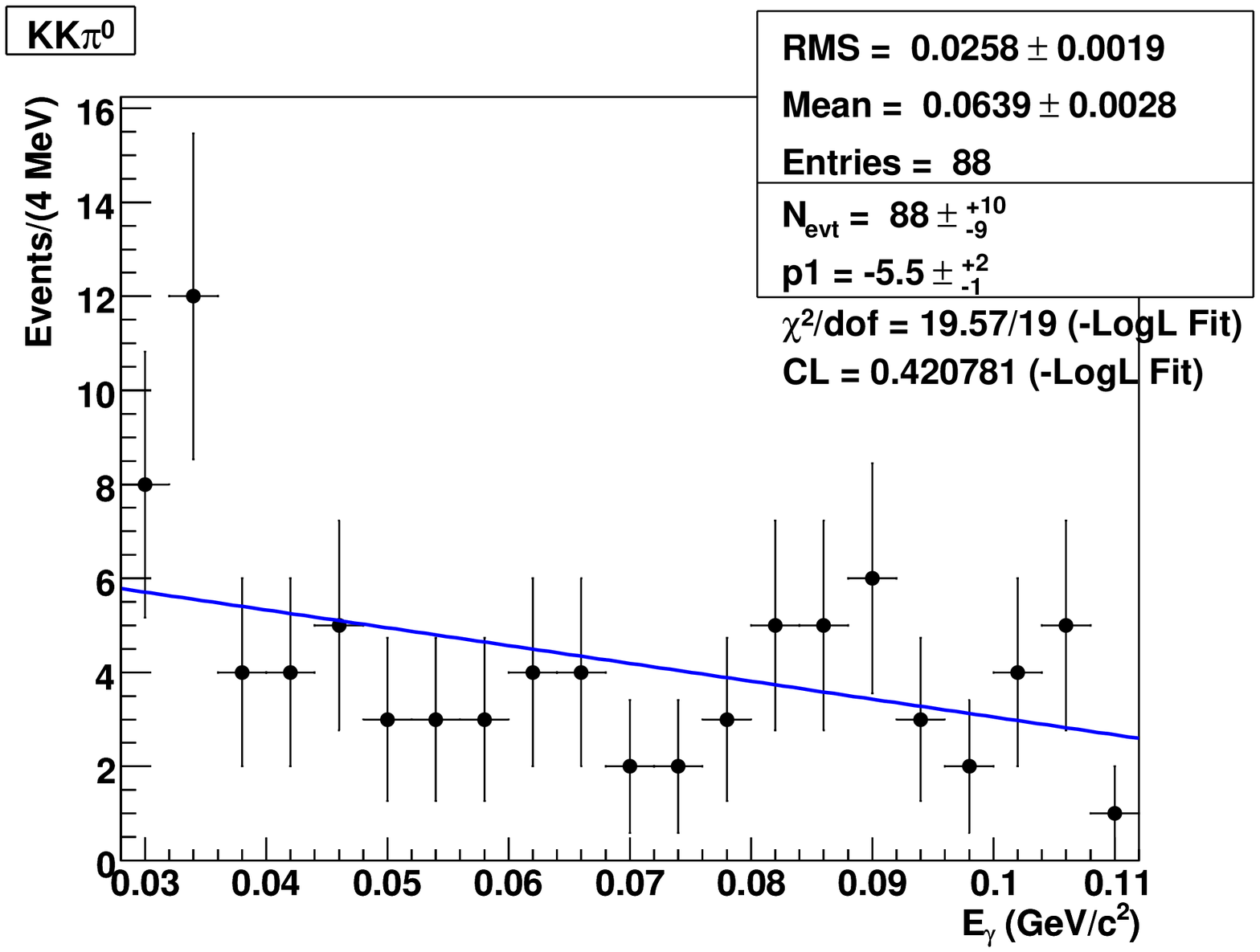}}
  \caption[Background fits of decay $J/\psi \to KK\pi^{0}$]
    {Background fits of the measured photon-energy distribution for
    $\psi(2S) \to \pi^{+} \pi^{-} J/\psi$ events with $J/\psi \to KK\pi^{0}$ 
    and a low energy shower. \ The upper left plot is the fit of 
    data to a single MC histogram. \ The upper right plot is the fit of data 
    to two MC histograms with independent parameters, one (dashed) histogram 
    excluding splitoff showers and the other (dotted) including only splitoff 
    showers. \ The lower left plot is the fit of data to a single MC 
    histogram with a tighter cut of full event 4-C fit $\chi^{2}/{\rm dof} < 3.0$. \ 
    The lower right plot is the fit of data with a linear function.}
  \label{fig:jpsi_KKPiPiPi0}
\end{figure}

\begin{figure}[htbp]
  \centering
  \subfigure
    {\includegraphics[width=.49\textwidth]{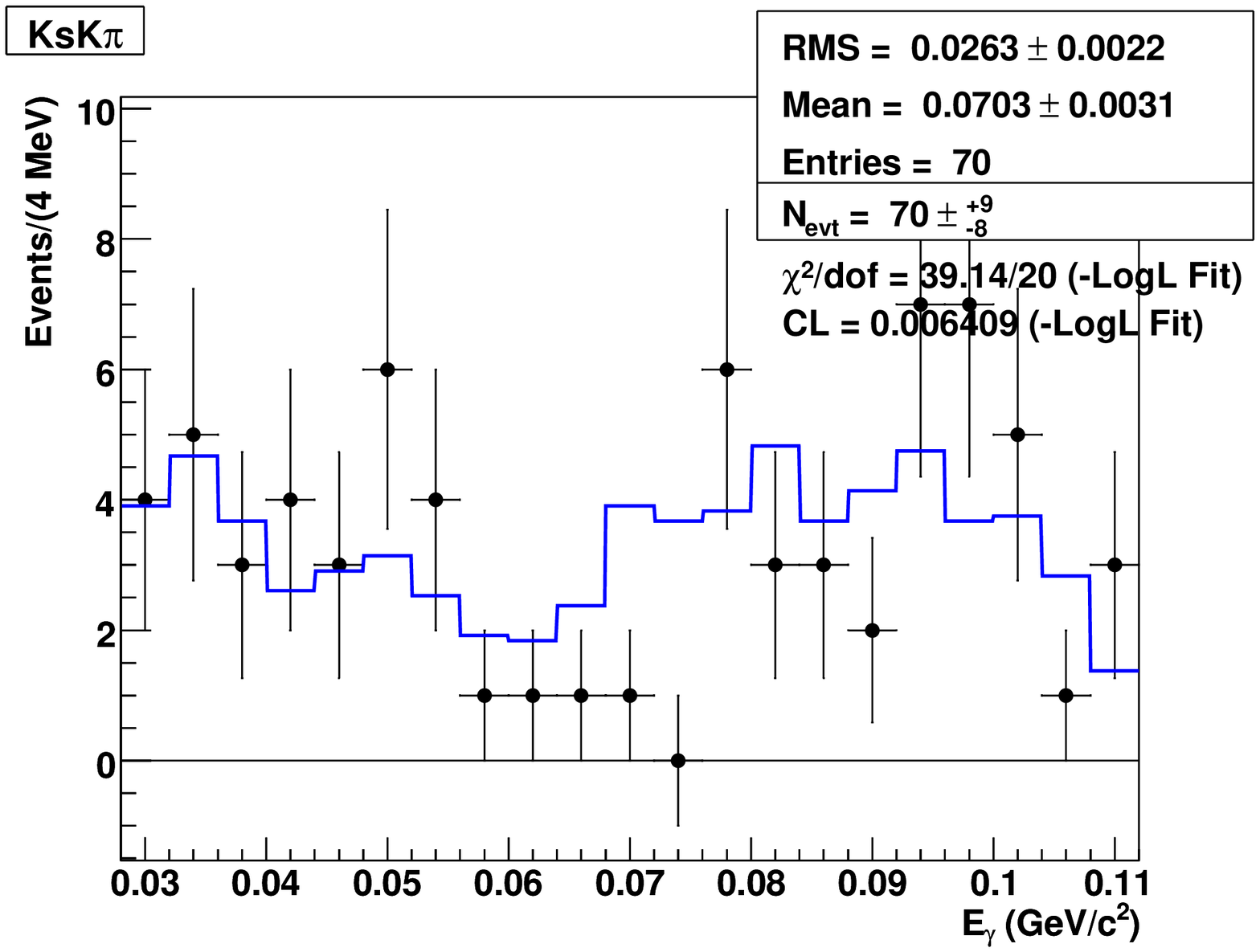}}
  \subfigure
    {\includegraphics[width=.49\textwidth]{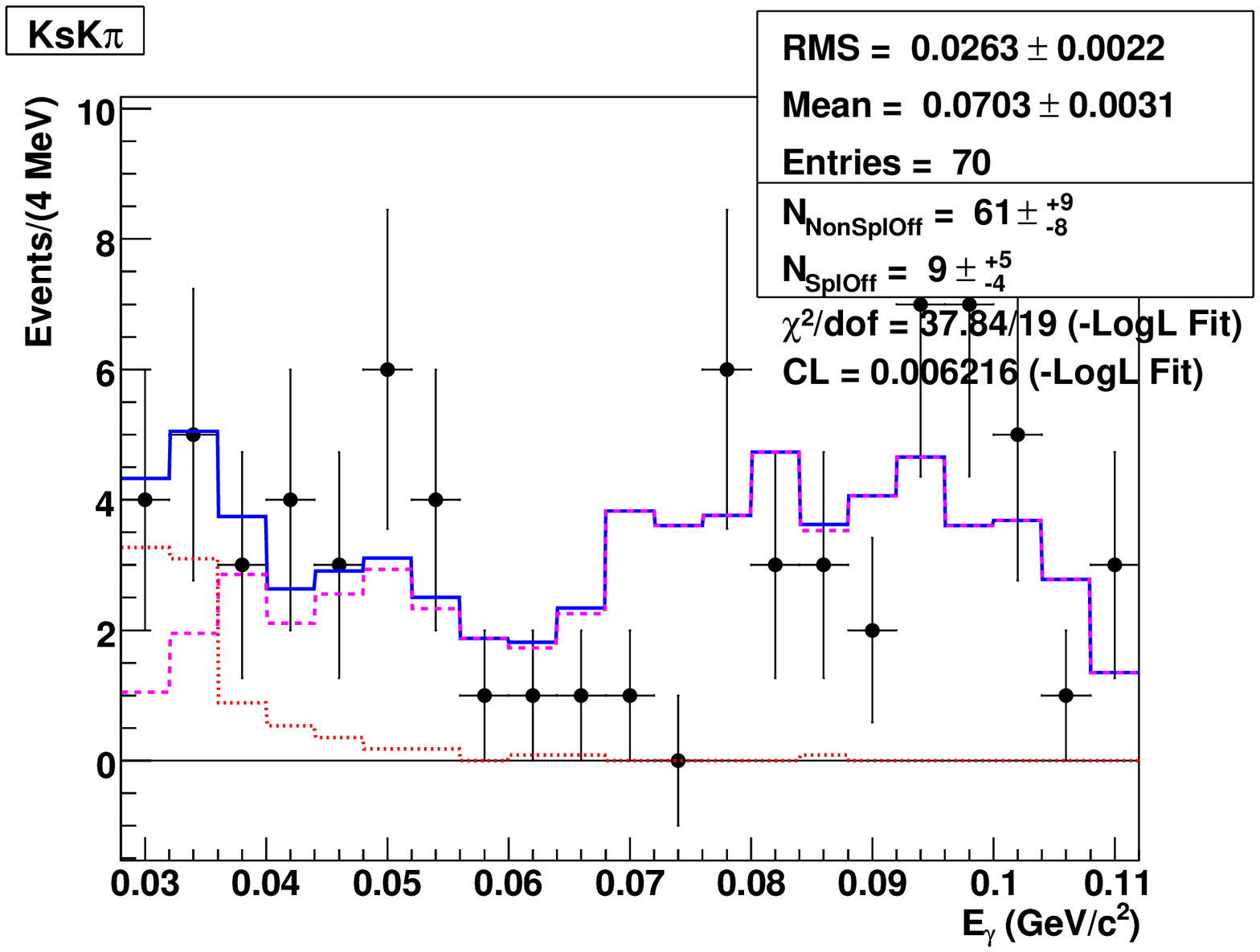}}
  \subfigure
    {\includegraphics[width=.49\textwidth]{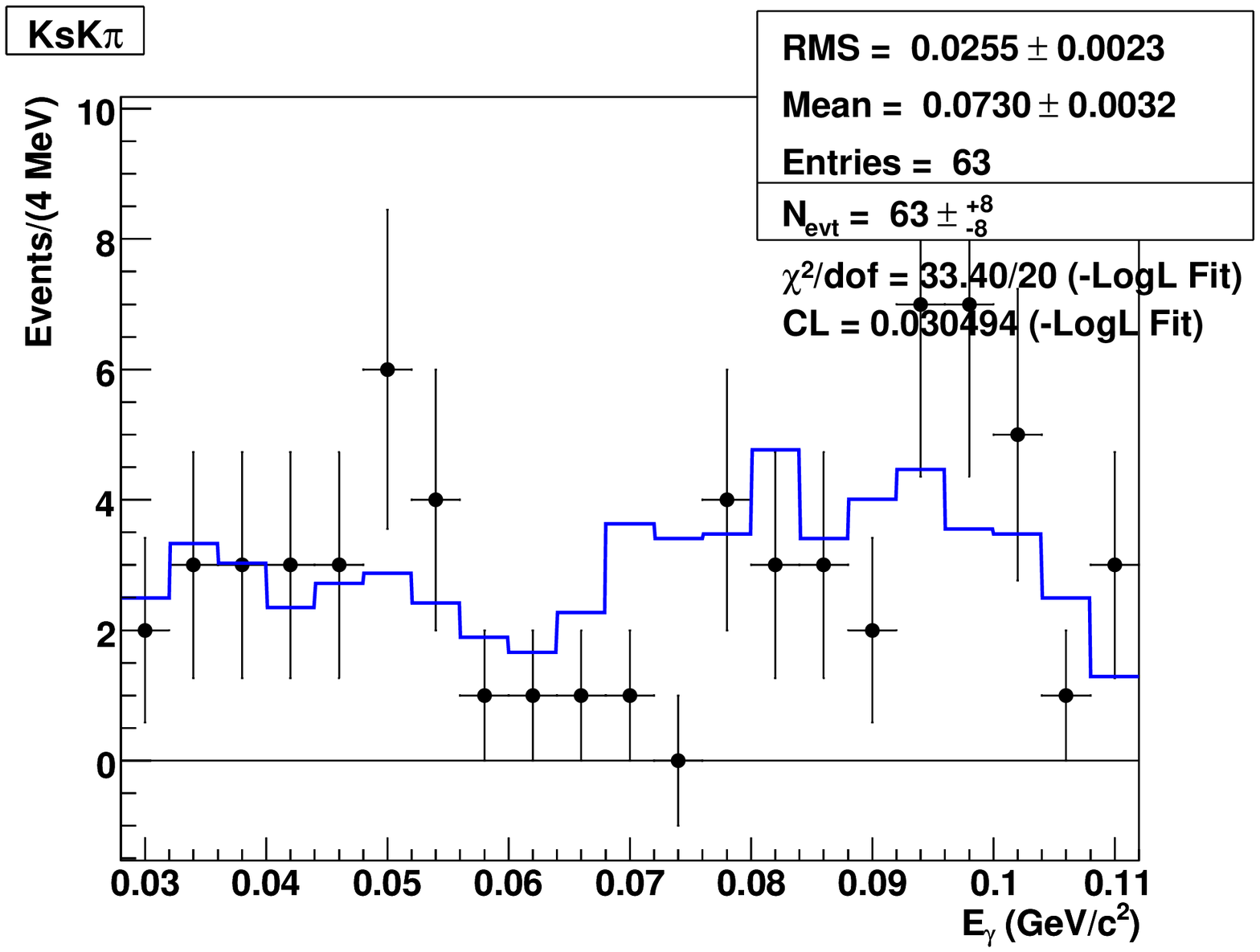}}
  \subfigure
    {\includegraphics[width=.49\textwidth]{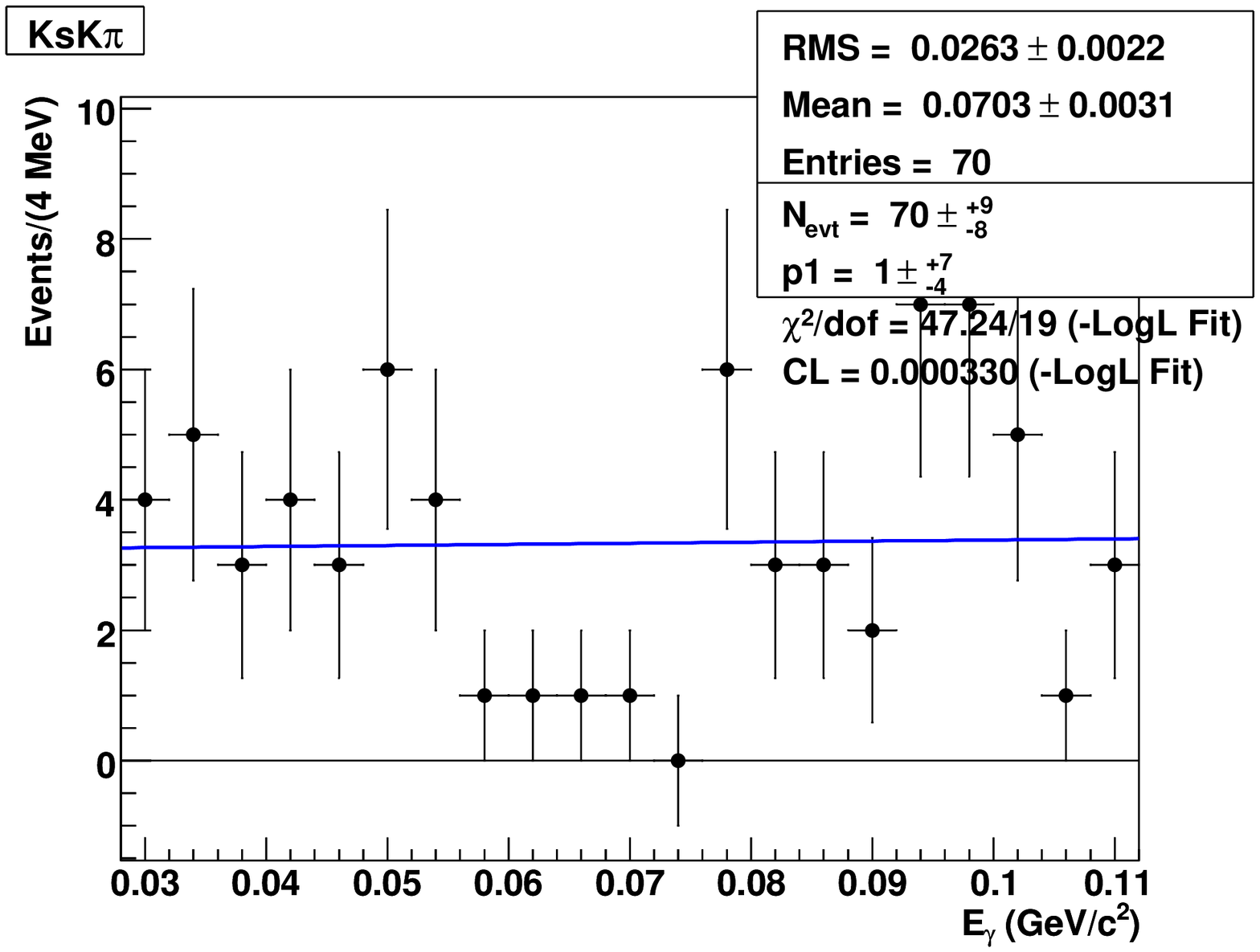}}
  \caption[Background fits of decay $J/\psi \to K_{S}K\pi$]
    {Background fits of the measured photon-energy distribution for
    $\psi(2S) \to \pi^{+} \pi^{-} J/\psi$ events with $J/\psi \to K_{S}K\pi$ 
    and a low energy shower. \ The upper left plot is the fit of 
    data to a single MC histogram. \ The upper right plot is the fit of data 
    to two MC histograms with independent parameters, one (dashed) histogram 
    excluding splitoff showers and the other (dotted) including only splitoff 
    showers. \ The lower left plot is the fit of data to a single MC 
    histogram with a tighter cut of full event 4-C fit $\chi^{2}/{\rm dof} < 3.0$. \ 
    The lower right plot is the fit of data with a linear function.}
  \label{fig:jpsi_KsK3Pi}
\end{figure}

The two primary sources of low energy shower backgrounds are splitoff 
showers and 
FSR. \ 
The splitoff showers are generally very close to the track associated with them. \ 
Most splitoff showers are removed by 
the cut on the transition photon candidate distance to the nearest 
track, but some pass through the cut. \ Additional splitoff showers
can be removed with a tighter cut on the full event 
4-constraint kinematic fit $\chi^{2}/{\rm dof}$. \ Figure~\ref{fig:jpsi_6Pi} 
shows the fits of the measured photon energy distribution for data in 
$J/\psi \to 4\pi$ decays with an accompanying 
(background) photon candidate. \ The upper left plot in the figure shows a 
fit using the 
5 times luminosity generic 
$\psi(2S)$ and continuum MC samples with one free parameter 
(normalization) for the background. \ The upper right plot shows a two 
parameter 
fit to the separate distributions for non-splitoff showers (mostly FSR) 
and for splitoff showers.  A shower was MC-truth tagged as a splitoff shower 
if its parent was not a photon.  \ 
("MC-truth tagging" uses
detailed information stored at the time the MC was created to select only
those particles that come from a particular process.) \ 
The lower left plot shows a 
histogram (one parameter) fit for events when the 4-constraint kinematic fit 
$\chi^{2}/{\rm dof}$ has been tightened to be less than 3. \ The lower right plot shows a fit 
with a linear background function. \ Figures~\ref{fig:jpsi_KK4Pi} through 
\ref{fig:jpsi_KsK3Pi} show similar plots for the other three $J/\psi$ 
decay modes. \ The fits show that the low energy shower distribution of 
the background MC sample represents the data quite well. \ Fitting with the two 
sources of backgrounds separately does not lead to obvious improvement 
of background fitting based on the resulting $\chi^{2}$. \ 
Table~\ref{table:jpsi_background_fit} shows the $\chi^{2}$ results of the four 
types of fits for all four $J/\psi$ decay modes studied.
\begin{table}[htbp]
\caption[Background study with $J/\psi$ decay modes]
{\label{table:jpsi_background_fit}Results of $\chi^{2}/{\rm dof}$ for 
background 
fits of the $J/\psi$ decay modes. \ In this table, ``1 hist'' means 
the results of fitting data with MC histogram; ``2 hist'' means the 
results of fitting data with two MC histograms, one for non-splitoff 
histogram and one for splitoff histogram; ``tight $\chi^{2}/{\rm dof}$'' 
means the results of fitting data with MC histogram when a tighter 4-C 
fit $\chi^{2}/{\rm dof}$ cut was applied; ``linear'' means the results of 
fitting data with a linear function.}
\begin{center}
\begin{tabular}{|l|c|c|c|c|}
  \hline
  $J/\psi$ & \multicolumn{4}{c|}{$\chi^{2}/{\rm dof}$} 
  \\ \cline{2-5}
  Mode &
  1 hist & 2 hist & tight $\chi^{2}$ & linear \\ \hline
  $4\pi$
    & 1.25 & 1.31 & 1.92 & 1.27 \\ \hline
  $KK\pi\pi$
    & 0.59 & 0.61 & 0.74 & 0.99 \\ \hline
  $KK\pi^{0}$
    & 0.59 & 0.52 & 0.96 & 0.99 \\ \hline
  $K_{S}K\pi$
    & 1.99 & 2.05 & 1.79 & 2.48 \\ \hline
\end{tabular}
\end{center}
\end{table}

This study showed that a linear background function provides a reasonable 
alternative to the MC-predicted distributions but one that is not based 
on the physics of $\psi(2S)$ decays or the CLEO-c detector. \ Based on 
this investigation, we chose the single-histogram as our standard 
background shape fit in studying the $\psi(2S)\to\gamma\eta_{c}(2S)$,
$\eta_{c}(2S)\to X$ decays. \ The linear fit is included as an 
alternative for evaluating the systematic uncertainty in the fits.

\subsection{Signal Fit}
\label{subsec:signalfit}

The measured, or unconstrained, shower energy distribution was fit for 
a signal using
a non-relativistic Breit-Wigner function (Cauchy distribution) 
\cite{Breit:1936zzb,Breit} convolved with the Crystal Ball 
resolution function. \ The parameters of the Crystal Ball function, 
$n$, $\alpha$, and $\sigma$ were obtained from the resolution fit and 
were fixed in the signal fits. \ The Breit-Wigner function is defined 
as follows:
\begin{equation}
f(x;\bar{x},\Gamma) \sim \frac{1}{(x - \bar{x})^{2}+(\Gamma/2)^{2}},
\end{equation}
where $\Gamma$ is the width of the resonance and $\bar{x}$ is the mean 
of the variable $x$. \ In our signal fits, $x$ represents the measured 
photon energy.

For the signal fits, the parameters 
are extracted with a binned likelihood 
fit, minimizing the negative logarithm of likelihood using RooFit 
\cite{roofit}. \ For each mode the histogram background shape was the 
sum of events from the 10 times generic $\psi(2S)$ and 5 times continuum 
MC sample which passed our selection criteria. \ We fit the data with the 
above signal function and background histogram.

To verify the fitting procedure, the decays of
$\psi(2S) \to \gamma\chi_{c2}, \chi_{c2}\to X$ were studied. 
 \ The details of this study can be found in Section~\ref{sec:chic2}.

\section{Test of Fit Procedure with $\psi(2S) \to \gamma \chi_{c2}$ Decays}
\label{sec:chic2}

The study of the decay channel $\psi(2S) \to \gamma \chi_{c2}$ was 
used to verify the analysis procedures for the decay 
$\psi(2S)\to\gamma\eta_{c}(2S)$.

\subsection{Introduction}

The radiative decay of $\psi(2S)\to\gamma\chi_{c2}$ is similar to the 
decay $\psi(2S)\to\gamma\eta_{c}(2S)$. \ 
The $\chi_{c2}$ is the P-wave spin triplet charmonium state with total 
angular momentum $J=2$ 
mass $3556.20 \pm 0.09$~MeV, and width $2.05\pm0.12$~MeV \cite{PDBook2006}. 
\ For a $\psi(2S)$ meson nearly at rest, 
the transition photon from the decay $\psi(2S)\to\gamma\chi_{c2}$ 
should have energy around $127.60$~MeV. \ This is 80 MeV higher than the 
energy of the transition photon in the decay 
$\psi(2S)\to\gamma\eta_{c}(2S)$. \ While the 
photon energy range and background parametrization 
for the $\chi_{c2}$ study are different from the $\eta_{c}(2S)$ study,
the other event features and analysis procedures are very similar.

\subsection{Analysis Procedures}

Since the purpose of studying this decay channel is to evaluate the 
analysis procedure of $\psi(2S)\to\gamma\eta_{c}(2S)$, similar procedures 
were applied to $\psi(2S)\to\gamma\chi_{c2}$ decays. \ The results were 
obtained from similar resolution and signal fits on signal MC and data samples.

Same as $\eta_{c}(2S)$ decays, the data sample for this study was 
the full sample of Data 32 and Data 42 of CLEO-c data, 25.9 M 
$\psi(2S)$ decays. \ As described in Section~\ref{subsec:signalmc}, 
the $\chi_{c2}$ signal MC sample 
was generated to replicate the full CLEO-c $\psi(2S)$ sample, similar 
to the $\eta_{c}(2S)$ sample.

To validate the event selection criteria for 
$\psi(2S)\to\gamma\eta_{c}(2S)$ decays, the same criteria 
for $\psi(2S)\to\gamma\eta_{c}(2S)$ decays were applied to 
$\psi(2S)\to\gamma\chi_{c2}$ decays, except that 
$\Delta M = M_{\psi(2S)} - M_{\rm had} = (0,100)~{\rm MeV}$
was removed and $E_{\gamma} = (30,110)~{\rm MeV}$ was replaced with
$E_{\gamma} = (90,145)~{\rm MeV}$.

\subsection{Results}

For each mode, the Crystal Ball function \cite{Gaiser:1982yw} was used to determine 
the detector resolution of the measured photon energy from signal MC. \ 
The parameter $n$ of the Crystal
Ball function was set at a fixed value of $140$ in order to obtain 
successful fits. \ From the resolution fits, the value of 
$\sigma$ and $\alpha$ were determined. \ These values were then fixed and 
applied in fitting the measured photon energy distribution of data. 

In the signal fit of each mode, we used a Breit-Wigner distribution 
convolved with a Crystal Ball resolution as the signal function. \ In 
addition to fixed $n$, $\alpha$ and $\sigma$, the width ($\Gamma$) 
was fixed to $2.05$~MeV. \ Unlike the 
$\psi(2S)\to\gamma\eta_{c}(2S)$ study, the background shape was either 
a linear function, constant, or no background, depending on the fitting 
status of the linear background fit. \ For all modes, a fit with a linear 
background and mean left free was performed first. \ If the linear fit 
did not converge, and the signal mean was within $\pm 1~{\rm MeV}$ of 
the nominal mean 
(127.60~MeV), then a fit with a constant background and mean left free 
was attempted. \ If the linear fit did not converge and the mean was not 
within $\pm 1~{\rm MeV}$ of the nominal value, then a fit with a constant
background and the mean fixed to 127.60~MeV was attempted.  If the linear fit 
resulted in a background with a negative yield, then a fit with no background 
was performed. \ According to these rules,
the modes $4\pi$, $6\pi$, $KK\pi\pi$, $KK4\pi$ and $K_{S}K3\pi$ were fit with a 
linear background; the modes $KK\pi^{0}$, $\pi\pi\eta(\gamma\gamma)$, 
$\pi\pi\eta(\pi\pi\pi^{0})$ and $KK\pi\pi\pi^{0}$ were fit with a
constant background; the modes $KK\eta(\gamma\gamma)$ and 
$\pi\pi\eta^{\prime}$ with $\eta^{\prime}\to\pi\pi\eta(\gamma\gamma)$ were
fit with a constant background and fixed mean; and the modes $K_{S}K\pi$ and
$KK\eta(\pi\pi\pi^{0})$ were fit without a background.

The efficiencies and the resolution fit results are listed 
in Table~\ref{table:chic2_result_resfit}. \ The branching fractions of 
the decays $\chi_{c2} \to X$ are given by
\begin{equation}
\mathcal{B}(\chi_{c2} \to X) = \frac{N_{\rm sig}}{\epsilon~N_{\psi(2S)}~
\mathcal{B}(\psi(2S) \to \gamma \chi_{c2})}
\end{equation}
where $N_{\rm sig}$ is the number of signal events (yields) determined by 
the signal 
fits, $\epsilon$ is the efficiency, $N_{\psi(2S)}$ is the total number of 
$\psi(2S)$ decays, and $\mathcal{B}(\psi(2S) \to \gamma \chi_{c2})$ is the 
branching fraction of $\psi(2S) \to \gamma \chi_{c2}$. \ When calculating 
the branching fractions of $\chi_{c2}\to X$, we 
used $N_{\psi(2S)} = 25.9 \times 10^{6}$ \cite{cbx07-4} and 
$\mathcal{B}(\psi(2S) \to \gamma \chi_{c2}) = 
(8.1 \pm 0.4) \%$ \cite{PDBook2006}. \ 
The results of the signal fits 
and the corresponding $\psi(2S)\to\gamma\chi_{c2}$ branching fractions are 
listed in 
Table~\ref{table:chic2_result_sigfit}. \ The data fits for all thirteen modes 
are shown in 
Figures~\ref{fig:chic2_result_4Pi} through \ref{fig:chic2_result_KK4Pi_KsK3Pi}, 
while the 
resolution fits are shown in Appendix ~\ref{appendix:resfnctfits}.

\begin{table}[htbp]
\caption[Efficiencies and resolution function parameters for 
$\psi(2S)\to\gamma\chi_{c2}$ decays]
{\label{table:chic2_result_resfit}
Efficiencies and resolution function parameters for $\psi(2S)\to\gamma\chi_{c2}$ decays.}
\begin{center}
\begin{tabular}{|l|c|c|c|}
  \hline
  Mode &
  $\epsilon$ (\%) &
  Res $\sigma$ (MeV) &
  XBall $\alpha$ \\ \hline
  $4\pi$
    & $26.49\pm0.17$ & $6.23\pm0.04$ & $1.26\pm0.02$ \\ \hline
  $6\pi$
    & $18.43\pm0.15$ & $6.40\pm0.05$ & $1.33\pm0.03$ \\ \hline
  $KK\pi\pi$
    & $24.91\pm0.16$ & $6.30\pm0.04$ & $1.42\pm0.03$ \\ \hline
  $KK\pi^{0}$
    & $25.48\pm0.16$ & $6.12\pm0.05$ & $1.10\pm0.02$ \\ \hline
  $K_{S}K\pi$
    & $25.46\pm0.16$ & $6.38\pm0.04$ & $1.56\pm0.04$ \\ \hline
  $\pi\pi\eta(\gamma\gamma)$
    & $16.68\pm0.14$ & $5.97\pm0.05$ & $1.33\pm0.04$ \\ \hline
  $\pi\pi\eta(\pi\pi\pi^{0})$
    & $11.77\pm0.12$ & $6.24\pm0.06$ & $1.40^{+0.05}_{-0.04}$ \\ \hline
  $\pi\pi\eta^{\prime}$
    & $12.86\pm0.15$ & $6.13\pm0.08$ & $1.26^{+0.05}_{-0.04}$ \\ \hline
  $KK\eta(\gamma\gamma)$
    & $18.86\pm0.15$ & $6.12\pm0.05$ & $1.16\pm0.03$ \\ \hline
  $KK\eta(\pi\pi\pi^{0})$
    & $14.01\pm0.13$ & $6.06\pm0.06$ & $1.05\pm0.03$ \\ \hline
  $KK\pi\pi\pi^{0}$
    & $11.93\pm0.12$ & $6.28\pm0.06$ & $1.69^{+0.08}_{-0.07}$ \\ \hline
  $KK4\pi$
    & $12.97\pm0.13$ & $6.37\pm0.06$ & $1.45\pm0.05$ \\ \hline
  $K_{S}K3\pi$
    & $14.21\pm0.13$ & $6.37\pm0.06$ & $1.49^{+0.05}_{-0.04}$ \\ \hline
\end{tabular}
\end{center}
\end{table}

\begin{table}[htbp]
\caption[Result of decay $\psi(2S)\to\gamma\chi_{c2}$]
{\label{table:chic2_result_sigfit}
Results for $\psi(2S)\to\gamma\chi_{c2}$ study. \ The value 
${\cal B}(\psi(2S)\to\gamma\chi_{c2}) = ( 8.1 \pm 0.4 )\%$ was used to 
determine 
$B = {\cal B}(\chi_{c2}\to X)$. \ Only statistical errors are listed. 
\ Note that the PDG value for $K_{S}K\pi$ listed here is corrected, 
because the PDG did not 
convert $K_{S}K\pi$ to $K^{0}K\pi$ when listing BES-II result 
\cite{PDBook2006,Ablikim:2006vm}. \ The mean in the 
$\pi\pi\eta(\pi\pi\pi^{0})$ and $\pi\pi\eta^{\prime}$ fits was fixed to
127.60~MeV, and all others are free.}
\begin{center}
\begin{tabular}{|l|c|c|c|c|}
  \hline
  Mode &
  Mean (MeV) &
  $N_{\rm sig}$ &
  $B$ ($10^{-3}$) &
  $B_{\rm PDG}$ ($10^{-3}$) \\ \hline
  $4\pi$
    & $127.23\pm0.10$ & $7215\pm119$ & $13.0\pm0.2$ & $12.5\pm1.6$ \\ \hline
  $6\pi$
    & $127.09\pm0.11$ & $6083^{+113}_{-112}$ & $15.7\pm0.3$ & $8.7\pm1.8$ \\ \hline
  $KK\pi\pi$
    & $127.20\pm0.12$ & $4717\pm95$ & $9.0\pm0.2$ & $10.0\pm2.6$ \\ \hline
  $KK\pi^{0}$
    & $127.71\pm0.58$ & $219\pm17$ & $0.41\pm0.03$ & $0.36\pm0.09$ \\ \hline
  $K_{S}K\pi$
    & $127.87\pm0.43$ & $294\pm17$ & $0.80\pm0.05$ & $0.71\pm0.11$ \\ \hline
  $\pi\pi\eta(\gamma\gamma)$
    & $128.15^{+0.95}_{-0.97}$ & $97\pm12$ & $0.70\pm0.10$ & $0.56\pm0.15$ \\ \hline
  $\pi\pi\eta(\pi\pi\pi^{0})$
    & $126.86^{+2.10}_{-1.82}$ & $31\pm7$ & $0.55\pm0.13$ & $0.56\pm0.15$ \\ \hline
  $\pi\pi\eta^{\prime}$
    & $127.60$ & $3.7\pm5.2$ & $0.08\pm0.11$ & $0.59\pm0.22$ \\ \hline
  $KK\eta(\gamma\gamma)$
    & $127.60$ & $29\pm8$ & $0.19\pm0.05$ & $<0.4$ \\ \hline
  $KK\eta(\pi\pi\pi^{0})$
    & $127.96^{+2.21}_{-2.61}$ & $17\pm5$ & $0.26\pm0.08$ & $<0.4$ \\ \hline
  $KK\pi\pi\pi^{0}$
    & $127.39\pm0.14$ & $3197\pm62$ & $12.8\pm0.3$ & - \\ \hline
  $KK4\pi$
    & $127.24\pm0.18$ & $2249\pm68$ & $8.3\pm0.3$ & - \\ \hline
  $K_{S}K3\pi$
    & $127.49\pm0.23$ & $1453^{+53}_{-54}$ & $7.2\pm0.3$ & - \\ \hline
\end{tabular}
\end{center}
\end{table}

\begin{figure}[htbp]
\begin{center}
\subfigure
{\includegraphics[width=.80\textwidth]{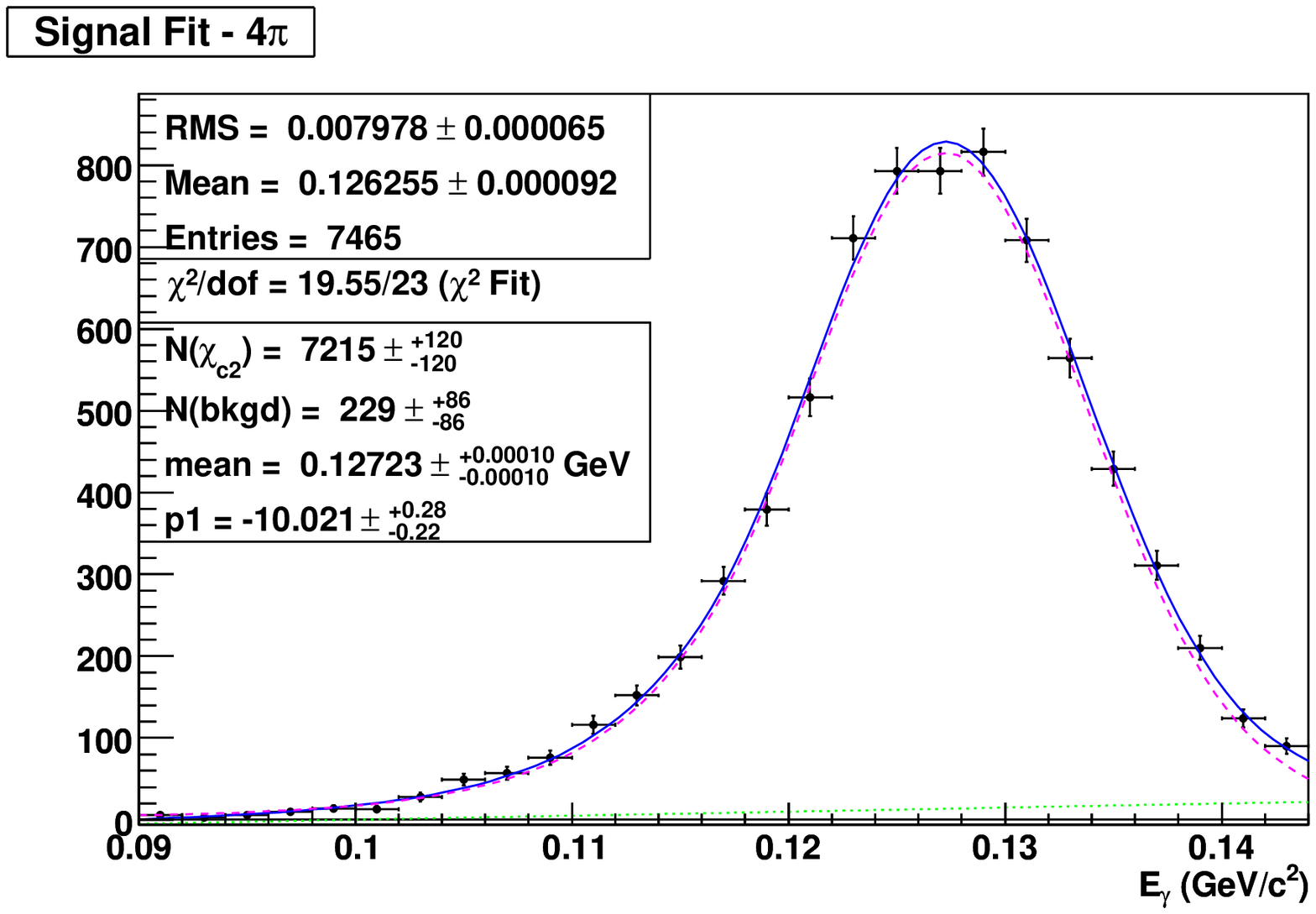}}
\end{center}
\caption[Measured photon energy fit for $\psi(2S)\to\gamma\chi_{c2}$, 
$\chi_{c2}\to 4\pi$ mode.]
{\label{fig:chic2_result_4Pi}
{Measured photon energy for the decay mode $\psi(2S)\to\gamma\chi_{c2}, 
\chi_{c2}\to 4\pi$.
 \ The points are from the 25.9 M $\psi(2S)$ data sample. \ The dashed line 
is the result of the signal fit, the dotted line is the background fit, and 
the solid line is the sum of the signal and background fits.}}
\end{figure}

\begin{figure}[htbp]
\begin{center}
\subfigure
{\includegraphics[width=.80\textwidth]{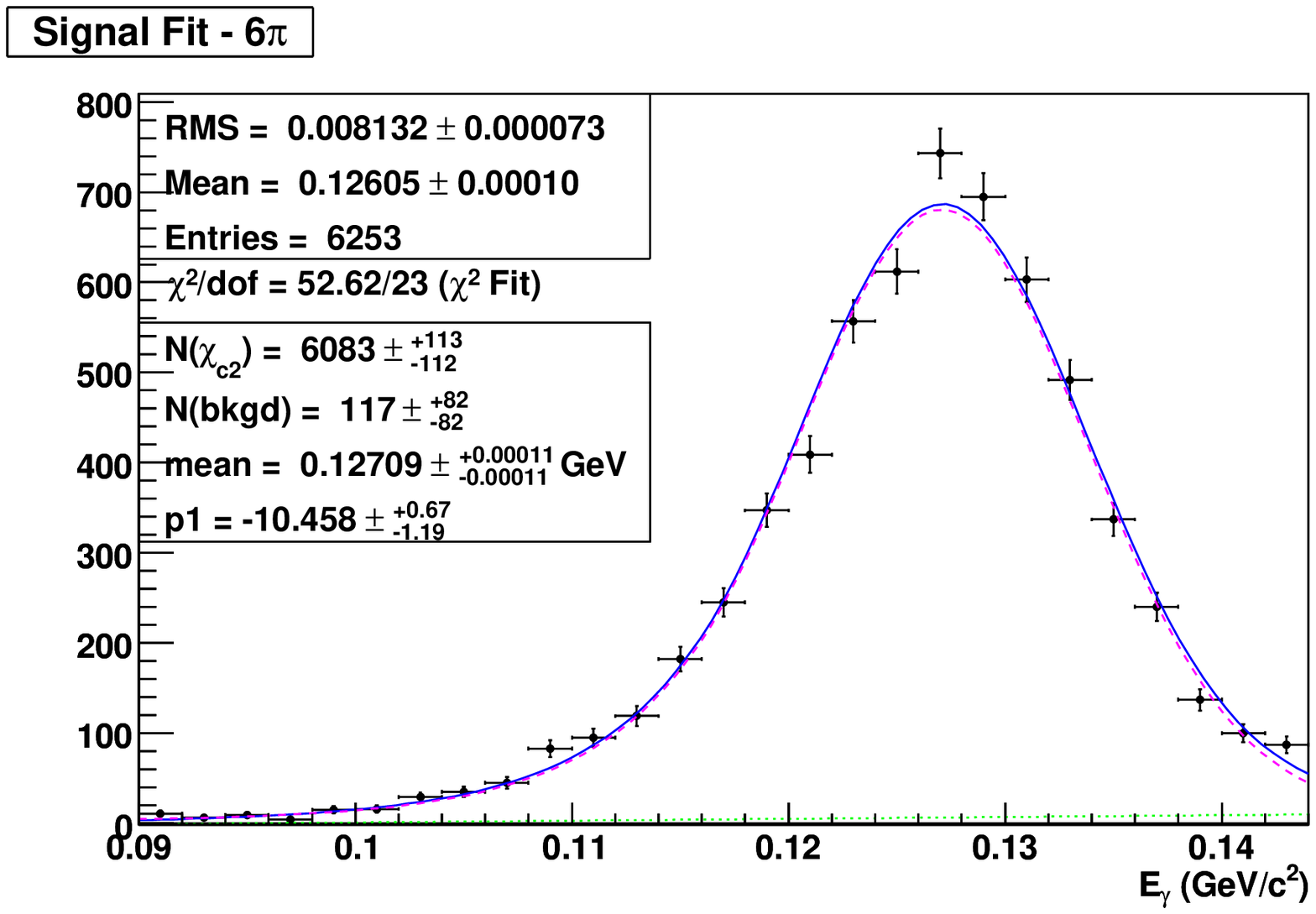}}
\subfigure
{\includegraphics[width=.80\textwidth]{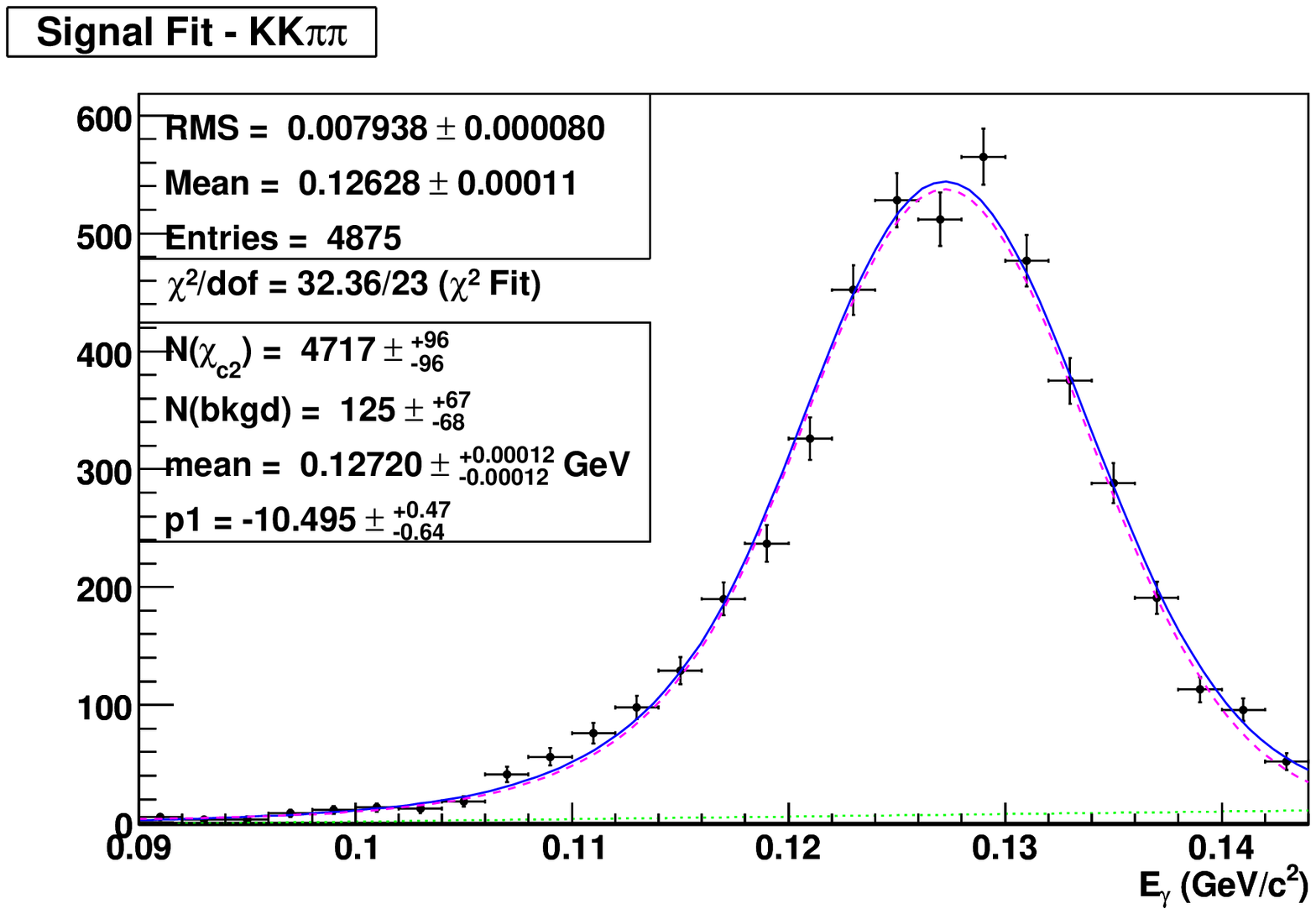}}
\end{center}
\caption[Measured photon energy fits for $\psi(2S)\to\gamma\chi_{c2}$, 
$\chi_{c2}\to 6\pi$ and $KK\pi\pi$ modes.]
{\label{fig:chic2_result_6Pi_KKPiPi}
{Measured photon energy for the decay modes 
$\psi(2S)\to\gamma\chi_{c2}, \chi_{c2}\to 6\pi$ (top) 
and $KK\pi\pi$ (bottom).  
 \ The points are from the 25.9 M $\psi(2S)$ data sample. \ The dashed line 
is the result of the signal fit, the dotted line is the background fit, and 
the solid line is the sum of the signal and background fits.}}
\end{figure}

\begin{figure}[htbp]
\begin{center}
\subfigure
{\includegraphics[width=.80\textwidth]{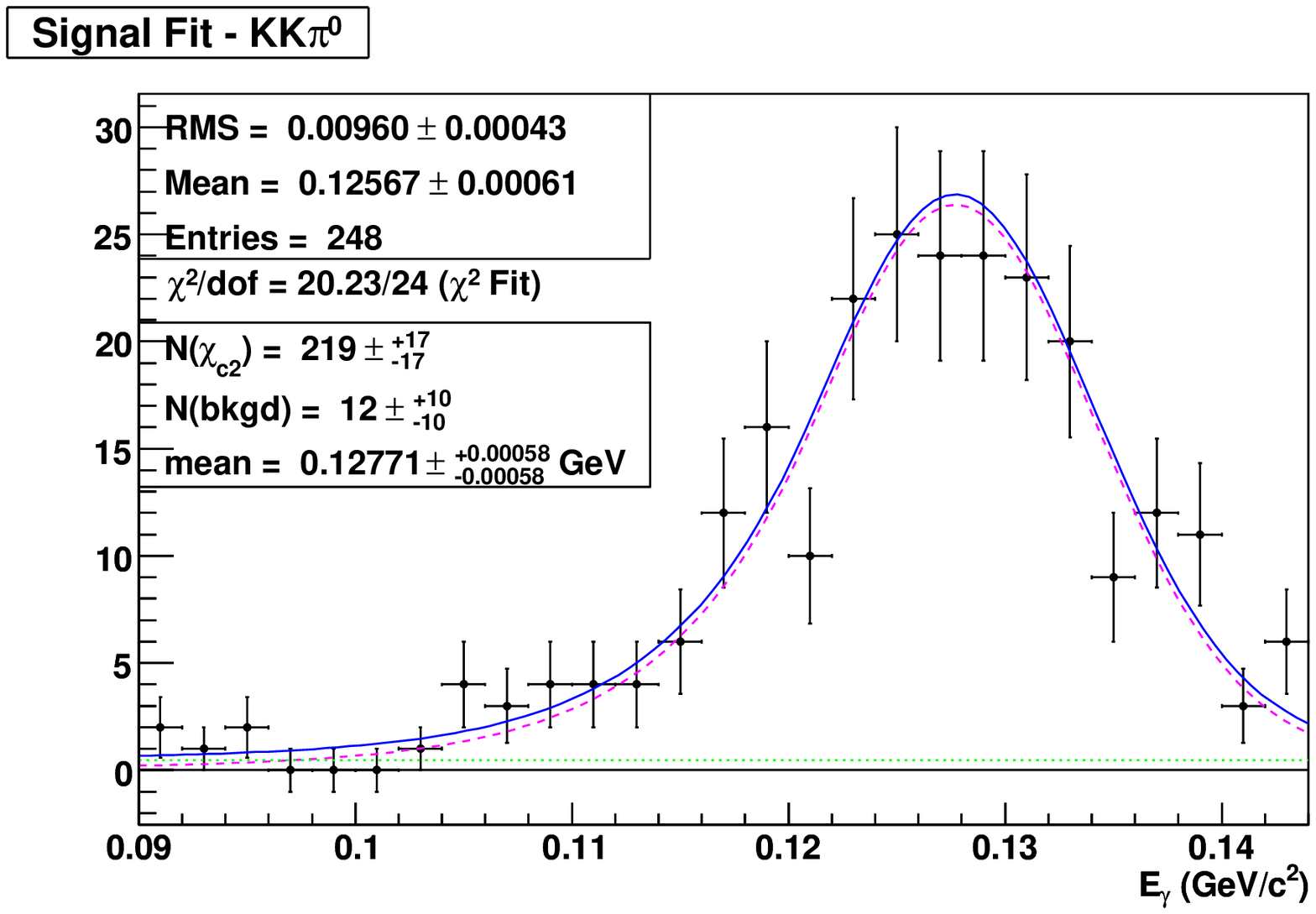}}
\subfigure
{\includegraphics[width=.80\textwidth]{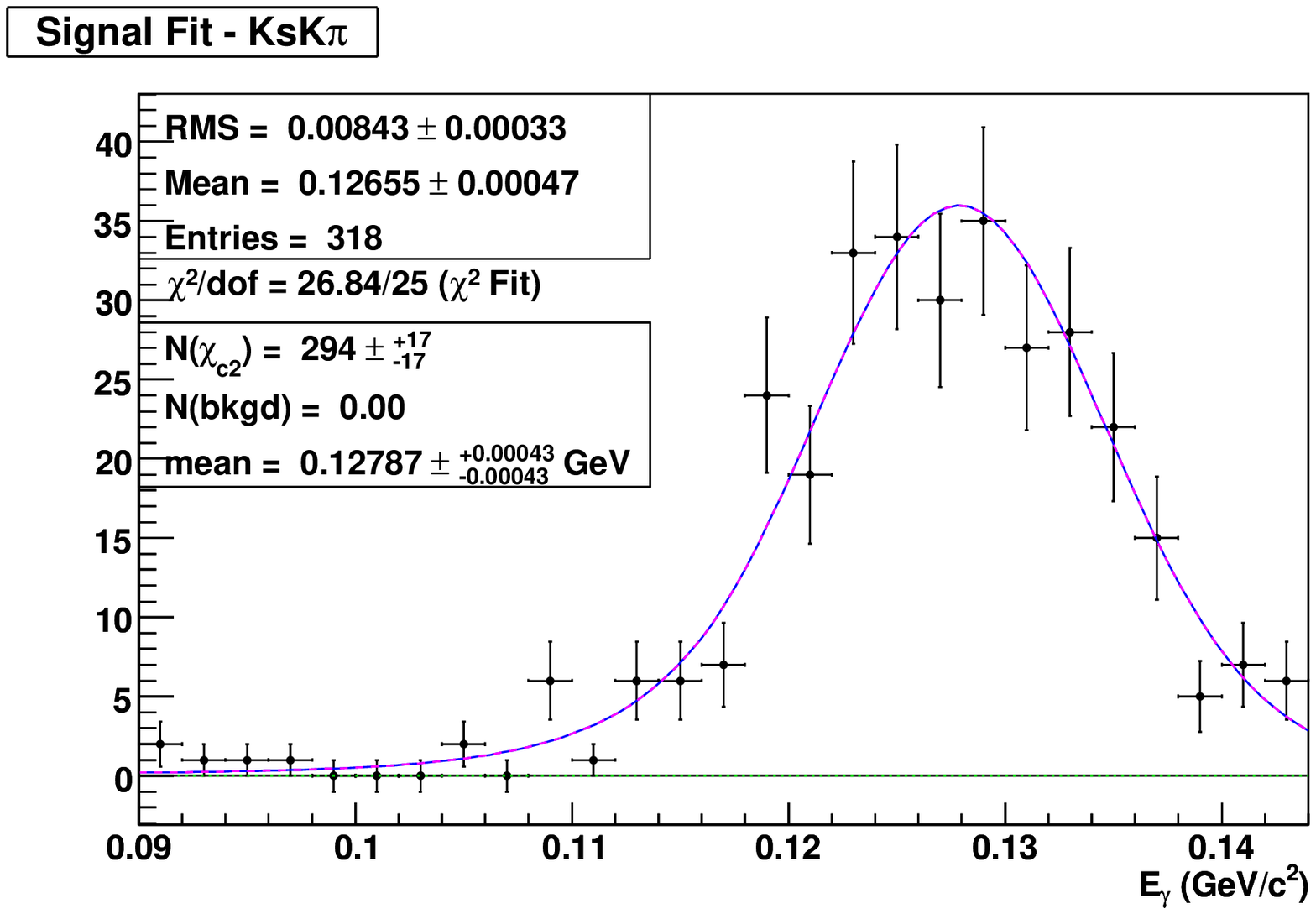}}
\end{center}
\caption[Measured photon energy fits for $\psi(2S)\to\gamma\chi_{c2}$, 
$\chi_{c2}\to KK\pi^{0}$ and $K_{S}K\pi$ modes.]
{\label{fig:chic2_result_KKPi0_KsKPi}
{Measured photon energy for the decay modes $\psi(2S)\to\gamma\chi_{c2}, 
\chi_{c2}\to  KK\pi^{0}$ (top) and $K_{S}K\pi$ (bottom).
 \ The points are from the 25.9 M $\psi(2S)$ data sample. \ The dashed line 
is the result of the signal fit, the dotted line is the background fit, and 
the solid line is the sum of the signal and background fits.}}
\end{figure}

\begin{figure}[htbp]
\begin{center}
\subfigure
{\includegraphics[width=.80\textwidth]{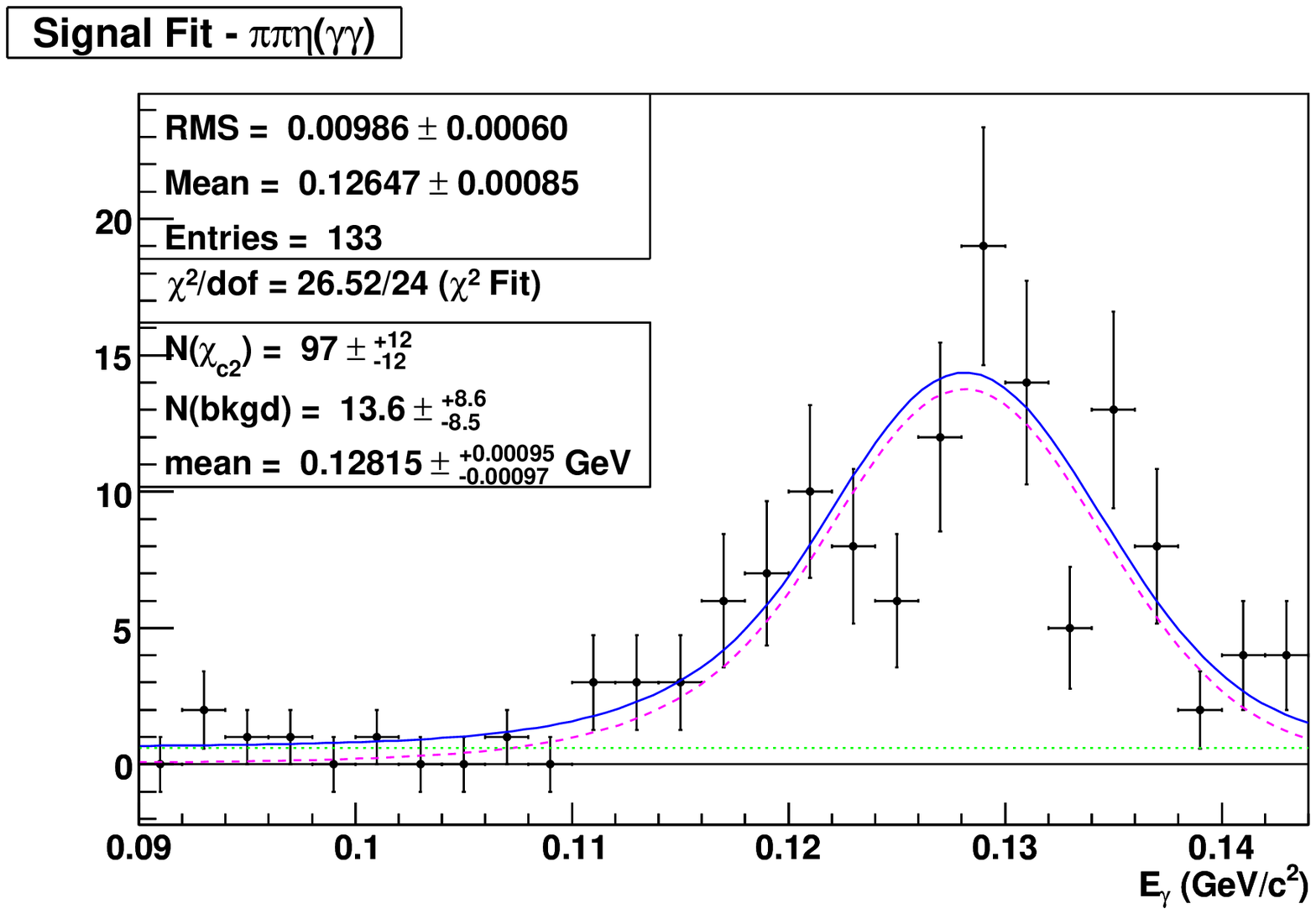}}
\subfigure
{\includegraphics[width=.80\textwidth]{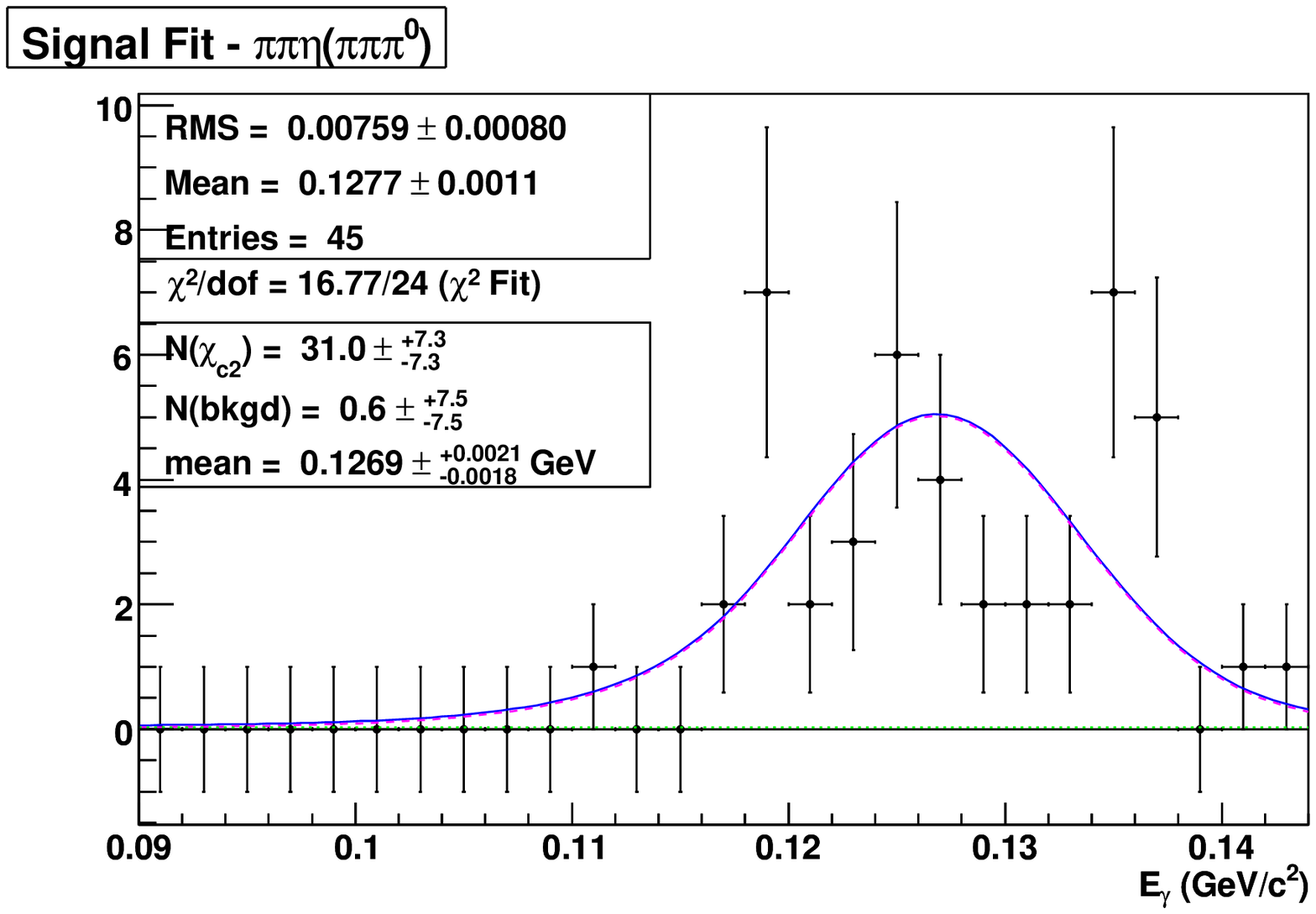}}
\end{center}
\caption[Measured photon energy fits for $\psi(2S)\to\gamma\chi_{c2}$, 
$\chi_{c2}\to \pi\pi\eta(\gamma\gamma)$ and $\pi\pi\eta(\pi\pi\pi^{0})$ modes.]
{\label{fig:chic2_result_PiPiEtaGG_PiPiEtaPiPiPi0}
{Measured photon energy for the decay modes $\psi(2S)\to\gamma\chi_{c2}, 
\chi_{c2}\to \pi\pi\eta(\gamma\gamma)$ (top) and $\pi\pi\eta(\pi\pi\pi^{0})$ (bottom).  
 \ The points are from the 25.9 M $\psi(2S)$ data sample. \ The dashed line 
is the result of the signal fit, the dotted line is the background fit, and 
the solid line is the sum of the signal and background fits.}}
\end{figure}

\begin{figure}[htbp]
\begin{center}
\subfigure
{\includegraphics[width=.80\textwidth]{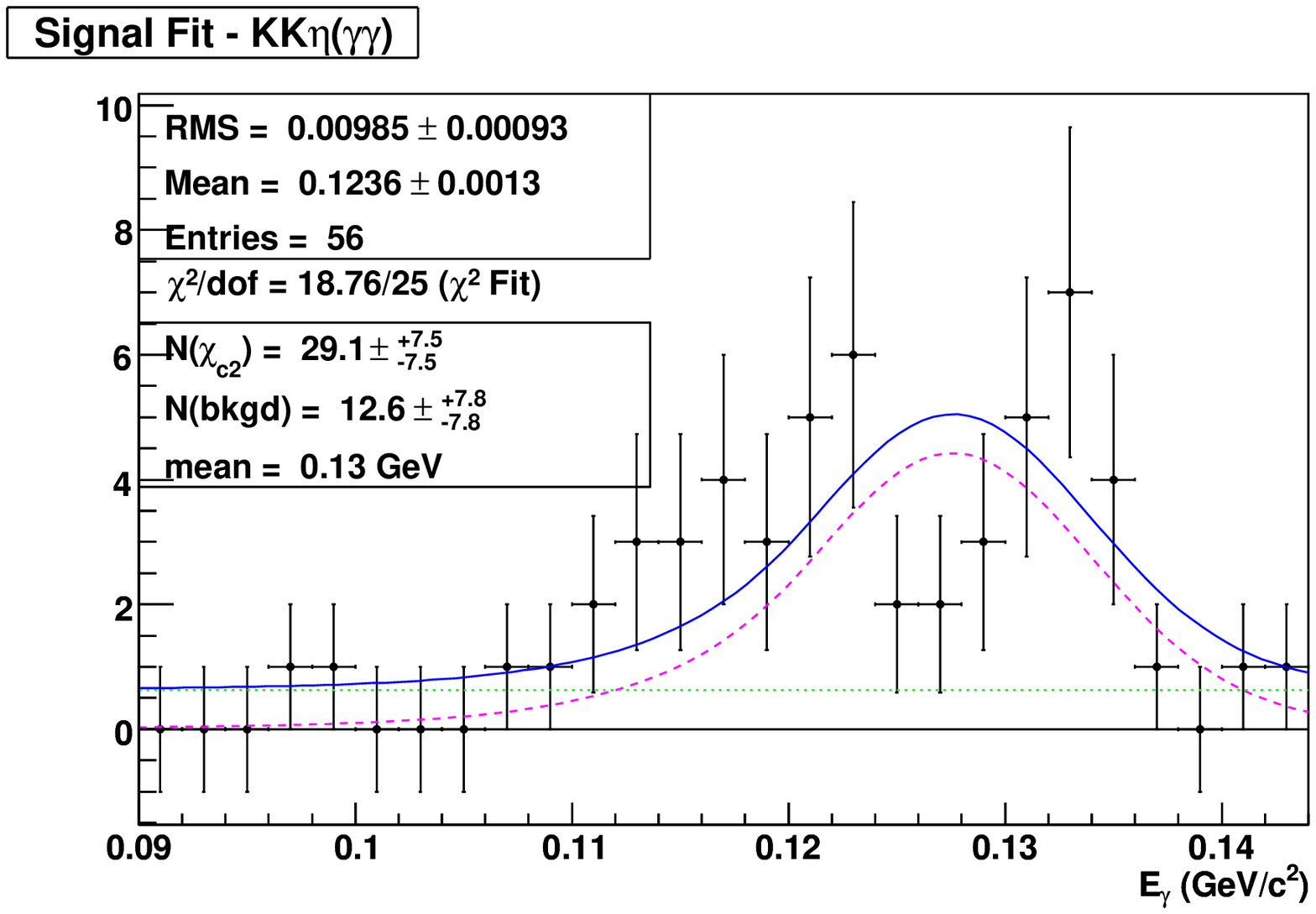}}
\subfigure
{\includegraphics[width=.80\textwidth]{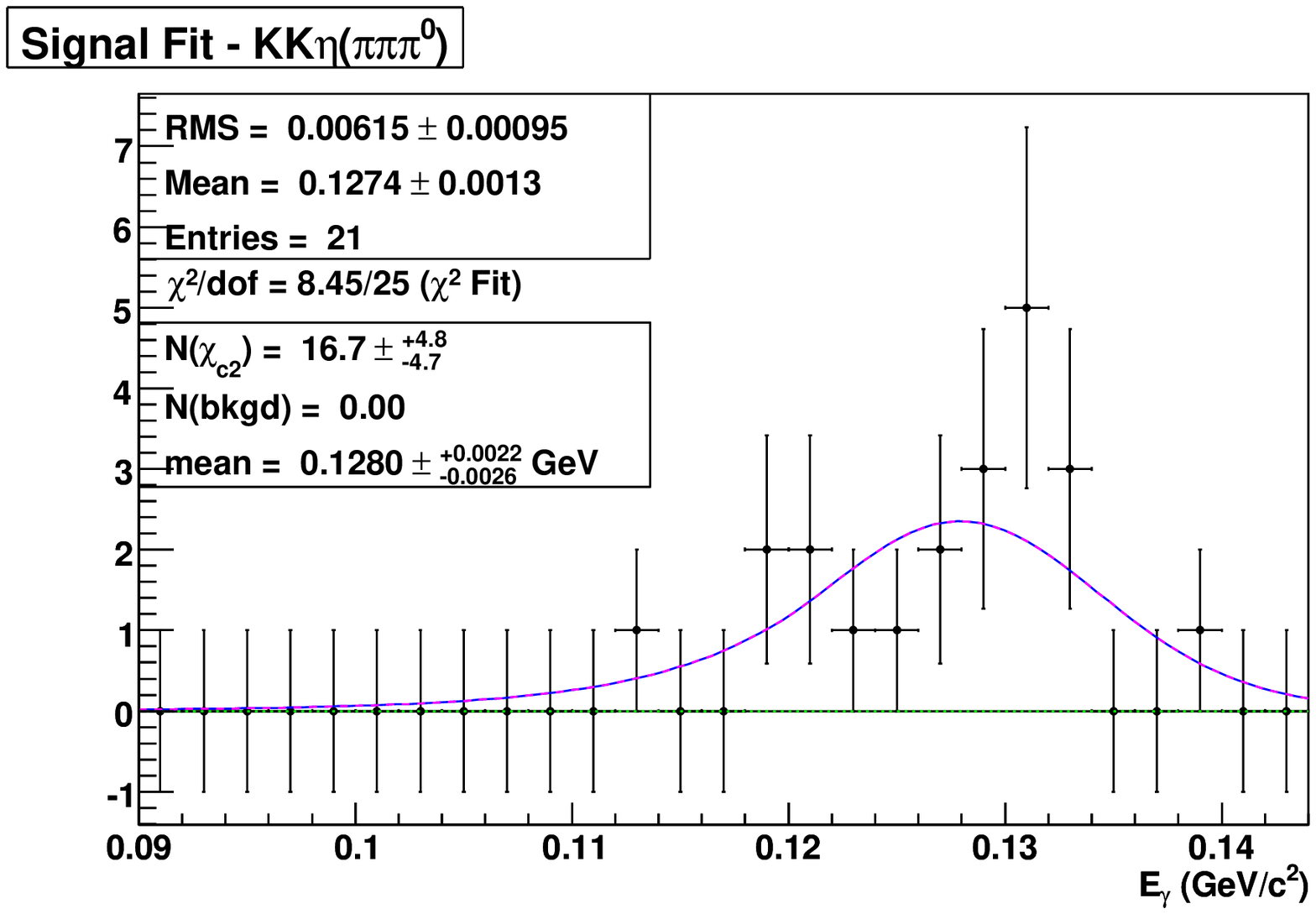}}
\end{center}
\caption[Measured photon energy fits for $\psi(2S)\to\gamma\chi_{c2}$, 
$\chi_{c2}\to KK\eta(\gamma\gamma)$ and $KK\eta(\pi\pi\pi^{0})$ modes.]
{\label{fig:chic2_result_KKEta_KKEtaPiPiPi0}
{Measured photon energy for the decay modes $\psi(2S)\to\gamma\chi_{c2}, 
\chi_{c2}\to KK\eta(\gamma\gamma)$ (top) and $KK\eta(\pi\pi\pi^{0})$ (bottom).  
 \ The points are from the 25.9 M $\psi(2S)$ data sample. \ The dashed line 
is the result of the signal fit, the dotted line is the background fit, and 
the solid line is the sum of the signal and background fits.}}
\end{figure}

\begin{figure}[htbp]
\begin{center}
\subfigure
{\includegraphics[width=.80\textwidth]{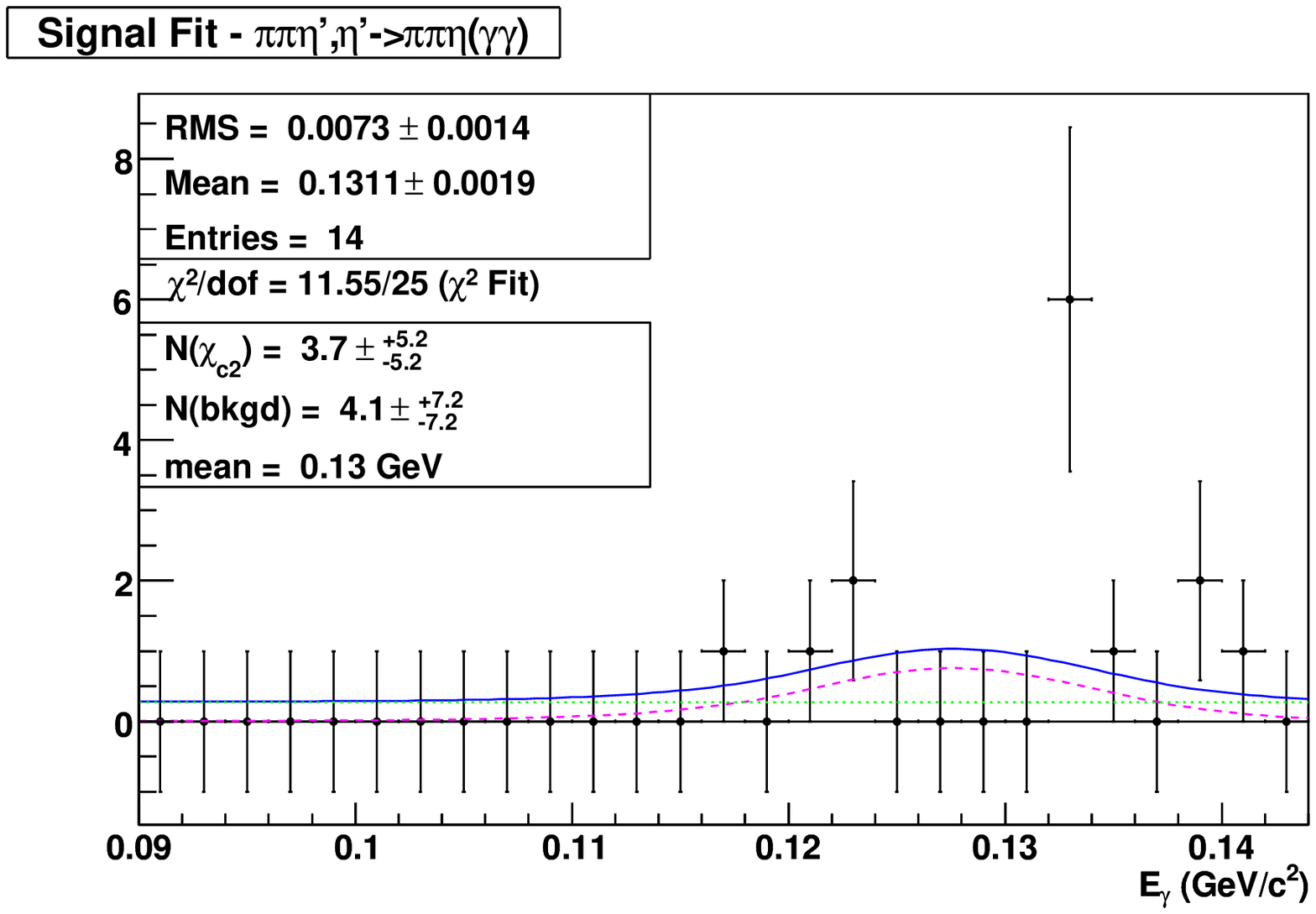}}
\subfigure
{\includegraphics[width=.80\textwidth]{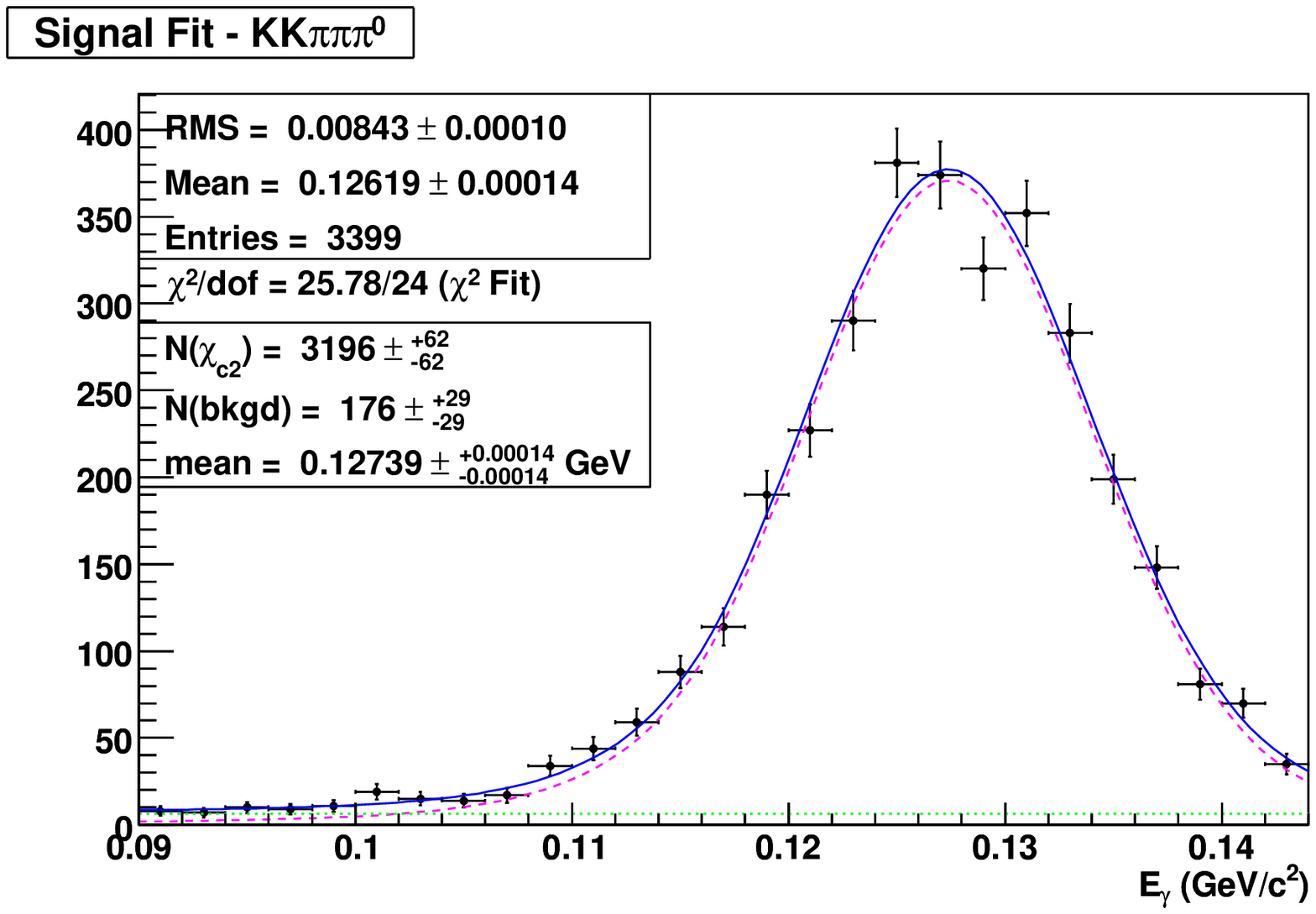}}
\end{center}
\caption[Measured photon energy fits for $\psi(2S)\to\gamma\chi_{c2}$, 
$\chi_{c2}\to \pi\pi\eta^{\prime}$ 
and $KK\pi\pi\pi^{0}$ modes.]
{\label{fig:chic2_result_PiPiEtaPPiPiEtaGG_KKPiPiPi0}
{Measured photon energy for the decay modes $\psi(2S)\to\gamma\chi_{c2}, 
\chi_{c2}\to \pi\pi\eta^{\prime}, \eta^{\prime}\to\pi\pi\eta(\gamma\gamma)$ (top) 
and $KK\pi\pi\pi^{0}$ (bottom).  
 \ The points are from the 25.9 M $\psi(2S)$ data sample. \ The dashed line 
is the result of the signal fit, the dotted line is the background fit, and 
the solid line is the sum of the signal and background fits.}}
\end{figure}

\begin{figure}[htbp]
\begin{center}
\subfigure
{\includegraphics[width=.80\textwidth]{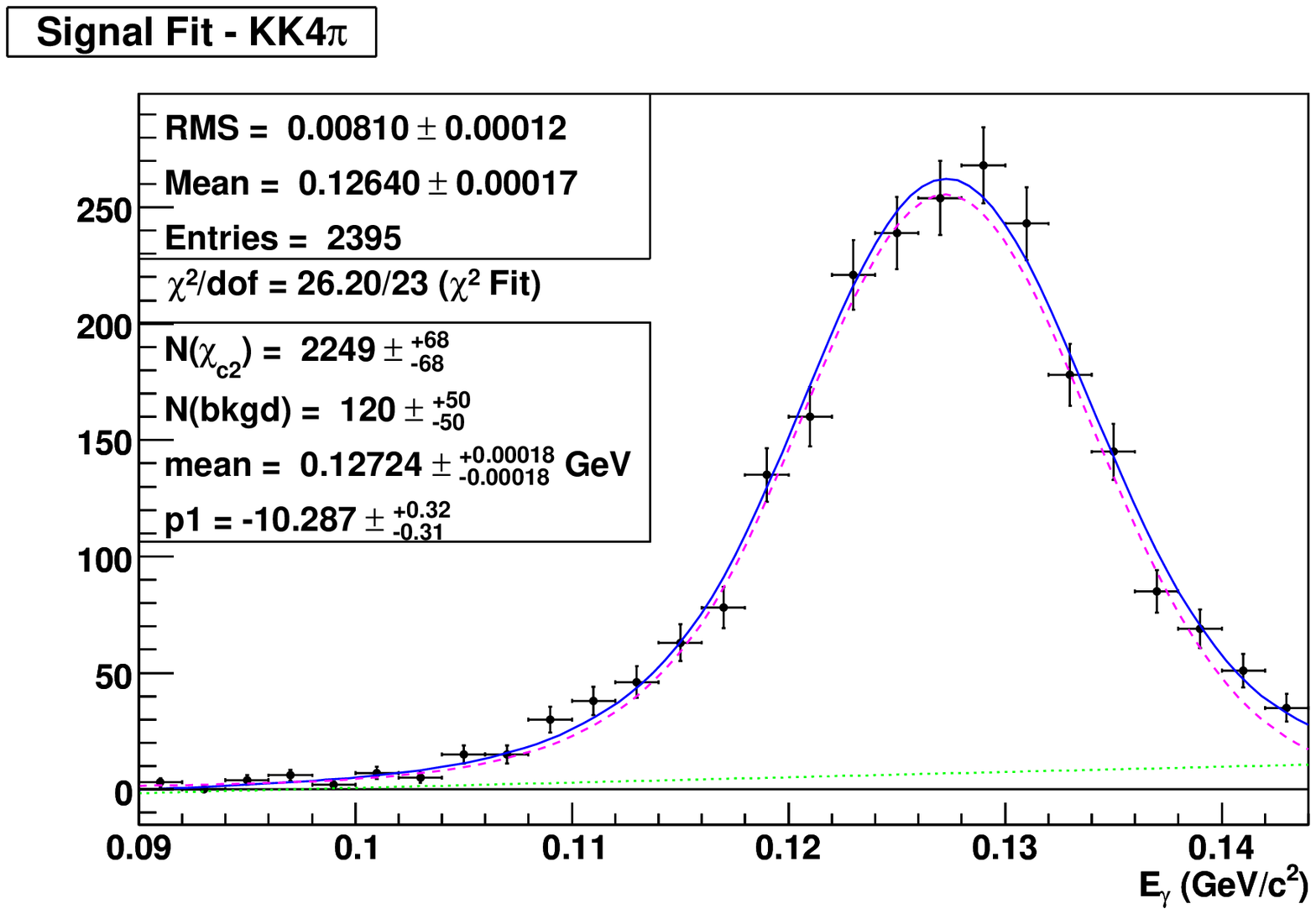}}
\subfigure
{\includegraphics[width=.80\textwidth]{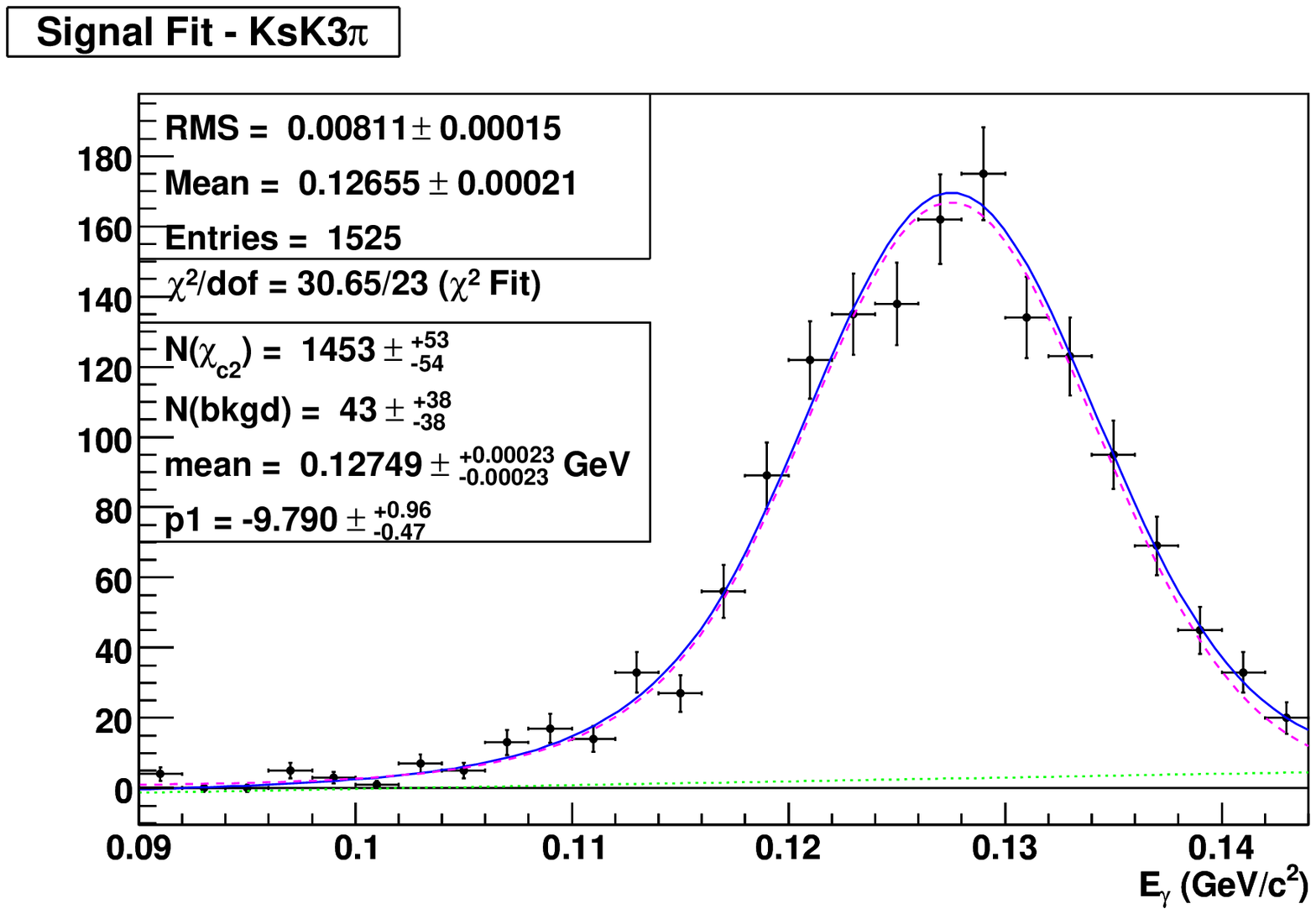}}
\end{center}
\caption[Measured photon energy fits for $\psi(2S)\to\gamma\chi_{c2}$, 
$\chi_{c2}\to KK4\pi$ and $K_{S}K3\pi$ modes.]
{\label{fig:chic2_result_KK4Pi_KsK3Pi}
{Measured photon energy for the decay modes $\psi(2S)\to\gamma\chi_{c2}, 
\chi_{c2}\to KK4\pi$ (top) and $K_{S}K3\pi$ (bottom).  
 \ The points are from the 25.9 M $\psi(2S)$ data sample. \ The dashed line 
is the result of the signal fit, the dotted line is the background fit, and 
the solid line is the sum of the signal and background fits.}}
\end{figure}

\subsection{Discussion and Conclusion for the $\chi_{c2}$ Study}

From the results we can see that for most modes the results are consistent
with the world averages. \ However, two modes show inconsistency with the 
PDG values, $6\pi$ and 
$\pi\pi\eta^{\prime}$.

The previous CLEO measurement \cite{athar:032002, cbx06-22} shows the result of
the branching fractions of $\chi_{c2} \to X$ decays to be 
$(0.49\pm0.12\pm0.05\pm0.03)\times 10^{-3}$ for $\pi\pi\eta$ mode,
$< 0.33 \times 10^{-3}$ for $KK\eta$ mode,
$(0.51\pm0.18\pm0.05\pm0.03)\times 10^{-3}$ for $\pi\pi\eta^{\prime}$ mode,
and 
$(0.31\pm0.07\pm0.03\pm0.02)\times 10^{-3}$ for $KK\pi^{0}$ mode. \ 
Our results are consistent with this measurement except for 
$\chi_{c2} \to \pi\pi\eta^{\prime}$ mode. \ The cuts in our measurements 
are not optimized for $\chi_{c2}$ studies.

The PDG branching fraction average for $\chi_{c2}\to 6\pi$ is dominated by 
measurement from the BES-I collaboration \cite{PhysRevD.60.072001}. \ As can 
be seen in Table~\ref{table:chic2_result_detail}, the BES-I results are 
systematically 
lower than our results, while our measurement of $\chi_{c2} \to 4\pi$ 
is completely consistent with the PDG average.

\begin{table}[htbp]
\caption[Detailed comparison of decay 
$\chi_{c2} \to X$ results]
{\label{table:chic2_result_detail}
Detailed comparison of decay $\psi(2S)\to\gamma\chi_{c2}$ results between 
our results, BES-I \cite{PhysRevD.60.072001}, and the PDG.  \ 
The PDG branching fraction for $\chi_{c2}\to 4\pi$ mode is from a 28 
parameter fit using properties of the 
$\chi_{c0}$, $\chi_{c1}$, $\chi_{c2}$, $\psi(2S)$ \cite{PDBook2006}. \ 
The BES-I measurement of $\chi_{c2}\to 6\pi$ dominates the PDG average. \  
All branching fractions listed in units of $10^{-3}$.} 
\begin{center}
\begin{tabular}{|l|c|c|c|c|}
  \hline
  Mode &
  PDG &
  This result &
  BES-I &
  This Result/BES-I \\ \hline
  $4\pi$
    & $12.5\pm1.6$ & $13.00\pm0.22$ & $9.2\pm2.4$ & $1.4\pm0.4$ \\ \hline
  $6\pi$
    & $8.7\pm1.8$ & $15.75\pm0.29$ & $8.7\pm1.9$ & $1.8\pm0.4$ \\ \hline
  $KK\pi\pi$
    & $10.0\pm2.6$ & $9.03\pm0.18$ & $7.6\pm1.9$ & $1.2\pm0.3$ \\ \hline
\end{tabular}
\end{center}
\end{table}

For the mode $\pi\pi\eta^{\prime}$,
we were not able to explain for the discrepancy other than
a statistical fluctuation in either our or the previous CLEO measurement
\cite{athar:032002, cbx06-22}.

Some of the decay modes like $KK\pi\pi\pi^{0}$, $KK4\pi$ and $K_{S}K3\pi$
have not been measured before. \ For the decay mode 
$KK\eta$, 
only upper limits were
given in PDG based on previous measurements. \ In this study, 
the branching fractions of $\chi_{c2}$ to these modes were determined. \
Even though the decays were studied with the optimized criteria for 
$\psi(2S)\to\gamma\eta_{c}(2S)$ decays, the branching fractions results 
are consistent with values listed by the PDG. \ If the decays were studied 
with the optimized criteria for $\chi_{c2}$ decays, the efficiencies would 
have been higher and the statistical error would have been reduced.

Overall, the results of the study of $\psi(2S)\to\gamma\chi_{c2}$, 
$\chi_{c2} \to X$ decays 
establish the validity of the analysis procedures for 
$\psi(2S)\to\gamma\eta_{c}(2S)$. \ There is no explanation for the small, 
but statistically significant, discrepancy observed in the 
$\pi\pi\eta^{\prime}$ mode, but since the analysis of this mode is not 
appreciably different from the higher-statistics modes, we assume that 
it too has no fundamental flaw.

\section{Results for $\psi(2S)\to \gamma \eta_c(2S)$}
\label{sec:results}

\begin{figure}[htbp]
\begin{center}
\subfigure
{\includegraphics[width=.80\textwidth]{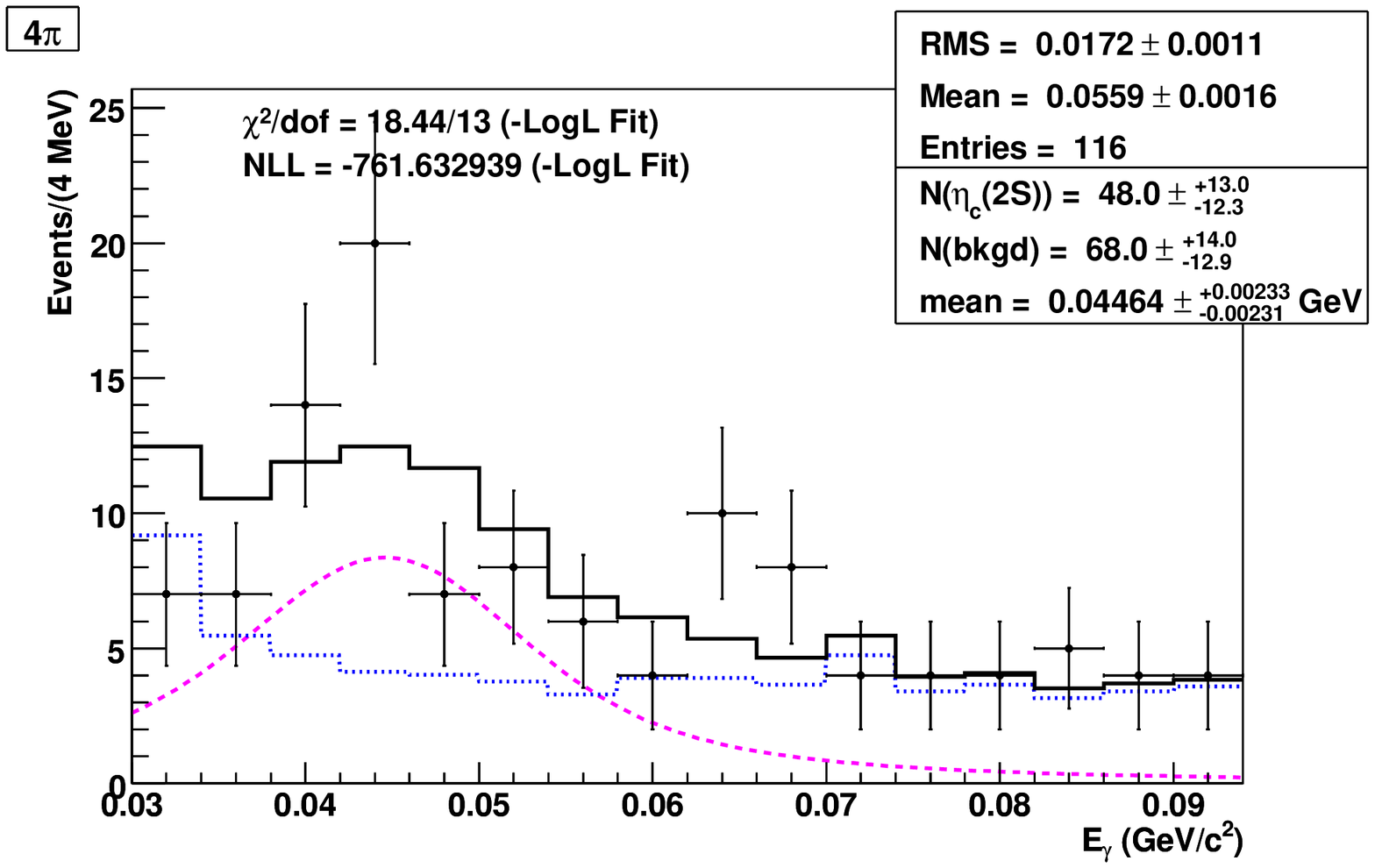}}
\subfigure
{\includegraphics[width=.80\textwidth]{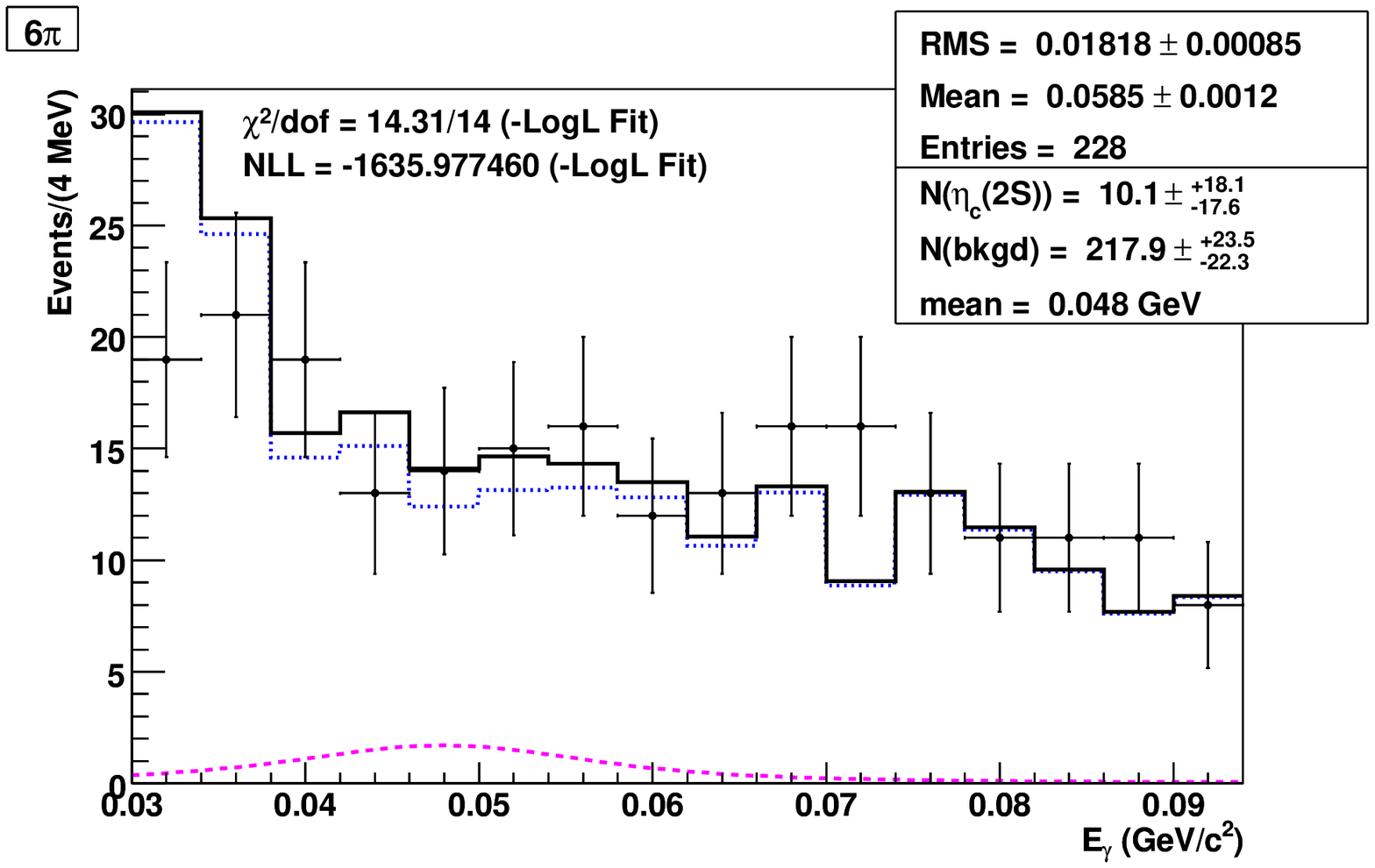}}
\end{center}
\caption[Measured photon energy for the final states $\gamma 4\pi$ and $\gamma 6\pi$ 
in the $\eta_c(2S)$ signal region.]
{\label{fig:etac2s_result_4Pi_6Pi}
{Measured photon energy for the final states $\gamma 4\pi$ (top) and 
$\gamma 6\pi$ (bottom) in the $\eta_c(2S)$ signal region.
 \ The points are from the 25.9~M $\psi(2S)$ data sample. \
The dotted lines are a Breit-Wigner convoluted with the Crystal Ball 
resolution signal shape. \ The dashed lines are the background histogram 
determined from the 10 times luminosity generic $\psi(2S)$ and 5 times 
luminosity continuum MC samples. \ The solid lines are the sum of the 
signal and background.}}
\end{figure}

\begin{figure}[htbp]
\begin{center}
\subfigure
{\includegraphics[width=.80\textwidth]{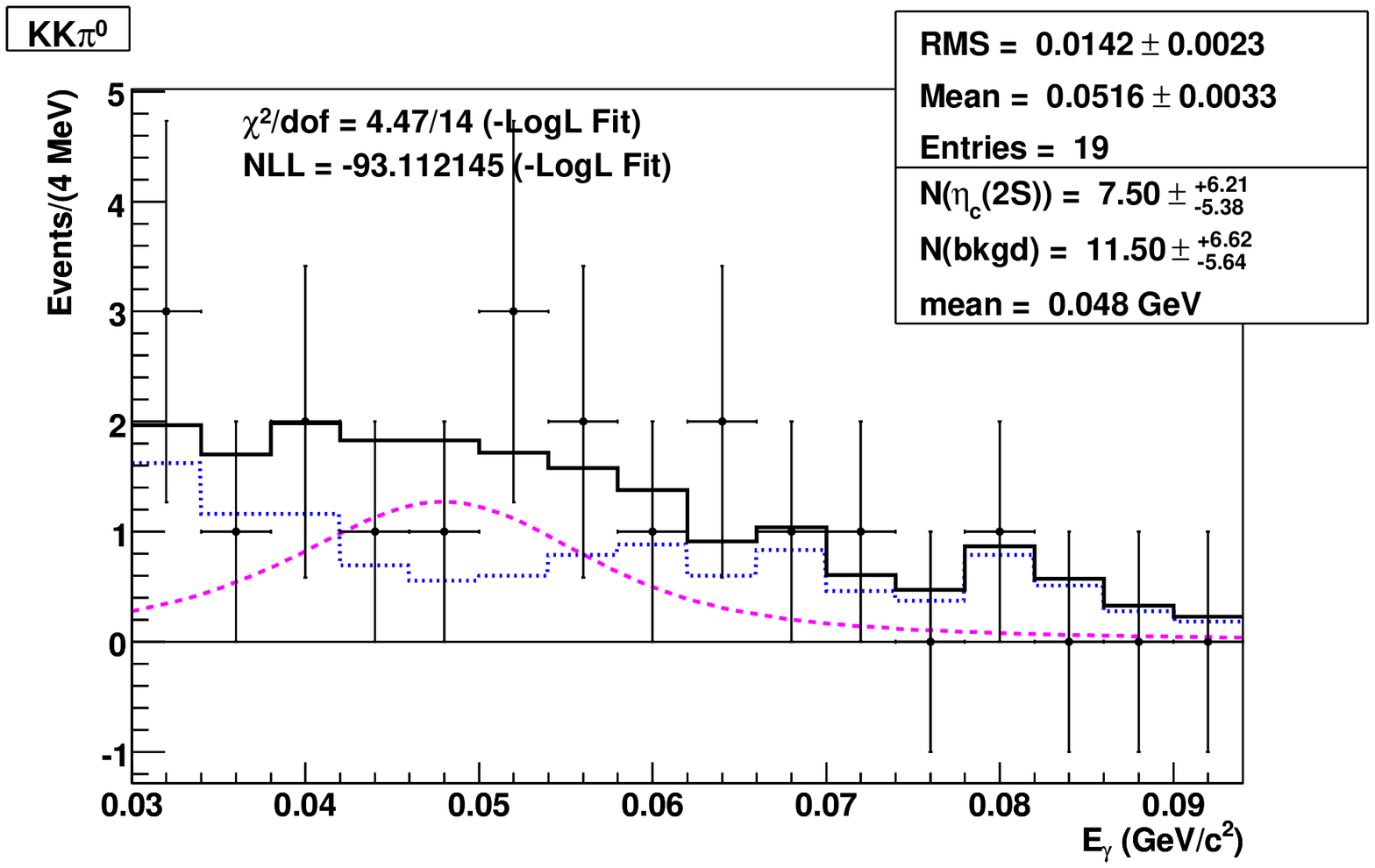}}
\subfigure
{\includegraphics[width=.80\textwidth]{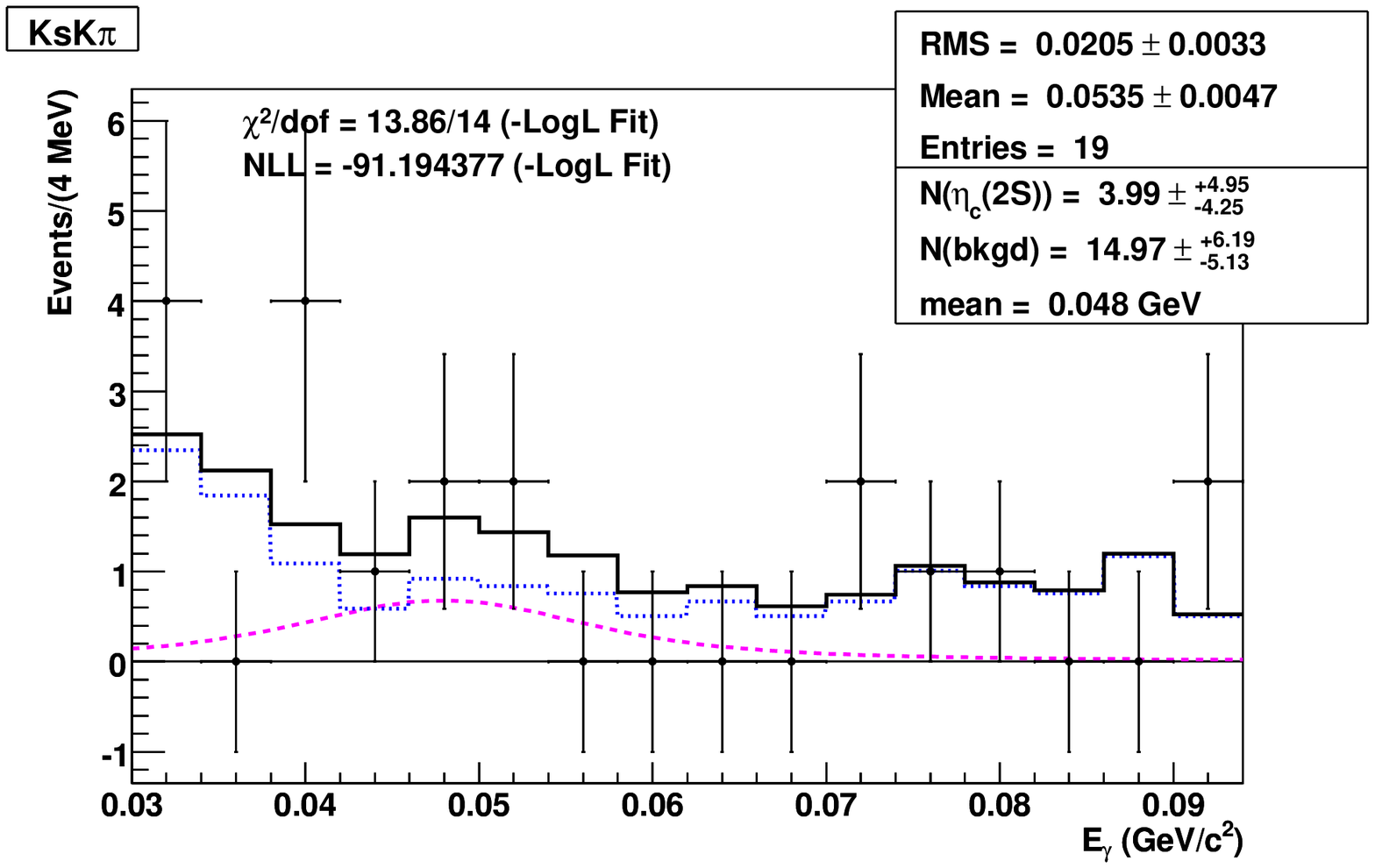}}
\end{center}
\caption[Measured photon energy for the final states $\gamma KK\pi^{0}$ and $\gamma K_{S}K\pi$ 
in the $\eta_c(2S)$ signal region.]
{\label{fig:etac2s_result_KKPi0_KsKPi}
{Measured photon energy for the final states $\gamma KK\pi^{0}$ (top) 
and $\gamma K_{S}K\pi$ (bottom) in the $\eta_c(2S)$ signal region.  
 \ The points are from the 25.9~M $\psi(2S)$ data sample. \
The dashed lines are a Breit-Wigner convoluted with the Crystal Ball 
resolution signal shape. \ The dotted lines are the background histogram 
determined from the 10 times luminosity generic $\psi(2S)$ and 5 times 
luminosity continuum MC samples. \ The solid lines are the sum of the 
signal and background.}}
\end{figure}

\begin{figure}[htbp]
\begin{center}
\subfigure
{\includegraphics[width=.80\textwidth]{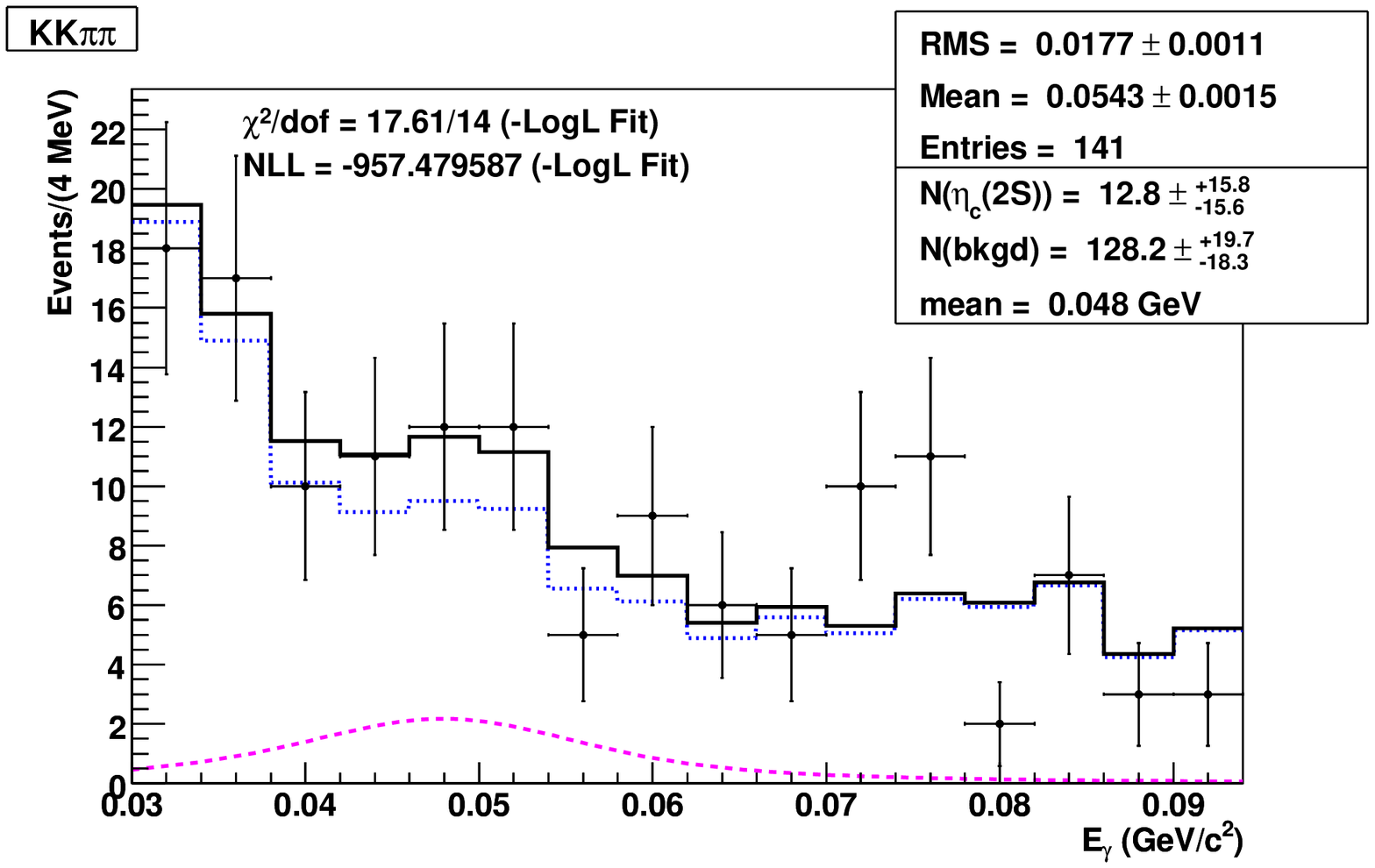}}
\subfigure
{\includegraphics[width=.80\textwidth]{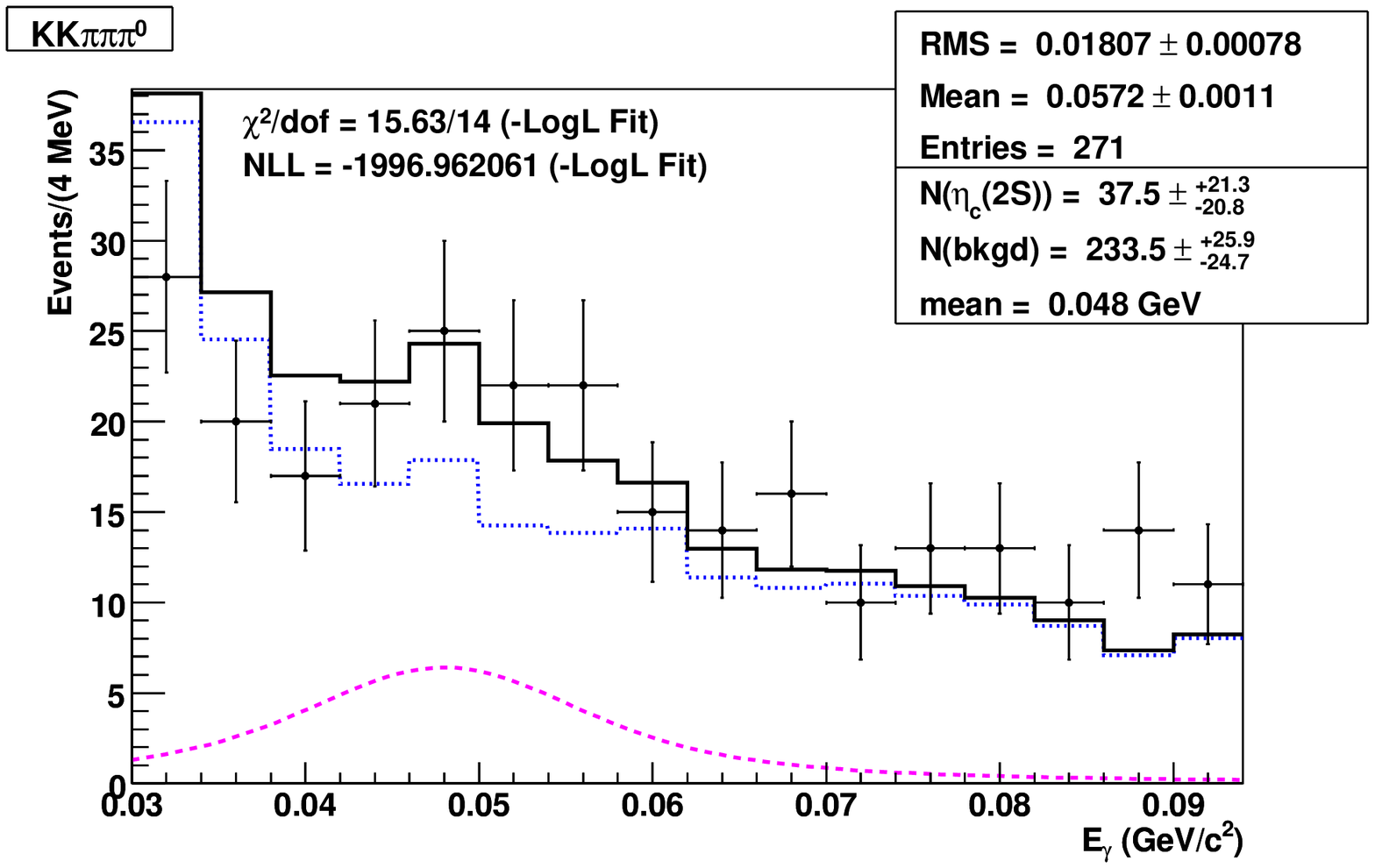}}
\end{center}
\caption[Measured photon energy for the final states $\gamma KK\pi\pi$ 
and $\gamma KK\pi\pi\pi^{0}$ in the $\eta_c(2S)$ signal region.]
{\label{fig:etac2s_result_KKPiPi_KKPiPiPi0}
{Measured photon energy for the final states $\gamma KK\pi\pi$ (top) 
and $\gamma KK\pi\pi\pi^{0}$ (bottom) in the $\eta_c(2S)$ signal region.  
 \ The points are from the 25.9~M $\psi(2S)$ data sample. \
The dashed lines are a Breit-Wigner convoluted with the Crystal Ball 
resolution signal shape. \ The dotted lines are the background histogram 
determined from the 10 times luminosity generic $\psi(2S)$ and 5 times 
luminosity continuum MC samples. \ The solid lines are the sum of the 
signal and background.}}
\end{figure}

\begin{figure}[htbp]
\begin{center}
\subfigure
{\includegraphics[width=.80\textwidth]{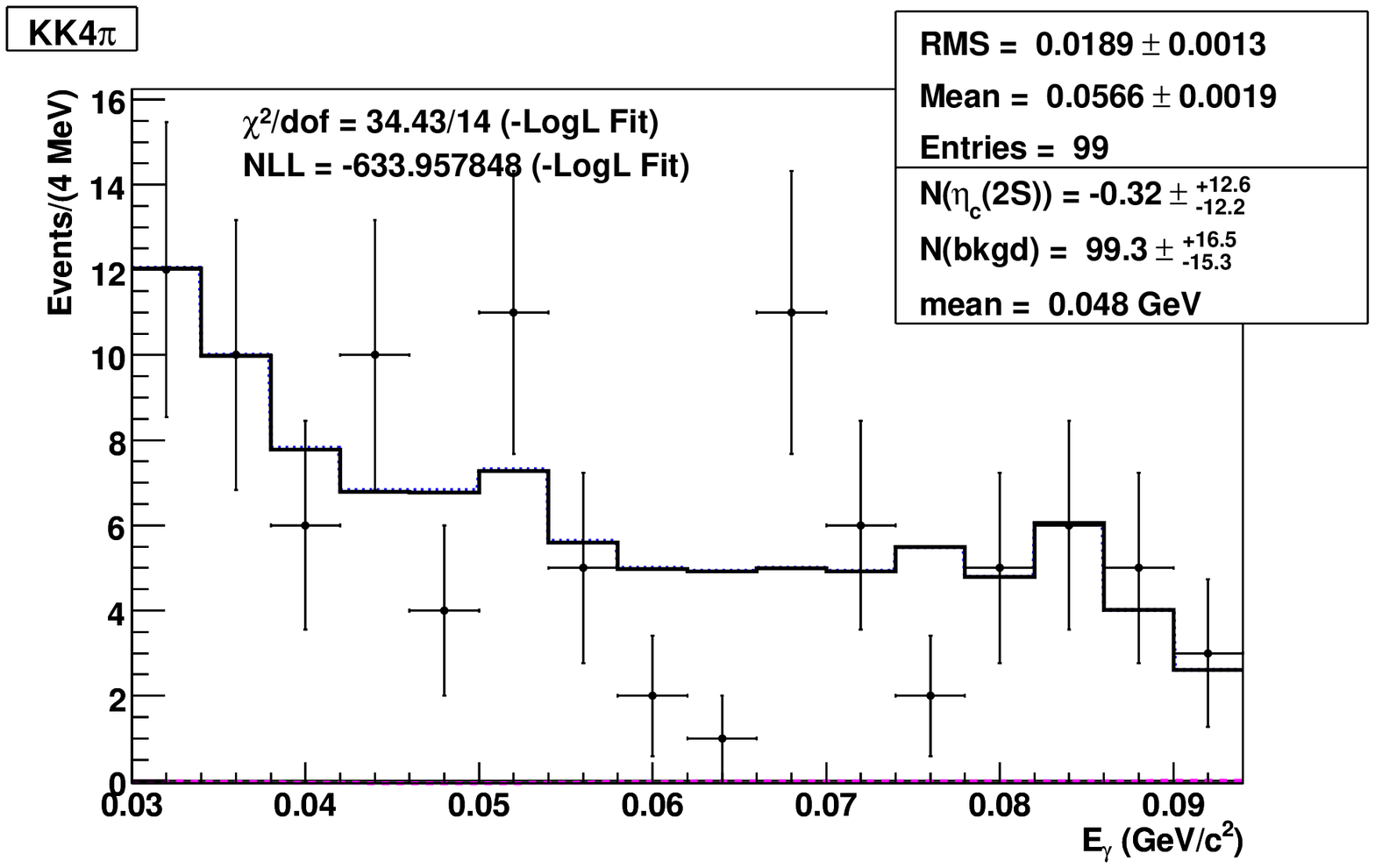}}
\subfigure
{\includegraphics[width=.80\textwidth]{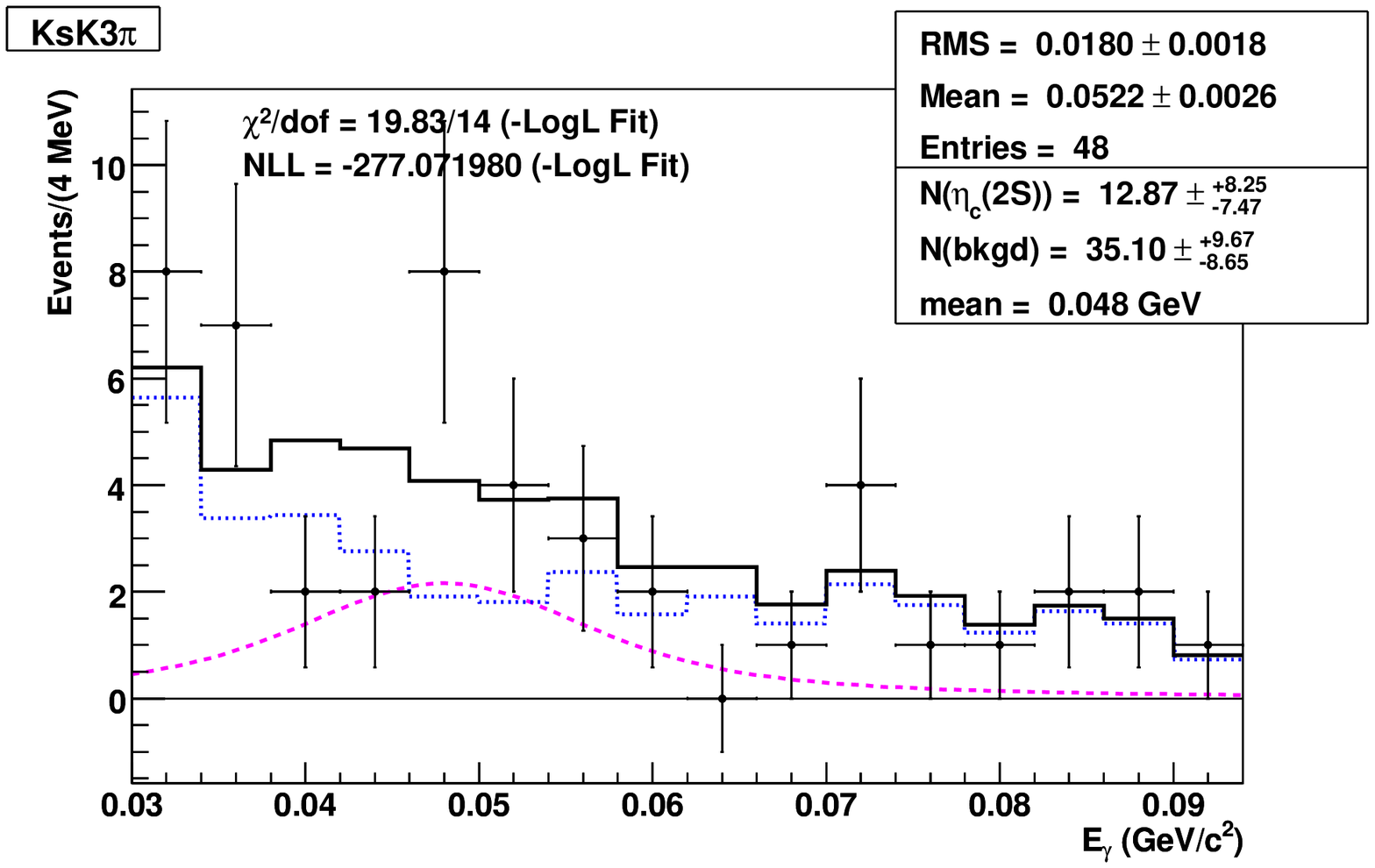}}
\end{center}
\caption[Measured photon energy for the final states $\gamma KK4\pi$ 
and $\gamma K_{S}K3\pi$ in the $\eta_c(2S)$ signal region.]
{\label{fig:etac2s_result_KK4Pi_KsK3Pi}
{Measured photon energy for the final states $\gamma KK4\pi$ (top) 
and $\gamma K_{S}K3\pi$ (bottom) in the $\eta_c(2S)$ signal region.  
 \ The points are from the 25.9~M $\psi(2S)$ data sample. \
The dashed lines are a Breit-Wigner convoluted with the Crystal Ball 
resolution signal shape. \ The dotted lines are the background histogram 
determined from the 10 times luminosity generic $\psi(2S)$ and 5 times 
luminosity continuum MC samples. \ The solid lines are the sum of the 
signal and background.}}
\end{figure}

\begin{figure}[htbp]
\begin{center}
\subfigure
{\includegraphics[width=.61\textwidth,height=0.23\textheight]{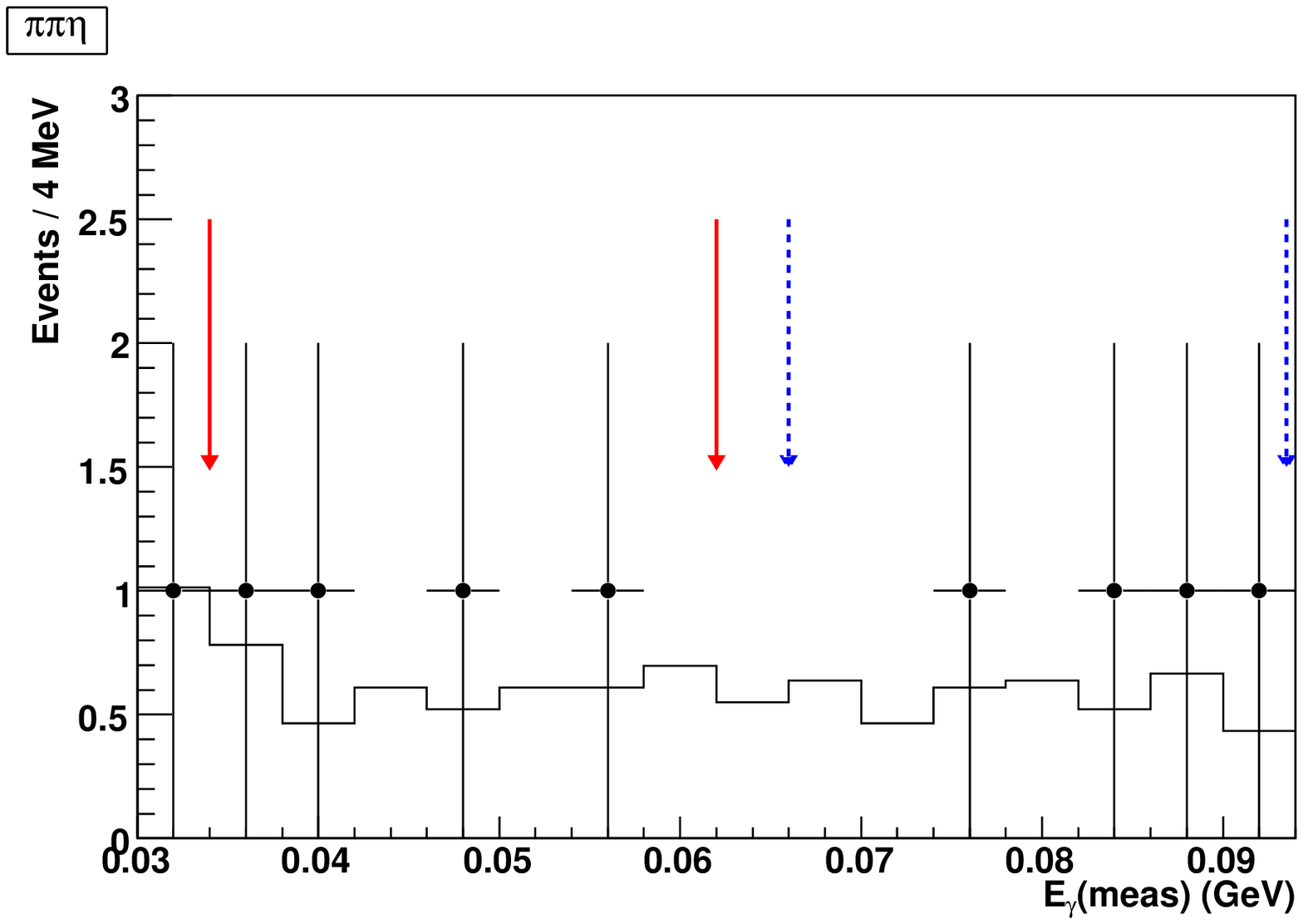}}
\subfigure
{\includegraphics[width=.61\textwidth,height=0.23\textheight]{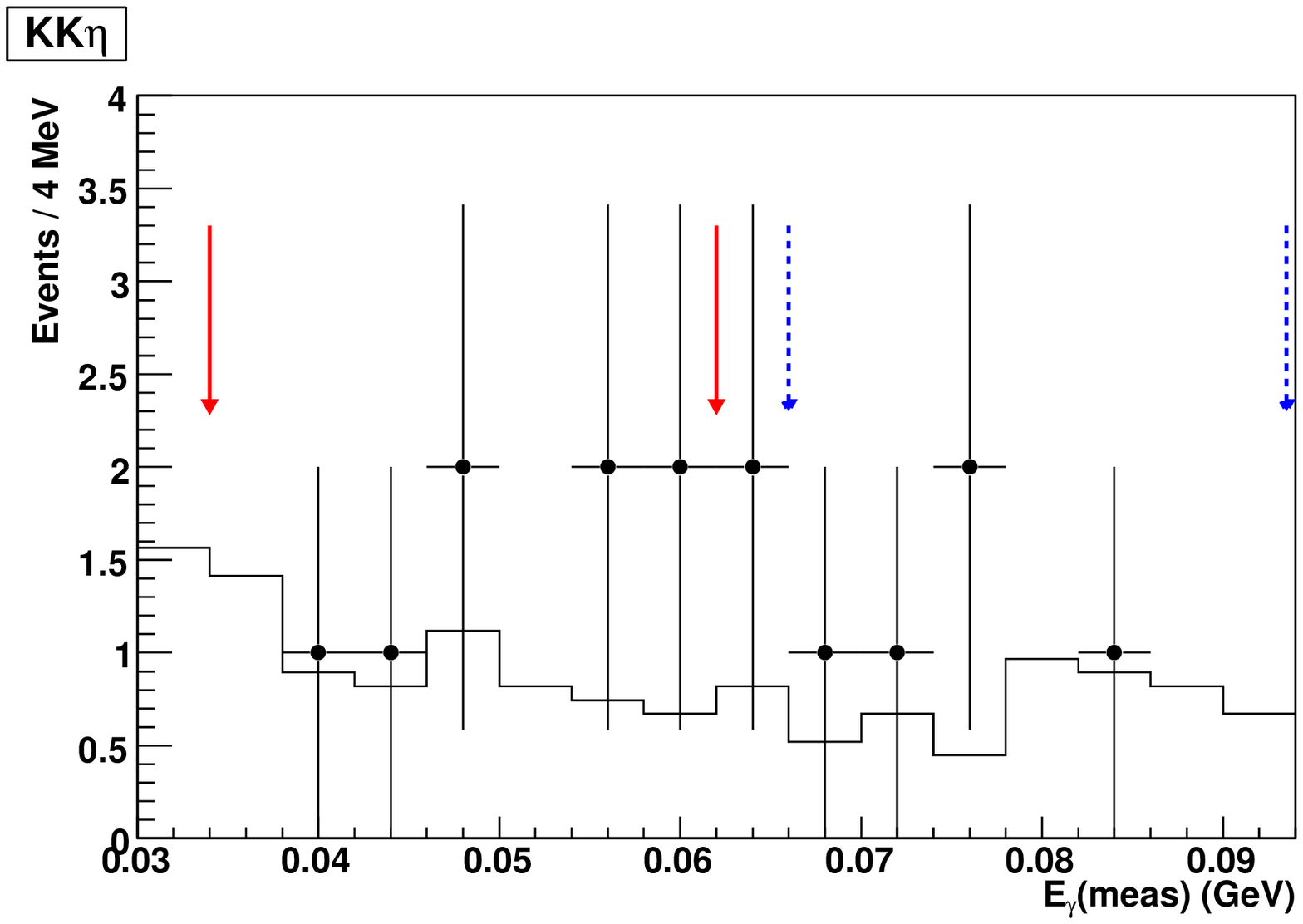}}
\subfigure
{\includegraphics[width=.61\textwidth,height=0.23\textheight]{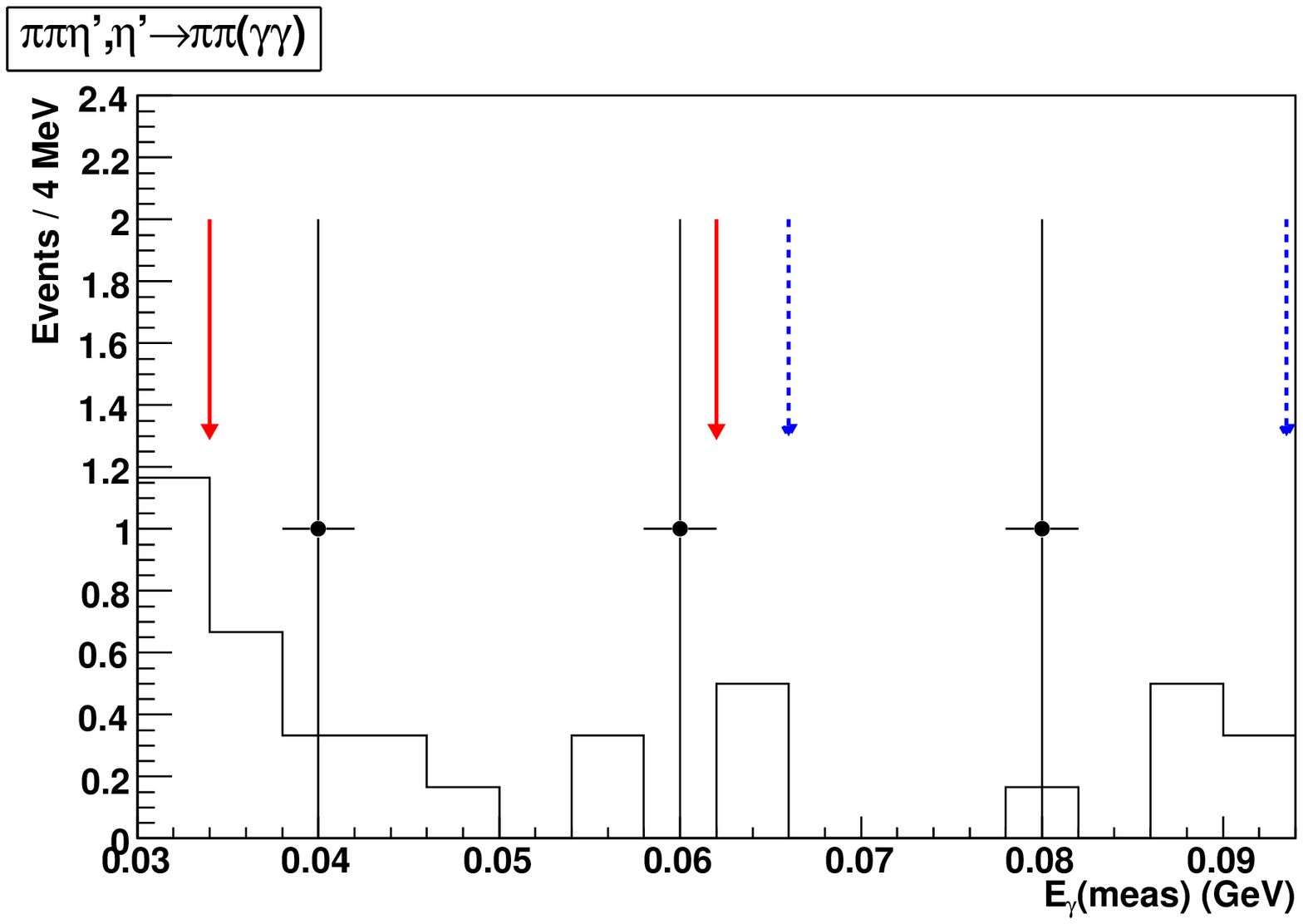}}
\end{center}
\caption[Measured photon energy for the final states $\gamma \pi\pi\eta$, $\gamma KK\eta$, 
and $\gamma \pi\pi\eta^{\prime}$
in the $\eta_c(2S)$ signal region.]
{\label{fig:etac2s_result_EtaModes}
{Measured photon energy for the final states $\gamma \pi\pi\eta$ (top) 
and $\gamma KK\eta$ (middle) and 
$\gamma \pi\pi\eta^{\prime}$
 (bottom) 
in the $\eta_c(2S)$ signal region. \ The points are from the 25.9~M $\psi(2S)$ 
data sample. \ 
The histograms are from the 10 times luminosity generic $\psi(2S)$ 
and 5 times luminosity continuum MC samples. \ The solid red arrows enclose 
the signal region, while the dashed blue arrows enclose the sideband region.}}
\end{figure}

The measured photon energy results for the $\eta_c(2S)$ signal region 
are listed in Table~\ref{table:etac2s_result_fit}. \ 
The resolution function fits are shown in Appendix~\ref{appendix:resfnctfits} and the 
corresponding values for $\sigma$ and $\alpha$ are fixed in the fits of the 
data sample. \ 
The $\eta_c(2S)$ resonance parameters $M = 3638~{\rm MeV}$ 
($E_{\gamma} = 48~{\rm MeV}$) and 
$\Gamma = 14~{\rm MeV}$ are also fixed in the fits. \ The photon energy distributions for all eleven modes 
are shown in Figures~\ref{fig:etac2s_result_4Pi_6Pi} through \ref{fig:etac2s_result_EtaModes}.

\begin{figure}[htbp]
\begin{center}
\subfigure
{\includegraphics[width=.80\textwidth]{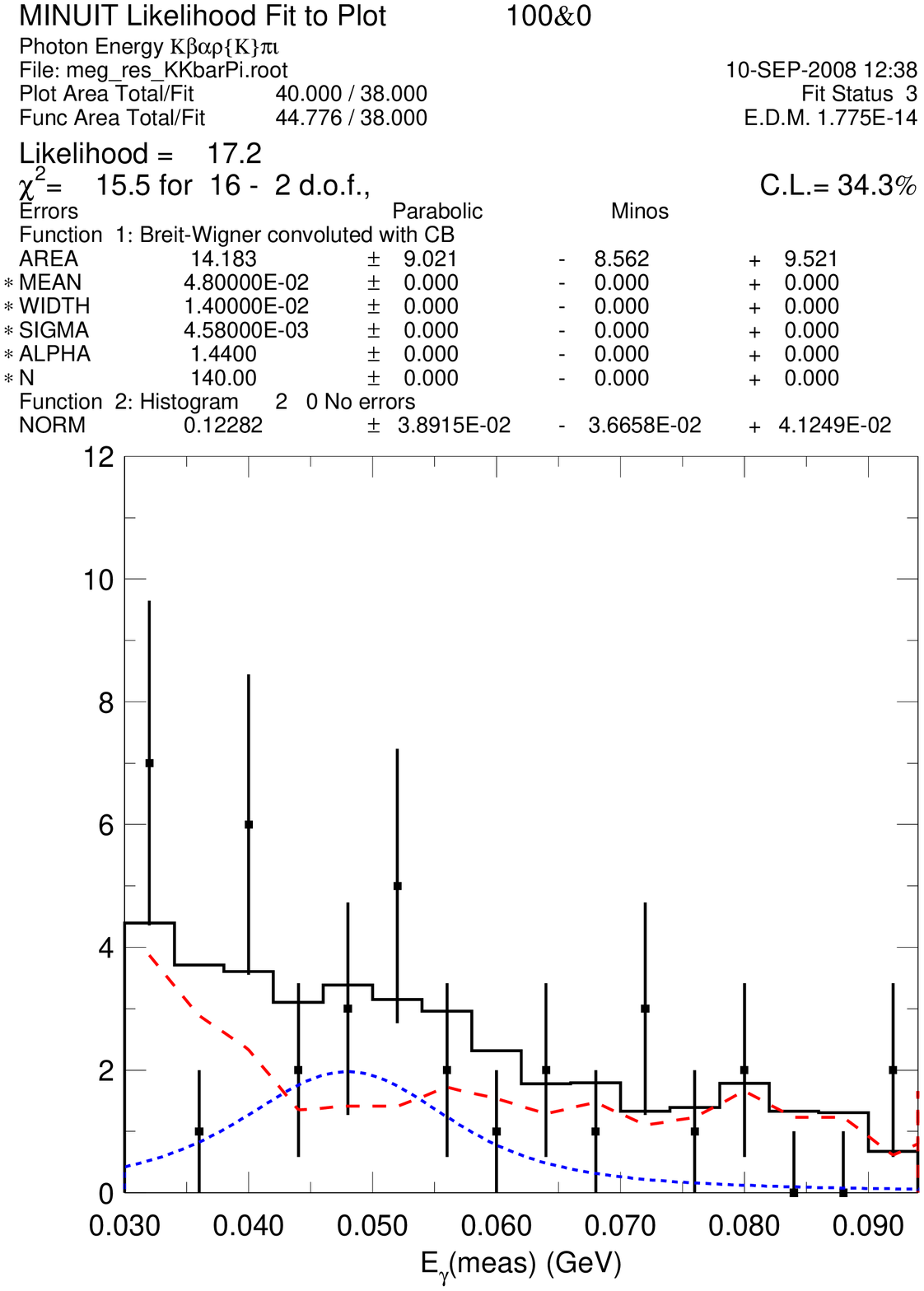}}
\end{center}
\caption[Measured photon energy for the final state $\gamma K\bar{K}\pi$ in the $\eta_c(2S)$ signal region.]
{\label{fig:PKZetac2s_result_KKbarPi}
{Measured photon energy for the final state $\gamma K\bar{K}\pi$ in the $\eta_c(2S)$ signal region.  
 \ The points are from the 25.9~M $\psi(2S)$ data sample. \
The dashed lines are a Breit-Wigner convoluted with the Crystal Ball 
resolution signal shape. \ The dotted lines are the background histogram 
determined from the 10 times luminosity generic $\psi(2S)$ and 5 times 
luminosity continuum MC samples. \ The solid line is the sum of the 
signal and background.}}
\end{figure}

Two methods are employed for determining the 90\% confidence level 
upper limits on the yields listed in 
Table~\ref{table:etac2s_result_fit}. \ 
For modes which 
do not contain $\eta$ decays, we have no apparent signals and substantial 
backgrounds. \
Therefore, we determine an upper limit by integrating the distribution
defined from the nominal yield result (a bifurcated Gaussian) up to
90\% of the area in the positive physical range. \ We use toy MC studies to 
verify
that we get consistent results with the nominal fit integration. \ For 
modes with 
$\eta$ decays, a very limited number of events pass our selection criteria, 
either signal or background. \ Therefore, we use the Feldman and Cousins 
method \cite{PhysRevD.57.3873}. \ The procedure of determining the upper 
limits is described in more detail in Appendix~\ref{appendix:ul}.

\begin{table}[htbp]
\caption[Result of decay $\psi(2S)\to\gamma\eta_{c}(2S)$]
{\label{table:etac2s_result_fit}
Results for $\psi(2S)\to\gamma\eta_{c}(2S)$, $\eta_{c}(2S)\to X$. \ 
The photon energy resolutions were determined by fitting the resolution 
function from the $\eta_c(2S)$ signal MC sample. \ 
Other parameters were from signal fits of data. \ 
The statistical significance, defined as $\sqrt{-2\Delta{\rm ln}{\cal L}}$, 
where is ${\cal L}$ is the likelihood, and the difference is computed between
the standard fit with both signal and background and an alternative
background-only fit. \ 
The product $B_{1} \times B_{2}$ is defined as 
${\cal B}(\psi(2S) \to \gamma~\eta_c(2S)) \times {\cal B}(\eta_c(2S)\to X)$. \ 
All upper limits are at 90\% confidence levels and only statistical errors are shown.}
\scriptsize{
\begin{center}
\begin{tabular}{|l|c|c|c|c|c|c|c|}
  \hline
  Mode& $\epsilon$ & Res $\sigma$ & XBall $\alpha$ & Sgnf & $N_{\rm sig}$ & $N_{\rm sig}$ & $B_{1} \times B_{2}$ \\
  ($X$)& (\%) & (MeV) & & ($\sigma$) & $(N_{\rm obs}/N_{\rm bg})$ & & ($10^{-6}$) \\ \hline
  $4\pi$
    & $20.94\pm0.16$ & $4.68\pm0.04$ & $1.50\pm0.04$ & $4.15$ & $48.0^{+13.0}_{-12.3}$ & $<64.8$ & $<12.0$ \\ \hline
  $6\pi$
    & $14.71\pm0.14$ & $4.70\pm0.05$ & $1.43^{+0.05}_{-0.04}$ & $0.56$ & $10.1^{+18.1}_{-17.6}$ & $<36.6$ & $<9.6$ \\ \hline
  $KK\pi\pi$
    & $20.15\pm0.16$ & $4.57\pm0.04$ & $1.45\pm0.04$ & $0.82$ & $12.8^{+15.8}_{-15.6}$ & $<35.2$ & $<6.7$\\ \hline
  $KK\pi^{0}$
    & $18.82\pm0.15$ & $4.54\pm0.04$ & $1.31^{+0.04}_{-0.03}$ & $1.44$ & $7.5^{+6.2}_{-5.4}$ & $<16.0$ & $<3.3$ \\ \hline
  $K_{S}K\pi$
    & $20.69\pm0.15$ & $4.61\pm0.04$ & $1.56^{+0.05}_{-0.04}$ & $0.94$ & $4.0^{+5.0}_{-4.3}$ & $<11.0$ & $<3.0$ \\ \hline
  $K\bar{K}\pi$
    & $7.89\pm0.04$* & $4.58\pm0.03$ & $1.56\pm0.03$ & $1.73$ & $11.7^{+7.8}_{-7.0}$ & $<21.9$ & $<10.7$ \\ \hline
  $\pi\pi\eta$
    & $6.00\pm0.05$* & -- & -- & -- & $4~/~4.3$ & $<4.3$ & $<2.3$ \\ \hline
  $\pi\pi\eta^{\prime}$
    & $8.63\pm0.11$ & -- & -- & -- & $2~/~1.8$ & $<4.1$ & $<10.5$ \\ \hline
  $KK\eta$
    & $6.90\pm0.05$* & -- & -- & -- & $8~/~6.6$ & $<7.4$ & $<3.5$ \\ \hline
  $KK\pi\pi\pi^{0}$
    & $9.42\pm0.12$ & $4.56\pm0.05$ & $1.82^{+0.11}_{-0.09}$ & $1.82$ & $37.5^{+21.3}_{-20.8}$ & $<65.4$ & $<27.4$ \\ \hline
  $KK4\pi$
    & $10.38\pm0.12$ & $4.64\pm0.06$ & $1.58^{+0.08}_{-0.07}$ & -- & $-0.3^{+12.6}_{-12.2}$ & $<20.6$ & $<7.7$ \\ \hline
  $K_{S}K3\pi$
    & $11.65\pm0.13$ & $4.82\pm0.05$ & $1.76^{+0.09}_{-0.08}$ & $1.78$ & $12.9^{+8.3}_{-7.5}$ & $<23.9$ & $<11.4$ \\ \hline
\end{tabular}
\end{center}
}
*Submode-decay branching fractions are included in efficiencies.
\end{table}

The modes $KK\pi^0$ and $K_SK\pi$ are actually specific final states of 
$K\bar{K}\pi$. \ 
Therefore, we also determined the product branching fraction for the decay 
$\psi(2S) \to \gamma \eta_c(2S), \eta_c(2S) \to K\bar{K}\pi$. \ 
Figure~\ref{fig:PKZetac2s_result_KKbarPi} shows the fit result of the 
combined final states. \ 
The results are also listed in Table~\ref{table:etac2s_result_fit}. 

The statistical significance, defined as $\sqrt{-2\Delta{\rm ln}{\cal L}}$,
where is ${\cal L}$ is the likelihood and the difference is computed between
the standard fit with both signal and background and an alternative
background-only fit. \ 
Excluding the 
$\eta_{c}(2S) \to \pi\pi\eta^{\prime}$ mode, the product of branching fractions 
${\cal B}(\psi(2S) \to \gamma~\eta_c(2S)) \times {\cal B}(\eta_c(2S)\to X)$
($B_{1} \times B_{2}$ in Table~\ref{table:etac2s_result_fit})
was obtained from
\begin{equation}
{\mathcal B}(\psi(2S) \to \gamma~\eta_c(2S))\times{\mathcal B}(\eta_c(2S)\to X)
= \frac{N_{\rm sig}}{\epsilon N_{\psi(2S)}} 
\end{equation}
where $N_{\rm sig}$ is the number of signal events (yields) determined by fits
for non-$\eta$ modes and Feldman-Cousins table for modes contain 
$\eta$'s, $\epsilon$ is the efficiency, $N_{\psi(2S)}$ is the total number of 
$\psi(2S)$ decays. \ For $\eta_{c}(2S) \to \pi\pi\eta^{\prime}$ mode additional 
factors of ${\mathcal B}(\eta^{\prime}\to\pi\pi\eta)$ and 
${\mathcal B}(\eta \to \gamma\gamma)$ should be considered and the 
corresponding equation becomes
\begin{equation}
{\mathcal B}(\psi(2S) \to \gamma~\eta_c(2S))\times
{\mathcal B}(\eta_c(2S)\to \pi\pi\eta^{\prime})
= \frac{N_{\rm sig}}{\epsilon N_{\psi(2S)} 
{\mathcal B}(\eta^{\prime}\to\pi\pi\eta) 
{\mathcal B}(\eta \to \gamma\gamma)} .
\end{equation}
When calculating the upper limits of the branching fraction product we 
used $N_{\psi(2S)} = 25.9 \times 10^{6}$ \cite{cbx07-4}. \ The results of 
the upper limits of branching fraction products at 90\% confidence level 
are also listed in Table~\ref{table:etac2s_result_fit}. 

No evidence for the decay $\psi(2S)\to\gamma\eta_c(2S)$ is observed in any 
of the these modes. \ The only mode in which an excess above background is present 
with a statistical significance greater than 2$\sigma$ is the 
4$\pi$ mode. \ We have investigated other aspects of the events in the 
signal region to determine if the 
excess constitutes evidence of a signal or is just a statistical fluctuation.  

\begin{figure}[htbp]
\begin{center}
\subfigure
{\includegraphics[width=.80\textwidth]{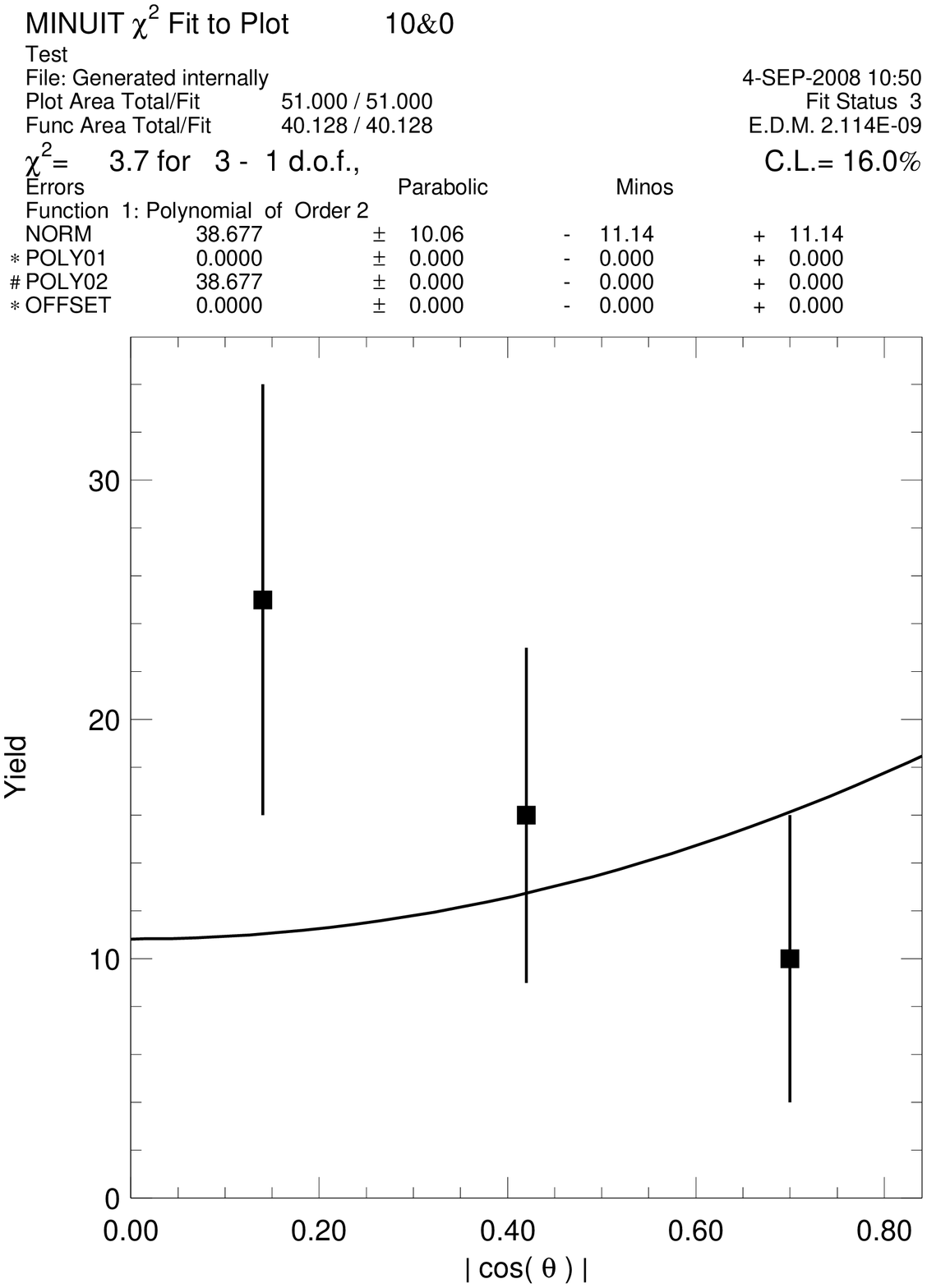}}
\end{center}
\caption[The $\cos\theta$ distribution of the 4$\pi$ mode.]
{\label{fig:PKZangdist_4Pi}
{The $\cos\theta$ distribution of the photon candidate for the 4$\pi$ 
mode. \ The points are the fit yields after requiring the photon candidate 
to lie with a particular $\cos\theta$ bin. \ The line is the result of 
fitting the points to 1 + $\cos^2\theta$ with only the normalization 
left free.}}
\end{figure}

Figure~\ref{fig:PKZangdist_4Pi} shows the $\cos\theta$ distribution of the 
photon for the 4$\pi$ mode. \ 
The points are determined by the measured photon energy yields after requiring 
the photon candidate to be within a given $\cos\theta$ region. \ 
This distribution is fit to the expected $1 + \cos^2 \theta$ dependence,
giving a $\chi^2/{\rm dof}$ of 3.7/2, which corresponds to a confidence 
level of
only 16\%.

\begin{figure}[htbp]
\begin{center}
\subfigure
{\includegraphics[width=.80\textwidth]{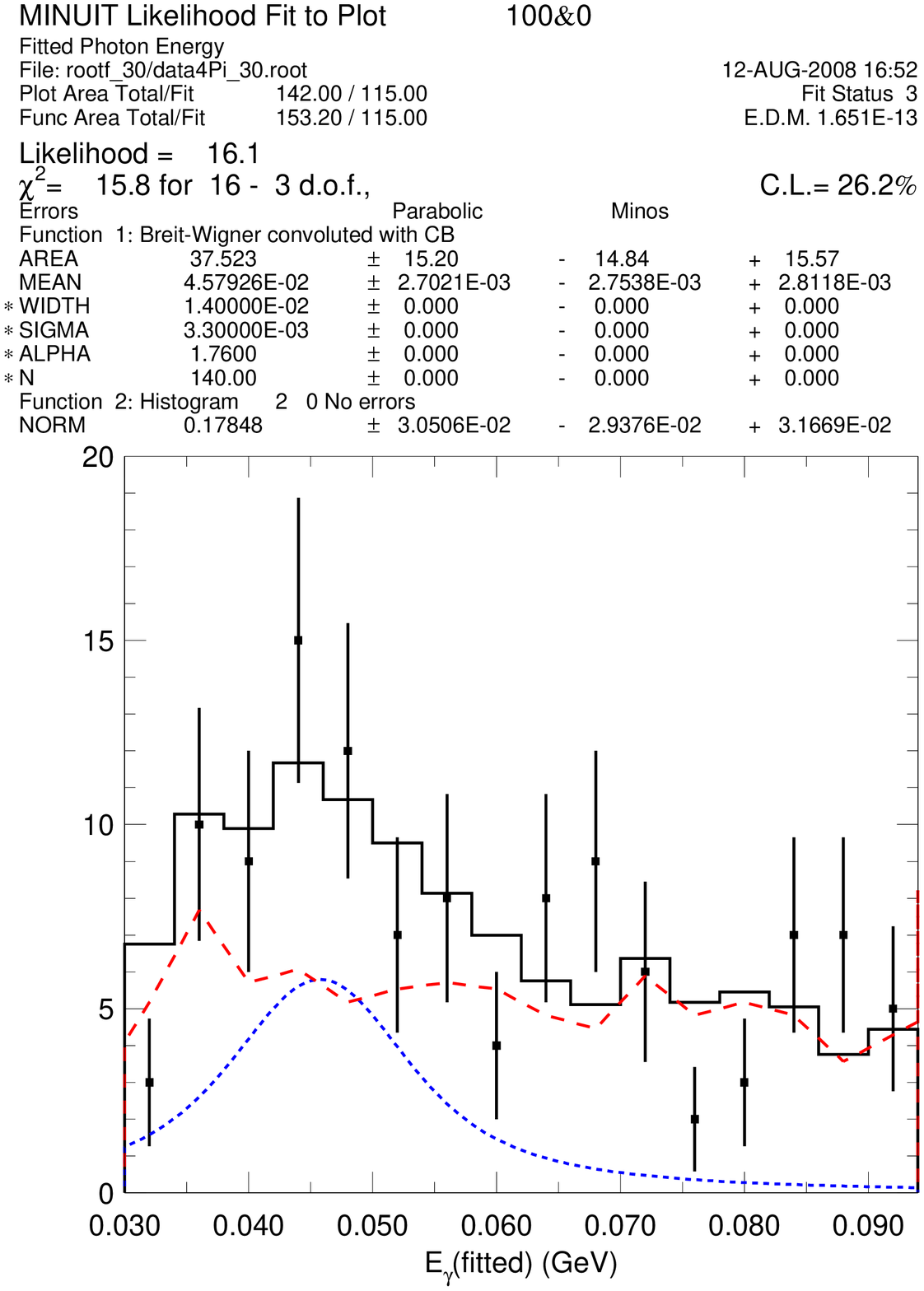}}
\end{center}
\caption[Constrained photon energy for the final state $\gamma 4\pi$ in 
the $\eta_c(2S)$ signal region.]
{\label{fig:PKZfittedEgamma_4Pi}
{Constrained photon energy for the final state $\gamma 4\pi$ in 
the $\eta_c(2S)$ signal region. \ 
The points are from the 25.9 M $\psi(2S)$ data sample. \ 
The dotted line is the Breit-Wigner convoluted with the Crystal Ball resolution 
signal shape. \ 
The dashed line is the background histogram determined from the 10 times 
luminosity generic $\psi(2S)$ 
and 5 times luminosity continuum MC samples. \ The solid line is the sum 
of the signal and background.}}
\end{figure}

Another test was to fit the distribution in the constrained photon energy,
the value returned by the 4C kinematic fit to the $\psi(2S) \rightarrow \gamma 
\eta_c(2S)$, $\eta_c(2S) \rightarrow 4 \pi$ hypothesis. \ If the events in
the excess are true signal, we would expect the significance to be
increased. \ In fact it is reduced to only $2.6 \sigma$. \ 
Figure~\ref{fig:PKZfittedEgamma_4Pi} 
displays the constrained photon energy for the $4\pi$  mode.

These tests suggest that the 
excess of events in the $\eta_c(2S)$ signal region for the $4\pi$ mode is 
caused by an upward fluctuation of the background rather than true signal. 

\begin{figure}[htbp]
\begin{center}
\subfigure
{\includegraphics[width=.95\textwidth]{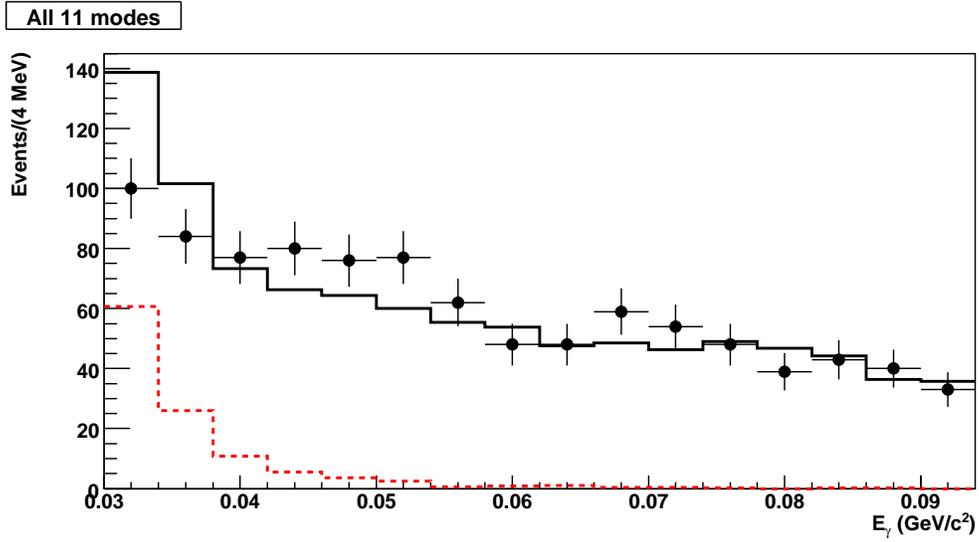}
}
\subfigure
{\includegraphics[width=.95\textwidth]{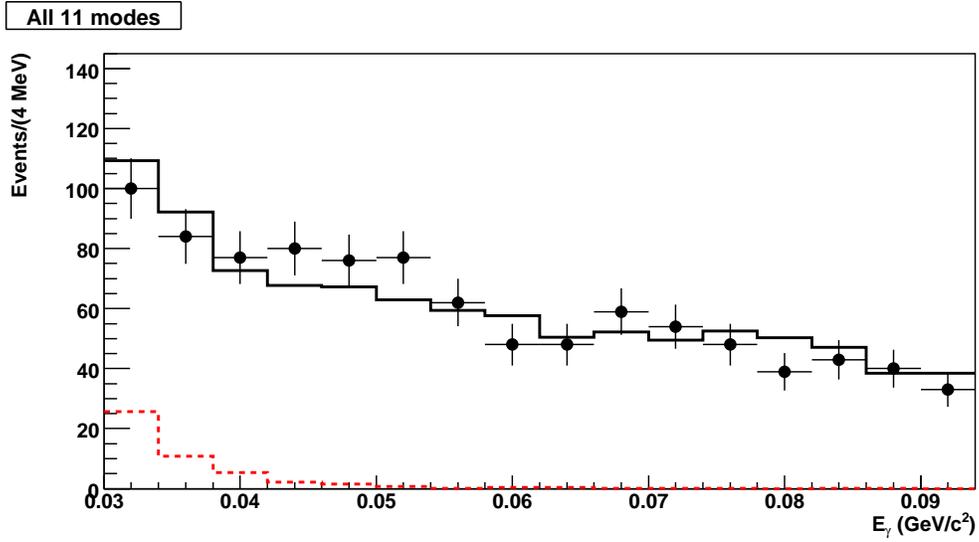}
}
\end{center}
\caption[Measured photon energy for all 11 modes in the $\eta_c(2S)$ signal region.]
{\label{fig:PKZetac2s_result_all11modes}
{Measured photon energy for all 11 modes in the $\eta_c(2S)$ signal region.  
The points are from the 25.9 M $\psi(2S)$ data sample.  
The histogram is the 10 times luminosity generic $\psi(2S)$ and 5 times 
luminosity continuum MC samples, 
with the contribution of each of the 11 final
states determined by its individual 
1-parameter (top) and 2-paramter (bottom) background-only fits. \ 
The red dashed histogram shows the events with transition photon candidates
tagged as splitoff showers in MC. \ Details about the fits can be found
in text.}}  
\end{figure}

Our conclusion is that there is no statistically significant excess
attributable to $\psi(2S) \rightarrow \eta_c(2S)$  in any of the 11 modes
considered. \ It is reasonable to ask whether the sum over all exclusive
modes shows evidence of an excess, even though the interpretation of such an
excess would be complicated. \ 
Figure~\ref{fig:PKZetac2s_result_all11modes}~(top) shows the measured photon
energy distribution summed over all 11 modes (points). \  It can be compared
with a ``background-only'' hypothesis constructed as the sum of the
distributions from mode-by-mode fits to the MC-determined background
histograms with no signal component included (solid histogram). \ A small
excess is visible in the signal region, but it is clearly not statistically
significant.

Figure~\ref{fig:PKZetac2s_result_all11modes}~(top) shows a clear 
discrepancy between the fitted background and the
observed yield in the lowest photon-energy bins. \ As was discussed
previously, this region has a significant contribution from splitoff
showers. \ In our $J/\psi$ study (Figure~\ref{fig:jpsi_6Pi} through 
\ref{fig:jpsi_KsK3Pi}), we tested the
reliability of the background simulation in events that are similar to
$\eta_{c}(2S)$ signal events. \ 
In addition to parameterizing the background with a single MC-predicted
distribution, we considered the alternative of fitting to two separated
background components: splitoff showers and showers from all other sources. \ 
While there was some improvement in fit quality for the fits with two
independent background parameters compared to fits with a single background
parameter, it was not statistically compelling and we adopted fits with one
background parameter for our standard procedure. \ The discrepancy observed
in Figure~\ref{fig:PKZetac2s_result_all11modes}~(top) may demonstrate that 
for the full set of 11 signal modes
the standard MC does not provide a satisfactory description of the
background. \ This led us to reconsider whether fitting with two background
parameters would give better agreement.

Figure~\ref{fig:PKZetac2s_result_all11modes}~(bottom) shows the same data 
points with the summed histogram obtained
from background-only fits with two parameters, one controlling the
contribution of splitoff showers and one for other showers. \ The separate
background components were constrained to be positive.
\ Because of
insufficient statistics to determine a second background parameter, the
treatment  of the 
$\eta$ modes was left unmodified. \ The sum of the
background distributions for the modified fits in 
Figure~\ref{fig:PKZetac2s_result_all11modes}~(bottom) shows clear
improvement in the agreement with the data, suggesting a deficiency in the
MC that is at least  partially addressed by the fits with two background
parameters. \ Our conclusion that there is no statistically significant
excess attributable to $\eta_{c}(2S)$ stands, but this exercise provides a
reminder of possible systematic sensitivity in our results to the treatment
of the background.

From Figure~\ref{fig:PKZetac2s_result_all11modes}, because of the low 
statistics, we cannot find an explanation other than statistical 
fluctuation for the possible discrepancy between data and MC around 
70~MeV.

\section{Systematic Uncertainties}
\label{sec:syst}

While no evidence for the $\psi(2S)\to\gamma\eta_{c}(2S)$ 
transition was observed in 
any of the 11 modes we studied, 
systematic uncertainties were evaluated and incorporated into the branching 
fraction upper limits.

The systematic uncertainties due to various cuts were studied in the 
following environments. \ Some systematic uncertainties have been performed in 
other CLEO analyses, and the others can be estimated by studying decays 
of $\psi(2S)\to\gamma\chi_{c2}$ or $\psi(2S)\to\gamma\eta_{c}(2S)$ fits. \ 
We used $\psi(2S)\to\gamma\chi_{c2}$, $\chi_{c2}\to X$ decays to study 
the selection criteria, although sufficient statistics were available for only 
six modes of 
$\chi_{c2}$ decays: $4\pi$, $6\pi$, $KK\pi\pi$, $KK\pi\pi\pi^{0}$, $KK4\pi$
and $K_{S}K3\pi$. \ For some other modes, we estimated systematic uncertainties
by relating to modes with similar combinations of final state particles, e.g.,  
for the $KK\pi^{0}$ mode we applied the corresponding systematic uncertainties 
from the $\chi_{c2}\to KK\pi\pi\pi^{0}$ studies. \ 
We studied the $\eta_{c}(2S)$ signal region to assess uncertainties in the 
$\eta_{c}(2S)$ signal fitting method.

\subsection{Systematic Uncertainties That Apply to All Studied Modes}

Some of the systematic uncertainties were applied to all modes. \
The uncertainty on the total number of $\psi(2S)$ was 2\% 
\cite{cbx07-4}. \ The systematic errors on the trigger, which arise from 
uncertainties in the efficiency of trigger decisions for different final 
states, were taken to be 1\% for all modes \cite{cbx07-9, :2007ppa}. \ The systematic 
error in the efficiency for detecting the transition photon was 2\% \cite{cbx07-9, :2007ppa}. \ 
The results of mode dependent systematic uncertainties that were determined 
by studying $\psi(2S)\to\gamma\chi_{c2}$ decays are listed in 
Table~\ref{table:systmodeindep_chic2}, while the results determined by studying 
the $\eta_c(2S)$ signal region are listed in Table~\ref{table:systmodeindep_etac2s}. \

\subsubsection{Global Event Selection Systematics}

We conservatively take the respective systematic uncertainties associated with 
$\chi^{2}/{\rm dof}$ from the full event vertex fit and the 
$\chi^{2}/{\rm dof}$ from the full event 4C fit to be the difference between 
the results obtained with and without those cuts applied. \ 
An uncertainty of $1.3\%$ was assigned to the $\chi^{2}/{\rm dof}$ from the full event 
vertex fit for all modes. \ For modes without (with) a $\pi^0,\eta \to \gamma\gamma$ decay 
we assigned a $2.2\%$ ($4.0\%$) uncertainty due to the $\chi^{2}/{\rm dof}$ from the full 
event 4C fit. \ 
The per mode results can be found in Table~\ref{table:systmodeindep_chic2}.

\subsubsection{Resolution Function Systematics}

Resolution function systematics were studied by individually adjusting the 
Crystal Ball parameters
$\sigma$ and $\alpha$ by one standard deviation and finding the changes in the 
results of branching fractions from $\psi(2S)\to\gamma\chi_{c2}$ 
decays. \ We also performed the same procedures for the 
$\psi(2S)\to\gamma\eta_{c}(2S)$ decay candidates and obtained consistent 
but smaller results. \  We assigned $0.8\%$ and $0.6\%$ systematic uncertainties 
due to the uncertainties in $\sigma$ and $\alpha$, respectively. \  The per mode results 
from the $\chi_{c2}$ study are listed in Table~\ref{table:systmodeindep_chic2}.

\subsubsection{Signal Region Systematics}

We studied the effect of the lower boundary of the $E_{\gamma}$ signal 
region through studying $\psi(2S)\to\gamma\chi_{c2}$ decays by 
increasing the lower boundary of the $E_{\gamma}$ signal region from 
$90~{\rm MeV}$ to 104-110~MeV, which is about $18~{\rm MeV}$ 
below the 
mean of the $\chi_{c2}$. \ This amount is the same as the difference
between the lower boundary in the $\eta_{c}(2S)$ signal region and the
mean of the $\eta_{c}(2S)$. \ 
We assigned a systematic uncertainty of $3.2\%$ to all modes. \ 
The per mode results can be found in Table~\ref{table:systmodeindep_chic2}.

The systematic uncertainties due to the high side boundary of the 
signal region were studied directly in the $\eta_{c}(2S)$ signal region. \ 
The changes of the upper limits of branching fraction products were 
determined by varying the upper boundary of the $E_{\gamma}$ signal region by 
$\pm 8~{\rm MeV}$. \ The results can be found in Table~\ref{table:systmodeindep_etac2s}.

\subsubsection{Hadronic Invariant Mass $\Delta M$ Systematics}

The systematics due to the hadronic invariant mass $\Delta M$ cut were 
estimated by removing the cut from the $\psi(2S)\to\gamma\eta_{c}(2S)$ 
decay search and observing the effect it has on the upper limits of 
branching fraction products. \ This actually studied the change 
in the background. \ The results are listed in 
Table~\ref{table:systmodeindep_etac2s}. \ 

We also studied the $\psi(2S)\to\gamma\chi_{c2}$ decays by adding a 
requirement of $\Delta M < 175~{\rm MeV}$ to simulate the effect of adding a cut of 
$\Delta M$ at the value $30~{\rm MeV}$ 
below to the lower boundary of the $\eta_{c}(2S)$ signal region. \ 
Results are shown in Table~\ref{table:systmodeindep_chic2}, 
and the sensitivity is found to 
be smaller than in the $\eta_{c}(2S)$ signal region.

\subsubsection{Background Shape Systematics}

The systematic uncertainties associated with the background shape were
estimated using the $\eta_{c}(2S)$ signal region. \ In non-$\eta$ decay 
modes, the uncertainties were obtained by repeating the signal fits but 
replacing the histogram background with linear or constant functions. \ 
For $\eta$ modes, in which we do not use 
fits for upper limit estimates, 
we use the same number of observed events but take the background in the signal 
region down by $1 \sigma$, where $\sigma$ is the 
square root of the number of data events in the sideband region. \ The 
difference between this and the nominal result was assigned as the  
systematic uncertainty. \ The results can be found in
Table~\ref{table:systmodeindep_etac2s}.

\begin{table}[htbp]
\caption[Systematic uncertainties determined from $\psi(2S)\to\gamma \chi_{c2}$ decays.]
{\label{table:systmodeindep_chic2}Mode dependent systematic uncertainties 
determined from $\psi(2S)\to\gamma \chi_{c2}$ decays.
 \ The uncertainties are due to the following cuts:
(A) $\chi^{2}/{\rm dof}$ from full event vertex fit;
(B) $\chi^{2}/{\rm dof}$ from full event 4C fit;
(C) resolution function;
(D) lower boundary of $E_{\gamma}$ signal region ranging from $E_{\gamma} = (104,110)~{\rm MeV}$;
(E) hadronic invariant mass $\Delta M$.
 \ The relative uncertainties are listed in percent.}
\begin{center}
\begin{tabular}{|l|c|c|c|c|c|c|}
  \hline
  Mode & A & B & \multicolumn{2}{c|}{C} & D & E \\ \cline{4-7}
       &   &   & $\sigma$ & $\alpha$ & $E_{\gamma}\sim 110~{\rm MeV}$ & $\Delta M < 175$~MeV \\ \hline
  $4\pi$
    & $ 0.8$   & $0.5$ & $0.5$ & $0.4$ & $3.2$ & $0.05$ \\ \hline
  $6\pi$
    & $ 1.3$   & $2.2$ & $0.6$ & $0.5$ & $3.0$ & $0.03$ \\ \hline
  $KK\pi\pi$
    & $ 1.1$   & $0.6$ & $0.5$ & $0.4$ & $3.0$ & $0.2$ \\ \hline
  $KK\pi\pi\pi^{0}$
    & $\sim 0$ & $4.0$ & $0.2$ & $0.3$ & $0.7$ & $0.3$ \\ \hline
  $KK4\pi$
    & $ 0.5$   & $0.2$ & $0.7$ & $0.6$ & $1.8$ & $0.1$ \\ \hline
  $K_{S}K3\pi$
    & $ 1.1$   & $1.1$ & $0.8$ & $0.5$ & $2.5$ & $0.04$ \\ \hline
\end{tabular}
\end{center}
\end{table}

\begin{table}[htbp]
\caption[Systematic uncertainties applied to all modes determined in the $\eta_{c}(2S)$ signal region]
{\label{table:systmodeindep_etac2s}
Mode dependent systematic uncertainties determined from studying the $\eta_{c}(2S)$ signal region.  
The labels A, B, and C refer to studies of the 
(A) high side boundary of $E_{\gamma}$ signal region;
(B) hadronic invariant mass $\Delta M$;
(C) background shape. \ The relative uncertainties are listed in percent.}
\begin{center}
\begin{tabular}{|l|c|c|c|}
  \hline
  Mode & A & B & C \\ \hline
  $4\pi$
    & $11.2$ & $2.5 $ & $6.1 $ \\ \hline
  $6\pi$
    & $11.7$ & $16.6$ & $20.4$ \\ \hline
  $KK\pi\pi$
    & $13.4$ & $ 7.7$ & $32.6$ \\ \hline
  $KK\pi^{0}$
    & $11.0$ & $26.8$ & $36.5$ \\ \hline
  $K_{S}K\pi$
    & $9.1 $ & $ 1.9$ & $22.2$ \\ \hline
  $K\bar{K}\pi$
    & $13.0 $ & $15.2$ & $16.9$ \\ \hline
  $\pi\pi\eta$
    & --     & $ 6.1$ & $46.5$ \\ \hline
  $\pi\pi\eta^{\prime}$
    &  --    & $ 8.9$ & $24.5$ \\ \hline
  $KK\eta$
    &  --    & $ 9.2$ & $29.8$ \\ \hline
  $KK\pi\pi\pi^{0}$
    & $24.3$ & $22.7$ & $15.2$ \\ \hline
  $KK4\pi$
    & $10.5$ & $ 0.8$ & $6.8 $ \\ \hline
  $K_{S}K3\pi$
    & $4.6 $ & $13.2$ & $17.8$ \\ \hline
\end{tabular}
\end{center}
\end{table}

\subsection{Systematic Uncertainties That Apply Only to Specific Modes}

Other systematic uncertainties apply to only specific modes. \ 
Table~\ref{table:systmodedep} shows a summary of these uncertainties. 

\begin{table}[htbp]
\caption[Systematic uncertainties applied only to specific modes]
{\label{table:systmodedep}Systematic uncertainties applied only to specific 
modes. \ The uncertainties are due to the following selection criteria:
(A) pion and kaon track finding;
(B) pion and kaon PID;
(C) $\pi^{0}$ and $\eta$ finding;
(D) mass ranges for $\eta\to 3 \pi\pi\pi^0$ and 
$\eta^{\prime} \to \pi\pi\eta(\gamma\gamma)$ decays;
(E) $K_{S}$ finding; 
(F) ${\rm Rec}~M(\pi\pi)$ for $J/\psi$ rejection;
(G) $M(X - \pi\pi)$ for $J/\psi$ rejection; 
(H) $\cos \theta_{\gamma,\pi}$;
(I) $d_{\gamma, trk}$;
(J) ${\rm Rec}~M(\pi^{0}~{\rm or}~\eta~{\rm child})$;
(K) ${\rm Rec}~M(\eta)$ for $J/\psi$ rejection. 
\ The relative uncertainties are listed in percent.}
\begin{center}
\begin{tabular}{|l|c|c|c|c|c|c|c|c|c|c|c|}
  \hline
  Mode & A & B & C & D & E & F & G & H & I & J & K \\ \hline
  $4\pi$
    & 1.2 & 0.8 & -   & -      & - & $0.08$ & -        & $2.4 $ & -     & -     & -     \\ \hline
  $6\pi$
    & 1.8 & 1.2 & -   & -      & - & $2.4 $ & $1.0 $   & -      & -     & -     & -     \\ \hline
  $KK\pi\pi$
    & 1.8 & 1.0 & -   & -      & - & $0.2 $ & $0.07$   & -      & $0.8$ & -     & -     \\ \hline
  $KK\pi^{0}$
    & 1.2 & 0.6 & 3.5 & -      & - & -      & -        & -      & $0.8$ & $2.3$ & -     \\ \hline
  $K_{S}K\pi$
    & 0.9 & 0.5 & -   & -      & 2 & -      & -        & -      & -     & -     & -     \\ \hline
  $\pi\pi\eta(\gamma\gamma)$
    & 0.6 & 0.4 & 3.5 & -      & - & -      & -        & -      & -     & $2.3$ & $1.0$ \\ \hline
  $\pi\pi\eta(\pi\pi\pi^{0})$
    & 1.2 & 0.8 & 3.5 & $14.4$ & - & -      & -        & $2.4$ & -     & -     & $0.8$ \\ \hline
  $\pi\pi\eta^{\prime}$
    & 1.2 & 0.8 & 3.5 & $7.6 $ & - & -      & -        & -      & -     & -     & -     \\ \hline
  $KK\eta(\gamma\gamma)$
    & 1.2 & 0.6 & 3.5 & -      & - & -      & -        & -      & -     & $2.3$ & -     \\ \hline
  $KK\eta(\pi\pi\pi^{0})$
    & 1.8 & 1.0 & 3.5 & $1.6 $ & - & -      & -        & -      & -     & -     & -     \\ \hline
  $KK\pi\pi\pi^{0}$
    & 1.8 & 1.0 & 3.5 & -      & - & $2.2 $ & $\sim 0$ & -      & -     & -     & -     \\ \hline
  $KK4\pi$
    & 2.4 & 1.4 & -   & -      & - & $1.3 $ & $1.5 $   & -      & -     & -     & -     \\ \hline
  $K_{S}K3\pi$
    & 1.5 & 0.9 & -   & -      & 2 & $0.3 $ & $0.9 $   & -      & -     & -     & -     \\ \hline
\end{tabular}
\end{center}
\end{table}

\subsubsection{Tracking and PID Systematics}

For tracking uncertainties we applied 0.3\% for pions and 0.6\% for kaons 
based on studies of hadronic $D$ decays, which used the full
$818~{\rm pb}^{-1}$ $\psi(3770)$ data and standard DTag track quality 
cuts \cite{cbx2008-040}.

Based on the study of the first $281~{\rm pb}^{-1}$ $\psi(3770)$ data \cite{cbx05-43}, 
the pion (kaon) PID efficiency determined from MC was 0.55\% (1.06\%)
higher than determined from data. \ Therefore, we decreased the 
MC-determined efficiencies by these amounts per track. \ We 
conservatively assign 0.2\% (0.3\%) uncertainty to the pion (kaon) PID.

The uncertainties due to the same type of tracks are combined with 
full correlation. \  The uncertainties due to different type of 
tracks are considered uncorrelated, since they are determined through
independent studies using specially selected event samples that have 
minimal overlap with our sample.

\subsubsection{$\pi^{0}$ and $\eta \to \gamma \gamma$ Finding Systematics}

The uncertainty for $\pi^{0}$ finding was determined by using the results 
of the $\pi^{0}$ finding study which used the full $818~{\rm pb}^{-1}$ 
$\psi(3770)$ data sample \cite{cbx2008-029}. \ 

Our $\pi^0$ selection criteria are slightly different from 
the nominal criteria used in that study since we use the CCFIX. \ The CCFIX makes 
two types of changes to standard photon reconstruction. \ It increases the 
photon resolution in MC by 20\% and increases the reported errors of the photon 
uncertainty in data and MC by 20\% \cite{ccfix}. \ 
Using the result in Ref.~\cite{cbx2008-029} with a loose $\pi^0$ pull-mass cut of $\pm6$ 
is then similar to using the CCFIX.  Defining 
$\Delta\epsilon \equiv \epsilon_{\rm data}/\epsilon_{\rm MC} - 1$, 
the result for $\Delta\epsilon$ with a $\pm6$ $\pi^0$ pull mass and 
E9/E25 OK cuts applied to the $\pi^0$ children is $(-3.5 \pm 2.5)\%$. \ Interestingly, 
$\Delta\epsilon = (-3.5 \pm 3.3)\%$ with a 
$\pm6$ $\pi^0$ pull mass and no E9/E25 OK cuts applied to the $\pi^0$ children, 
suggesting the E9/E25 OK cuts do not effect $\pi^0$ finding. 

The uncertainty due to $\eta\to\gamma \gamma$ finding was determined 
based on the studies of $\psi(2S) \to \eta J/\psi$ decays in the 27.4~M $\psi(2S)$ data 
sample \cite{cbx07-15}. \ This study found $\Delta\epsilon = (-6.5 \pm 1.3)\%$ 
without the CCFIX but applying the E9/E25 OK cuts, 
$\Delta\epsilon = (-5.6 \pm 1.3)\%$ without the CCFIX and the E9/E25 OK cuts, 
$\Delta\epsilon = (-3.5 \pm 1.3)\%$ with the CCFIX and without the E9/E25 OK cuts.  
This suggests that the MC-determined efficiency overestimates 
the $\eta\to\gamma\gamma$ finding efficiency by $3.5\%$ and that the E9/E25 OK cut 
does not effect $\eta\to\gamma\gamma$ finding. 

Based on the interpretations of the studies above, we decreased the 
MC-determined efficiencies by 3.5\% 
and conservatively assigned systematic uncertainties of 3.5\% for both $\pi^{0}$ and 
$\eta \to \gamma \gamma$ finding. \ The systematic uncertainty for 
$\eta \to \pi\pi\pi^{0}$ decays was determined by applying the assigned 
$\pi$ and $\pi^{0}$ systematic uncertainties. \ Similarly, the systematic uncertainty for 
$\eta^{\prime} \to \pi \pi \eta (\gamma \gamma)$ decays were determined 
by applying the assigned $\pi$ and $\eta \to \gamma \gamma$ systematic uncertainties.

\subsubsection{$K_{S}$ Finding Systematics}

The systematic uncertainty associated with $K_{S}$ finding has been studied 
in several CLEO analyses \cite{cbx2008-041, cbx07-28, cbx06-22, cbx05-7, cbx05-3}. \ 
We applied an uncertainty of $2\%$ based on the analyses which apply selection 
criteria similar to ours \cite{cbx06-22, cbx05-7}. 

\subsubsection{$\eta \to \pi \pi \pi^0$ and $\eta^{\prime}$ Selection Systematics}

The uncertainties due to the mass ranges used to select $\eta\to\pi\pi\pi^{0}$ 
and $\eta^{\prime}\to\pi\pi\eta$ decays were determined by 
changing the corresponding $\eta$ or $\eta^{\prime}$ mass ranges to ranges 
in which the inefficiencies double with respect to our nominal range. \ 
These inefficiencies correspond to the mass ranges of 
$|M_{\pi\pi\pi^{0}} - M_{\eta}| < 8.5~{\rm MeV}$ 
for $\eta\to\pi\pi\pi^{0}$ decays and 
$|M_{\pi\pi\pi^{0}} - M_{\eta^{\prime}}| < 8.5~{\rm MeV}$ 
for the $\eta^{\prime}\to\pi\pi\eta$ mode.

\subsubsection{$\pi\pi$ Recoil Mass Systematics}

The uncertainties associated with $J/\psi$ rejection based on the $\pi\pi$ recoil mass 
criteria were determined by studying $\psi(2S)\to\gamma\chi_{c2}$, $\chi_{c2}\to X$ decays 
with the cuts removed and observing the change of the branching fractions.

\subsubsection{$M(X - \pi\pi)$ Systematics}

The uncertainties due to the cut on the hadronic invariant mass determined while 
excluding a pion pair $M(X - \pi\pi)$ for $J/\psi$ rejection were determined based 
on decays of $\psi(2S)\to\gamma\chi_{c2}, \chi_{c2}\to X$ by 
removing the cuts and observing the change of the branching fractions.

\subsubsection{$\cos \theta_{\gamma,\pi}$ and $d_{\gamma, trk}$ Systematics}

The systematics due to the cuts on the
angle between the momenta of the transition photon and the closest
pion ($\cos \theta_{\gamma,\pi}$), and
the distance between the transition photon and the  nearest track 
($d_{\gamma, trk}$) were determined by removing the cut and observing the 
change of the final results. \ For the $4\pi$ and $KK\pi\pi$ modes 
we studied $\psi(2S)\to\gamma\chi_{c2}$, $\chi_{c2}\to X$ decays.
For $KK\pi^{0}$ and $\pi\pi\eta(\pi\pi\pi^{0})$ modes, 
we applied the results from the $\psi(2S)\to\gamma\chi_{c2}$, 
$\chi_{c2}\to X$ decay studies for the $KK\pi\pi$ and $4\pi$ modes, respectively.

\subsubsection{Photon Recoil Mass Systematics}

The uncertainties associated with the photon recoil mass cuts, 
which rejects $\psi(2S)\to\gamma\chi_{cJ}$ events where the transition 
photon was used as a child in a reconstructed $\pi^0$ or $\eta \to \gamma \gamma$ 
decay (${\rm Rec}~M(\pi^{0}~{\rm or}~\eta~{\rm child})$), 
were determined based on the decay of 
$\psi(2S)\to\gamma\chi_{c2}$, $\chi_{c2}\to KK\pi\pi\pi^{0}$ by 
applying the cut to this mode and observing the change of the branching fraction. \ 
We applied this uncertainty to the 
$KK\pi^{0}$, $\pi\pi\eta(\gamma\gamma)$, and $KK\eta(\gamma\gamma)$ modes.

\subsubsection{$\eta$ Recoil Mass Systematics}

The systematic uncertainties due to the $\eta$ recoil mass cuts (${\rm Rec}~M(\eta)$), 
which were used for $J/\psi$ rejection, were determined by changing 
the recoil mass range to double the inefficiency 
with respect to the nominal cut. \ 
These inefficiencies correspond to mass range
$|{\rm Rec} M(\eta) - M(J/\psi)| \geq 85~{\rm MeV}$ 
for the $\pi\pi\eta(\gamma\gamma)$ mode and
$|{\rm Rec} M(\eta) - M(J/\psi)| \geq 35~{\rm MeV}$ 
for the $\pi\pi\eta(\pi\pi\pi^{0})$ mode.

\subsection{Summary}

Consideration of the full set of systematic uncertainties listed in 
Tables~\ref{table:systmodeindep_chic2} through \ref{table:systmodedep},
as well as those due to the trigger, the number of 
$\psi(2S)$, and transition photon finding, reveals that the majority of the 
overall systematic uncertainty is due to the treatment of the background. \ 
These factors are listed in Column~C of Table~\ref{table:systmodeindep_etac2s}. \ 
The $E_{\gamma}$ signal region (Column~A of Table~\ref{table:systmodeindep_etac2s}) 
also shows significant uncertainty associated with the high $E_{\gamma}$ limit. \ 
However, this sensitivity is due to the fact that 
changing the upper boundary of the $E_{\gamma}$ signal region changes the 
background shape because of the low statistics in the 
$E_{\gamma} = (66,94)~{\rm MeV}$ region. \ 
Therefore, the results listed in Column~A of
Table~\ref{table:systmodeindep_etac2s} are mostly redundant with the 
background shape uncertainties (Column~C of
Table~\ref{table:systmodeindep_etac2s}). \ 
Table~\ref{table:systerr} lists the total systematic uncertainty applied to 
each mode. \  In general, the individual uncertainties are treated as uncorrelated 
and are combined in quadrature to obtain the overall 
systematic uncertainties in our product branching fraction measurements.  

\begin{table}[htbp]
\caption[Systematic uncertainties]
{\label{table:systerr}Systematic uncertainties of all modes.}
\begin{center}
\begin{tabular}{|l|c|}
  \hline
  Mode & Syst Err (\%) \\ \hline
  $4\pi$ & $14.3$ \\ \hline
  $6\pi$ & $29.4$ \\ \hline
  $KK\pi\pi$ & $36.5$ \\ \hline
  $KK\pi^{0}$ & $47.2$ \\ \hline
  $K_{S}K\pi$ & $24.7$ \\ \hline
  $K\bar{K}\pi$ & $26.9$ \\ \hline
  $\pi\pi\eta$ & $48.0$ \\ \hline
  $\pi\pi\eta^{\prime}$ & $28.1$ \\ \hline
  $KK\eta$ & $32.1$ \\ \hline
  $KK\pi\pi\pi^{0}$ & $37.4$ \\ \hline
  $KK4\pi$ & $14.0$ \\ \hline
  $K_{S}K3\pi$ & $23.4$ \\ \hline
\end{tabular}
\end{center}
\end{table}

\chapter{Conclusions and Discussions}
\label{chap:conclusions}

\section{Summary of Results}

We do not observe the transition $\psi(2S) \to \gamma \eta_{c}(2S)$ 
in any of the decay modes studied. \ Table~\ref{table:etac2s_resultsummary} 
lists the results  of the product branching fractions 
${\cal B}(\psi(2S) \to \gamma~\eta_c(2S)) \times {\cal B}(\eta_c(2S)\to X)$ using 
the nominal value of the $\eta_c(2S)$ full width 
($\Gamma(\eta_c(2S)) = 14 \pm 7~{\rm MeV}$ \cite{PDBook2008}). \ 
Even though we may have an excess of events in the $4\pi$ mode, we did not find any 
strong evidence for the $\psi(2S) \to \gamma~\eta_c(2S)$ decay. \ 

\begin{table}[htbp]
\caption[Final results of decay $\psi(2S)\to\gamma\eta_{c}(2S)$]
{\label{table:etac2s_resultsummary}
Summary for $\psi(2S)\to\gamma\eta_{c}(2S)$, $\eta_{c}(2S)\to X$ 
product branching fraction results using $\Gamma(\eta_c(2S)) = 14~{\rm MeV}$. \  
The product $B_{1} \times B_{2}$ is defined as 
${\cal B}(\psi(2S) \to \gamma~\eta_c(2S)) \times {\cal B}(\eta_c(2S)\to X)$ 
and are at the $10^{-6}$ level. \  
The first three columns only include statistical uncertainties, while the last 
column includes both statistical and systematic uncertainties. \
All upper limits are at 90\% confidence level.}
\begin{center}
\begin{tabular}{|l|c|c|c|c|}
  \hline
  Mode & $N_{\rm sig}$ & Corrected  & $B_{1} \times B_{2}$ & $B_{1} \times B_{2}$ \\ 
       & (90\% C.L.)   & $\epsilon$ (\%) & (stat only) & (stat+syst) \\ \hline
  $4\pi$                & $<64.8$ & $20.49\pm0.16$ & $<12.2$ & $<14.0$ \\ \hline
  $6\pi$                & $<36.6$ & $14.22\pm0.14$ & $<9.9$  & $<12.9$ \\ \hline
  $KK\pi\pi$            & $<35.2$ & $19.49\pm0.15$ & $<7.0$  & $<9.5 $ \\ \hline
  $KK\pi^{0}$           & $<16.0$ & $17.76\pm0.14$ & $<3.5$  & $<5.2 $ \\ \hline
  $K_{S}K\pi$           & $<11.0$ & $20.40\pm0.15$ & $<3.0$  & $<3.8 $ \\ \hline
  $K\bar{K}\pi$         & $<21.9$ & $7.63\pm0.04$* & $<11.1$ & $<14.1$ \\ \hline
  $\pi\pi\eta$          & $<4.3$  & $5.68\pm0.05$* & $<2.9$  & $<4.3 $ \\ \hline
  $\pi\pi\eta^{\prime}$ & $<4.1$  & $8.14\pm0.10$  & $<11.1$ & $<14.2$ \\ \hline
  $KK\eta$              & $<7.5$  & $6.47\pm0.05$* & $<4.4$  & $<5.8 $ \\ \hline
  $KK\pi\pi\pi^{0}$     & $<65.4$ & $8.74\pm0.11$  & $<29.2$ & $<40.2$ \\ \hline
  $KK4\pi$              & $<20.6$ & $9.93\pm0.11$  & $<8.0$  & $<9.1 $ \\ \hline
  $K_{S}K3\pi$          & $<23.9$ & $11.39\pm0.13$ & $<11.7$ & $<14.4$ \\ \hline
\end{tabular}
\end{center}
*Submode-decay branching fractions are included in efficiencies.
\end{table}

The effect of the $\eta_c(2S)$ full width uncertainty was determined as follows. \ Separate signal 
MC samples with $\Gamma(\eta_c(2S)) =$ 7 and 21 MeV were generated in the same manner as the 
nominal MC sample described in Section~\ref{subsec:signalmc}. \ The measured 
photon energy distributions were fit in the same manner as the nominal procedure, but with the 
resolution functions determined from these MC samples and the full width of the signal shape 
adjusted according to the full width being investigated. \ The results using 
$\Gamma(\eta_c(2S)) =$ 7 and 21 MeV, and the linear extrapolation of the 
product branching fraction as a function of $\Gamma(\eta_c(2S))$, are listed in 
Table~\ref{table:etac2s_resultsummary}. 

\begin{table}[htbp]
\caption[Summary of results for $\psi(2S)\to\gamma\eta_{c}(2S)$ 
with $\Gamma(\eta_c(2S)) = 7,21~{\rm MeV}$]
{\label{table:etac2swidth_7and21MeV}
Summary of results for $\psi(2S)\to\gamma\eta_{c}(2S)$ 
with $\Gamma(\eta_c(2S)) = 7,21~{\rm MeV}$. \ 
All upper limits are at 90\% confidence level. \ 
The efficiencies include PID, $\pi^0$ finding, and $\eta$ finding corrections. \ 
The product $B_{1} \times B_{2}$ is defined as 
${\cal B}(\psi(2S) \to \gamma~\eta_c(2S)) \times {\cal B}(\eta_c(2S)\to X)$ 
and are at the $10^{-6}$ level. \  
The $B_{1} \times B_{2}$ results include statistical and systematic uncertainties. \ 
The $y$-intercept and slope variables $a$ and $b$ are determined by 
$B_{1} \times B_{2} < a + b*\Gamma(\eta_c(2S))$.}
\begin{center}
\small{
\begin{tabular}{|l|c|c|c|c|c|c|c|c|}
  \hline
  Mode & \multicolumn{3}{c|}{$\Gamma(\eta_c(2S)) = 7~{\rm MeV}$} 
    & \multicolumn{3}{c|}{$\Gamma(\eta_c(2S)) = 21~{\rm MeV}$} & a & b \\ \cline{2-7}
       & \scriptsize{$N_{\rm sig}$} & \scriptsize{$\epsilon$ (\%)} & \scriptsize{$B_{1} \times B_{2}$} 
    & \scriptsize{$N_{\rm sig}$} & $\epsilon$ (\%) & \scriptsize{$B_{1} \times B_{2}$} 
    & \scriptsize{($10^{-6}$)} & \scriptsize{($10^{-6}~{\rm MeV}^{-1}$)} \\ \hline
  $4\pi$                & $<53.1$ & $22.06$  & $<10.6$ & $<77.5$ & $19.41$  & $<17.7$ & 7.04 & 0.505 \\ \hline
  $6\pi$                & $<26.4$ & $14.71$  & $<9.0 $ & $<49.8$ & $13.03$  & $<19.1$ & 3.88 & 0.727 \\ \hline
  $KK\pi\pi$            & $<25.6$ & $20.44$  & $<6.6 $ & $<45.7$ & $17.72$  & $<13.6$ & 3.10 & 0.500 \\ \hline
  $KK\pi^{0}$           & $<12.0$ & $19.15$  & $<3.6 $ & $<19.5$ & $16.88$  & $<6.6 $ & 2.08 & 0.217 \\ \hline
  $K_{S}K\pi$           & $<9.7 $ & $21.78$  & $<3.1 $ & $<12.4$ & $19.53$  & $<4.4 $ & 2.43 & 0.095 \\ \hline
  $K\bar{K}\pi$         & $<17.2$ & $8.21 $* & $<10.3$ & $<26.7$ & $7.31 $* & $<17.9$ & 6.48 & 0.542 \\ \hline
  $\pi\pi\eta$          & $<4.3 $ & $6.79 $* & $<3.6 $ & $<4.3 $ & $4.97 $* & $<4.9 $ & 2.95 & 0.095 \\ \hline
  $\pi\pi\eta^{\prime}$ & $<4.1 $ & $9.46 $  & $<12.3$ & $<4.1 $ & $6.98 $  & $<16.6$ & 10.1 & 0.309 \\ \hline
  $KK\eta$              & $<7.5 $ & $7.72 $* & $<5.0 $ & $<7.5 $ & $5.68 $* & $<6.7 $ & 4.08 & 0.127 \\ \hline
  $KK\pi\pi\pi^{0}$     & $<49.4$ & $9.47 $  & $<28.0$ & $<83.9$ & $8.16 $  & $<55.2$ & 14.4 & 1.95  \\ \hline
  $KK4\pi$              & $<17.0$ & $10.50$  & $<7.1 $ & $<24.6$ & $9.37 $  & $<11.6$ & 4.91 & 0.317 \\ \hline
  $K_{S}K3\pi$          & $<20.2$ & $12.00$  & $<11.6$ & $<27.4$ & $10.23$  & $<18.4$ & 8.19 & 0.486 \\ \hline
\end{tabular}
}
\end{center}
*Submode-decay branching fractions are included in efficiencies.
\end{table}

The branching fractions of $\eta_{c}(2S) \to X$ may be 
estimated from $\eta_{c}(1S) \to X$ branching fractions and the ratio of the full width 
of the $\eta_{c}(2S)$ and $\eta_{c}(1S)$ if the matrix elements governing the 
decays of the $\eta_c(2S)$ are similar to those of the $\eta_c(1S)$. \ 
Since the studies of 
$\psi(2S) \to \gamma~\chi_{c2}, \chi_{c2} \to X$ built confidence 
in this analysis procedure, 
the most likely explanation of the non-observation is that the branching 
fraction assumptions were overly optimistic. \ Our original 
objectives for measuring the resonance properties of the $\eta_c(2S)$ cannot 
be achieved with this data sample. \ The upper limits of the branching 
fractions were obtained instead.

\section{Comparison with Published Results}

The recent CLEO measurement for ${\cal B}(J/\psi \to \gamma~\eta_c(1S))$ 
requires a reestimate of ${\cal B}(\psi(2S) \to \gamma~\eta_c(2S))$, as computed using 
Equation~\ref{widthjpsitogetac} in Section~\ref{ExpNumProdEvt}. \   
Reaveraging ${\cal B}(J/\psi \to \gamma~\eta_c(1S))$, after including the CLEO result of 
${\cal B}(J/\psi \to \gamma~\eta_c(1S)) = (1.98 \pm 0.09 \pm 0.30)\%$ \cite{Mitchell:CLEOjpsitoetac}, 
leads to ${\cal B}(J/\psi \to \gamma~\eta_c(1S)) = (1.72 \pm 0.25)\%$. \ 
Using the PDG 2008 values for $M(\eta_c(1S))$, $\Gamma(J/\psi)$, $M(\eta_c(2S))$, 
and $\Gamma(\psi(2S))$  \cite{PDBook2008}, along 
with this new value for ${\cal B}(J/\psi \to \gamma~\eta_c(1S))$, replaces the 
prediction given in Equation~\ref{widthpsi2s} with 
${\cal B}(\psi(2S) \to \gamma~\eta_c(2S)) = (3.9 \pm 1.1)\times 10^{-4}$, 
which is 50\% larger.

The BaBar collaboration recently measured the 
branching fraction for $\eta_c(2S) \to K\bar{K}\pi$ to be 
${\mathcal B}(\eta_{c}(2S) \to K{\bar K}\pi) = (1.9\pm0.4\pm0.5\pm1.0)\%$ 
\cite{Aubert:2008kp}, where the first two errors are statistical and 
systematic, respectively, and the last error is from the measurement of
${\mathcal B}(B^{+} \to K^{+} \eta_{c}(2S)~+~{\rm c.c.}) = (3.4 \pm 1.8) 
\times 10^{-4}$ \cite{InclEtac2S}. \ 
Using the central value of the BaBar $\mathcal{B}(\eta_c(2S) \rightarrow K
\bar{K} \pi)$ result with our 90\% confidence level upper limit of
$\mathcal{B}(\psi(2S) \rightarrow \gamma \eta_c(2S)) \times \mathcal{B}(\eta_c(2S)
\rightarrow K \bar{K} \pi)$ leads to an upper limit of $\mathcal{B}(\psi(2S)
\rightarrow \gamma \eta_c(2S)) < 7.4 \times 10^{-4}$. \ This limit is roughly
twice the revised phenomenological prediction given in the last paragraph.

The BaBar measurement of $\mathcal{B}(\eta_c(2S) \rightarrow K \bar{K} \pi)$
can also be used to obtain an improved estimate of the two-photon partial
width of the $\eta_c(2S)$. \ CLEO \cite{cleo2004fusion} previously measured 
the ratio
\begin{eqnarray}
R(\eta_{c}(2S)/\eta_{c}(1S)) & \equiv &
\frac{\Gamma_{\gamma\gamma}(\eta_{c}(2S)) \times 
\mathcal{B}(\eta_{c}(2S) \to K{\bar K}\pi)}
{\Gamma_{\gamma\gamma}(\eta_{c}(1S)) \times 
\mathcal{B}(\eta_{c}(1S) \to K{\bar K}\pi)} \nonumber \\
& = & 0.18 \pm 0.05 \pm 0.02 ,
\end{eqnarray}
and assumed equal branching fractions to $K \bar{K} \pi)$ for $\eta_c(1S)$
and $\eta_c(2S)$ to obtain a first estimate of $\Gamma_{\gamma
\gamma}(\eta_c(2S))=(1.3 \pm 0.6)$~keV. \ Using BaBar's $\mathcal{B}(\eta_c(2S)
\rightarrow K \bar{K} \pi)$, the PDG value of $\Gamma_{\gamma
\gamma}(\eta_c(1S))=(7.2 \pm 2.1)$~keV, and $\mathcal{B}(\eta_c(1S) \rightarrow
K \bar{K} \pi)=7.0 \pm 1.2$\% \cite{PDBook2008}, the CLEO measurement of
$R(\eta_c(2S)/\eta_c(1S))$ can be reinterpreted to give $\Gamma_{\gamma
\gamma}(\eta_c(2S))=(4.8 \pm 3.7)$~keV.

A recent analysis from the Belle collaboration found 
$\Gamma_{\gamma\gamma}(\eta_{c}(2S))\mathcal{B}(\eta_{c}(2S) \to 4\pi) < 6.5~{\rm eV}$ 
and 
$\Gamma_{\gamma\gamma}(\eta_{c}(2S))\mathcal{B}(\eta_{c}(2S) \to KK\pi\pi) < 5.0~{\rm eV}$ 
\cite{Uehara2008}, both at $90\%$ confidence level. \
Using the $\Gamma_{\gamma\gamma}(\eta_c(2S))$ result above, 
these two results give upper limits of $\mathcal{B}(\eta_{c}(2S) \to 4\pi) < 0.14\%$ 
and $\mathcal{B}(\eta_{c}(2S) \to KK\pi\pi) < 0.10\%$. 

Our results for the $4\pi$ and $KK\pi\pi$ modes can be used to determine comparable upper limits. \ 
Using $\mathcal{B}(\psi(2S) \to \gamma \eta_{c}(2S)) = (3.9 \pm 1.1) \times 10^{-4}$ leads 
to upper limits of $\mathcal{B}(\eta_{c}(2S) \to 4\pi) < 3.6\%$ and
$\mathcal{B}(\eta_{c}(2S) \to KK\pi\pi) < 2.4\%$.  Our upper limit results are 
an order of magnitude larger than the corresponding upper limits determined from the 
Belle results.

The branching fraction measurement for $\eta_{c}(2S) \to K{\bar K}\pi$ and 
the upper limits for $\eta_{c}(2S) \to 4\pi, KK\pi\pi$ from the Belle collaboration 
are all lower than predictions made by assuming the partial widths of the these 
decays are the same as for the $\eta_c(1S)$. \ Using the PDG 2008 
values for the $\eta_c(1S)$ and $\eta_c(2S)$ full widths, 
$\Gamma(\eta_c(1S)) = 26.7 \pm 3.0~{\rm MeV}$ and 
$\Gamma(\eta_c(2S)) = 14 \pm 7~{\rm MeV}$, and assuming the partial widths for the 
$\eta_c(1S)$ and $\eta_c(2S)$ decays are identical, leads to all $\eta_c(1S)$ 
branching fractions being increased by $1.9 \pm 1.0$ for $\eta_c(2S)$. \ 
Using the PDG 2008 values for the $\eta_{c}(1S)$ branching fractions
${\mathcal B}(\eta_{c}(1S) \to K{\bar K}\pi) = (7.0 \pm 1.2)\%$,
${\mathcal B}(\eta_{c}(1S) \to 4\pi) = (1.20 \pm 0.30)\%$, and
${\mathcal B}(\eta_{c}(1S) \to KK\pi\pi) = (1.5 \pm 0.6)\%$,
leads to
${\mathcal B}(\eta_{c}(2S) \to K{\bar K}\pi) = (13.4 \pm 2.3 \pm 6.9)\%$,
${\mathcal B}(\eta_{c}(2S) \to 4\pi) = (2.3 \pm 0.6 \pm 1.2)\%$ and
${\mathcal B}(\eta_{c}(2S) \to KK\pi\pi) = (2.9 \pm 1.1 \pm 1.5)\%$,
where the last error is from the full width scale factor. \
The ${\mathcal B}(\eta_{c}(2S) \to K{\bar K}\pi)$ measurement is $7.1\times$
smaller than the partial width scaling assumption but, after
considering the uncertainties, they only differ by $1.6~\sigma$. \ 
The $4\pi$ and $KK\pi\pi$ upper limit results are $16.4\times$ 
and $21\times$ smaller, respectively, but only differ by 1.6 and 1.5~$\sigma$. \
Higher precision measurements of the $\eta_{c}(2S)$ full width and 
${\mathcal B}(B^{+} \to K^{+} \eta_{c}(2S)~+~{\rm c.c.})$ are needed 
in order to determine if the partial width scaling assumption is correct 
for the $\eta_{c}(1S,2S)$ states.  It should be noted that this may not 
be true since, for example, we see large differences in the decay rates of 
$J/\psi$ and $\psi(2S)$ to $\rho\pi$.

Is the experimental measurements for the $\eta_c(2S)$ full width correct? \ 
Comparing the $J/\psi$ and $\psi(2S)$ full widths would lead us to think the 
$\eta_c(2S)$ full width should be larger than the $\eta_c(1S)$. \ Experimentally, 
this is not true. \ Is there something unusual with the full widths of the 
$J/\psi$ or $\psi(2S)$? \ Table~\ref{table:1s2swidthratio} compares the 
full width and $e^+e^-$ partial widths for the first two $S$-wave spin 
triplet states in the $c\bar{c}$ and $b\bar{b}$ systems, along with the 
full widths and $\gamma\gamma$ partial widths of the $\eta_c(1S)$ and 
$\eta_c(2S)$. \ The full width ratio of $\eta_c(2S)/\eta_c(1S)$ agrees with 
$\Upsilon(2S)/\Upsilon(1S)$ and not with $\psi(2S)/J/\psi$, while the 
$e^+e^-$ and $\gamma\gamma$ partial widths for all three agree. \ This 
seems to suggest that the $\psi(2S)$ full width is the odd character, 
possibly related to $D\bar{D}$ threshold effects.

\begin{table}[htbp]
\caption[Ratio of widths between $2S$ and $1S$ states]
{\label{table:1s2swidthratio}Ratio of widths between $2S$ and $1S$ states
in quarkonia. \ The ``$x$'' in the table denotes the type of partial widths, 
which is $\gamma \gamma$ for $\eta_{c}(1S)$, $\eta_{c}(2S)$ and $e^{+}e^{-}$ 
for $J/\psi$, $\psi(2S)$, $\Upsilon(1S)$, and $\Upsilon(2S)$. \ 
All values are derived from the PDG 2008 \cite{PDBook2008}.}
\begin{center}
\begin{tabular}{c|ccc}
  & $\eta_{c}(2S)/\eta_{c}(1S)$ & $\psi(2S)/J/\psi$ & $\Upsilon(2S)/\Upsilon(1S)$ \\ \hline
  $\Gamma(2S)/\Gamma(1S)$ & $0.52\pm0.27$ & $3.40\pm0.12$ & $0.59\pm0.05$ \\
  $\Gamma_{x}(2S)/\Gamma_{x}(1S)$ & $0.67\pm0.55$ & $0.429\pm0.013$ & $0.457\pm0.010$ \\
\end{tabular}
\end{center}
\end{table}

\section{Conclusions and Future Prospects}

In order to experimentally observe decays of $\psi(2S) \to \gamma \eta_{c}(2S)$, 
our results show that a much larger data sample is needed. \ 
CESR-c and CLEO-c have ceased operation, but the more powerful collider 
BEPC-II and its detector BES-III are now beginning operation in Beijing, 
China. \ BES-III is expected to collect 
billion $\psi(2S)$ events in one year of running \cite{Asner:BES3YellowBook}, 
forty times the CLEO-c $\psi(2S)$ sample. \ 
Our results are an important input for the development of future 
analyses of $\psi(2S) \to \gamma \eta_{c}(2S)$.

\appendix
\chapter{Resolution Function Fits}
\label{appendix:resfnctfits}

\begin{figure}[htbp]
\begin{center}
\subfigure
{\includegraphics[width=.95\textwidth]{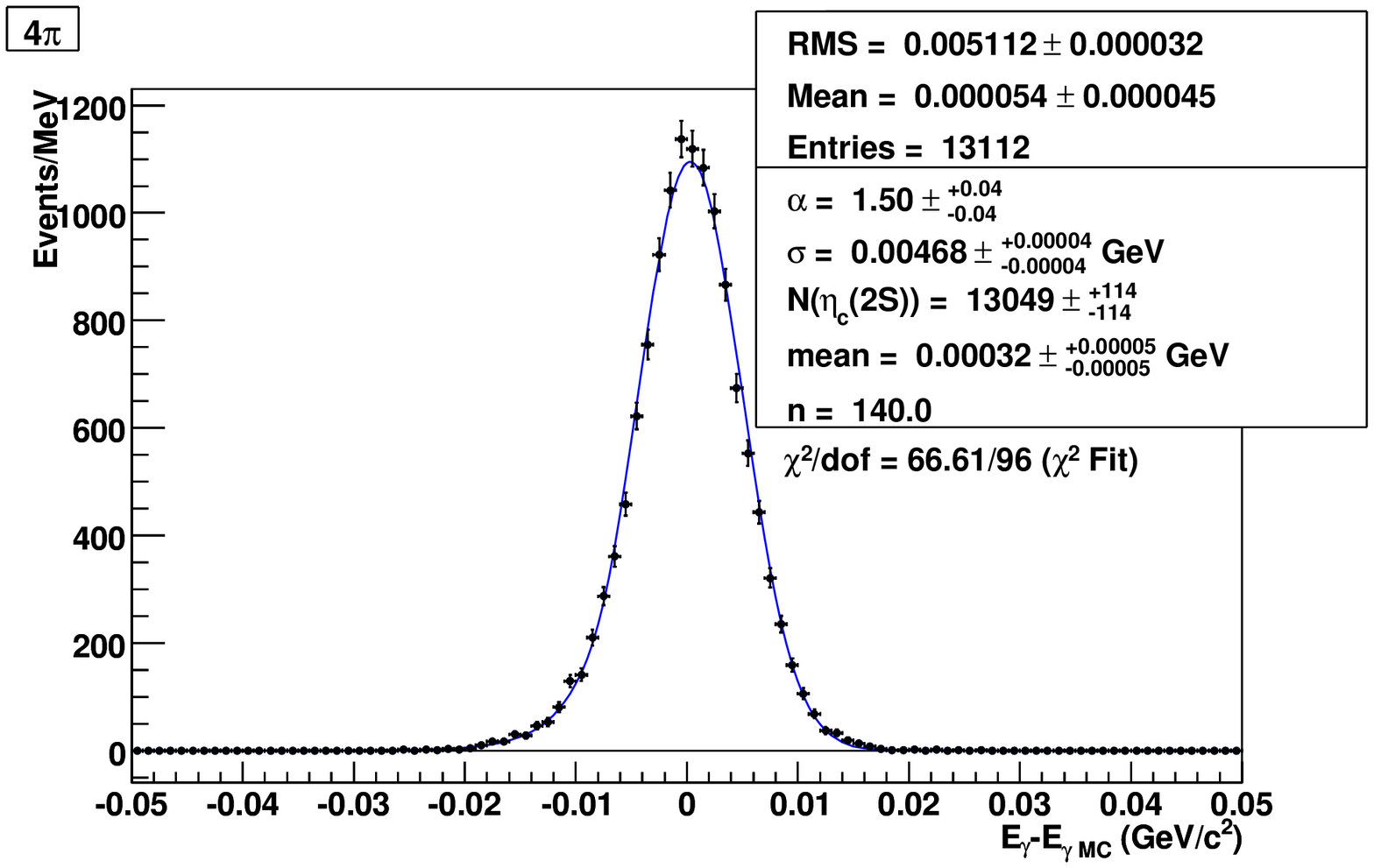}}
\end{center}
\caption[Resolution function for the $\psi(2S)\to\gamma\eta_c(2S),\eta_c(2S)\to 4\pi$ mode.]
{\label{fig:etac2s_resfnct_4Pi}
{Resolution function for the decay mode $\psi(2S)\to\gamma\eta_c(2S), 
\eta_c(2S)\to 4 \pi$.
 \ The points are from the $\eta_c(2S)$ signal MC and 
the solid line is the result of the fit to the Crystal Ball function.}}
\end{figure}

\begin{figure}[htbp]
\begin{center}
\subfigure
{\includegraphics[width=.85\textwidth]{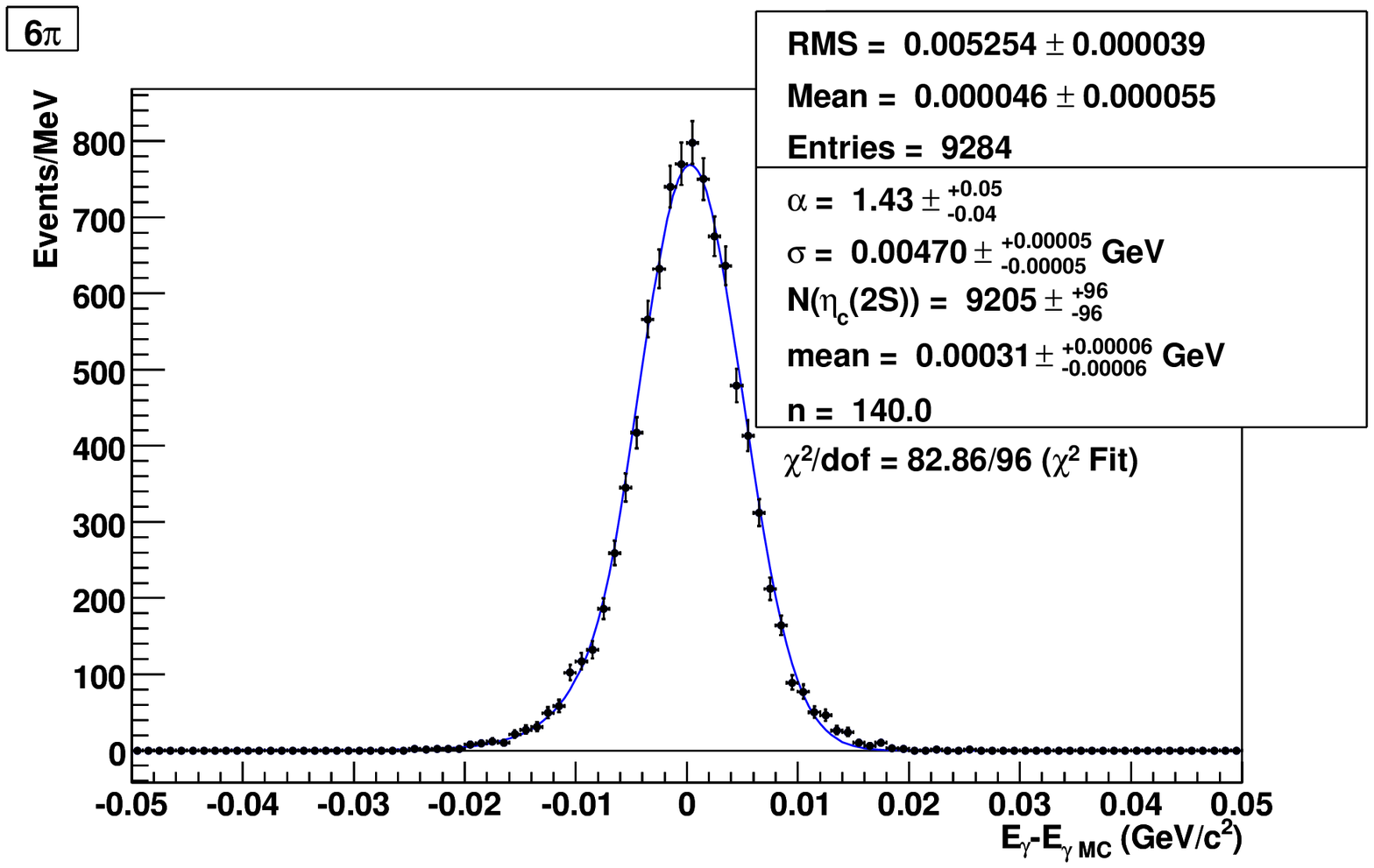}}
\subfigure
{\includegraphics[width=.85\textwidth]{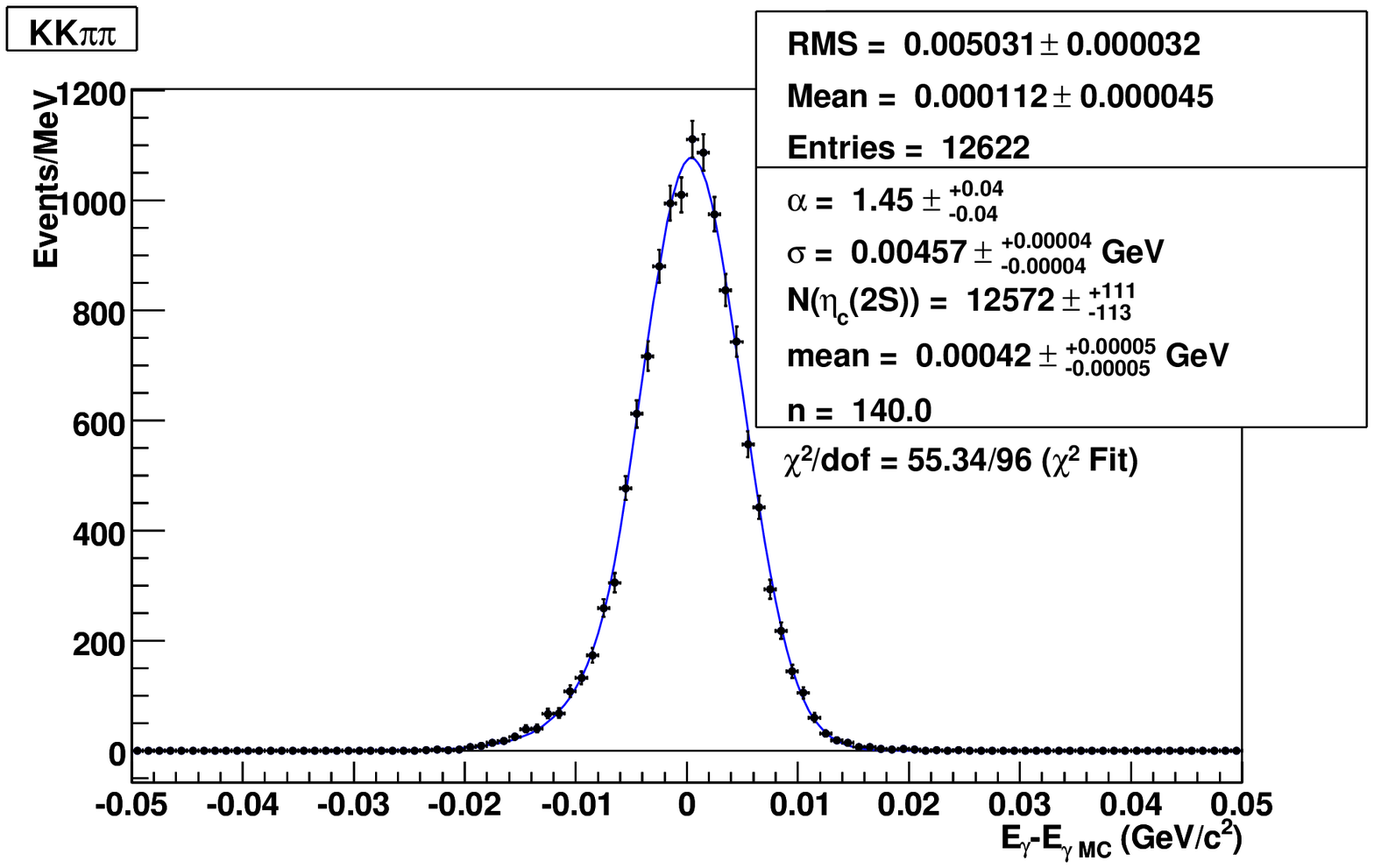}}
\end{center}
\caption[Resolution function for the $\psi(2S)\to\gamma\eta_c(2S),\eta_c(2S)\to 6\pi$
and $KK\pi\pi$ modes.]
{\label{fig:etac2s_resfnct_6Pi_KKPiPi}
{Resolution functions for the decay modes 
$\psi(2S)\to\gamma\eta_c(2S),\eta_c(2S)\to 6\pi$  (top) 
and $KK\pi\pi$ (bottom).
 \ The points are from the $\eta_c(2S)$ signal MC and 
the solid line is the result of the fit to the Crystal Ball function.}}
\end{figure}

\begin{figure}[htbp]
\begin{center}
\subfigure
{\includegraphics[width=.85\textwidth]{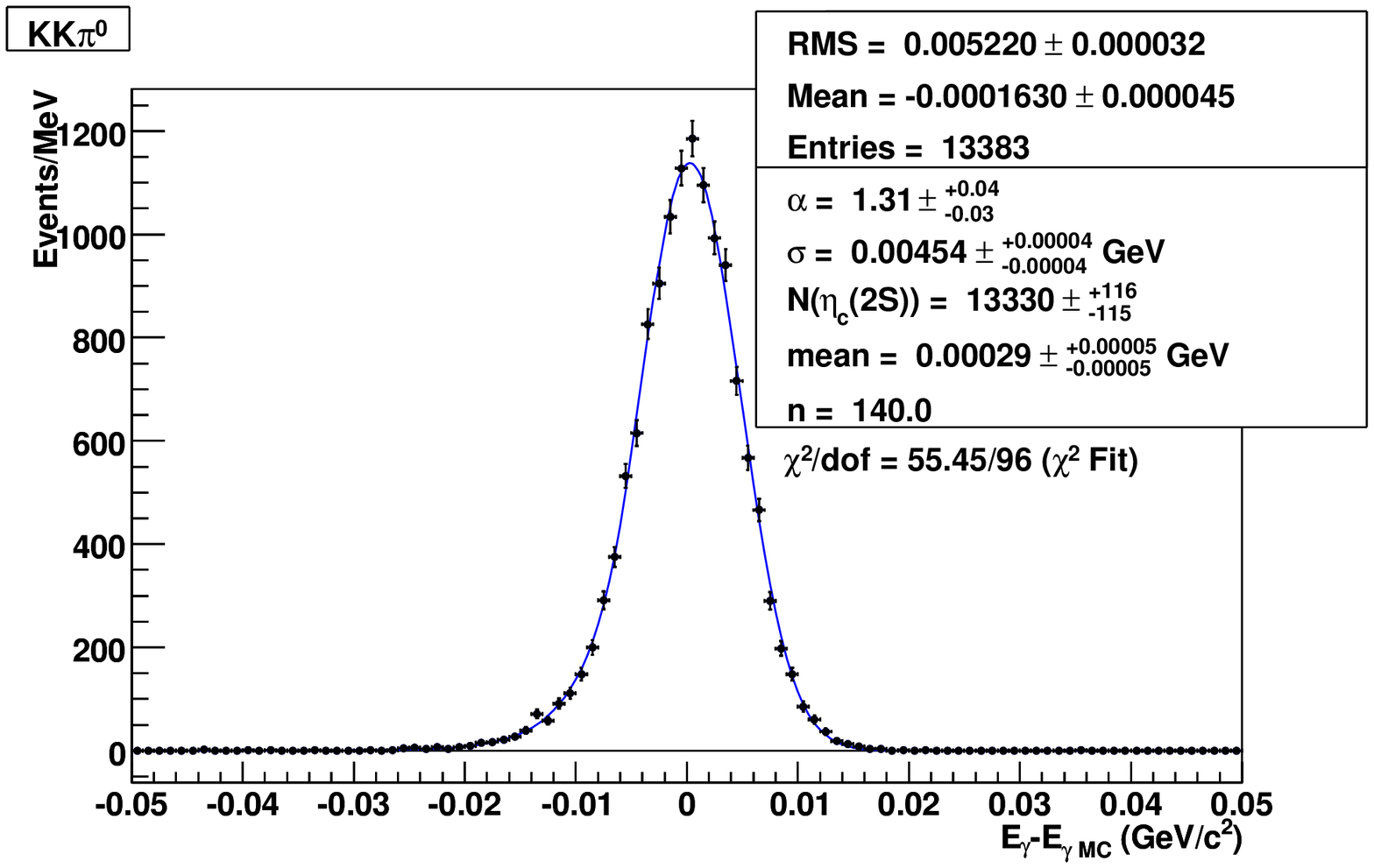}}
\subfigure
{\includegraphics[width=.85\textwidth]{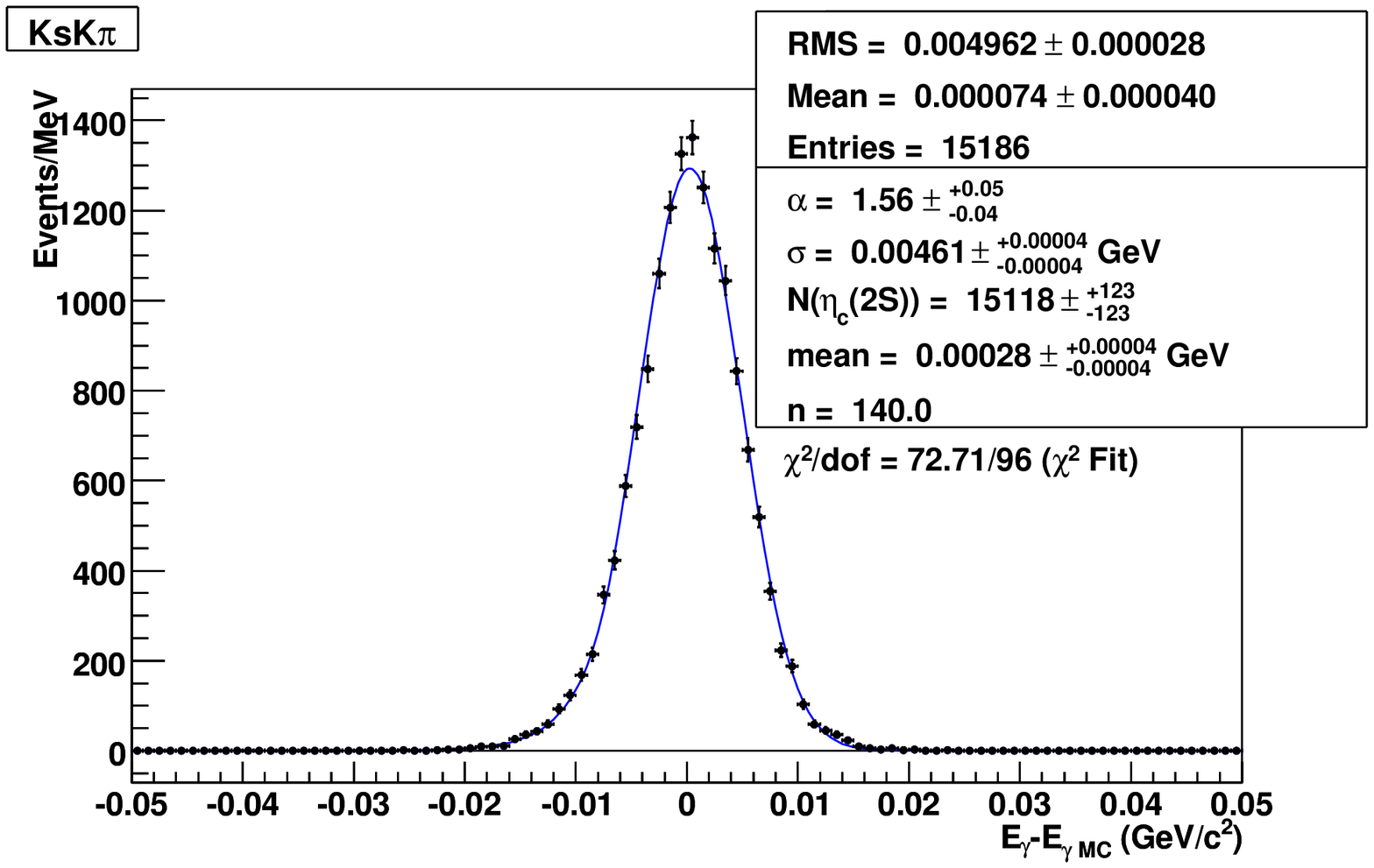}}
\end{center}
\caption[Resolution functions for the 
$\psi(2S)\to\gamma\eta_c(2S),\eta_c(2S)\to KK\pi^0$ 
and $K_{S}K\pi$ modes.]
{\label{fig:etac2s_resfnct_KKPi0_KsPi}
{Resolution functions for the decay modes 
$\psi(2S)\to\gamma\eta_c(2S),\eta_c(2S)\to KK\pi^0$  (top) 
and $K_{S}K\pi$ (bottom).
 \ The points are from the $\eta_c(2S)$ signal MC and 
the solid line is the result of the fit to the Crystal Ball function.}}
\end{figure}

\begin{figure}[htbp]
\begin{center}
\subfigure
{\includegraphics[width=.85\textwidth]{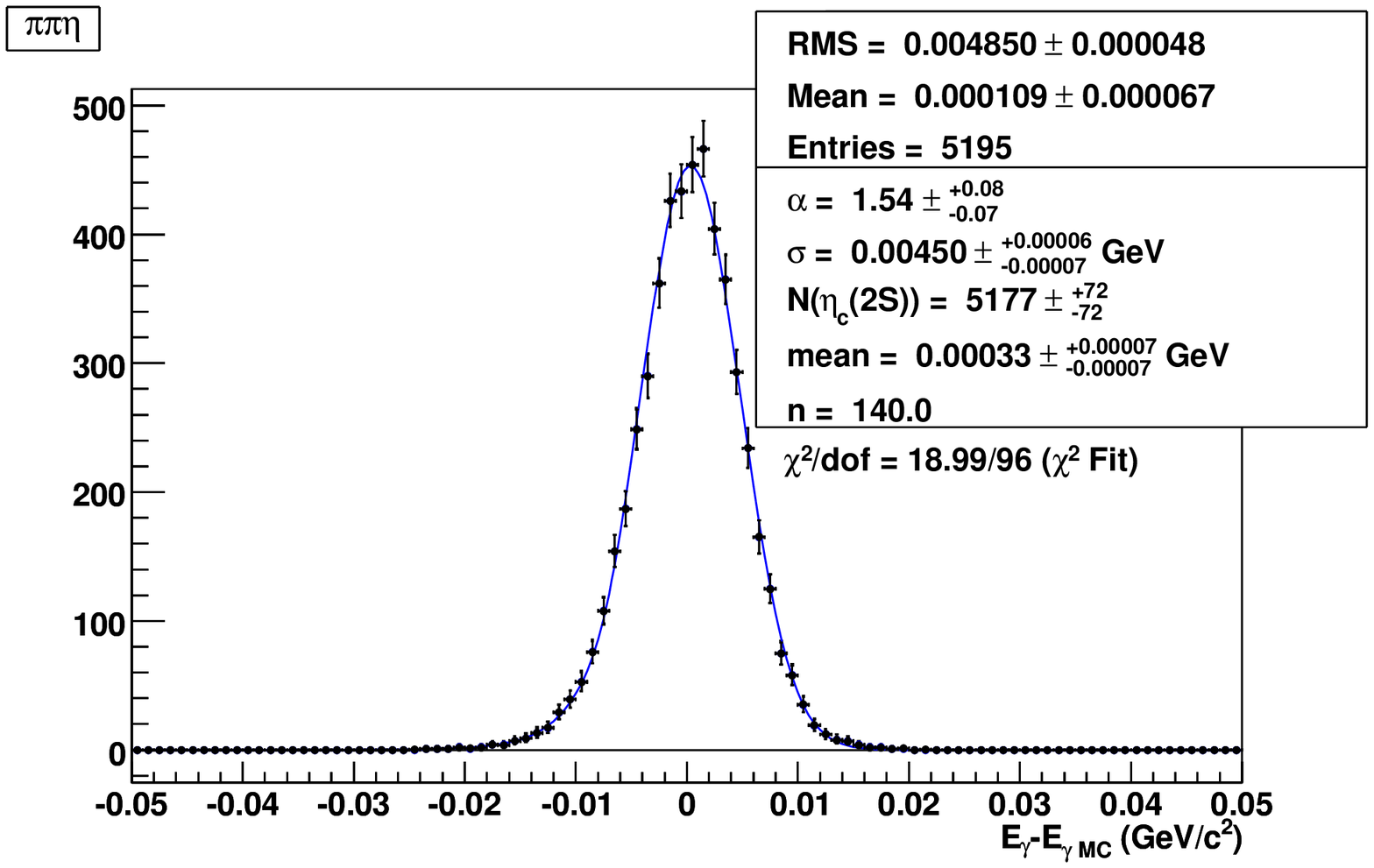}}
\subfigure
{\includegraphics[width=.85\textwidth]{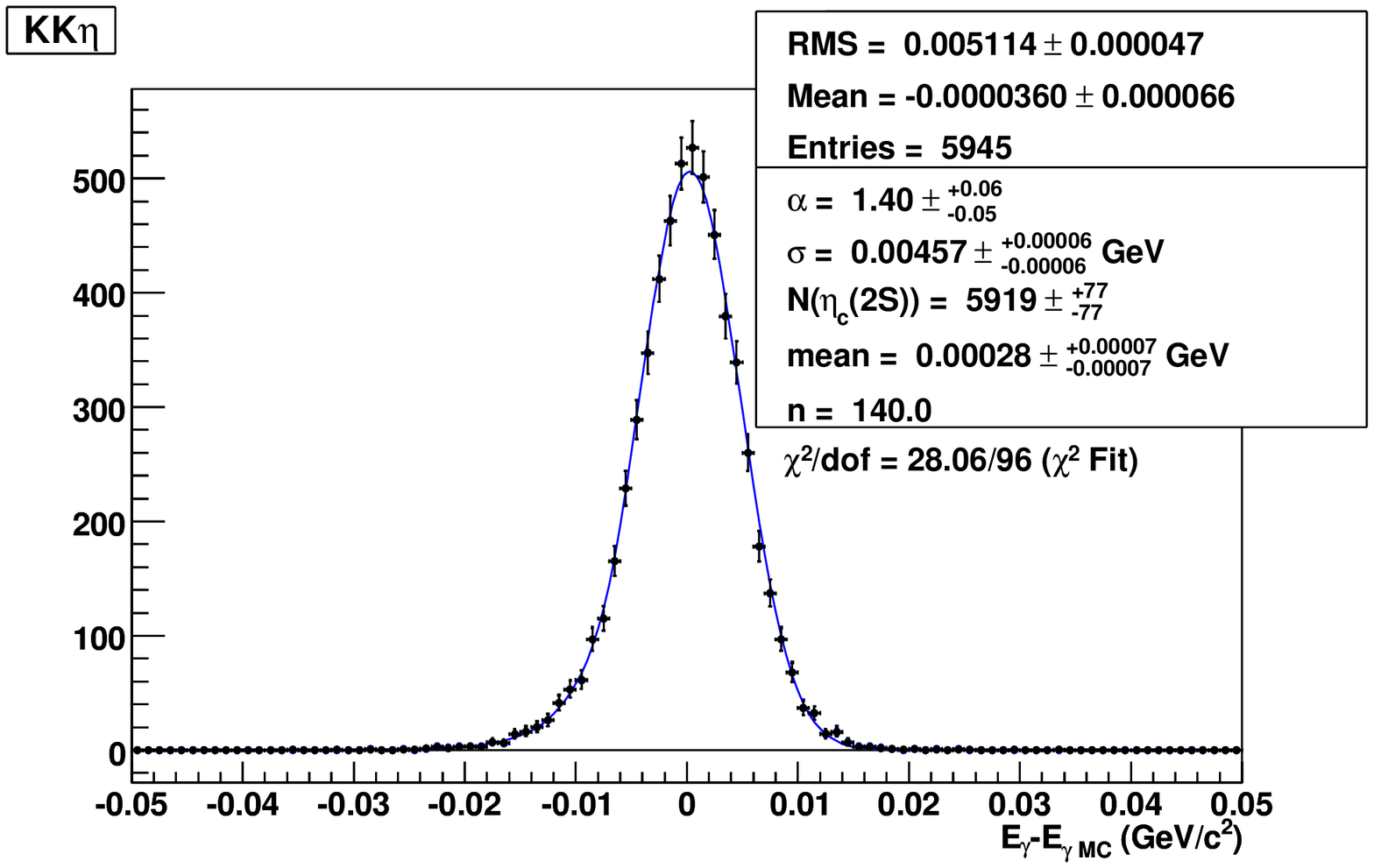}}
\end{center}
\caption[Resolution function for the 
$\psi(2S)\to\gamma\eta_c(2S),\eta_c(2S)\to \pi\pi\eta$
and $KK\eta$ modes.]
{\label{fig:etac2s_resfnct_PiPiEta_KKEta}
{Resolution functions for the decay modes 
$\psi(2S)\to\gamma\eta_c(2S),\eta_c(2S)\to \pi\pi\eta$  (top) 
and $KK\eta$ (bottom).
 \ The points are from the $\eta_c(2S)$ signal MC and 
the solid line is the result of the fit to the Crystal Ball function.}}
\end{figure}

\begin{figure}[htbp]
\begin{center}
\subfigure
{\includegraphics[width=.85\textwidth]{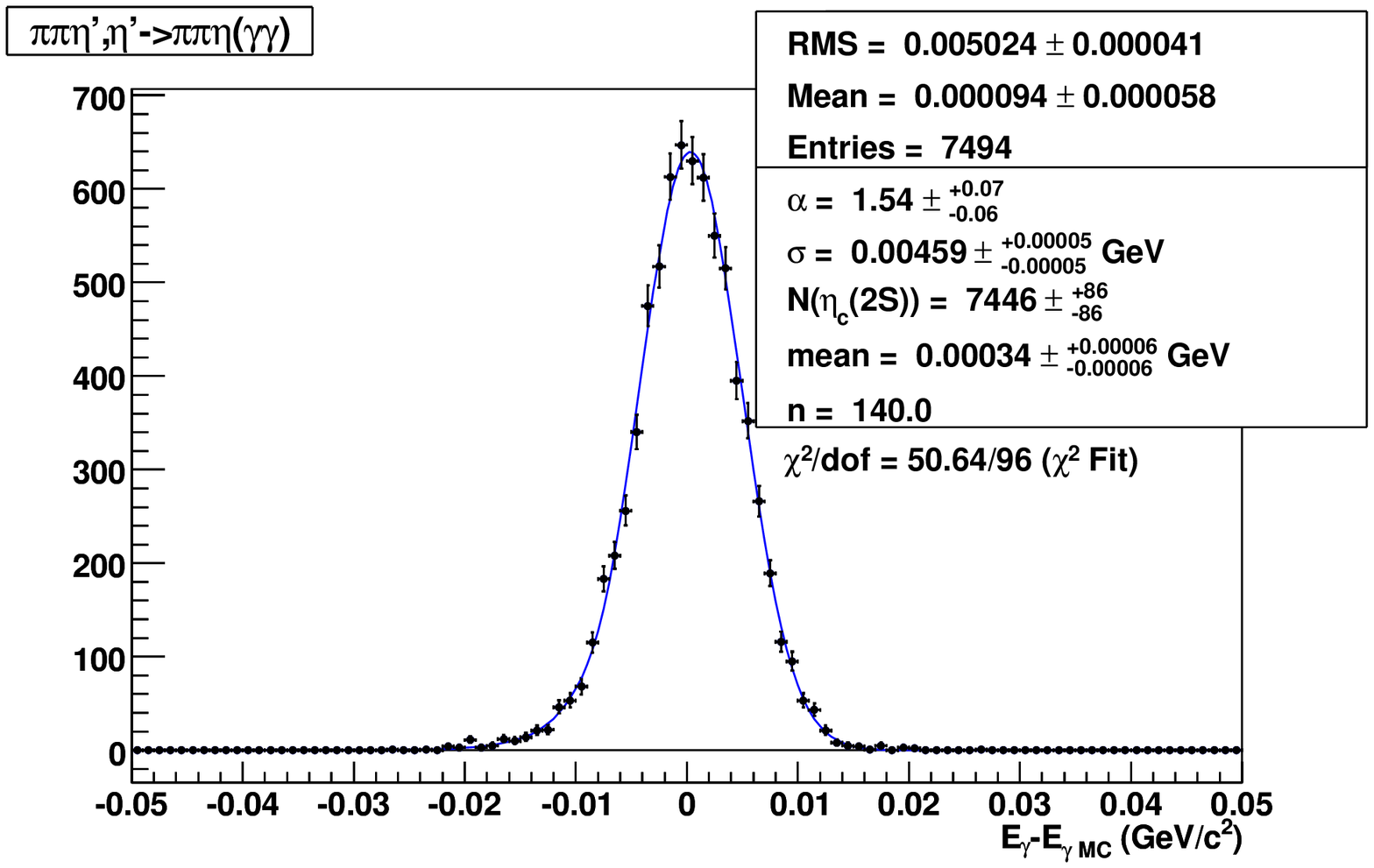}}
\subfigure
{\includegraphics[width=.85\textwidth]{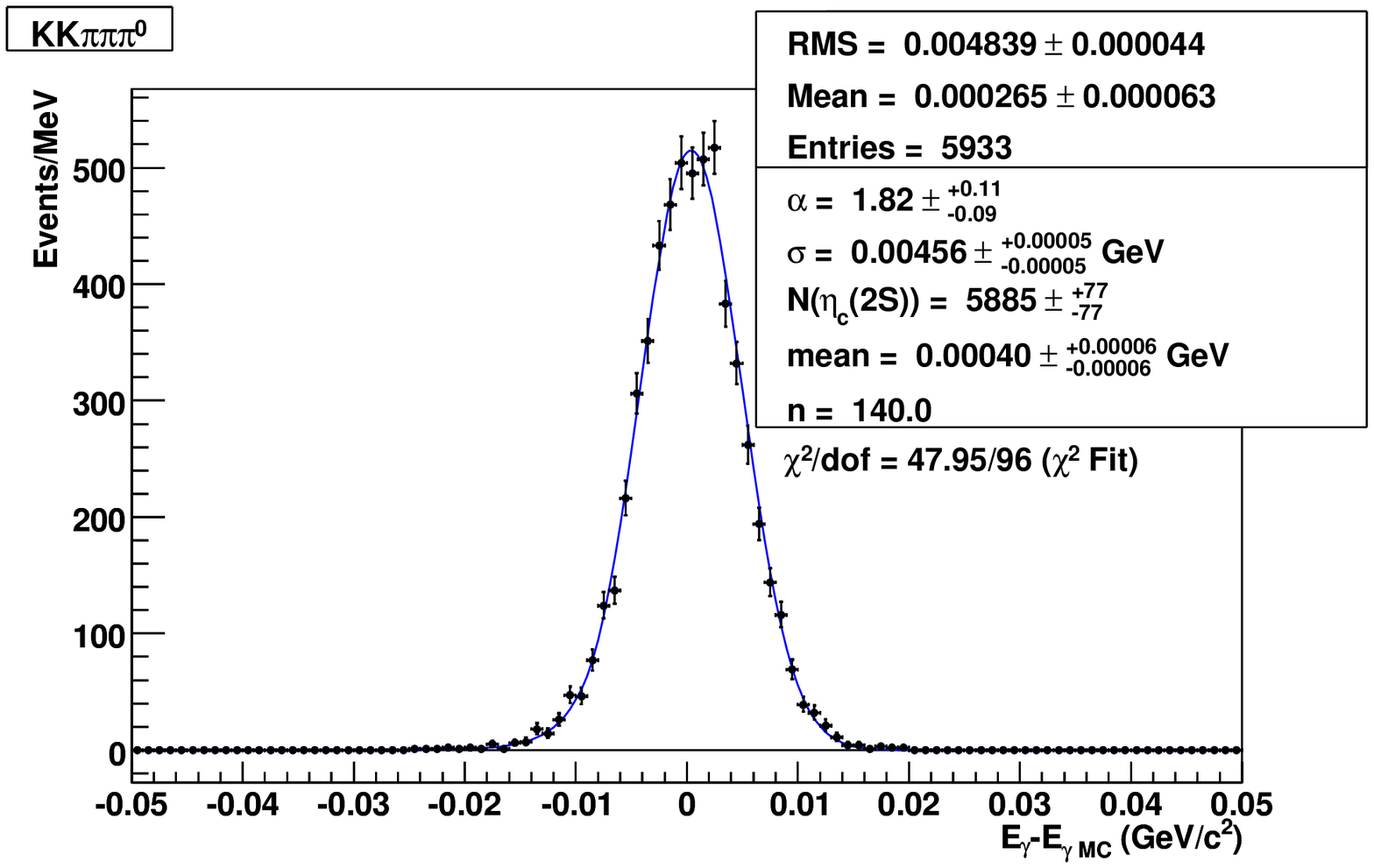}}
\end{center}
\caption[Resolution functions for the $\psi(2S)\to\gamma\eta_c(2S),
\eta_c(2S)\to \pi\pi\eta^{\prime}$ mode] 
{\label{fig:etac2s_resfnct_PiPiEtaPPiPiEtaGG_KKPiPiPi0}
{Resolution functions for the decay modes 
$\psi(2S)\to\gamma\eta_c(2S)$, $\eta_c(2S)\to \pi\pi\eta^{\prime}$,
$\eta^{\prime}\to\pi\pi\eta(\gamma\gamma)$ (top) and $KK\pi\pi\pi^0$ 
(bottom).
 \ The points are from the $\eta_c(2S)$ signal MC and 
the solid line is the result of the fit to the Crystal Ball function.}}
\end{figure}

\begin{figure}[htbp]
\begin{center}
\subfigure
{\includegraphics[width=.85\textwidth]{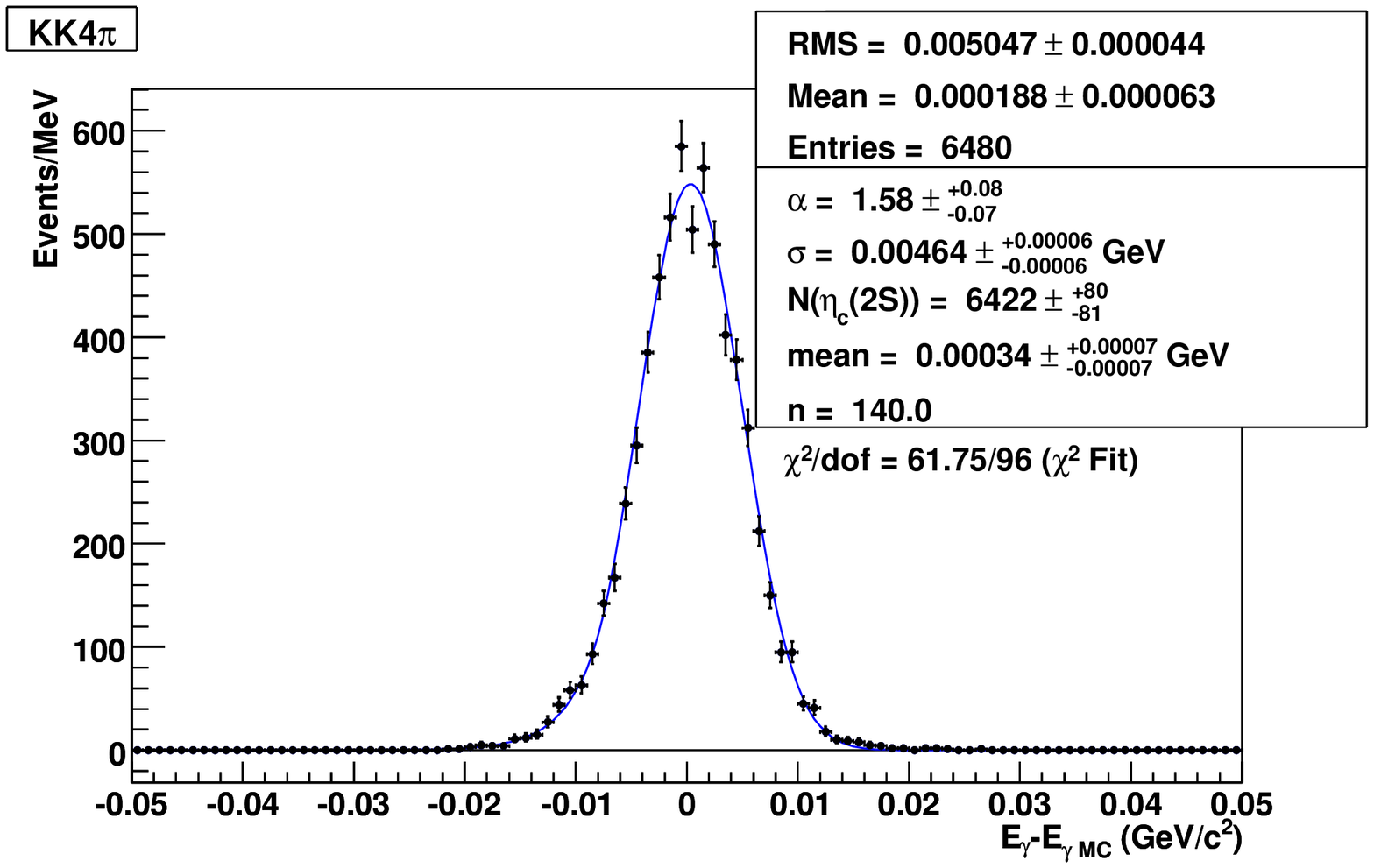}}
\subfigure
{\includegraphics[width=.85\textwidth]{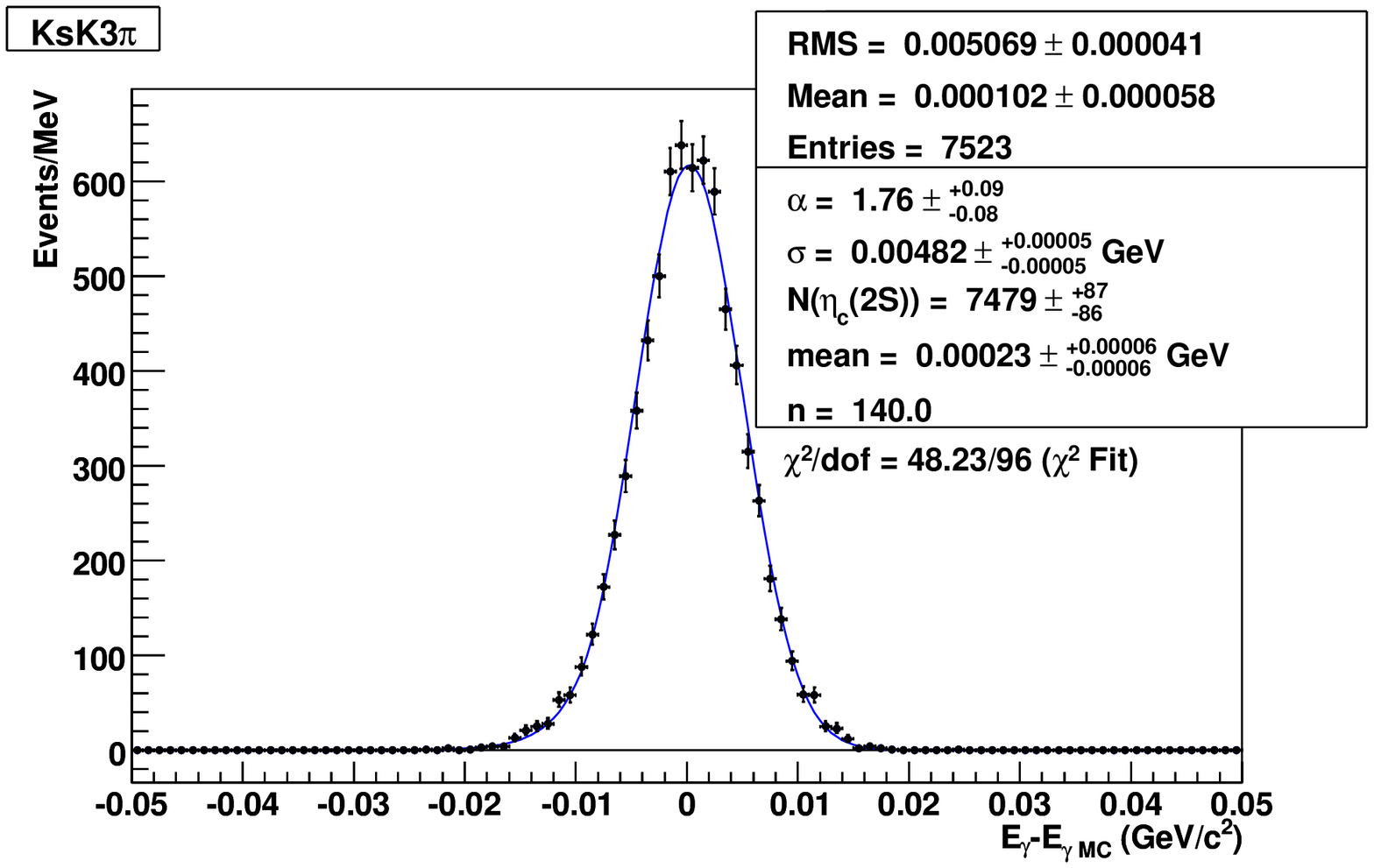}}
\end{center}
\caption[Resolution functions for the 
$\psi(2S)\to\gamma\eta_c(2S),\eta_c(2S)\to KK4\pi$  
and $K_{S}K3\pi$ modes.]
{\label{fig:etac2s_resfnct_KK4Pi_KsK3Pi}
{Resolution functions for the decay modes 
$\psi(2S)\to\gamma\eta_c(2S),\eta_c(2S)\to  KK4\pi$  (top) 
and $K_{S}K3\pi$ (bottom).
 \ The points are from the $\eta_c(2S)$ signal MC and 
the solid line is the result of the fit to the Crystal Ball function.}}
\end{figure}

\begin{figure}[htbp]
\begin{center}
\subfigure
{\includegraphics[width=.95\textwidth]{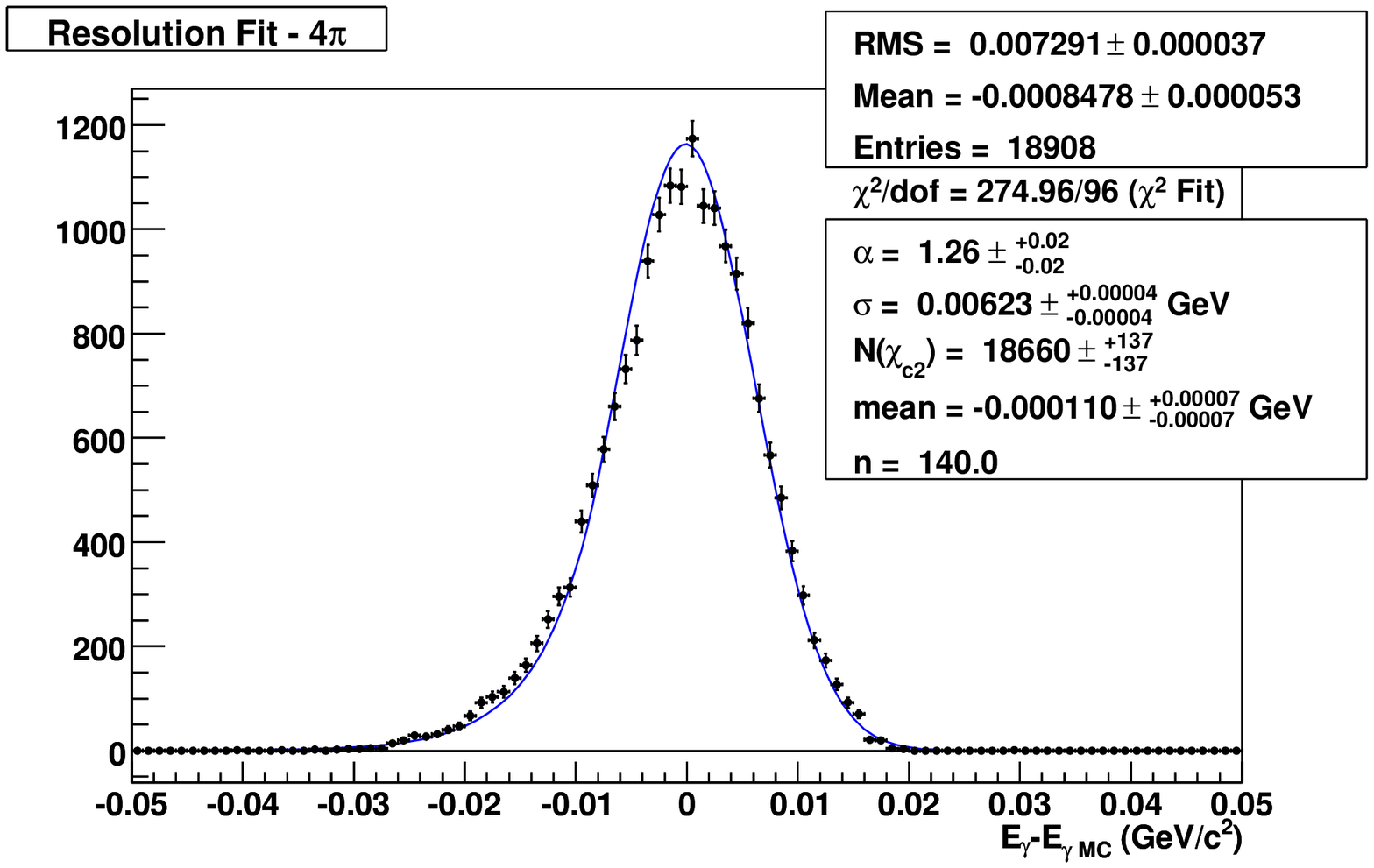}}
\end{center}
\caption[Resolution function for the 
$\psi(2S)\to\gamma\chi_{c2},\chi_{c2}\to 4\pi$ mode.]
{\label{fig:chic2_resfnct_4Pi}
{Resolution function for the decay mode $\psi(2S)\to\gamma\chi_{c2}, 
\chi_{c2}\to 4 \pi$.
 \ The points are from the $\chi_{c2}$ signal MC and the solid line 
is the result of the fit to the Crystal Ball function.}}
\end{figure}

\begin{figure}[htbp]
\begin{center}
\subfigure
{\includegraphics[width=.85\textwidth]{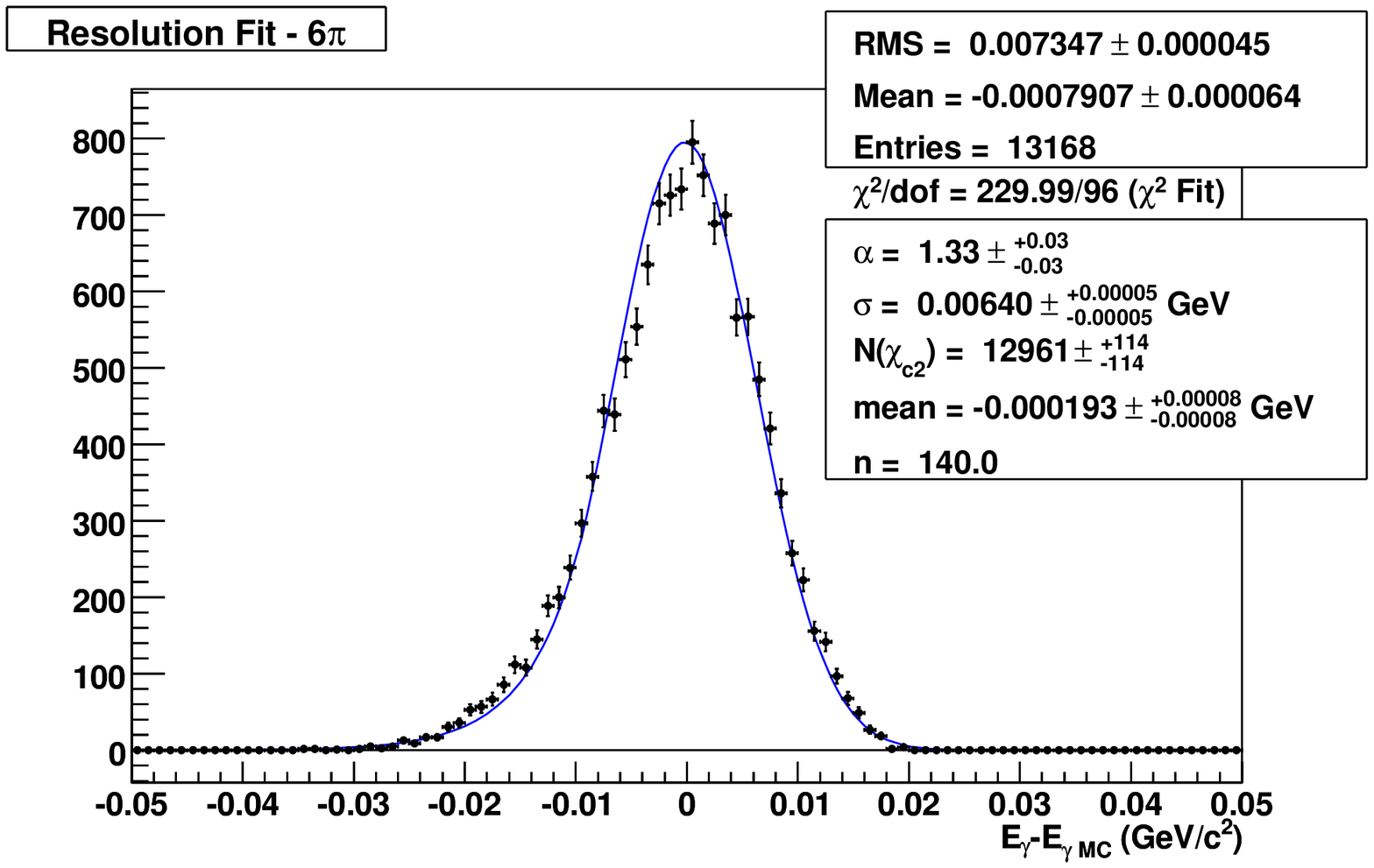}}
\subfigure
{\includegraphics[width=.85\textwidth]{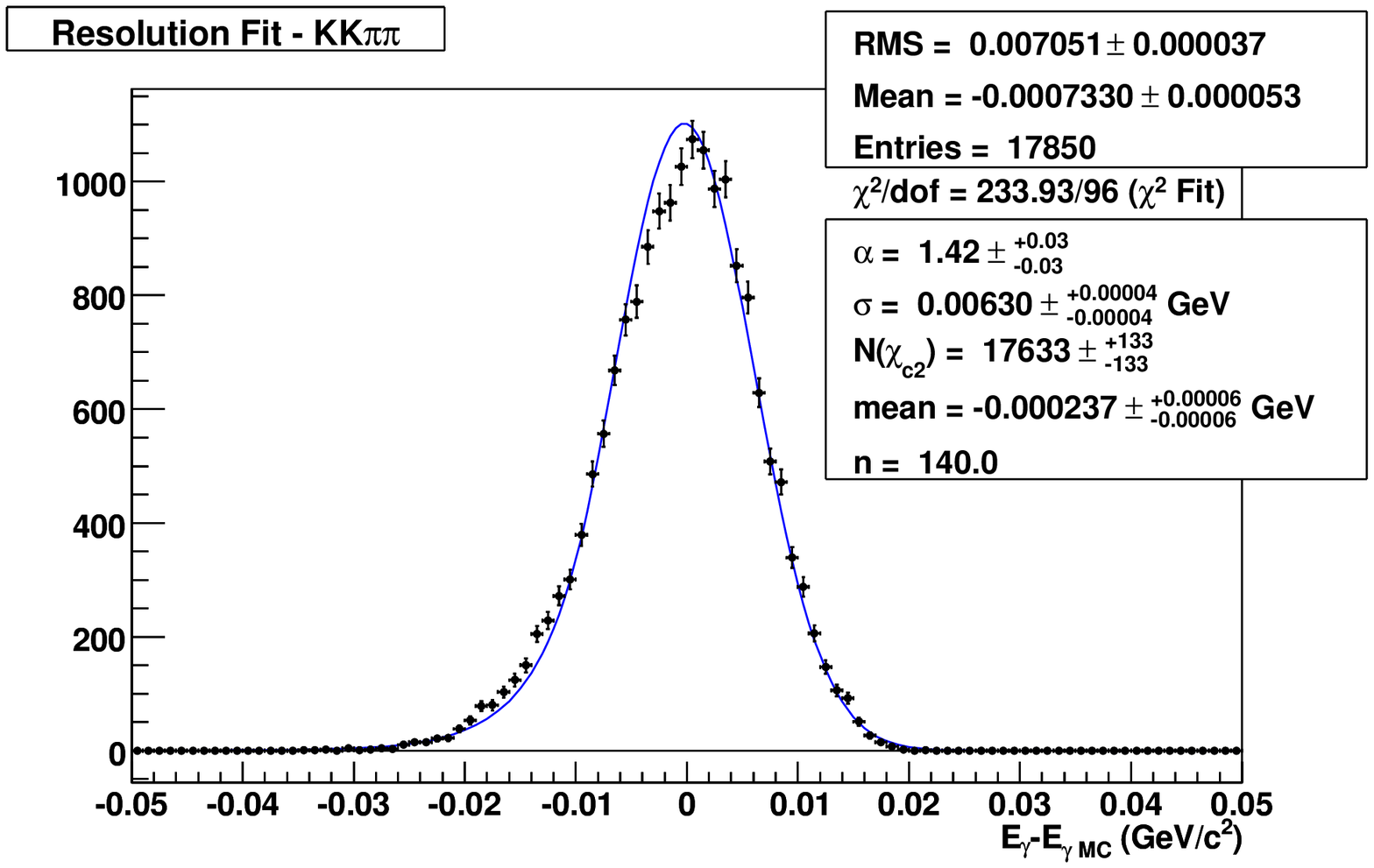}}
\end{center}
\caption[Resolution function for the 
$\psi(2S)\to\gamma\chi_{c2},\chi_{c2}\to 6\pi$  
and $KK\pi\pi$  modes.]
{\label{fig:chic2_resfnct_6Pi_KKPiPi}
{Resolution functions for the decay modes 
$\psi(2S)\to\gamma\chi_{c2},\chi_{c2}\to 6\pi$  (top) 
and $KK\pi\pi$ (bottom).
 \ The points are from the $\chi_{c2}$ signal MC and the solid line 
is the result of the fit to the Crystal Ball function.}}
\end{figure}

\begin{figure}[htbp]
\begin{center}
\subfigure
{\includegraphics[width=.85\textwidth]{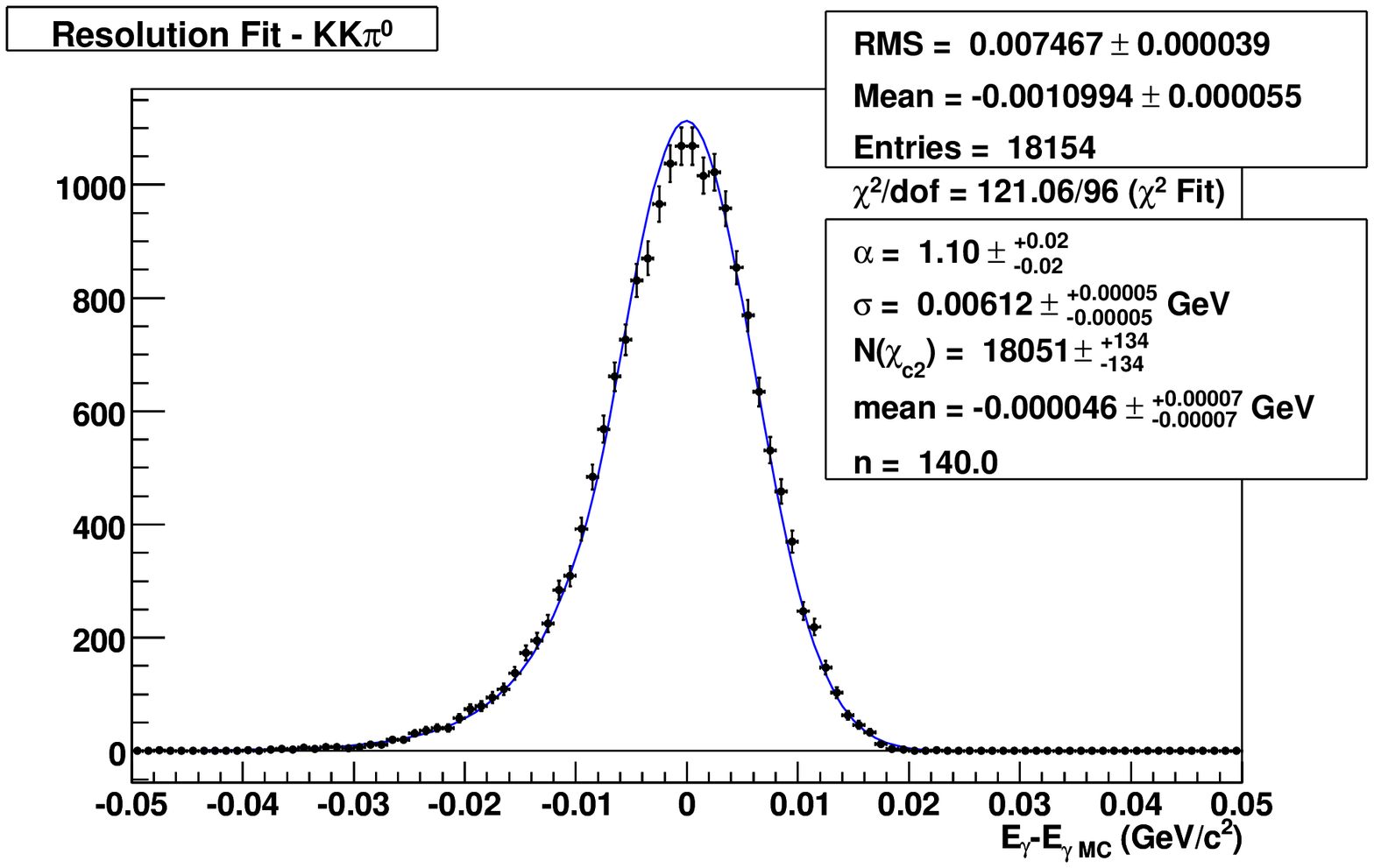}}
\subfigure
{\includegraphics[width=.85\textwidth]{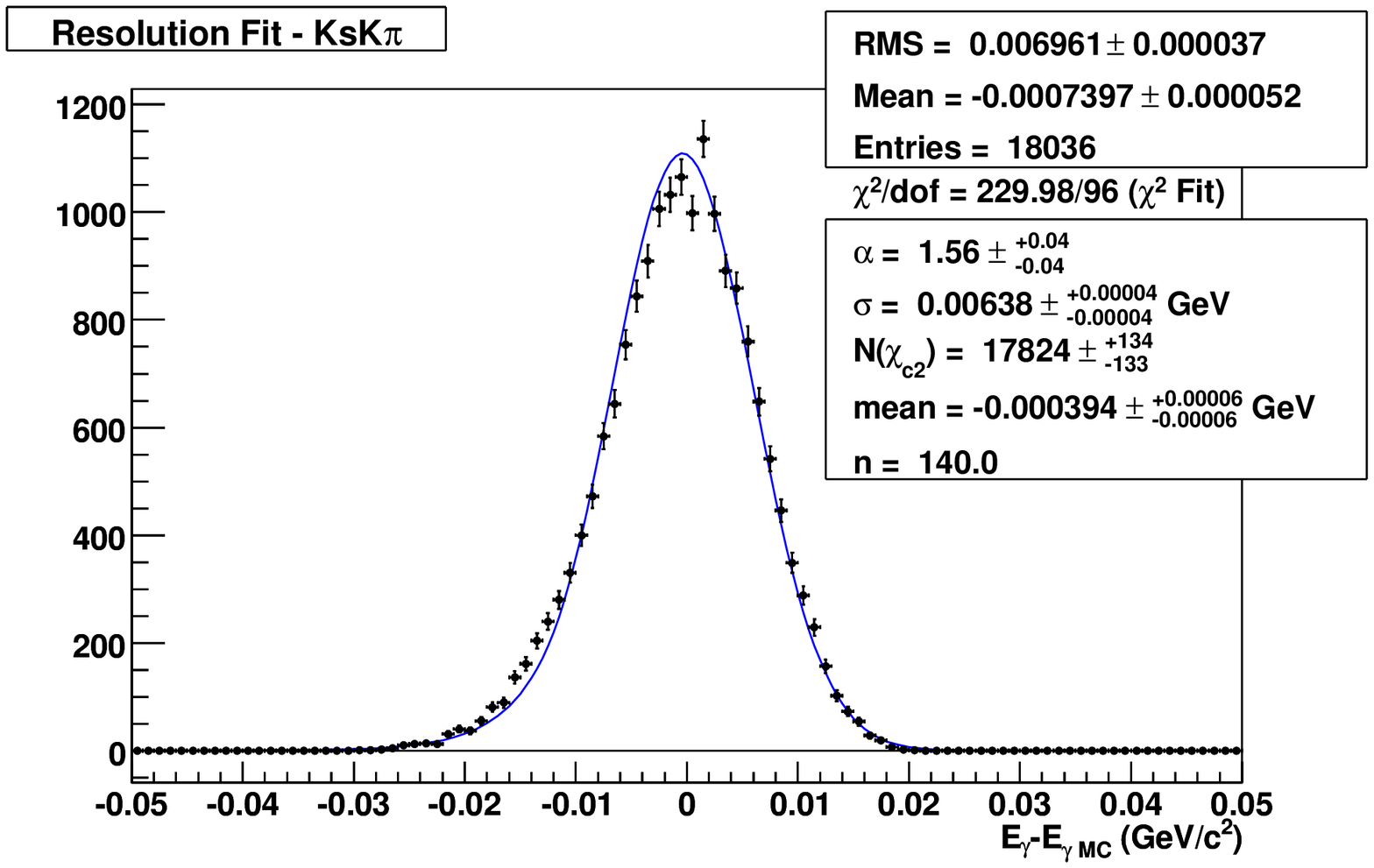}}
\end{center}
\caption[Resolution functions for the 
$\psi(2S)\to\gamma\chi_{c2},\chi_{c2}\to KK\pi^0$  
and $K_{S}K\pi$  modes.]
{\label{fig:chic2_resfnct_KKPi0_KsPi}
{Resolution functions for the decay modes 
$\psi(2S)\to\gamma\chi_{c2},\chi_{c2}\to KK\pi^0$  (top) 
and $K_{S}K\pi$ (bottom).
 \ The points are from the $\chi_{c2}$ signal MC and the solid line 
is the result of the fit to the Crystal Ball function.}}
\end{figure}

\begin{figure}[htbp]
\begin{center}
\subfigure
{\includegraphics[width=.85\textwidth]{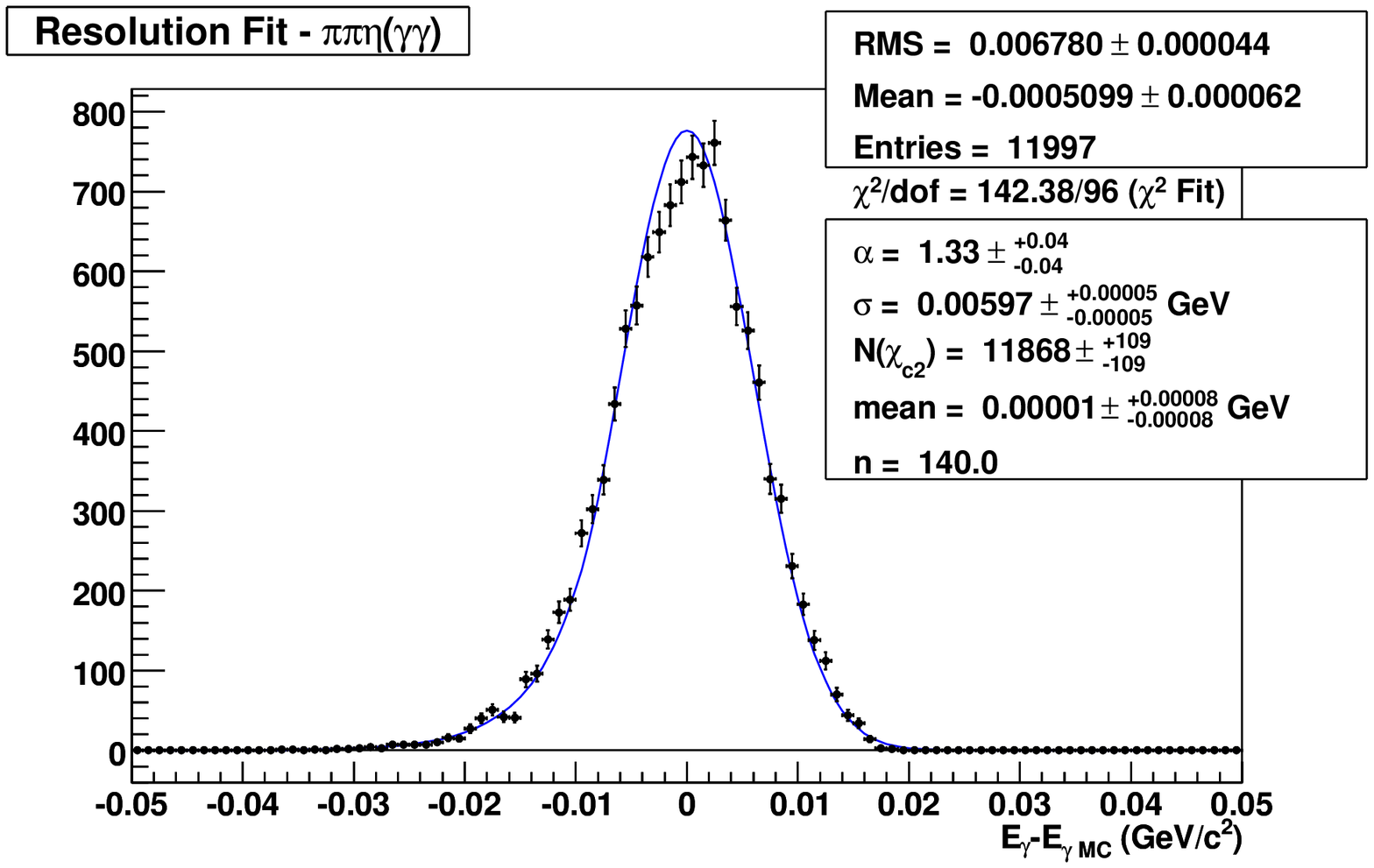}}
\subfigure
{\includegraphics[width=.85\textwidth]{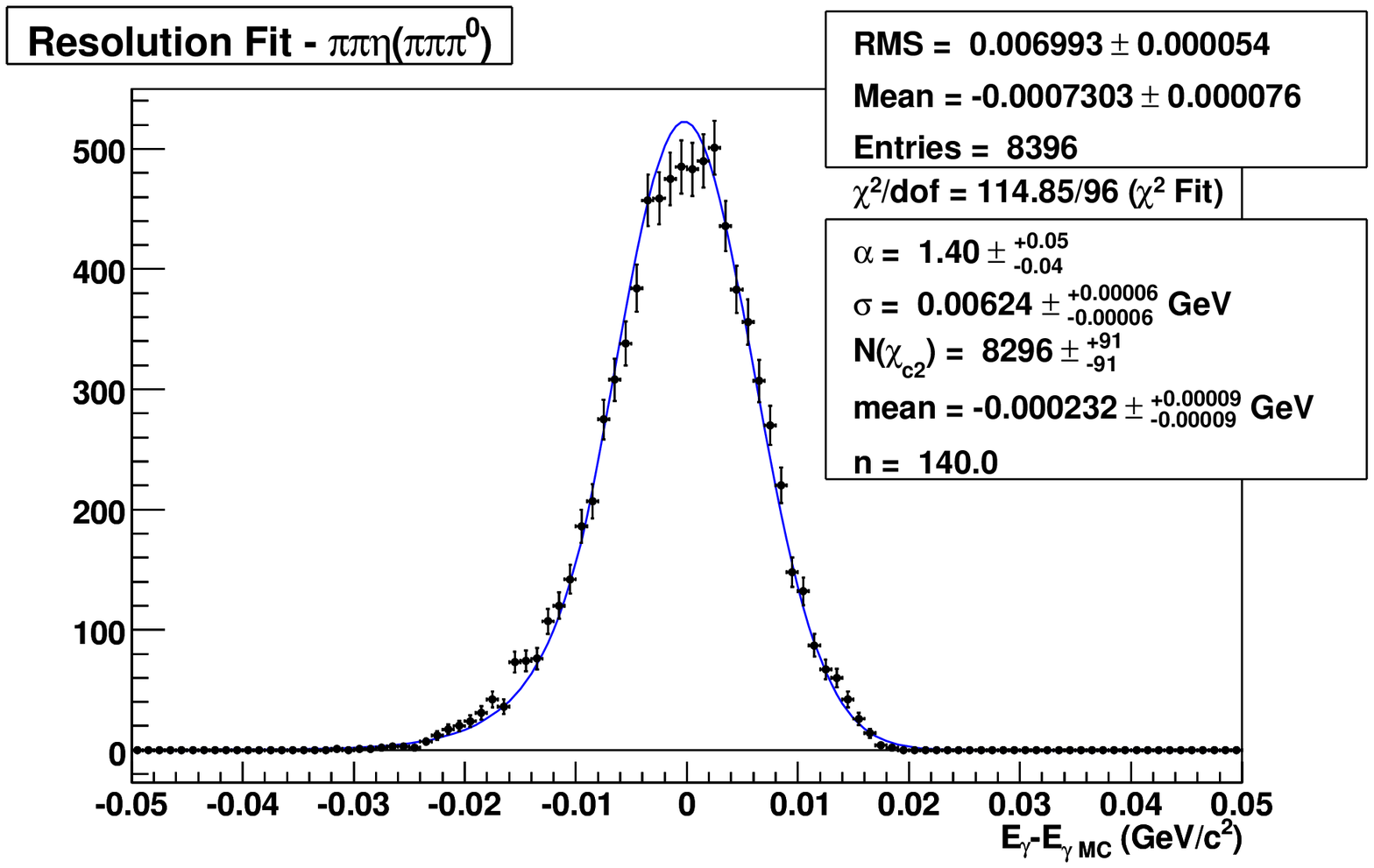}}
\end{center}
\caption[
Resolution function for the 
$\psi(2S)\to\gamma\chi_{c2},\chi_{c2}\to \pi\pi\eta(\gamma\gamma)$ 
and $\pi\pi\eta(\pi\pi\pi^0)$ modes.]
{\label{fig:chic2_resfnct_PiPiEtaGG_PiPiEtaPiPiPi0}
{Resolution functions for the decay modes 
$\psi(2S)\to\gamma\chi_{c2},\chi_{c2}\to \pi\pi\eta(\gamma\gamma)$  (top) 
and $\pi\pi\eta(\pi\pi\pi^0)$ (bottom).
 \ The points are from the $\chi_{c2}$ signal MC and the solid line 
is the result of the fit to the Crystal Ball function.}}
\end{figure}

\begin{figure}[htbp]
\begin{center}
\subfigure
{\includegraphics[width=.85\textwidth]{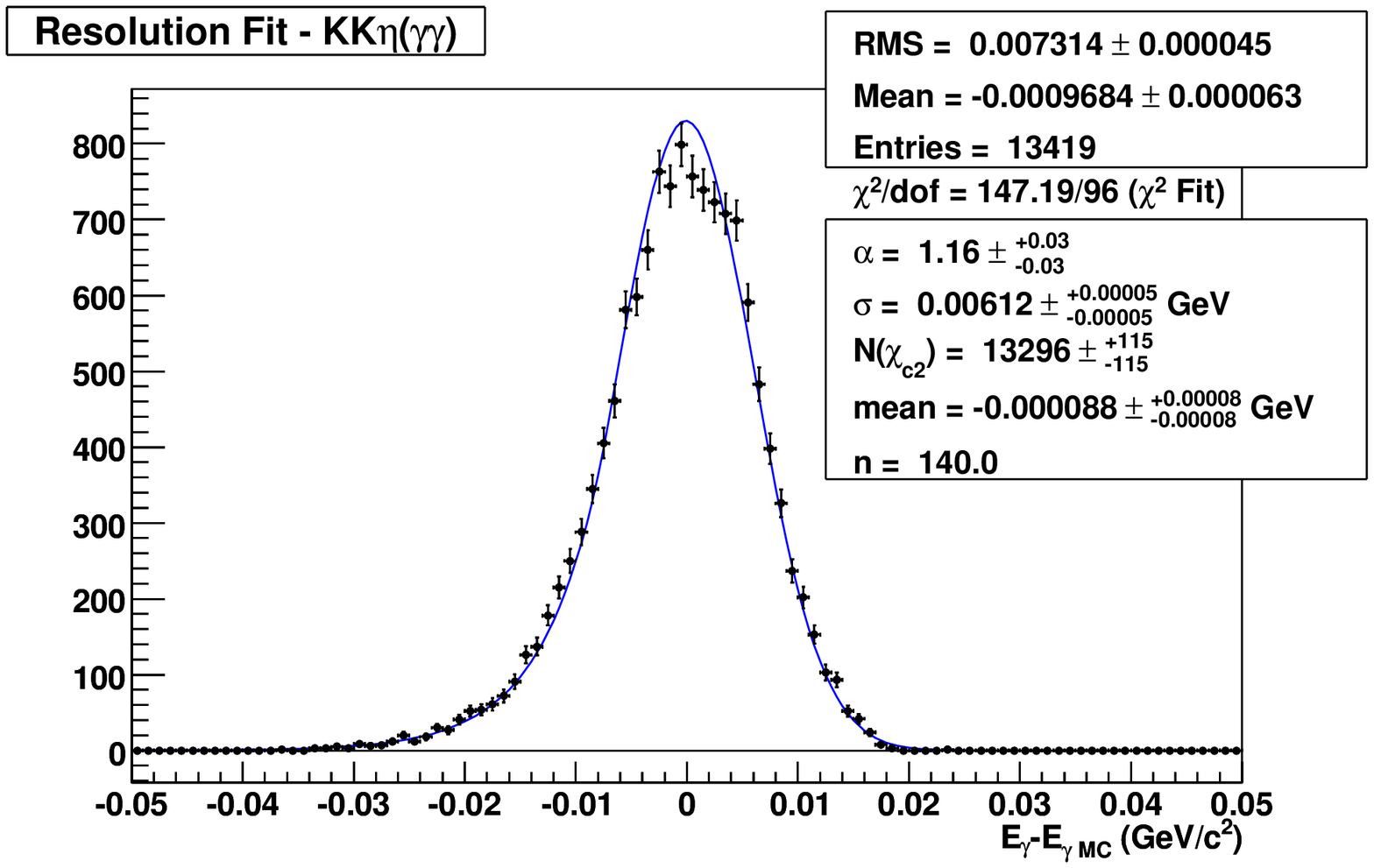}}
\subfigure
{\includegraphics[width=.85\textwidth]{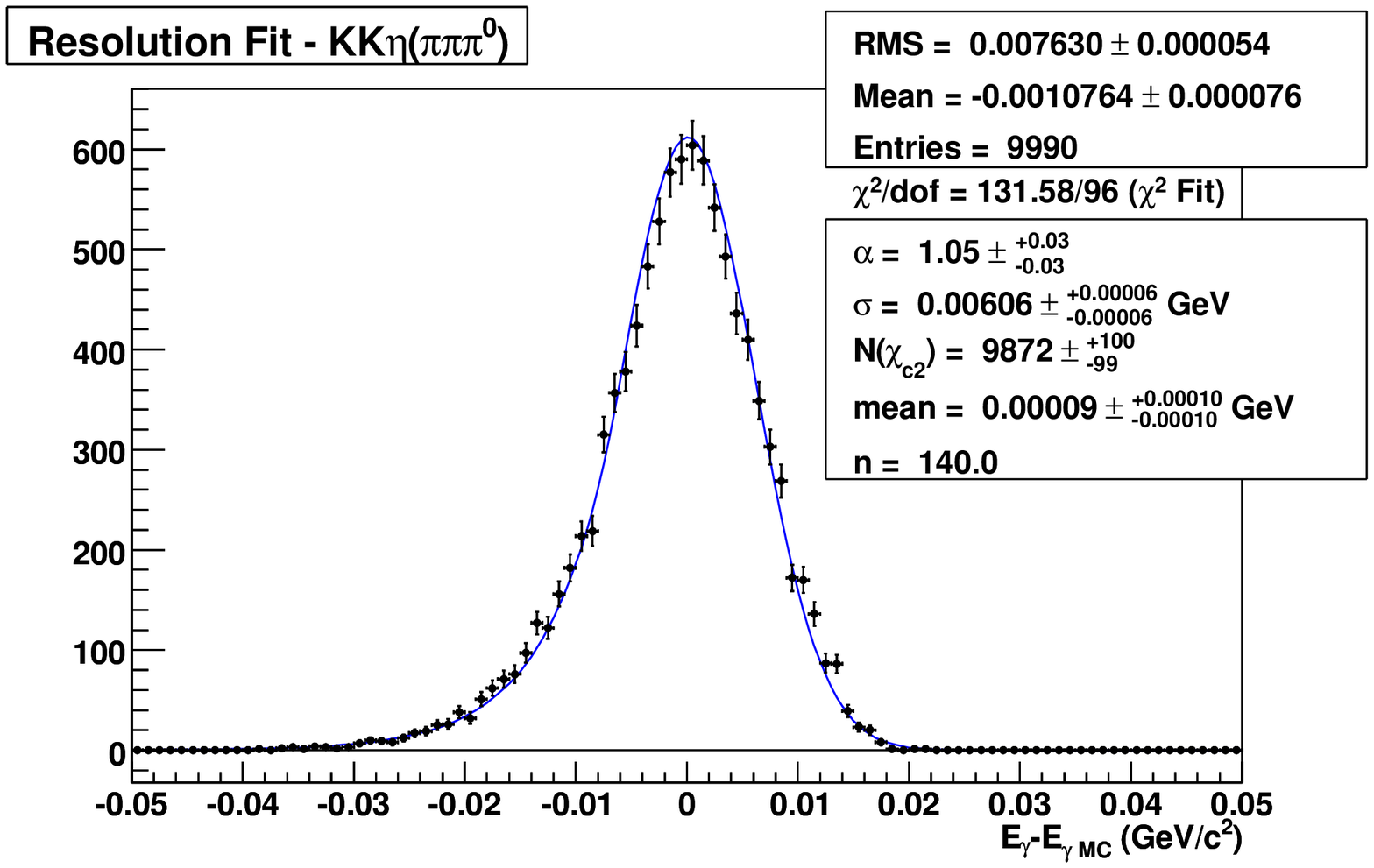}}
\end{center}
\caption[Resolution functions for the 
$\psi(2S)\to\gamma\chi_{c2},\chi_{c2}\to KK\eta(\gamma\gamma)$ 
and $KK\eta(\pi\pi\pi^0)$ modes.]
{\label{fig:chic2_resfnct_KKEtaGG_KKEtaPiPiPi0}
{Resolution functions for the decay modes 
$\psi(2S)\to\gamma\chi_{c2},\chi_{c2}\to KK\eta(\gamma\gamma)$  (top) 
and $KK\eta(\pi\pi\pi^0)$ (bottom).
 \ The points are from the $\chi_{c2}$ signal MC and the solid line 
is the result of the fit to the Crystal Ball function.}}
\end{figure}

\begin{figure}[htbp]
\begin{center}
\subfigure
{\includegraphics[width=.85\textwidth]{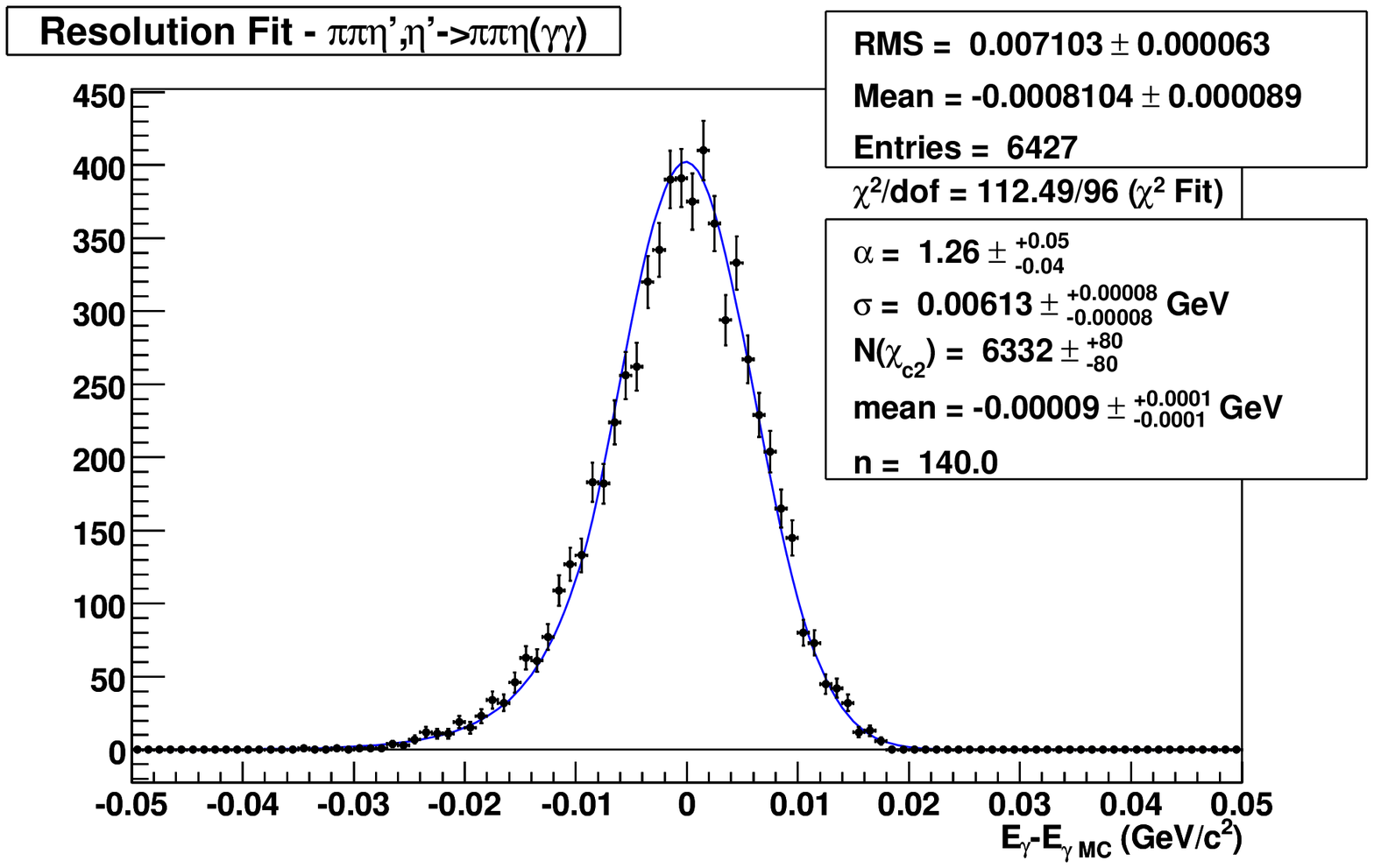}}
\subfigure
{\includegraphics[width=.85\textwidth]{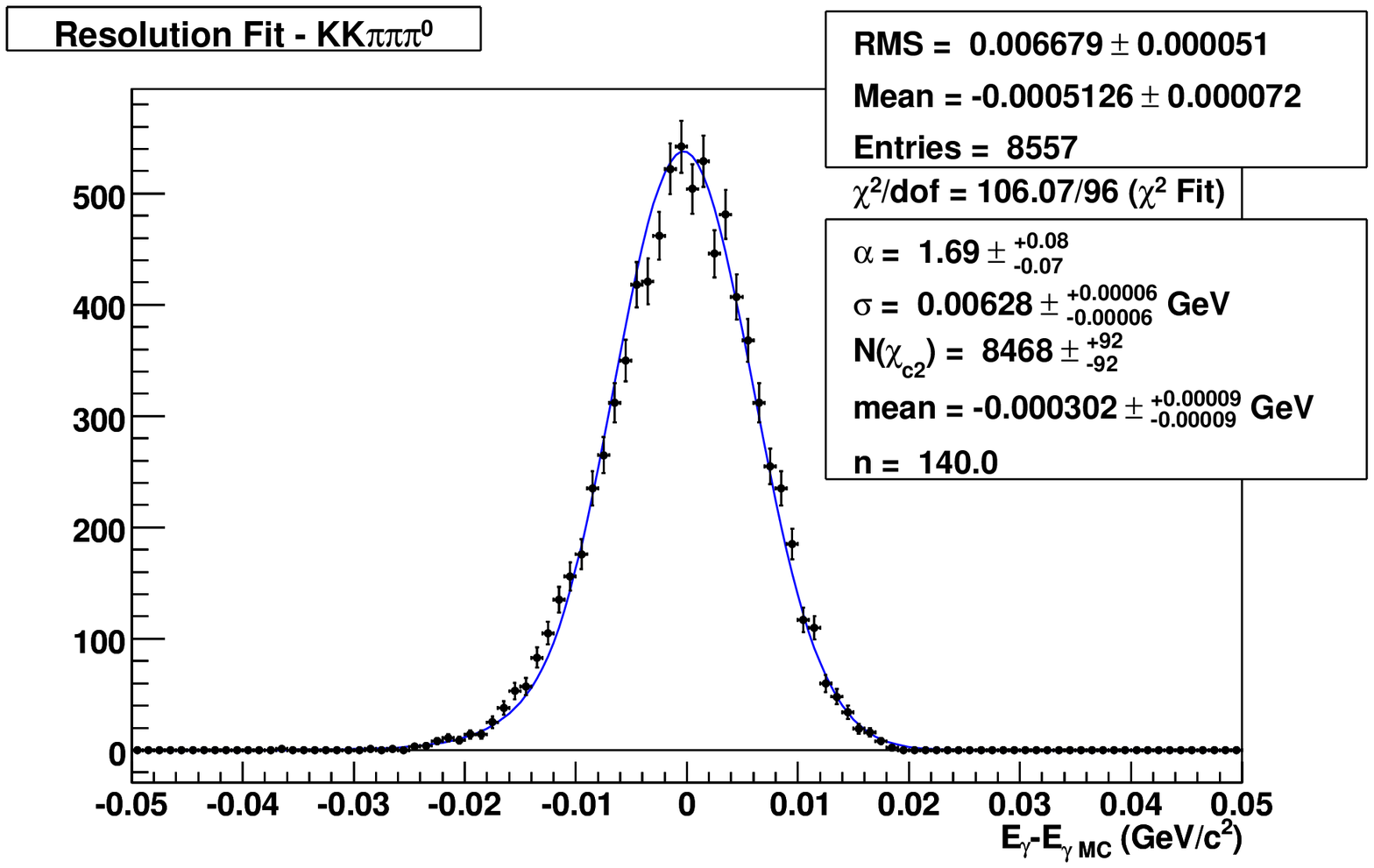}}
\end{center}
\caption[Resolution functions for the $\psi(2S)\to\gamma\chi_{c2},
\chi_{c2}\to \pi\pi\eta^{\prime}$ and 
$KK\pi\pi\pi^0$ modes.]
{\label{fig:chic2_resfnct_PiPiEtaPPiPiEtaGG_KKPiPiPi0}
{Resolution functions for the decay modes 
$\psi(2S)\to\gamma\chi_{c2}$, $\chi_{c2}\to \pi\pi \eta^{\prime}$, 
$\eta^{\prime}\to\pi\pi\eta(\gamma\gamma)$  (top) and $KK\pi\pi\pi^0$ 
(bottom).
 \ The points are from the $\chi_{c2}$ signal MC and the solid line 
is the result of the fit to the Crystal Ball function.}}
\end{figure}

\begin{figure}[htbp]
\begin{center}
\subfigure
{\includegraphics[width=.85\textwidth]{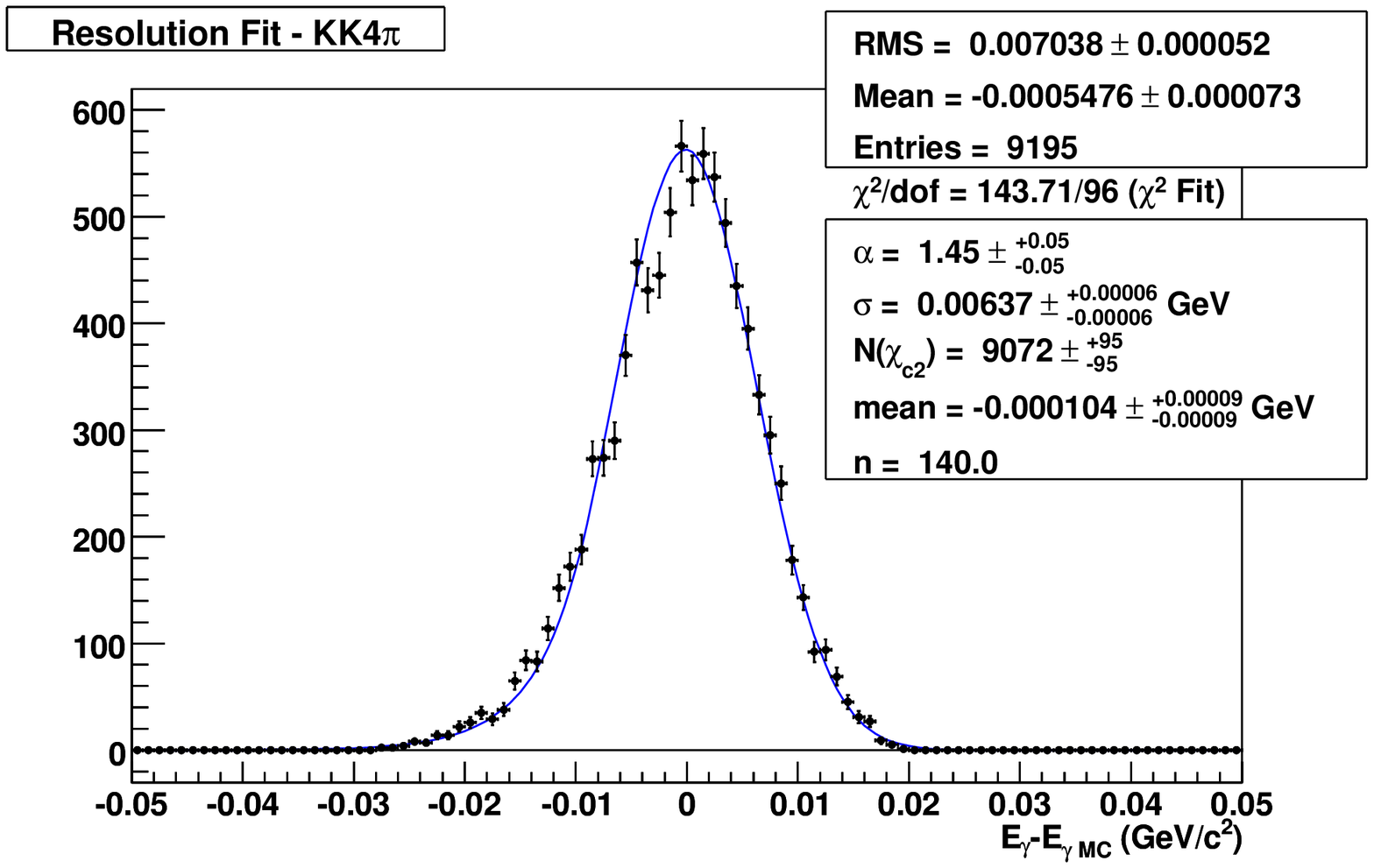}}
\subfigure
{\includegraphics[width=.85\textwidth]{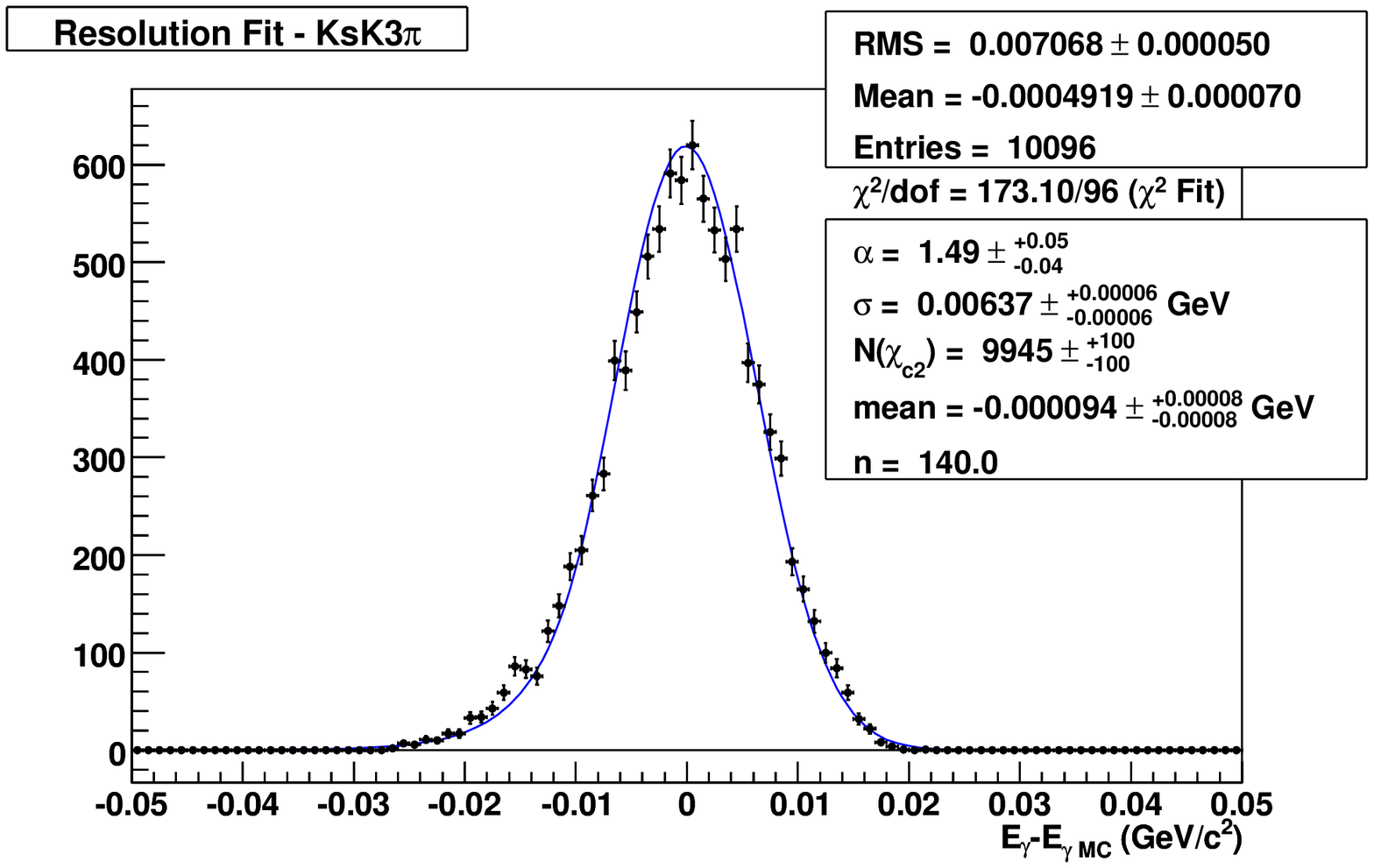}}
\end{center}
\caption[Resolution functions for the 
$\psi(2S)\to\gamma\chi_{c2},\chi_{c2}\to KK4\pi$  
and $K_{S}K3\pi$ modes.]
{\label{fig:chic2_resfnct_KK4Pi_KsK3Pi}
{Resolution functions for the decay modes 
$\psi(2S)\to\gamma\chi_{c2},\chi_{c2}\to  KK4\pi$  (top) 
and $K_{S}K3\pi$ (bottom).
 \ The points are from the $\chi_{c2}$ signal MC and the solid line 
is the result of the fit to the Crystal Ball function.}}
\end{figure}

The resolution function fits of the measured photon energy for 
the $\psi(2S)\to\gamma\chi_{c2}$ and $\psi(2S)\to\gamma\eta_c(2S)$ 
decays are given here. \ A Crystal Ball function 
\cite{Skwarnicki:1986xj, Gaiser:1982yw} 
was used to fit the resolution distribution from the signal MC samples 
as described in 
Section~\ref{subsec:signalmc}. \ The resolution distribution is the 
histogram of the difference 
between the measured and generator-level photon energy. \ The RooFit software 
package \cite{roofit} was used to perform the $\chi^{2}$ fits. \ 
The resolution function fits for 
$\psi(2S) \to \gamma \eta_c(2S), \eta_c(2S) \to X$ 
are shown in Figures~\ref{fig:etac2s_resfnct_4Pi} through 
\ref{fig:etac2s_resfnct_KK4Pi_KsK3Pi}. \ 
The resolution function fits for 
$\psi(2S) \to \gamma \chi_{c2}, \chi_{c2} \to X$ 
are shown in Figures~\ref{fig:chic2_resfnct_4Pi} through 
\ref{fig:chic2_resfnct_KK4Pi_KsK3Pi}. \ Some of the $\chi_{c2}$ fits give 
relatively poor values of $\chi^{2}$, reflecting the fact that the 
Crystal Ball function is an imperfect description of the photon energy
resolution. \ The systematic uncertainty related to this effect is 
negligible.

\chapter{Upper Limit Determination Procedures}
\label{appendix:ul}

Two methods were employed for determining the upper limits on the yields 
listed in Table~\ref{table:etac2s_result_fit}. \  
For modes which do not contain $\eta$ decays, we have no apparent signals 
and substantial backgrounds. \ 
Therefore, we determined a 90\% confidence level upper limit by integrating the distribution
defined from the nominal yield result up to 90\% of the area 
in the physical range (that is with the lower integration
limit of zero yield). \ 
We used toy MC studies (explained below) to verify
that we get consistent results with the nominal fit integration. \ 
For modes with 
$\eta$ decays, a very limited number of events pass our selection criteria, 
either signal or background.  Therefore, we use the Feldman and Cousins 
method \cite{PhysRevD.57.3873}, which is a standard procedure in high 
energy physics for setting Poisson upper limits in the presence of 
nonzero backgrounds that are separately estimated.

The distributions of the fitted number of signal events were best 
represented by bifurcated Gaussian distributions. \ However, to determine 
the 90\% confidence level upper limit for each distribution, we find the 
value of the number of signal events at which the area of the 
distribution between 0 and this value is 90\% of the area. \ 
This procedure is standard in high energy 
physics and avoids the artificially stringent limits that would result 
from a nonphysical central value.

\begin{figure}[htbp]
  \begin{center}
  \subfigure
    {\includegraphics[width=.49\textwidth]{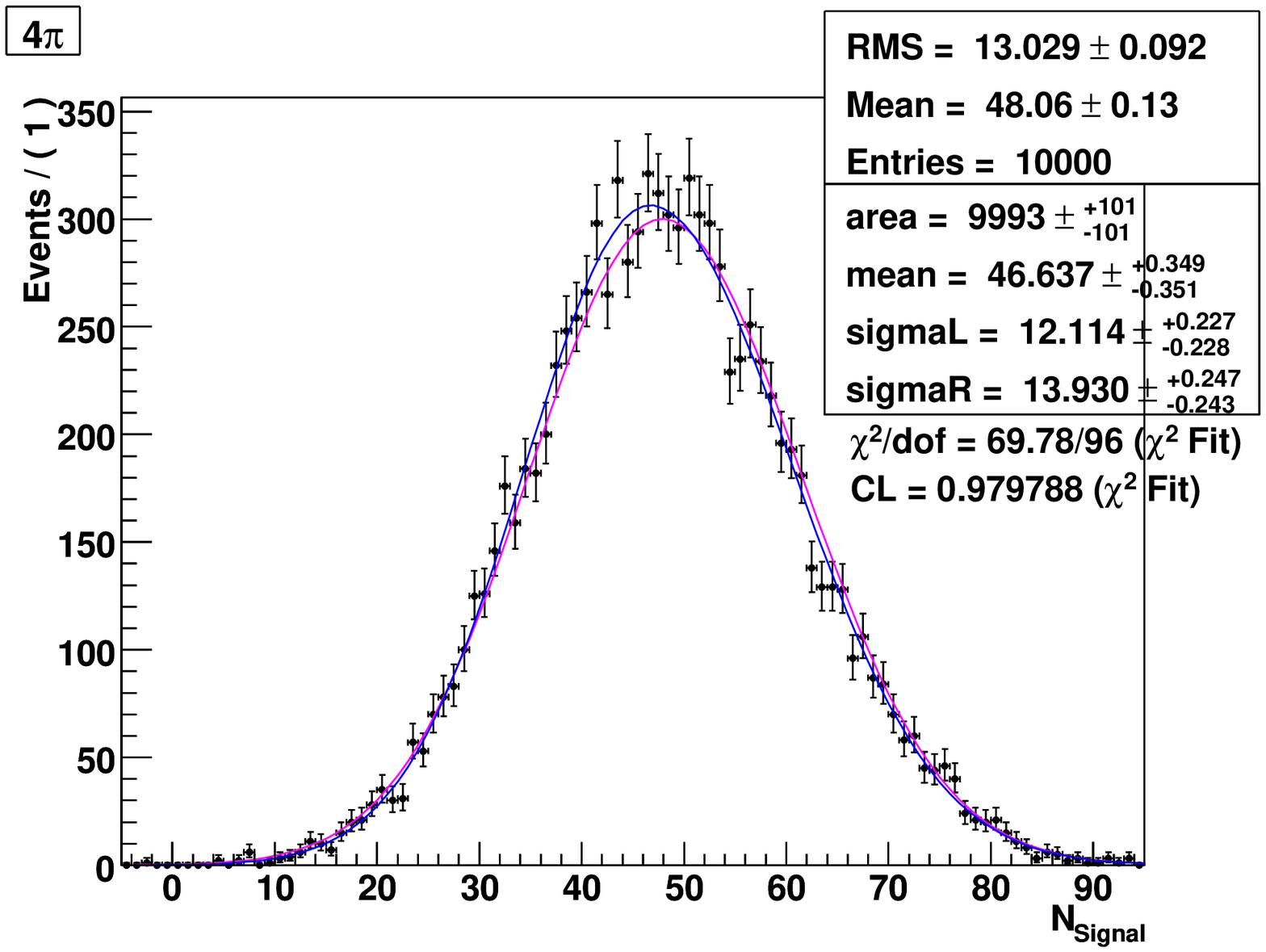}}
  \subfigure
    {\includegraphics[width=.49\textwidth]{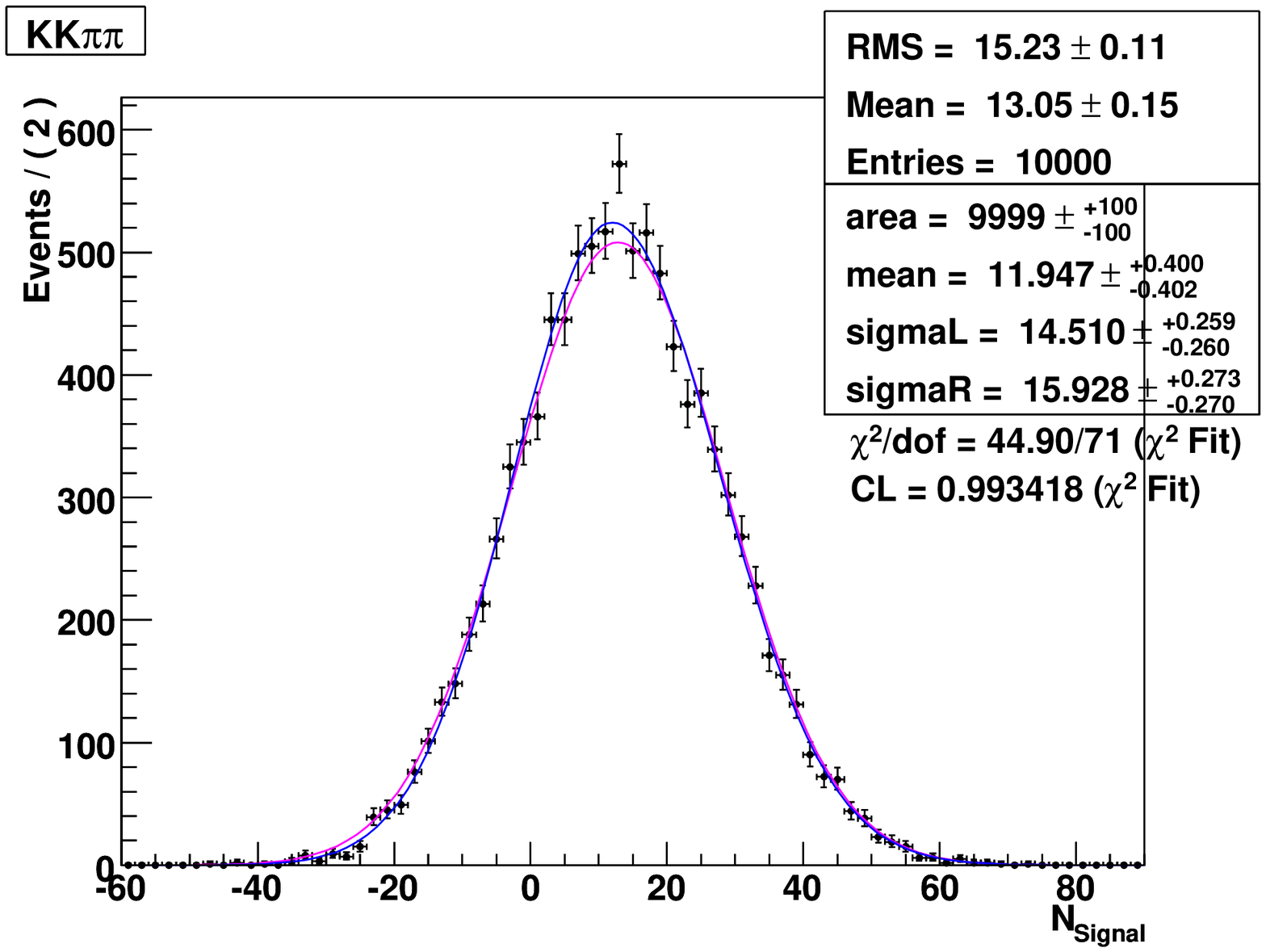}}
  \subfigure
    {\includegraphics[width=.49\textwidth]{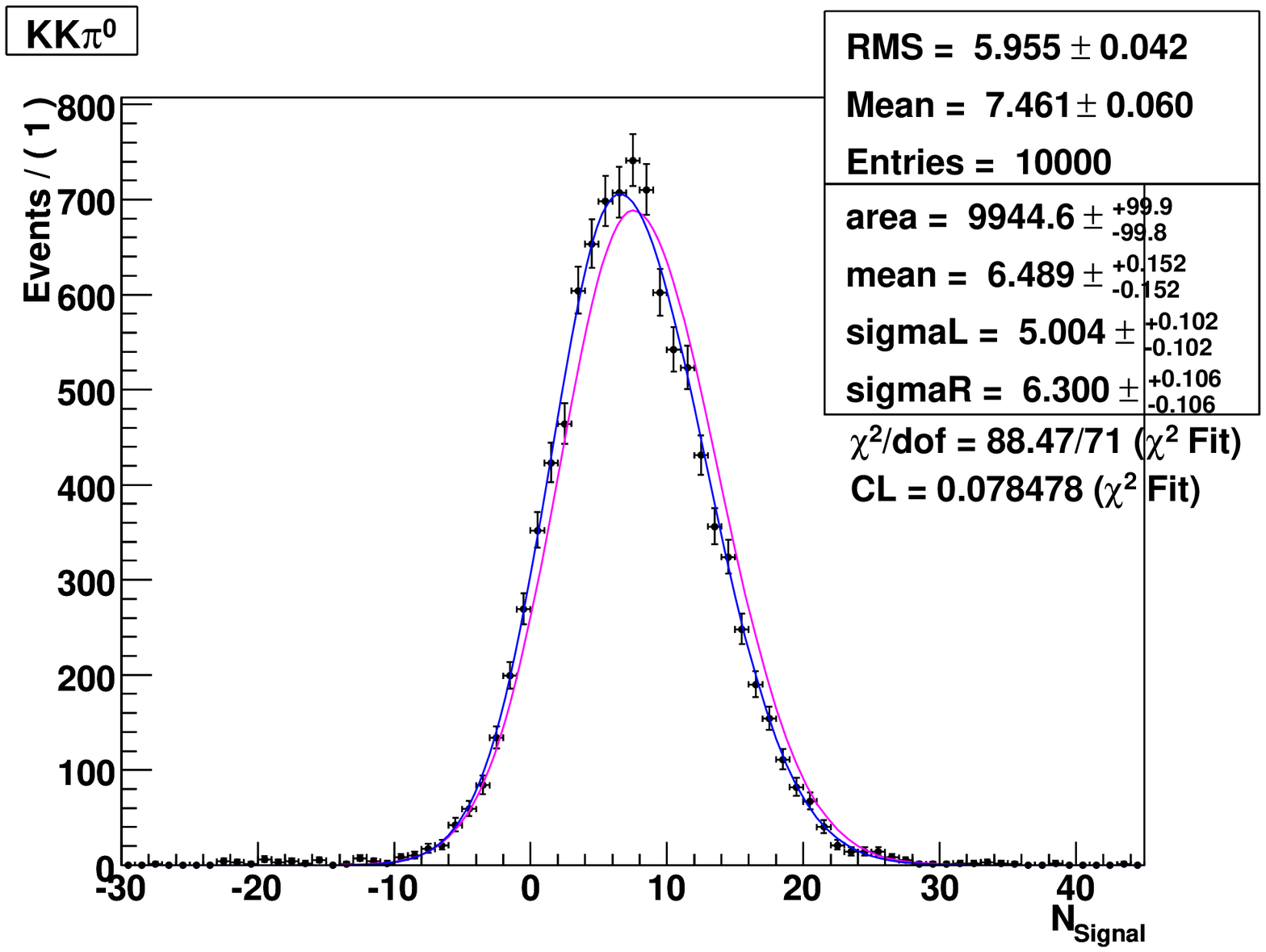}}
  \subfigure
    {\includegraphics[width=.49\textwidth]{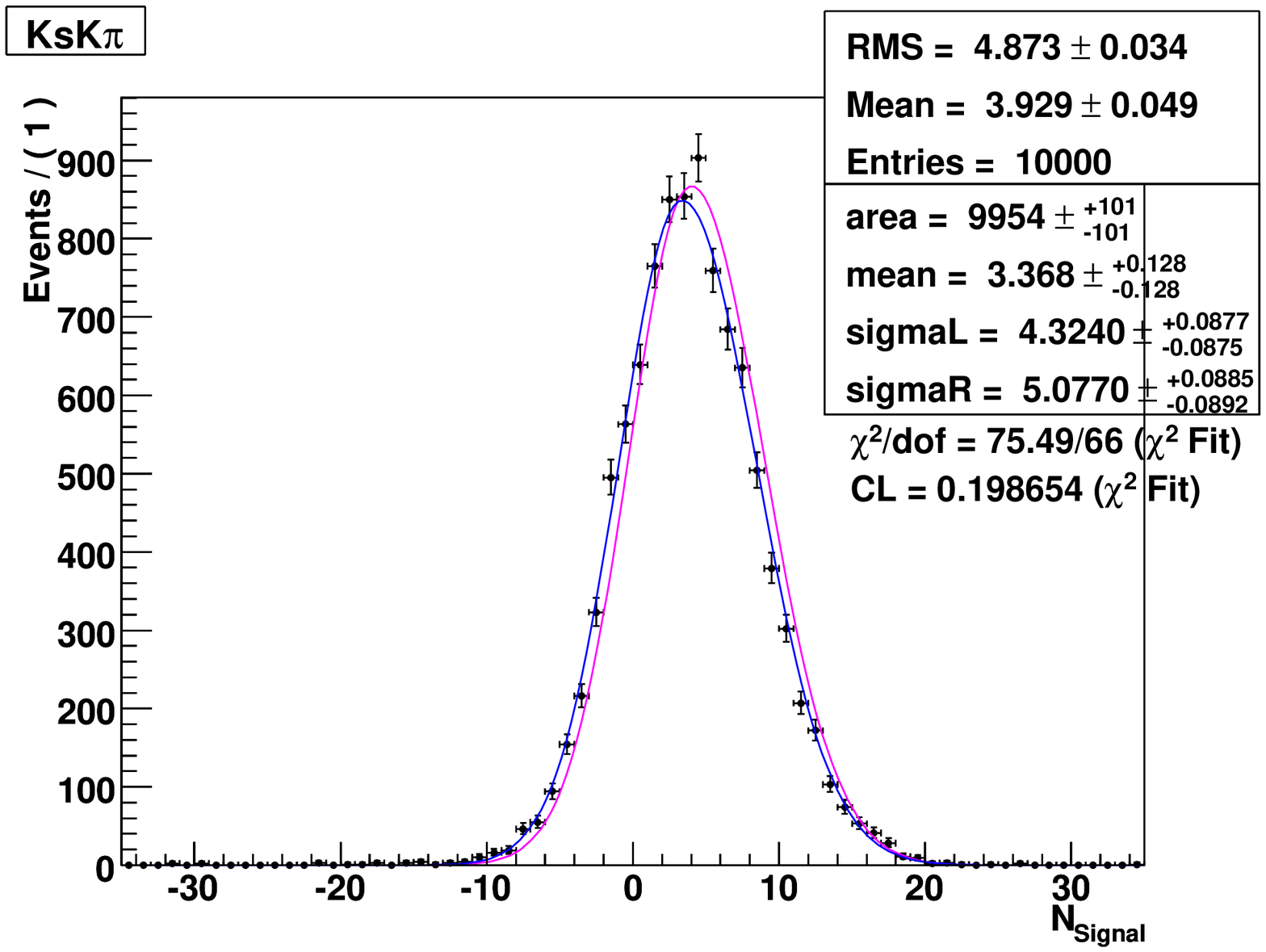}}
  \end{center}
  \caption[$N_{\rm Signal}$ distribution from toy MC samples (1)]
   {\label{fig:Nsig1}{$N_{\rm Signal}$ distributions based on the signal and background fit 
   for 10,000 toy MC samples for the modes $4\pi$ (top left), $KK\pi\pi$ (top right), 
   $KK\pi^0$ (bottom left), and $K_{S}K\pi$ (bottom right).}}
\end{figure}

\begin{figure}[htbp]
  \begin{center}
  \subfigure
    {\includegraphics[width=.49\textwidth]{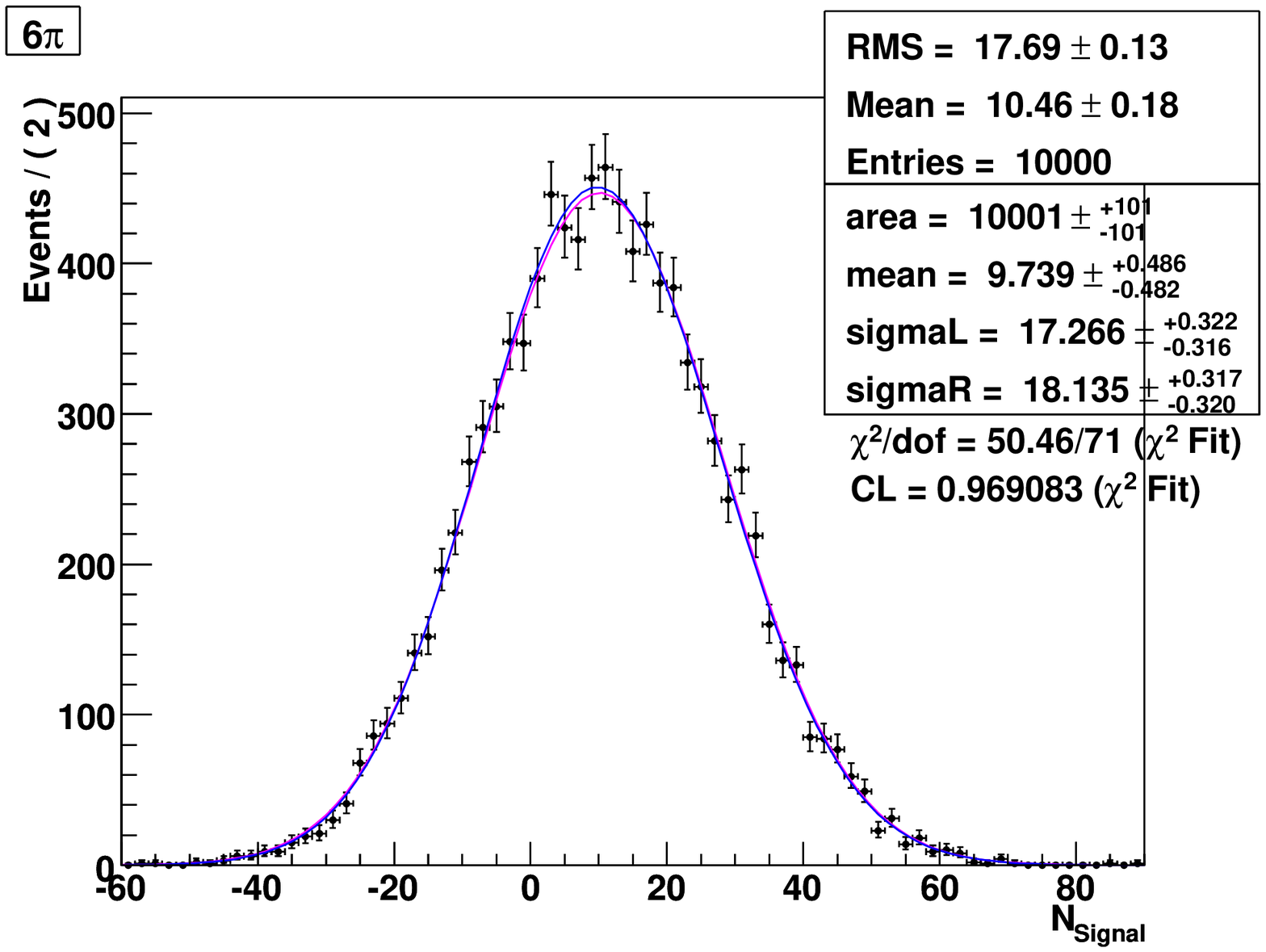}}
  \subfigure
    {\includegraphics[width=.49\textwidth]{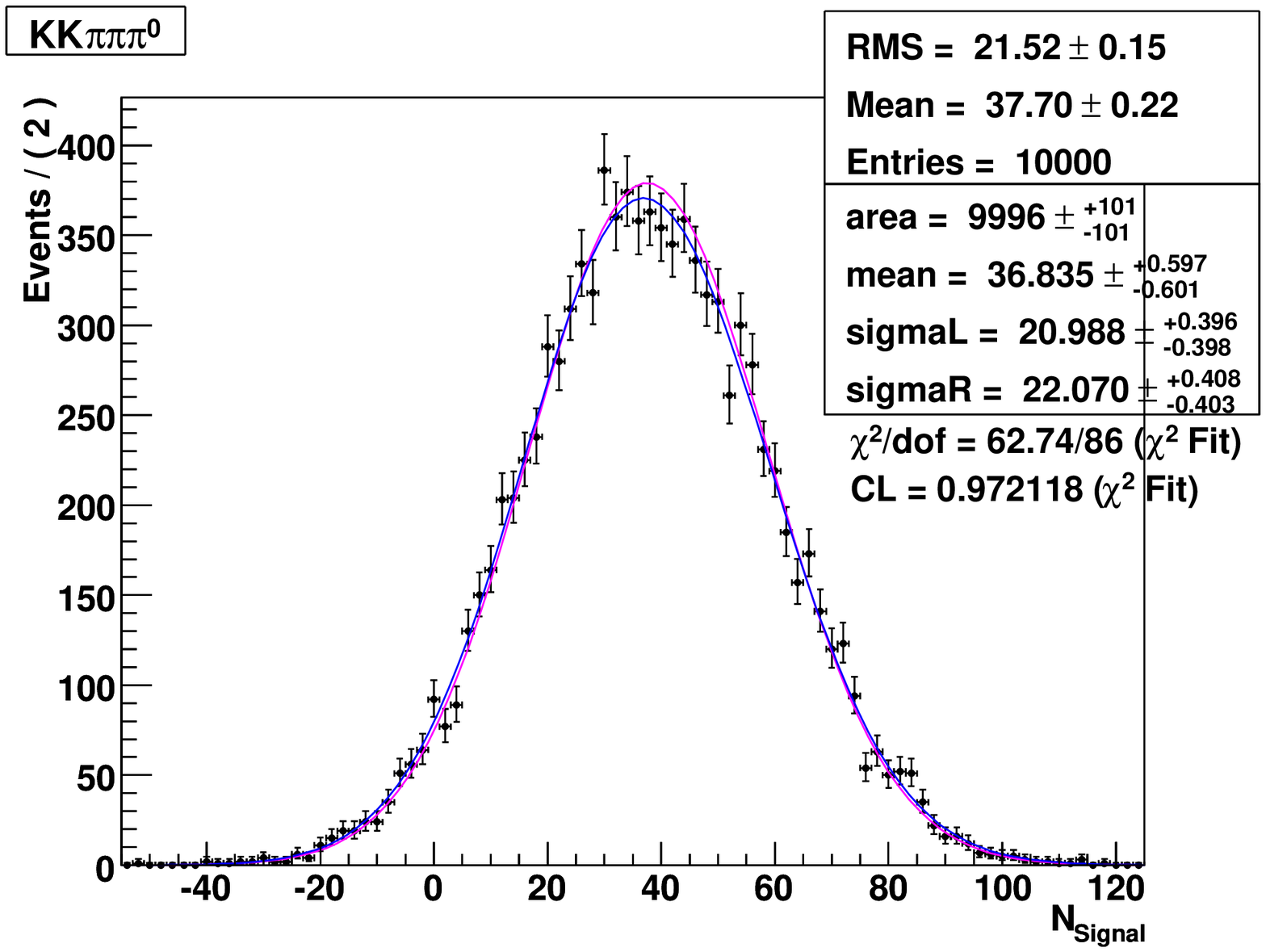}}
  \subfigure
    {\includegraphics[width=.49\textwidth]{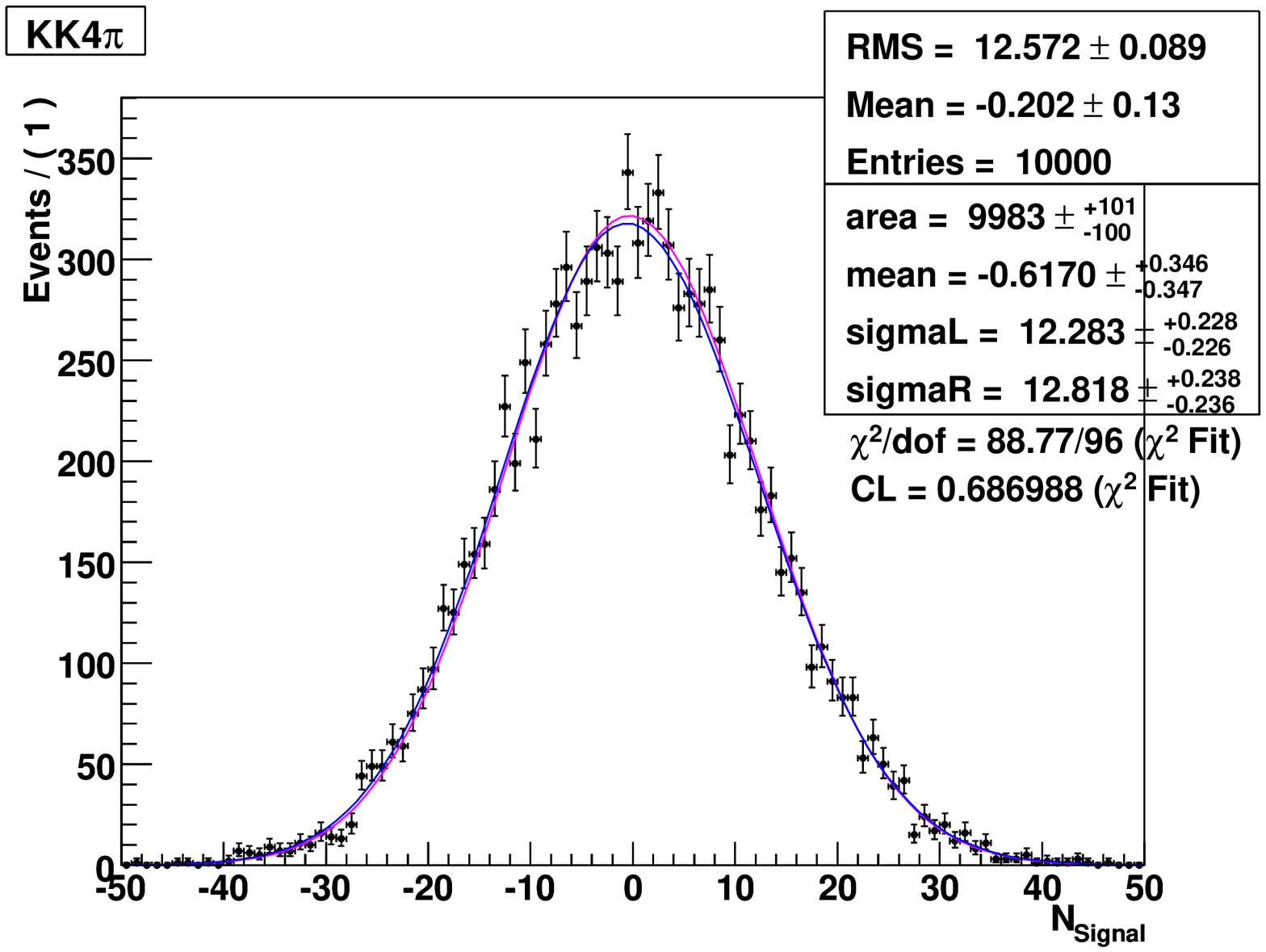}}
  \subfigure
    {\includegraphics[width=.49\textwidth]{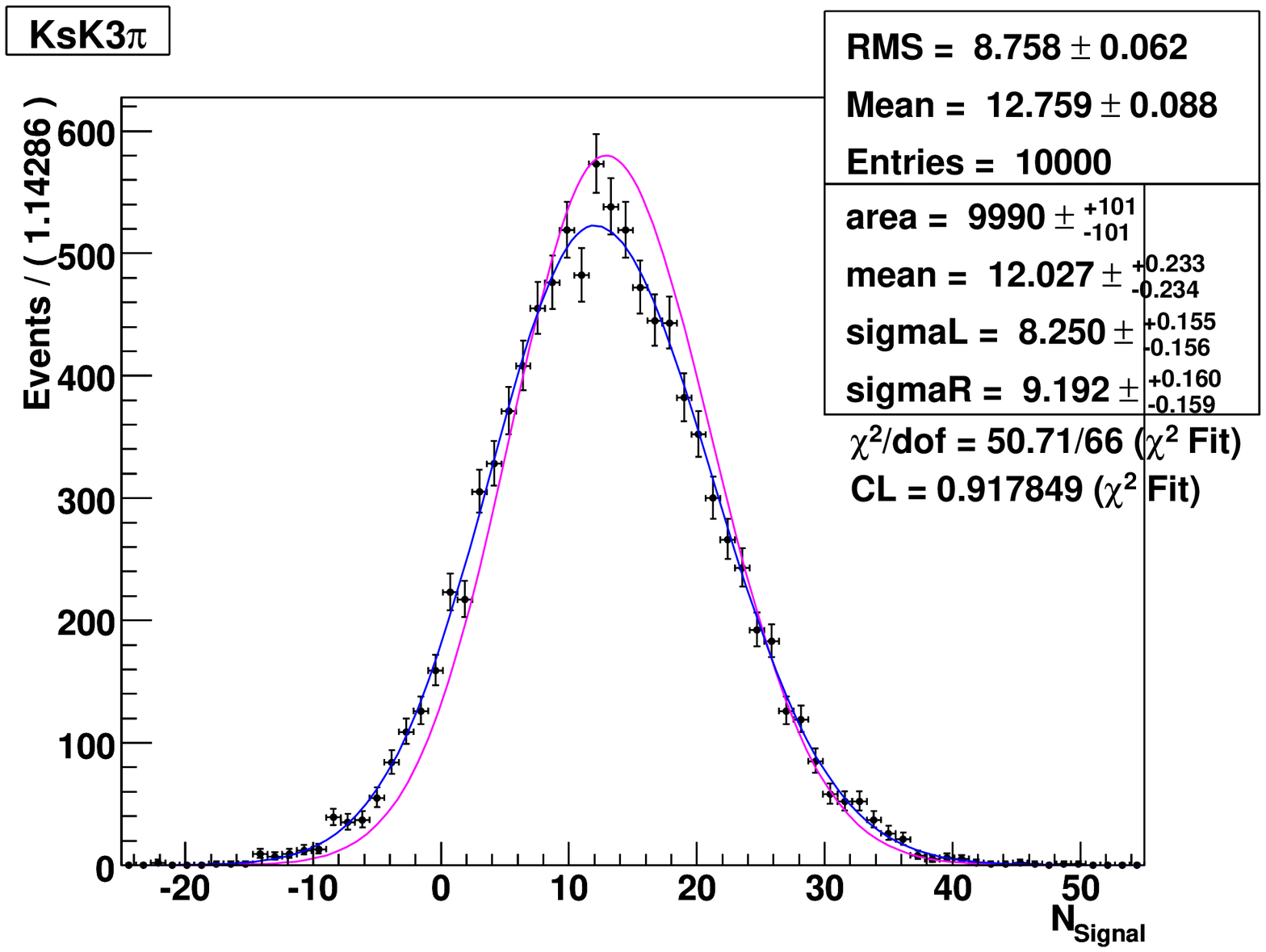}}
  \end{center}
  \caption[$N_{\rm Signal}$ distribution from toy MC samples (2)]
   {\label{fig:Nsig2}{$N_{\rm Signal}$ distributions based on the signal and background fit 
   for 10,000 toy MC samples for the modes $6\pi$ (top left), $KK\pi\pi\pi^0$ (top right), 
   $KK4\pi$ (bottom left), and $K_{S}K3\pi$ (bottom right).}}
\end{figure}

For the modes without $\eta$ decays, the measured photon energy were 
individually fit with the signal function and background histogram for 
all modes, as shown in Figures~\ref{fig:etac2s_result_4Pi_6Pi} through 
\ref{fig:etac2s_result_KK4Pi_KsK3Pi}. \ For each mode, we used the histogram 
from the fit result as the mean for generating toy MC samples. \
In a toy MC sample, for each bin we use a random number generator to 
randomly assign the bin content based on the Poisson 
probability distribution with the mean equal to the histogram bin value. \ 
After all bins of the toy MC sample
histogram were filled, a simulated data sample was produced. \ This 
simulated data sample was then fit with the same signal function and 
background histogram as the real data sample. \ 
The signal yield of this fit was recorded. \ For 
each mode, 10,000 toy MC samples were generated and fit, 
producing a distribution of the 10,000 yields. \ 
Figure~\ref{fig:Nsig1} and \ref{fig:Nsig2} show 
the yield distributions for the 10,000 toy MC 
samples for all non-$\eta$ modes. \ 
Table~\ref{table:compare_ul} compares the toy MC results with results from 
integrating the physical region of the yield from the nominal data fit.

\begin{table}[htbp]
\caption[Comparison of upper limit methods]
{\label{table:compare_ul}
Comparison of upper limit methods.  The column labeled ``Fit Integration'' 
is the result from 
determining the 90\% level of the distribution from the nominal fit result with 
only positive area. \ The column labeled ``Toy MC'' is the result 
determining the 
90\% level from the toy MC samples with fit yield $>$ 0.}
\begin{center}
\begin{tabular}{|l|c|c|}
  \hline
  Mode & Fit Integration & Toy MC \\  \hline
  $4\pi$ & $<64.8$ & $<65.1$ \\ \hline
  $6\pi$ & $<36.6$ & $<36.3$ \\ \hline
  $KK\pi\pi$ & $<35.2$ & $<34.6$\\ \hline
  $KK\pi^{0}$ & $<16.0$ & $<15.3$ \\ \hline
  $K_{S}K\pi$ & $<11.0$ & $<10.7$ \\ \hline
  $KK\pi\pi\pi^{0}$ & $<65.4$ & $<66.1$ \\ \hline
  $KK4\pi$ & $<20.6$ & $<20.8$ \\ \hline
  $K_{S}K3\pi$ & $<23.9$ & $<24.5$ \\ \hline
\end{tabular}
\end{center}
\end{table}

For the modes with $\eta$ decays, the following procedure was followed. \ 
The signal region was defined in the range $E_{\gamma} = [34,62]~{\rm MeV}$, 
corresponding to $\pm$ one full width ($\Gamma(\eta_c(2S)) = 14~{\rm MeV}$)
about the nominal central value. \ 
The number of observed events is counted. \ Next, the sideband region 
defined as $E_{\gamma} = [66,94]~{\rm MeV}$ was fit with only the 
background histogram to 
determine the background normalization over the $\eta_c(2S)$ signal 
region. \ Using this background normalization, the number of background 
events in the signal region is estimated. \ Using the number 
of observed and background events in the signal region, we 
applied the Feldman and Cousins procedure to determine the 90\% upper limit on 
the number of signal events. \ The efficiency for the signal region, 
as compared to the $E_{\gamma} = [30,94]~{\rm MeV}$ range for non-$\eta$ modes,
is then used to determine the upper limit on the product branching ratio.


\bibliography{thesis}

\end{document}